\documentclass[french,12pt,twoside,openright,a4paper]{book}
\pdfoutput=1 
\usepackage[T1]{fontenc}
\usepackage[utf8]{inputenc}  

\usepackage{natbib}
\usepackage[francais]{babel}

\usepackage{amsmath}
\usepackage{moreverb}
\usepackage{a4,epsfig,amssymb}

\usepackage{graphicx}
\usepackage{rotating}
\usepackage{graphics}
\usepackage{txfonts}
\usepackage{subfigure}
\usepackage{floatflt}
\usepackage{longtable}
\usepackage{ccaption}
\usepackage[final]{pdfpages}

\usepackage[top=2.5cm, bottom=2.5cm, left=2.cm , right=2.cm]{geometry}
\usepackage{fancyheadings}
\renewcommand{\sectionmark}[1]{\markright{\thesection\ #1}}
\newcommand{\chaptermarkfree}[1]{\markboth{#1}{}}
\newcommand{\chaptermarkannexe}[1]{\markboth{Annexe \thechapter:\ #1}{}}
\usepackage{pslatex}

\usepackage{lscape}

\usepackage{xspace} 
\usepackage{subfigure}

\newcommand{\FP}{Fabry-Perot}
\newcommand{\kms}{$\mathsf{~km.s^{-1}}$}
\newcommand{\degr}{\hbox{$^\circ$}}

\newcommand{\vdms}{velocity dispersion maps}

\newcommand{\vfs}{velocity fields}

\newcommand{\Vfs}{Velocity fields}
\newcommand{\rc}{rotation curve}
\newcommand{\rcs}{rotation curves}

\newcommand{\Ha} {H$\alpha$}

\usepackage[french]{minitoc}

\newcommand{\hl}{\\\hbox{\raisebox{0.4em}{\vrule depth 0pt height 0.4pt width \textwidth}}}

\usepackage[]{hyperref}

\def\jnl@style{\it}
\def\aaref@jnl#1{{\jnl@style#1}}

\def\aaref@jnl#1{{\jnl@style#1}}

\def\aj{\aaref@jnl{AJ}}                   
\def\araa{\aaref@jnl{ARA\&A}}             
\def\apj{\aaref@jnl{ApJ}}                 
\def\apjl{\aaref@jnl{ApJ}}                
\def\apjs{\aaref@jnl{ApJS}}               
\def\ao{\aaref@jnl{Appl.~Opt.}}           
\def\apss{\aaref@jnl{Ap\&SS}}             
\def\aap{\aaref@jnl{A\&A}}                
\def\aapr{\aaref@jnl{A\&A~Rev.}}          
\def\aaps{\aaref@jnl{A\&AS}}              
\def\azh{\aaref@jnl{AZh}}                 
\def\baas{\aaref@jnl{BAAS}}               
\def\jrasc{\aaref@jnl{JRASC}}             
\def\memras{\aaref@jnl{MmRAS}}            
\def\mnras{\aaref@jnl{MNRAS}}             
\def\pra{\aaref@jnl{Phys.~Rev.~A}}        
\def\prb{\aaref@jnl{Phys.~Rev.~B}}        
\def\prc{\aaref@jnl{Phys.~Rev.~C}}        
\def\prd{\aaref@jnl{Phys.~Rev.~D}}        
\def\pre{\aaref@jnl{Phys.~Rev.~E}}        
\def\prl{\aaref@jnl{Phys.~Rev.~Lett.}}    
\def\pasp{\aaref@jnl{PASP}}               
\def\pasj{\aaref@jnl{PASJ}}               
\def\qjras{\aaref@jnl{QJRAS}}             
\def\skytel{\aaref@jnl{S\&T}}             
\def\solphys{\aaref@jnl{Sol.~Phys.}}      
\def\sovast{\aaref@jnl{Soviet~Ast.}}      
\def\ssr{\aaref@jnl{Space~Sci.~Rev.}}     
\def\zap{\aaref@jnl{ZAp}}                 
\def\nat{\aaref@jnl{Nature}}              
\def\iaucirc{\aaref@jnl{IAU~Circ.}}       
\def\aplett{\aaref@jnl{Astrophys.~Lett.}} 
\def\apspr{\aaref@jnl{Astrophys.~Space~Phys.~Res.}}
\def\bain{\aaref@jnl{Bull.~Astron.~Inst.~Netherlands}} 
\def\fcp{\aaref@jnl{Fund.~Cosmic~Phys.}}  
\def\gca{\aaref@jnl{Geochim.~Cosmochim.~Acta}}   
\def\grl{\aaref@jnl{Geophys.~Res.~Lett.}} 
\def\jcp{\aaref@jnl{J.~Chem.~Phys.}}      
\def\jgr{\aaref@jnl{J.~Geophys.~Res.}}    
\def\jqsrt{\aaref@jnl{J.~Quant.~Spec.~Radiat.~Transf.}}
\def\memsai{\aaref@jnl{Mem.~Soc.~Astron.~Italiana}}
\def\nphysa{\aaref@jnl{Nucl.~Phys.~A}}   
\def\physrep{\aaref@jnl{Phys.~Rep.}}   
\def\physscr{\aaref@jnl{Phys.~Scr}}   
\def\planss{\aaref@jnl{Planet.~Space~Sci.}}   
\def\procspie{\aaref@jnl{Proc.~SPIE}}   

\begin{document}
\renewcommand{\bibname}{Bibliographie du texte en Français}
\renewcommand\labelitemi{\textbullet}

\thispagestyle{empty}
\begin{centering}
{\large\textsc{Université de Provence - Aix-Marseille I}}\\
{\large\textsc{\'Ecole Doctorale de Physique et Sciences de la Matière}}\\
{\large\textit{Laboratoire d'Astrophysique de Marseille}}\\
\vspace{1.3cm}
\LARGE{\textsc{\textbf{Thèse de Doctorat}}}\\
\vspace{0.6cm}
\normalsize{
\textit{présentée pour obtenir le grade de}\\
\vspace{0.6cm}
{\large Docteur de l'Université de Provence}\\
{\large Spécialité: Astrophysique}}\\
\vspace{0.4cm}
{\large \textit{par}}\\
\vspace{0.5cm}
{\LARGE \textbf{Benoît \textsc{Epinat}}}\\
\vspace{1.cm}

\rule{16.cm}{0.5pt}
\huge{\textsc{\textbf{Des galaxies proches aux galaxies lointaines}}}\\
\LARGE{\textsc{\textbf{\'Etudes cinématique et dynamique}}}\\
\rule{16.cm}{0.5pt}

\vspace{2.5cm}
\normalsize{
\textit{{\large Soutenue publiquement le \emph{jeudi 6 Novembre 2008}}\\
{\large Au Laboratoire d'Astrophysique de Marseille}\\
{\large Devant le Jury composé de}}\\
}
\vspace{1.cm}
\begin{tabular}{c r}
\hspace{7cm} \\
{\large Christian \textsc{Marinoni}~\hrulefill}&{\large Président} \\
{\large Jonathan \textsc{Bland-Hawthorn}~\hrulefill}&{\large Rapporteur} \\
{\large Thierry \textsc{Contini}~\hrulefill}&{\large Rapporteur} \\
{\large Claudia \textsc{Mendes de Oliveira}~\hrulefill}&{\large Examinatrice} \\
{\large \'Eric \textsc{Emsellem}~\hrulefill}&{\large Examinateur} \\
{\large Matthew \textsc{Lehnert}~\hrulefill}&{\large Examinateur} \\
{\large Philippe \textsc{Amram}~\hrulefill}&{\large Directeur} \\
{\large Chantal \textsc{Balkowski} ~\hrulefill}&{\large Co-directrice} \\
\end{tabular}

\end{centering}

\clearpage
\thispagestyle{empty}
\dominitoc

\frontmatter  

\addstarredchapter{Remerciements} 
\chapter*{Remerciements}
\chaptermarkfree{Remerciements}

\`A mon arrivée au Laboratoire d'Astrophysique de Marseille, au palais Longchamps, j'avais une idée assez vague de ce que serait mon travail. Petit à petit, les briques se sont assemblées pour ébaucher un ouvrage de plus en plus passionnant. Une thèse est un travail solitaire à bien des égards, et pourtant, les participations hétéroclites des uns et des autres ainsi que les échanges sont à la base de tout le travail qui a été effectué. Ce sont en quelques sortes les fondations de l'ouvrage.

Je tiens tout d'abord à remercier Chantal Balkowski qui m'a permis de faire cette thèse en insistant auprès de Philippe Amram pour qu'il propose un sujet. C'est aussi après l'avoir rencontrée à Paris que l'idée de travailler sur les galaxies a germé. Enfin, lors de cette rencontre, j'ai senti le courant passer, ce qui m'a mis en confiance pour arriver au LAM.


J'y ai été chaleureusement accueilli dans l'équipe ``Physique des Galaxies'' et j'en remercie vivement l'ensemble des membres. Je les remercie également pour toutes les discussions que nous avons eues, tant sur les plans technique avec Jean-Luc, Philippe, Olivier et Jacques que scientifique avec Philippe, Michel, Yvon, Georges, Véronique, Alessandro, Denis et Samuel. C'est cette dualité qui fait l'originalité de cette équipe.
L'éloge de l'humour de Michel Marcelin n'étant plus à faire (les remerciements de la thèse de mon propre directeur de thèse montraient déjà la couleur), je veux plutôt le remercier pour sa grande disponibilité, sa pédagogie, ses conseils et la considération qu'il m'a portée.

Je veux aussi remercier les personnes avec lesquelles j'ai partagé mon bureau. Tout d'abord Maxime, stagiaire de Michel, puis \'Elodie, stagiaire de Philippe avec qui j'ai passé de bon moments. J'ai eu plaisir à m'impliquer dans leurs travaux. Enfin, Henri Plana, ``ex'' thésard exilé au Brésil qui est venu quelques mois chaque année partager son ancien bureau avec moi et avec qui j'ai eu des discussions très intéressantes et très instructives.

Je remercie Françoise Herniou pour sa relecture finale du manuscript ainsi que pour d'autres aspects pratiques qui m'ont été utiles tout au long de ma thèse. 

\`A l'aube du vingt-et-unième siècle, l'informatique est un outil indispensable en astronomie, mais lorsqu'on a la poisse et qu'on est novice, comme c'est le cas pour moi, on est alors extrêmement reconnaissant à des personnes telles que Valérie Novak et Jean-Charle Lambert qui s'avèrent être des personnes très compétentes, très patientes, très disponibles et très sympathiques.

L'astronomie est pour moi indissociable de l'observation du ciel. C'est pourquoi les moments forts de ma thèse resteront sans aucun doute les missions d'observation. Débuter ma thèse par une mission à l'observatoire du mont Mégantic au Québec fut une chance d'autant plus grande que j'ai pu assister à une magnifique aurore boréale. Pourtant, ce fut un début un peu difficile puisque je n'avais qu'une vague idée de la physique que j'allais bientôt étudier. De plus, je ne connaissais alors personne. Je tiens donc à remercier Laurent Chemin qui m'a accueilli et m'a à proprement parlé fait débuter ma thèse, m'a présenté l'instrumentation Fabry-Perot, l'acquisition et la réduction des données. Un grand merci aussi à Olivier Daigle qui était avec moi pour la fin de cette mission et qui m'a fait partager ses programmes de réduction des données. Je remercie également Claude Carignan qui, à cette occasion, m'a accueilli au Laboratoire d'Astronomie Expérimentale de Montréal.

De plus, terminer ma thèse avec deux missions d'observations au WHT (La Palma, Espagne) fut un immense honneur et un réel plaisir. John Beckman est le principal acteur de la réussite de ces missions et je l'en remercie vivement. Je tiens à remercier également toute l'équipe de GH$\alpha$Fas, et tout particulièrement, Olivier Hernandez et Philippe Balard pour avoir eu la patience de nous apprendre, Marie-Maude et moi-même, à mettre en place l'instrument. Je veux aussi remercier Kambiz, Javier, Sébastien et Marie-Maude pour des parties de ping pong au sommet (de la montagne, bien sûr), pour leur bonne humeur et leur enthousiasme tant pour monter l'instrument que pour faire les observations.


Je remercie toutes les personnes avec lesquelles j'ai eu le plaisir de travailler pendant ma thèse et avec lesquelles, je l'espère, l'aventure continuera.

Je tiens à remercier l'ensemble de mon jury de thèse pour avoir accepté de remplir ce rôle.

Merci en particulier à Claudia Mendes de Oliveira pour m'avoir rassuré lors de ma première prestation orale en anglais à Marseille et pour m'avoir permis de travailler avec son étudiant en thèse, Sergio Torres.

Un grand merci supplémentaire à Thierry Contini pour m'avoir fait participer si naturellement à la collaboration MASSIV lors de ma dernière année de thèse, et pour me permettre de faire mon premier post-doc avec lui à Toulouse. Je tiens d'ailleurs à remercier l'ensemble de la collaboration MASSIV pour la confiance dont chacun des membres m'a gratifié.

Enfin, je ne remercierai probablement jamais assez la personne censée être la moins disponible et qui pourtant a su être là pour moi chaque fois que j'en ai eu besoin (j'ai appris à me jouer de son agenda), je pense bien entendu à mon directeur de thèse, Philippe Amram. Il a été à l'origine de tous mes résultats, de toutes mes collaborations, de mes missions, de mes enseignements, etc., bref, de tout ce qui m'a été profitable et qui a été mentionné dans toutes ces lignes. Il est pour moi bien plus qu'un directeur de thèse. Son sens du partage, son humanité, sa sympathie ont été pour moi un soutien extrêmement appréciable.

Les remerciements non officiels qui suivent me tiennent également à c\oe ur.

Un sujet source de sarcasmes à l'Observatoire de Longchamps, avant qu'on ne déménage au technopôle de Château-Gombert, fut la cantine. Quel casse-tête pour les Universitaires pour pouvoir se restaurer dans une cantine CNRS! Les étudiants ont subi le même problème. Je remercie donc le président ``Momo'' d'avoir bien voulu accepter les ``squatters'' à sa table, ainsi que Christine et Gaby qui étaient aux petits soins avec moi.

Je fais une dédicace toute spéciale (il faut savoir se jouer de sa réputation!) à mes monocycles sans qui ce travail de thèse n'aurait pu aboutir. C'est grâce à eux que je suis venu au laboratoire chaque jour pour avancer mes travaux. D'ailleurs j'avoue qu'ils m'en ont un peu voulu lorsque je leur ai annoncé que je prendrais le métro pour aller sur le site de Château Gombert. Ils ont donc insisté pour m'accompagner dans le métro et me reconduire chez moi de temps à autres.

Merci à l'équipe de monocyclistes de Marseille pour toutes les sorties qui font indéniablement du bien à l'esprit et en particulier à Yvon pour la ballade improvisée à Saint Michel de l'Observatoire (un peu impraticable d'ailleurs) lors de ma mission d'observation de la photométrie des galaxies GHASP.

Je remercie l'ensemble des thésards du LAM, en particulier Benoît pour son aide précieuse lors de la rédaction, ainsi qu'Arthur et Manu pour la détente entre les heures de travail.

Je remercie chaleureusement ma promotion de DEA et Jacqueline pour leur soutien, et plus particulièrement Xavier, Benoît et Marc que j'ai retrouvés avec plaisir lors de multiples conférences.

J'ai également pu compter sur mes amis pendant ces trois années pour passer de bons moments en dehors du labo. Un merci particulier à Guillaume qui m'a soutenu jusqu'au bout en me faisant répéter une ultime fois ma soutenance de thèse le matin même!

Céline, je te remercie pour tout ce que tu as fait pour moi. Ta patience, ton courage, ta relecture de ce manuscript qui était (du moins à première lecture!) pour toi assez obscur m'ont véritablement aidé à mener ce travail jusqu'au bout.

Je veux finir ces remerciements par ma famille: Maman, Papa, Guillaume, Virginie, Mylaine. Nous avons su nous serrer les coudes, être solidaires et présents les uns pour les autres.

\clearpage
\backmatter
\phantomsection
\addstarredchapter{Table des acronymes}
\chapter*{Table des acronymes}
\chaptermarkfree{TABLE DES ACRONYMES}

\begin{longtable}{ll}

\textbf{ADHOC} & Analyse et Dépouillement Homogène des Observations \textsc{Cigale}\\
\textbf{ADHOCw} & Analyse et Dépouillement Homogène des Observations \textsc{Cigale} windows\\
\textbf{BH$\alpha$Bar} & Big H-Alpha kinematical sample of Barred spiral galaxies\\
\textbf{BTFI} & Brazilian Tunable Filter Imager\\
\textbf{CCD} & Charge-Coupled Device\\
\textbf{CFHT} & Canada-France-Hawaï Telescope\\
\textbf{CIGALE} & CInématique des GALaxies\\
\textbf{CRAL} & Centre de Recherche Astronomique de Lyon\\
\textbf{DM} & Dark Matter\\
\textbf{ELT} & Extremely Large Telescope\\
\textbf{EAGLE} & Elt Ao for GaLaxy Evolution\\
\textbf{ESO} & European Southern Observatory\\
\textbf{FaNTOmM} & \FP~de Nouvelle Technologie pour l'Observatoire du mont  Mégantic\\
\textbf{FLAMES} & Fibre Large Array Multi Element Spectrograph\\
\textbf{G\'EPI} & Galaxies \'Etoiles Physique et Instrumentation\\
\textbf{GH$\alpha$FaS} & Galaxy H-Alpha \FP~System\\
\textbf{GHASP} & Gassendi H-Alpha survey of SPirals\\
\textbf{GIPSY} & Groningen Image Processing SYstem\\
\textbf{GIRAFFE} & High and Intermediate Resolution Spectrograph\\
\textbf{IAC} & Instituto de Astrof\'isica de Canarias\\
\textbf{IAG} & Instituto de Astronomia, Geof\'isica e Ciências Atmosféricas\\
\textbf{iBTF} & imaging Bragg Tunable Filter\\
\textbf{ICOS} & IC Optical System\\
\textbf{IDL} & Interactive Data Langage\\
\textbf{IFTS} & Imaging Fourier Transform Spectrometer\\
\textbf{IFU} & Integral Field Units\\
\textbf{IMAGES} & Intermediate MAss Galaxy Evolution Sequence\\
\textbf{INPE} & Instituto Nacional de Pesquisas Espaciais\\
\textbf{INT} & Isaac Newton Telescope\\
\textbf{IPCS} & Imaging Photo Counting Systems\\
\textbf{LAE} & Laboratoire d'Astrophysique Expérimentale\\
\textbf{LAM} & Laboratoire d'Astrophysique de Marseille\\
\textbf{LASER} & Light Amplification by the Stimulated Emission of Radiation\\
\textbf{LIRG} & Luminous Infra-Red Galaxie\\
\textbf{LTAO} & Laser Tomography Adaptive Optics\\
\textbf{MASSIV} & Mass ASsembly with \textsc{SInfoni} and \textsc{Vvds}\\
\textbf{MMTF} & Maryland-Magellan Tunable Filter\\
\textbf{MOAO} & Multi-Object Adaptive Optics\\
\textbf{MOND} & MOdified Newtonian Dynamics\\
\textbf{NTT} & New Technology Telescope\\
\textbf{OHP} & Observatoire de Haute Provence\\
\textbf{ON\'ERA} & Office National d'\'Etudes et de Recherches Aérospatiales\\
\textbf{OSIRIS} & OH-Suppressing Infra-Red Imaging Spectrograph\\
\textbf{PYTHEAS} & Prismes, Interféromètres et Trames de lentilles pour l'Holométrie et l'Endoscopie des\\
& Astres et des Sources\\
\textbf{PSF} & Point Spread Function en anglais, Réponse Impulsionnelle en français\\
\textbf{SAM} & Semi-Analytical method\\
\textbf{SESO} & Société Européenne de Systèmes Optiques\\
\textbf{SINFONI} & Spectrograph for INtegral Field Observations in the Near Infrared\\
\textbf{SPIE} & Society of Photo-Optical Instrumentation Engineers\\
\textbf{SOAR} & SOuthern Astrophysical Research\\
\textbf{TAURUS} & Fabry-Perot imaging spectrograph\\
\textbf{TTF} & Taurus Tunable Filter\\
\textbf{USP} & Universidade de S\~ao Paulo\\
\textbf{VLT} & Very Large Telescope\\
\textbf{VO} & Virtual Observatory\\
\textbf{VVDS} & \textsc{Vimos}-\textsc{Vlt} Deep Survey\\
\textbf{WFSpec} & Wide Field Spectrograph\\
\textbf{WHT} & William Herschel Telescope\\
\textbf{WMAP} & Wilkinson Microwave Anisotropy Probe

\end{longtable}

\backmatter

\clearpage
\phantomsection
\addstarredchapter{Table des matières}
\tableofcontents
\backmatter

\clearpage
\phantomsection
\addstarredchapter{Liste des tableaux}
\listoftables
\backmatter

\clearpage
\phantomsection
\addstarredchapter{Table des figures}
\listoffigures
\backmatter

\mainmatter  
\clearpage
\phantomsection
\addstarredchapter{Prélude}
\chapter*{Prélude}
\chaptermarkfree{Prélude}

La physique, et plus particulièrement l'astronomie, me fait depuis longtemps penser aux poupées gigognes, ces poupées semblables imbriquées les unes dans les autres: une petite poupée dans une poupée un peu plus grande, dans un poupée encore plus grande, etc. En effet, des quarks se regroupent pour former des nucléons qui s'associent entre eux ainsi qu'avec des électrons pour former des atomes. \`A leur tour, ces atomes s'agglomèrent pour former des molécules. Atomes et molécules forment des structures macroscopiques telles que les objets qui nous entourent ou bien à plus grande échelle les étoiles avec leurs planètes et leurs satellites. Ces systèmes se regroupent pour former des galaxies qui, à leur tour, se regroupent en amas, etc. Cette séquence fait penser à la création de l'Univers, telle qu'on l'imagine au début du vingt-et-unième siècle.
En passant d'une échelle à une autre, l'interaction change de nature, de l'interaction forte (nucléaire), à l'interaction faible (gravitation) en passant par l'interaction électromagnétique. Les lois de la physique sont différentes mais ont pourtant des caractéristiques communes.

\`A l'échelle de temps de l'espèce humaine, la compréhension des mécanismes responsables sur Terre de l'alternance périodique du jour et de la nuit et des saisons, rythmée sur une horloge céleste, a nécessité beaucoup de temps pour émerger et être acceptée.
Dès le troisième siècle avant notre ère, Aristarque de Samos avait émis l'hypothèse que les planètes tournent autour du Soleil, sa logique voulant que les plus petits objets tournent autour des plus gros: ``Pourquoi faire tourner la torche autour de la mouche ...''. Mais il fallut attendre le seizième siècle pour que cette idée soit admise, grâce aux travaux de Copernic qui lui permirent de comprendre que l'alternance entre le jour et la nuit était due à la rotation de la Terre sur elle-même à un rythme régulier et que les saisons résultaient du long trajet de la Terre autour de l'Astre Solaire.
Il fallut probablement tant de temps pour en arriver à cette conclusion à cause de la représentation géocentrique de l'Univers défendue par l'\'Eglise conjuguée aux difficultés observationnelles. En effet, comme le disait Descartes, ``nos sens nous trompent parfois''. Ainsi nos sens voient les astres posés sur une sphère céleste à deux dimensions alors qu'ils évoluent en réalité dans un espace à trois dimensions.
Kepler interpréta un siècle plus tard le mouvement des planètes autour du Soleil comme étant dû à une force d'intensité décroissante avec le carré de la distance, comme c'est le cas pour l'intensité de la lumière.
Galilée, grâce à sa lunette astronomique, apporta de nombreuses preuves observationnelles que la Terre n'est pas le centre du monde, en particulier avec les satellites de Jupiter et les phases de Vénus.
Enfin, à la fin du dix-septième siècle, Newton comprit que la force qui fait tourner les astres est due à la masse, et que les lois de gravitation sont universelles, c'est-à-dire qu'elles s'appliquent aussi bien pour expliquer le mouvement des planètes autour du Soleil, que celui de la Lune autour de la Terre ou encore des satellites de Jupiter.
Il inventa également le télescope composé de miroirs au lieu de lentilles comme c'est la cas de la lunette.
Ces progrès observationnels menèrent, petit à petit, à la découverte en 1924 des galaxies en tant que réels ``mondes à part'' par Edwin Hubble à partir d'observations au télescope de $2.5~m$ du mont Wilson.


Des milliards d'étoiles ainsi que de grandes quantités de gaz et de poussières constituent en apparence les galaxies. Chaque galaxie pouvant être considérée comme une entité en soi, il est primordial de comprendre quelle est l'origine de leur cohésion. \`A l'image des poupées gigognes, on entrevoit l'analogie avec le système solaire où les planètes sont retenues par l'attraction gravitationnelle due à la masse du Soleil.
Les galaxies proches sont le laboratoire idéal de l'astronome pour observer cette cohésion grâce, entre autres, à l'étude de leur cinématique. Les astronomes ne peuvent pas mesurer de vitesses dans les galaxies par l'observation des mouvements apparents, car ils sont trop petits pour être décelés ainsi depuis la Terre à l'échelle de temps d'une vie humaine. Ils mesurent directement la vitesse projetée le long de la ligne de visée en observant la lumière émise par le gaz et par les étoiles. Grâce à l'effet Doppler-Fizeau, le décalage spectral de raies d'absorption des atmosphères stellaires ou bien de raies d'émission du gaz (ou des étoiles) a permis dès les années 1930 de mesurer des vitesses dans des galaxies.
De même que pour le système solaire, il est alors nécessaire de représenter les galaxies dans l'espace pour interpréter les vitesses mesurées.
La grande résolution spatiale des observations des galaxies les plus proches permet dans certains cas d'affirmer que leur matière est confinée dans un plan.
L'interprétation naturelle est alors d'expliquer les mouvements dans ce plan par une rotation due au potentiel gravitationnel de la galaxie. Les premières courbes de rotation de galaxies ont ainsi été mesurées
par \citet{Burbidge:1960} par spectroscopie à longue fente placée le long du grand axe.
Les astronomes s'attendaient alors à observer des courbes de rotation avec une vitesse décroissante à grande échelle étant données la courbe de luminosité exponentielle des galaxies \citep{Kormendy:1977} et les lois de la gravitation universelle de Newton. Seulement, contre toute attente, les courbes de rotation restent plates à très grande échelle, jusqu'aux derniers points où des vitesses sont mesurables \citep{Bosma:1978,Rubin:1978}. Une explication avancée est que les galaxies possèdent de la matière qu'il est impossible de détecter, la matière sombre ou matière noire \citep{Gunn:1980} qui représenterait jusqu'à $90$\% de la masse totale des galaxies.
D'autres théories cherchent plutôt à expliquer cette observation par un changement des lois de la physique de Newton. C'est la théorie MOND
\citep{Milgrom:1983}.

Aujourd'hui encore, malgré les progrès observationnels, mais peut-être aussi grâce à eux, la cinématique de galaxies proches est source d'interrogations et d'investigations. De plus, avec la possibilité d'observer des galaxies lointaines, se posent de nouvelles questions sur l'évolution de la cinématique des galaxies et sur leur formation. C'est dans ce cadre que cette thèse s'est déroulée, pour apporter une petite pierre à un édifice en construction depuis bien des années...

\clearpage
\phantomsection
\addstarredchapter{Introduction}
\chapter*{Introduction}
\chaptermarkfree{Introduction}

L'étude de données cinématiques de galaxies proches et lointaines est indispensable à la compréhension de la formation, de l'évolution et de la stabilité des galaxies. En effet, la cinématique permet d'estimer la masse dynamique des galaxies, incluant leur matière sombre et leur matière visible. Elle permet également de qualifier le support dynamique et de quantifier le moment angulaire. La comparaison d'échantillons statistiques de galaxies proches et de galaxies lointaines (donc plus jeunes) permet alors de suivre l'évolution dynamique des galaxies à travers les âges. Cependant, la dimension angulaire des galaxies lointaines étant bien inférieure à celle des galaxies proches, il est nécessaire d'utiliser de nouvelles méthodes d'analyse comparative.

L'obtention de données cinématiques a fortement évolué depuis un demi-siècle. Ainsi, l'amélioration des techniques d'observation, plus particulièrement l'essor de l'imagerie électronique et l'augmentation du diamètre des télescopes, a permis le développement de la spectroscopie à champ intégral (appelée aussi spectroscopie 3D). En particulier, l'utilisation de radiotélescopes à synthèse d'ouverture (premières observations par \citealp{Gottesman:1975}) et l'utilisation d'interféromètres de Fabry-Perot dans le visible (premières observations par \citealp{Taylor:1980}) permettent d'obtenir des cartes de vitesses couvrant l'ensemble du champ des galaxies proches alors que les spectroscopes à longue fente limitent les observations à une unique direction spatiale.
La mesure des vitesses projetées en chaque point le long de la ligne de visée dans le disque des galaxies spirales ou lenticulaires permet de calculer la courbe de rotation des galaxies. Cette dernière conduit alors à la détermination de la distribution radiale du potentiel gravitationnel et de la masse galactique lumineuse et sombre.
La grande résolution spatiale et spectrale que l’on peut atteindre lors des études cinématiques à champ intégral des galaxies locales permet également de sonder les écarts à la rotation circulaire et d'observer des signatures cinématiques de structures telles que des bras spiraux \citep{Fathi:2008}, des barres \citep{Hernandez:2005a}, des interactions entre galaxies \citep{Amram:2007} ou encore des gauchissements de disques \citep{Christodoulou:1993}.

Les premières observations cinématiques de galaxies lointaines (décalage spectral entre $1$ et $2$) ont été obtenues par spectroscopie à longue fente et datent seulement de 1996 \citep{Vogt:1996}.
Jusqu'alors, pour des raisons observationnelles, les galaxies lointaines étaient vues comme des sources ponctuelles. En cosmologie, les galaxies étaient alors traditionnellement considérées comme des briques élémentaires traçant les grandes structures. L'étude des propriétés des galaxies permettait ainsi, à partir de grands relevés constitués de dizaines de milliers de galaxies, de contraindre les scénarios de formation de l'Univers par ses propriétés statistiques \citep{Cole:2001,Le-Fevre:2005}, sans pour autant avoir à regarder les propriétés dynamiques des galaxies en détail.
L'étude cinématique résolue à grand décalage spectral étant devenue possible, elle est désormais utilisée en cosmologie afin d'améliorer la compréhension des mécanismes d'agrégation de la matière et donc de formation des galaxies. Quand la formation d'étoiles fut-elle le plus intense? Où a-t-elle lieu? Les populations d'étoiles ont-elles évolué? Les galaxies se sont-elles formées par accrétion lente de matière ou bien sont-elles le résultat d'interactions et de fusions entre galaxies de faible masse? Les interactions favorisent-elles la formation d'étoiles? Comment le support dynamique des galaxies évolue-t-il? Ce support est-il ordonné (rotation) ou désordonné (mouvements aléatoires)? Quelle est la distribution de masse des galaxies?
\par
Si une galaxie est considérée comme un objet ponctuel, alors sa masse ne peut être déterminée qu'à partir du flux qu'elle émet ou de la largeur des raies de son spectre intégré, c'est-à-dire de la dispersion de vitesses globale qui est une estimation de la vitesse de rotation maximale projetée. Dans le premier cas, c'est la masse stellaire qui est estimée. Le rapport masse-luminosité étant mal contraint, l'incertitude est assez grande. Dans le second cas, il s'agit de la masse dynamique intégrée jusqu'à un rayon donné. Toutefois, l'incertitude reste grande car l'inclinaison est mal contrainte, la distribution de la matière responsable des raies est inconnue et le rayon à l'intérieur duquel est mesurée la masse est mal défini.
Les études sur la cinématique résolue à partir de spectrographie à longue fente puis à partir de spectroscopie à champ intégral depuis 2006 seulement \citep{Forster-Schreiber:2006,Flores:2006,Law:2007} permettent d'avoir une idée plus précise de la masse dynamique car la vitesse est alors mesurée à un rayon connu et car l'inclinaison et la distribution de matière peuvent être estimées. Ces observations permettent également de contrôler l'accord entre la morphologie et la cinématique et de mesurer des dispersions de vitesses locales et non globales. De plus, si on observe des galaxies en interaction ou en fusion, les observations spectroscopiques à champ intégral permettront dans certains cas de dissocier les composantes et ainsi de donner une estimation non biaisée de la masse.
Toutefois, la représentation dans l'espace reste une difficulté majeure pour les galaxies lointaines et ce pour plusieurs raisons:
\textit{(i)} la résolution spatiale reste faible pour ces galaxies malgré l'utilisation de l'optique adaptative;
\textit{(ii)} le signal provenant de ces galaxies est faible et noyé dans les raies d'émission du ciel nocturne;
\textit{(iii)} la géométrie de structures jeunes ou en formation peut être bien différente de celle des galaxies locales.
Il faudra attendre le développement des ELT (télescopes dont le diamètre sera de l'ordre de quarante mètres)
et d'une optique adaptative permettant d'atteindre leur résolution théorique pour obtenir des observations spectroscopiques à champ intégral de galaxies lointaines avec une résolution comparable à celle dont disposent actuellement les observations de galaxies proches. Ces technologies verront le jour dans une dizaine d'années et il est important de préparer ces futures observations dès à présent.

Le travail de cette thèse repose sur l'utilisation de cubes de données cinématiques pour un échantillon de $203$ galaxies locales observées dans le cadre du projet GHASP (Gassendi H-Alpha survey of SPirals). Ces cubes de données sont observés autour de la raie \Ha~du gaz d'hydrogène ionisé avec un Fabry-Perot à balayage.
L'étude cinématique d'un échantillon de galaxies proches est l'occasion d'affiner les méthodes d'extraction des paramètres cinématiques des disques optiques, de réaliser une étude statistique des propriétés cinématiques des galaxies locales, d'étudier les signatures de mouvements non circulaires, de contraindre la forme des halos de matière noire ainsi que de tester des théories de gravitation modifiée.
Par ailleurs, la constitution d'un échantillon de galaxies locales permet de définir des cubes de données cinématiques de référence qui seront utilisés pour étudier la cinématique des galaxies lointaines.
Afin de dissocier les biais observationnels des effets d'évolution cinématique, une solution consiste à simuler les galaxies locales telles qu'elles seraient observées si elles étaient aussi éloignées que les galaxies à fort décalage spectral, d'où l'intérêt plus particulier suscité par les données spectroscopiques à champ intégral
de galaxies locales.

Dans le premier chapitre, sont exposées les techniques d'observations adaptées à l'étude cinématique des galaxies proches. Leur grande taille angulaire nécessite d'utiliser des techniques permettant d'obtenir un grand champ de vue. L'accent est donc mis sur les techniques d'observations utilisant l'interféromètre de Fabry-Perot et, plus particulièrement, sur ma participation au développement en cours de l'instrument 3D-NTT, développé dans une collaboration entre le LAM (Marseille), le LAE (Montréal) et le G\'EPI (Paris) et qui sera sur le ciel fin 2009.
Une étude d'un concept d'instrument spectroscopique à grand champ (WFSpec) de première génération pour les ELT est également présentée.
Le deuxième chapitre est consacré au projet GHASP. Ce projet a permis d'obtenir l'échantillon de référence de galaxies locales le plus important observé avec les techniques de Fabry-Perot pour des études cinématiques et dynamiques. Ses objectifs scientifiques y sont présentés et les méthodes de dépouillement et d'analyse des données sont détaillées dans deux articles présentant l'ensemble des données GHASP \citep{Epinat:2008b,Epinat:2008a}\footnote{Epinat, B., Amram, P., et Marcelin, M. : 2008a, \textit{MNRAS}, \textbf{390}, 466}$^{,}$\footnote{Epinat, B., Amram, P., Marcelin, M., Balkowski, C., Daigle, O., Hernandez, O., Chemin, L., Carignan, C., Gach, J.-L., et Balard, P. : 2008b, \textit{MNRAS} \textbf{388}, 500}. Des études cinématiques et dynamiques de cet échantillon local sont également présentées dans ce deuxième chapitre sous forme d'un article en préparation \citep{Epinat:prep}\footnote{Epinat, B., Amram, P., Marcelin, M., et al. : 2009b, \textit{en préparation}}.
Le troisième chapitre concerne l'étude de données cinématiques de galaxies lointaines obtenues à partir d'instruments à champ intégral. L'analyse cinématique de sept nouvelles galaxies lointaines observées par SINFONI \citep{Epinat:2009}\footnote{Epinat, B., Contini, T., Le Fèvre, O., Vergani, D., Garilli, B., Amram, P., Queyrel, J., Tasca, L. et Tresse, L. : 2009c} y est présentée. Ces données ainsi que d'autres données de la littérature sont alors mises en regard avec l'échantillon de référence GHASP: dans un dernier article \citep{Epinat:2008c}\footnote{Epinat, B., Amram, P., Balkowski, C., Marcelin, M. : 2009a}, les galaxies de l'échantillon de référence local sont projetées au décalage spectral des données de la littérature mettant en évidence les biais observationnels ainsi que l'évolution cinématique des galaxies.
\par
L'Annexe \ref{annexe_instrumentation} contient des articles dans lesquels je suis impliqué. Ils approfondissent certains points des études instrumentales exposées au chapitre \ref{instrumentation} (\citealp{Marcelin:2008}~\footnote{Marcelin, M., Amram, P., Balard, P., Balkowski, C., Boissin, O., Boulesteix, J., Carignan, C., Daigle, O., de Denus-Baillargeon, M.-M., Epinat, B., Gach, J.-L., Hernandez, O., Rigaud, F., et Vallée, P. : 2008, in \textit{Ground-based and Airborne Instrumentation for Astronomy II, Ian S. McLean ; Mark M. Casali, Editors, 701455}, Vol. 7014 of \textit{Presented at the Society of PhotoOptical Instrumentation Engineers (SPIE) Conference}}; \citealp{Hernandez:2008}~\footnote{Hernandez, O., Fathi, K., Carignan, C., Beckman, J., Gach, J.-L., Balard, P., Amram, P., Boulesteix, J., Corradi, R. L. M., de Denus-Baillargeon, M.-M., Epinat, B., Relaño, M., Thibault, S., et Vallée, P. : 2008, \textit{PASP} \textbf{120}, 665}; \citealp{Moretto:2006}~\footnote{Moretto, G., Bacon, R., Cuby, J.-G., Hammer, F., Amram, P., Blais-Ouellette, S., Blanc, P.-E., Devriendt, J., Epinat, B., Fusco, T., Jagourel, P., Hernandez, O., Kneib, J.-P., Montilla, I., Neichel, B., Pécontal, E., Prieto, E., et Puech, M. : 2006, in \textit{Ground-based and Airborne Instrumentation for Astronomy. Edited by McLean, Ian S. ; Iye, Masanori. Proceedings of the SPIE, Volume 6269, pp. 62692G (2006)}., Vol. 6269 of \textit{Presented at the Society of Photo-Optical Instrumentation Engineers (SPIE) Conference}}). L'Annexe \ref{help_computeeverything} contient l'aide de l'outil de réduction développé pour dépouiller les données GHASP du chapitre \ref{ghasp_donnees}. L'Annexe \ref{spectre_vitesse} explique comment extraire les données cinématiques à partir d'un spectre. Elle expose les méthodes utilisées pour l'extraction des cartes cinématiques des galaxies locales et lointaines présentées respectivement dans les chapitres \ref{ghasp_donnees} et \ref{etudes_highz}. L'Annexe \ref{ghasp_images} illustre les données cinématiques (cartes cinématiques, courbes de rotations, profils de dispersion de vitesses et base de données) de l'échantillon GHASP (chapitre \ref{ghasp_donnees}). L'Annexe \ref{analyse_ghasp}, quant à elle, présente un article auquel j'ai participé \citep{Spano:2008}\footnote{Spano, M., Marcelin, M., Amram, P., Carignan, C., Epinat, B., et Hernandez, O. : 2008, \textit{MNRAS} \textbf{383}, 297} concernant la forme des halos de matière sombre des galaxies et qui résulte de l'exploitation des courbes de rotation de l'échantillon GHASP (chapitre \ref{ghasp_donnees}). Enfin, l'Annexe \ref{annexe_highz} expose une étude de la relation de Tully-Fisher à un décalage spectral proche de $0.6$ à partir de données GIRAFFE \citep{Puech:2006}\footnote{Puech, M., Flores, H., Hammer, F., Yang, Y., Neichel, B., Lehner t, M., Chemin, L., Nesvadba, N., Epinat, B., Amram, P., Balkowski, C., Cesarsky, C., Dannerbauer, H., di Serego Alighieri, S., Fuentes-Carrera, I., Guiderdoni, B., Kembhavi, A., Liang, Y. C., Östlin, G., Pozzetti, L., Ravikumar, C. D., Rawat, A., Vergani, D., Vernet, J., et Wozniak, H. : 2008, \textit{A\&A} \textbf{484}, 173} pour laquelle j'ai projeté plusieurs galaxies GHASP en utilisant la méthode décrite au chapitre \ref{etudes_highz} afin de vérifier les biais induits lors de la détermination de la vitesse maximale.


\chapter{La spectroscopie à champ intégral avec l'interféromètre de \FP}
\label{instrumentation}
\minitoc
\textit{Ce chapitre concerne les aspects instrumentaux de cette thèse. La technique de spectroscopie utilisant l'interféromètre de \FP~est présentée, ainsi que certains aspects de la réduction des données sur lesquels j'ai travaillé. Les études concernant la définition et l'utilisation des \FP~de l'instrument 3D-NTT auxquelles j'ai participé sont ensuite détaillées. Enfin, dans le cadre de WFSpec, projet de spectrographe à grand champ pour les ELT, je présente une étude que j'ai réalisée sur le facteur de mérite des trois concepts instrumentaux proposés ainsi que ma contribution à l'un de ces concepts, nommé iBTF.}
\hl

\section{Différentes méthodes pour faire de la spectro-imagerie}
\label{spectro}

Les études qui sont présentées dans cette thèse concernent principalement des données obtenues avec des instruments utilisant l'interféromètre de \FP. Une revue des différents types d'instruments de spectro-imagerie utilisés actuellement est l'occasion de montrer les domaines d'applications des diverses méthodes.

\subsection{Spectroscopie, spectrographie ou spectrométrie?}

Tout d'abord, commençons par clarifier un peu le vocabulaire de ce que l'on appelle couramment la ``spectro''.
Le préfixe \hbox{``spectro-''} désigne le spectre, soit, en optique, les différentes longueurs d'onde ou bien les fréquences qui composent un faisceau lumineux, en d'autres termes, les couleurs.
Le suffixe \hbox{``-scopie''} nous vient du grec et signifie ``examiner, observer''. La spectroscopie est donc l'examen, l'observation de spectres. Un spectroscope est donc un instrument qui permet d'observer des spectres.
Le suffixe \hbox{``-graphie''} nous vient également du grec et signifie ``écrire''. Il fait donc référence à une notion d'enregistrement. La spectrographie désigne donc l'enregistrement des spectres et un spectrographe est un instrument permettant d'enregistrer des spectres.
Le suffixe \hbox{``-métrie''} quant à lui, désigne la ``mesure''. La spectrométrie consiste donc en la mesure de spectres et un spectromètre est ainsi un instrument permettant la mesure des spectres.
\par
La différence entre chacun de ces termes est subtile.
Un spectroscope doit nous permettre d'observer un spectre à l'oeil, on a donc nécessairement besoin d'un système dispersif, c'est-à-dire qui étale la spectre selon une direction de l'espace, ce qui est le cas d'un prisme ou d'un réseau optique.
Pour obtenir un spectrographe, il faut ajouter un système d'acquisition, typiquement une caméra ou une plaque photo.
Pour obtenir un spectromètre, il faut une référence, donc un système permettant une calibration en longueur d'onde et/ou en intensité.
\par
La spectroscopie ayant aboutit le siècle dernier à la possibilité de faire de la spectro-imagerie\footnote{encore appelée spectro 3D ou spectroscopie à champ intégral}, c'est-à-dire obtenir un spectre pour chacun des points d'une image, la terminologie a elle aussi évolué. Ainsi aujourd'hui, un spectrographe désigne un instrument qui permet de faire l'acquisition de spectres sur la surface du récepteur, donc sur une image unique, alors qu'un spectromètre désigne un instrument qui permet la mesure de spectres au cours du temps, en fonction de créneaux de balayage, ce qui nécessite donc plusieurs lectures du récepteur.

\subsection{Les systèmes dispersifs}

Les systèmes dispersifs sont des spectrographes qui utilisent soit des prismes, soit des réseaux, soit les deux. Le phénomène de réfraction permet au prisme de disperser la lumière. C'est la variation de l'indice optique du prisme avec la longueur d'onde qui fait que les différentes composantes du spectre sont réfractées dans des directions différentes. Il y a deux types de réseaux, par réflexion et par transmission, mais dans les deux cas, c'est un phénomène interférométrique dû à sa structure périodique qui en fait un système dispersif. Un autre système optique dispersif peut être créé par l'utilisation d'un réseau accolé sur une face d'un prisme. Ce système s'appelle un Grism\footnote{\textit{Gr} pour \textit{gr}ating et \textit{ism} pour pr\textit{ism} (réseau et prisme en anglais)} et est utilisé afin de pouvoir utiliser un montage en ligne. En effet, la dispersion du prisme est calculée de manière à compenser la dispersion globale de l'ordre d'interférence utile du réseau.
Afin de permettre une mesure avec ces systèmes dispersifs, il est nécessaire de les coupler avec une fente.
L'observation ne possède donc qu'une dimension spatiale (le long de la fente) et le spectre est dispersé dans la direction perpendiculaire à la fente.

Cependant, les prismes ne permettent pas d'atteindre des résolutions spectrales très grandes et les réseaux n'ont pas une efficacité optimale car chaque longueur d'onde est dispersée dans plusieurs ordres d'interférence. Par ailleurs, lorsqu'on veut réaliser une cartographie contenant en chaque point une information spectrale, du fait que la dispersion se fait selon une direction spatiale, ces systèmes nécessitent soit plusieurs poses avec déplacement du champ de vue correspondant à la largeur de la fente, soit un couplage avec un système optique comme un découpeur d'image ou des fibres optiques, qui va découper le champ de vue et le réordonner le long d'une fente. Ce type d'instrument 3D est intéressant lorsqu'un grand domaine spectral est nécessaire sur un champ de vue modeste. Néanmoins, cela induit une réduction de données dont la difficulté principale est la reconstruction de l'image. Les systèmes à fibres optiques ont généralement une efficacité faible à cause des pertes dues au couplage avec les fibres. Les observations de galaxies lointaines utilisent ce genre d'instruments car le champ de vue nécessaire est petit.

\subsection{Les systèmes interférométriques à balayage}

Deux types de spectromètres basés sur le phénomène d'interférences optiques permettent naturellement de faire de la spectroscopie à champ intégral. Il s'agit du spectromètre à transformée de Fourier, dit IFTS,
et de l'interféromètre de \FP. Ces interféromètres ne font que moduler le flux et ne perturbent donc pas le caractère imageur des instruments dans lesquels ils sont placés.
\par
Un spectromètre à transformée de Fourier est un interféromètre de Michelson: ce type d'instrument est composé d'une lame semi-réfléchissante et de deux miroirs, dont un mobile, formant deux voies. La lumière va être décomposée par la lame en deux faisceaux envoyés dans chacune des voies et recomposée après réflexion sur les miroirs, à leur retour sur la lame: les ondes en provenance des deux voies interfèrent. En faisant varier la différence de chemin optique grâce au miroir mobile, il est possible d'enregistrer une série d'interférogrammes. Celle-ci contient la transformée de Fourier du spectre de la lumière en chaque point du champ de vue: pour chaque exposition, l'information enregistrée concerne donc l'ensemble du spectre.
\par
Un spectromètre de \FP~est un système dynamique composé d'une paire de lames semi-réfléchissantes parallèles. La lumière traversant ce système subit des réflexions multiples entre les lames qui interfèrent à la sortie de l'interféromètre. Pour chaque espacement des lames, seules certaines longueurs d'onde sont transmises par l'interféromètre. Le spectre sur l'ensemble du champ est donc obtenu en faisant varier séquentiellement l'espacement entre les lames. Ce type d'instrumentation est idéal pour étudier la cinématique d'objets étendus (aussi bien les nébuleuses que les galaxies) car le domaine de longueurs d'onde utile peut être ajusté à la dynamique en vitesse de l'objet observé et au bénéfice d'un champ de vue important.
\par
L'IFTS est donc un interféromètre à deux ondes alors que le \FP~est un interféromètre à ondes multiples. La fonction de transfert dans le cas de l'interféromètre à deux ondes est donc sinusoïdale alors que celle de l'interféromètre à ondes multiples est une fonction d'Airy (voir partie \ref{relations_fondamentales}) qui est plus sélective. L'utilisation du \FP~est une technique éprouvée pour des observations à faible flux du gaz ionisé galactique \citep{Russeil:2005} et extragalactique \citep{Marcelin:1987} alors que l'IFTS est actuellement principalement utilisé pour des applications galactiques stellaires ou planétaires (voir l'article de \citealp{Maillard:1996} et les références qui y sont attachées).
\par
Dans la suite, nous nous focalisons sur les instruments utilisant un interféromètre de \FP.

\section{Principes de la spectrométrie avec l'interféromètre de \FP}
\label{Principe du Fabry Perot}

\subsection{Historique de l'interféromètre de \FP}
\citet{Georgelin:1995} présentent un historique très détaillé de la découverte du \FP~et de ses premières
applications. Les lignes qui suivent indiquent le contexte scientifique de cette découverte ainsi que les développements et observations astronomiques qui en ont découlé.
\par
Le principe de cet interféromètre repose sur le caractère ondulatoire de la lumière qui fut principalement établit par Thomas Young (1773-1829). Le cadre théorique de l'optique ondulatoire est élaboré un peu plus tard par Augustin Fresnel (1788-1827) et l'expression mathématique de l'interféromètre de \FP~est déterminée en 1831 par Georges Airy (1801-1892) qui décrivit l'addition cohérente de réflexions multiples entre deux surfaces planes (voir partie \ref{relations_fondamentales}), grâce à la fonction éponyme. L'interféromètre de \FP~a été inventé en 1896 par Charles Fabry (1867-1945) et Alfred Perot (1863-1925) dans le domaine de la métrologie, afin de mesurer très précisément des variations d'épaisseur. Ils pressentent alors déjà la possibilité de placer leur interféromètre au foyer d'un télescope. Les applications de cet interféromètre s'avèrent nombreuses. En 1909, les inventeurs de cet interféromètre l'utilisent dans le but de comparer une lumière monochromatique au mètre étalon, afin d'améliorer la précision des mesures réalisées par Michelson en 1893.
\par
Nous devons les premières utilisations astronomiques de cet interféromètre à \citet{Buisson:1914} lorsqu'ils observèrent en 1914 la nébuleuse d'Orion à travers un interféromètre de \FP~monté sur le télescope de Foucault de $80~cm$ de l'Observatoire de Marseille. Ces observations leur permirent d'obtenir des informations quant à la cinématique du gaz ionisé de cette nébuleuse. La technique d'observation a été à nouveau utilisée une cinquantaine d'années plus tard par \citet{Courtes:1972} afin d'étudier les régions de gaz ionisé (régions HII, voir partie \ref{regions_h2}) de notre Galaxie mais également dans d'autres galaxies. Jusqu'alors, ces observations utilisaient un interféromètre de \FP~dont l'espacement entre les lames est fixe également appelé étalon.
Le premier \FP~à balayage date de 1970 et a été développé à l'Imperial College de Londres, aboutissant quelques années plus tard à un interféromètre contrôlé informatiquement et adapté à l'utilisation au foyer de télescopes \citep{Hicks:1976}.
Les premières utilisations d'un \FP~à balayage datent respectivement de 1980, avec le développement de l'instrument TAURUS initialement pour l'INT,
et de 1982 avec le développement de l'instrumentation CIGALE
par \citet{Boulesteix:1984} et ses collaborateurs pour le foyer Cassegrain du télescope de $3.6~m$ du CFHT.
Outre les observations avec étalon par \citet{de-Vaucouleurs:1974}, ces deux instruments conduisirent aux premiers champs de vitesses de galaxies \citep{Taylor:1980,Marcelin:1987}, prémices d'une longue pratique d'observations utilisant le \FP~à balayage au sein de l'équipe Interférométrie du LAM \citep{Boulesteix:1987,Laval:1987,Georgelin:1987,Marcelin:1987,Amram:1991,Le-Coarer:1992,Plana:1996,Russeil:1998,Garrido:thesis}.
Depuis, de nombreux instruments basés sur l'utilisation de l'interféromètre de \FP~ont été développés, notamment pour des applications à basse résolution spectrale \citep{Pogge:1995,Bland:1989}. Dans ce cas, l'interféromètre est nommé filtre accordable. Nous mentionnerons également les TTF
et le MMTF
qui, avec les instruments de haute résolution spectrale ont inspiré le développement de l'instrument versatile 3D-NNT
(voir partie \ref{3dntt}) par l'équipe Interférométrie du LAM en collaboration avec le G\'EPI
(observatoire de Paris) et le LAE
(Université de Montréal).
\par
Notons également que le principe de l'interféromètre de \FP~est utilisé pour réaliser les filtres interférentiels très sélectifs (quelques $nm$) qui sont d'ailleurs utilisés lors des observations astronomiques avec interféromètres de \FP~(voir Figure \ref{filtre_interferentiel}). Le \FP~est également utilisé pour réaliser des cavités laser.
Dans ce cas, les miroirs ne sont plus plans, mais concaves. Enfin le \FP~est largement utilisé dans le domaine des télécommunications afin de contrôler la longueur d'onde des signaux.

\subsection{Relations fondamentales}
\label{relations_fondamentales}

\begin{figure}[htbp]
\begin{center}
\includegraphics[width=12cm]{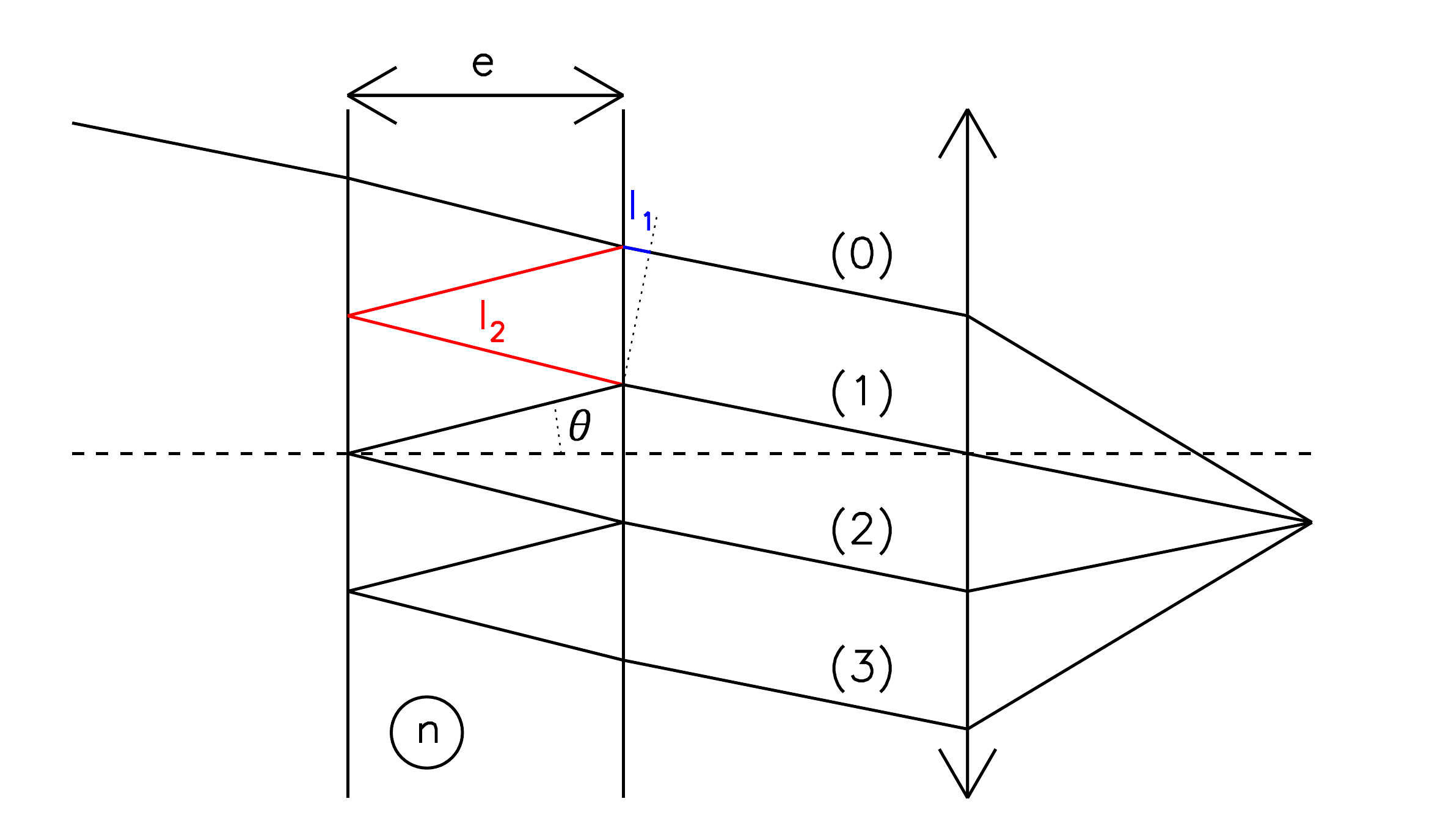}
\caption{Trajet de la lumière à l'intérieur de la cavité Fabry-Perot et construction de l'image à l'infini par l'objectif de la caméra.}
\label{FP_chemin_optique}
\end{center}
\end{figure}

\begin{figure}[h]
\begin{center}
\includegraphics[width=10cm]{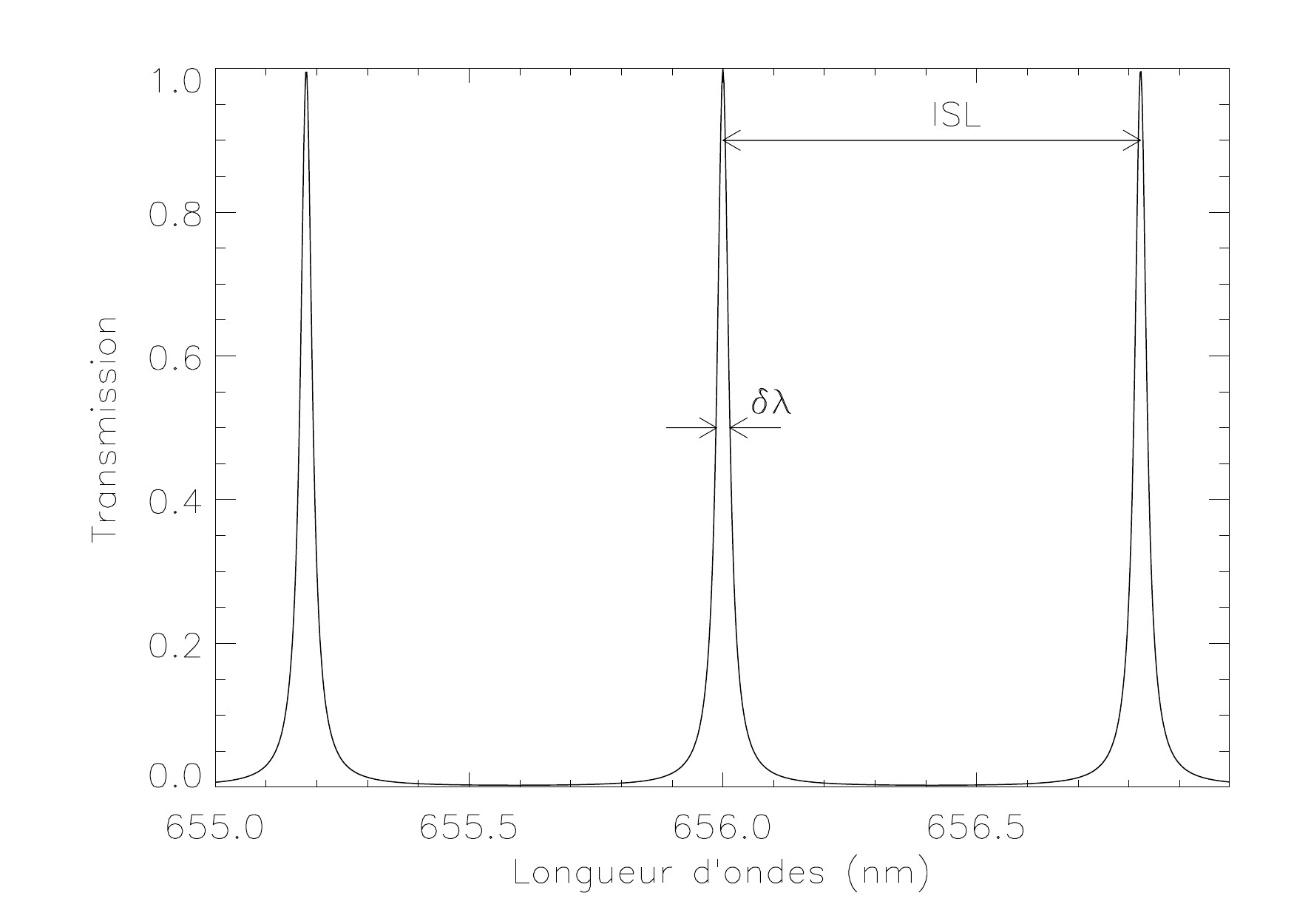}
\caption{Transmission d'un \FP~théorique (fonction d'Airy), possédant un facteur de réflexion de $0.9$ et fonctionnant à l'ordre 798 pour la raie \Ha~($656.3~nm$).}
\label{FP}
\end{center}
\end{figure}

Le \FP~est composé de deux lames semi-réfléchissantes dont on note les transmissions $\tau_1$ et $\tau_2$, et les réflectivités $\rho_1$ et $\rho_2$. On note $e$ l'écart entre les lames et $n$ l'indice du matériau entre les lames, qui est de l'air dans nos applications (donc $n=1$).
Considérons un faisceau collimaté d'inclinaison $\theta$ par rapport à l'axe optique. En entrée du \FP, le champ électrique de l'onde plane est
$$E(\lambda)=Ue^{-i\omega t + i\phi_0}$$
où $\omega=2\pi c/\lambda$ est la fréquence angulaire et $\phi_0$ est la phase à l'origine.
La différence de chemin optique entre deux réflexions successives de l'onde est (Figure \ref{FP_chemin_optique}):
$$\delta=l_1 - l_2 = 2ne \cos(\theta)$$
\\
Les ondes qui sortent du \FP~s'expriment en fonction de $k$, le nombre d'allers-retours dans la cavité:
$$E_k(\lambda,\theta)=U\tau_1 \tau_2 (\rho_1 \rho_2)^{k} e^{-i\omega t +i\phi_1 - 2i \pi \frac{k \delta}{ \lambda}}$$
Ces ondes vont interférer entre elles à l'infini.
L'intensité sortante est donc $$I(\lambda,\theta)=|\sum_{k=0}^{\infty} E_k(\lambda,\theta)|^2$$
Posons pour alléger les écritures:
\begin{itemize}
\item $\tau=\tau_1 \tau_2$
\item $\rho=\rho_1 \rho_2$
\item $I_0=|U|^2$
\item $\phi=2\pi \frac{2ne\cos(\theta)}{\lambda}$
\end{itemize}
Nous obtenons
$$I(\lambda,\theta)=I_0 \tau^2 |\sum_{k=0}^{\infty} (\rho e^{-i\phi})^k|^2=I_0 \tau \frac{1}{|1-\rho e^{-i\phi}|^2}$$
En remarquant que
$$|1-\rho e^{-i\phi}|^2=(1-\rho)^2(1+\frac{4\rho}{(1-\rho)^2}\sin^2(\phi/2))$$
et en posant la finesse
\begin{equation}
F=\frac{\pi \sqrt{\rho}}{1-\rho}
\label{finesse}
\end{equation}
on obtient l'expression de la fonction d'Airy (Figure \ref{FP}):
\begin{equation}
I(\lambda,\theta)=I_0 \left( \frac{\tau}{1-\rho} \right)^2 \frac{1}{1+\frac{4F^2}{\pi^2}\sin^2(\phi/2)}
\end{equation}
Cette fonction est $2\pi$-périodique en $\phi$ et maximale pour $\phi=0\left[2\pi\right]$ soit pour
\begin{equation}
\lambda p = 2ne\cos{\theta} \mathrm{~~~~,~~~~avec~~~~}p \in \mathbb{N}
\label{fp_fondamental}
\end{equation}

\begin{figure}[h]
\begin{center}
\begin{tabular}{c c}
\includegraphics[width=8cm]{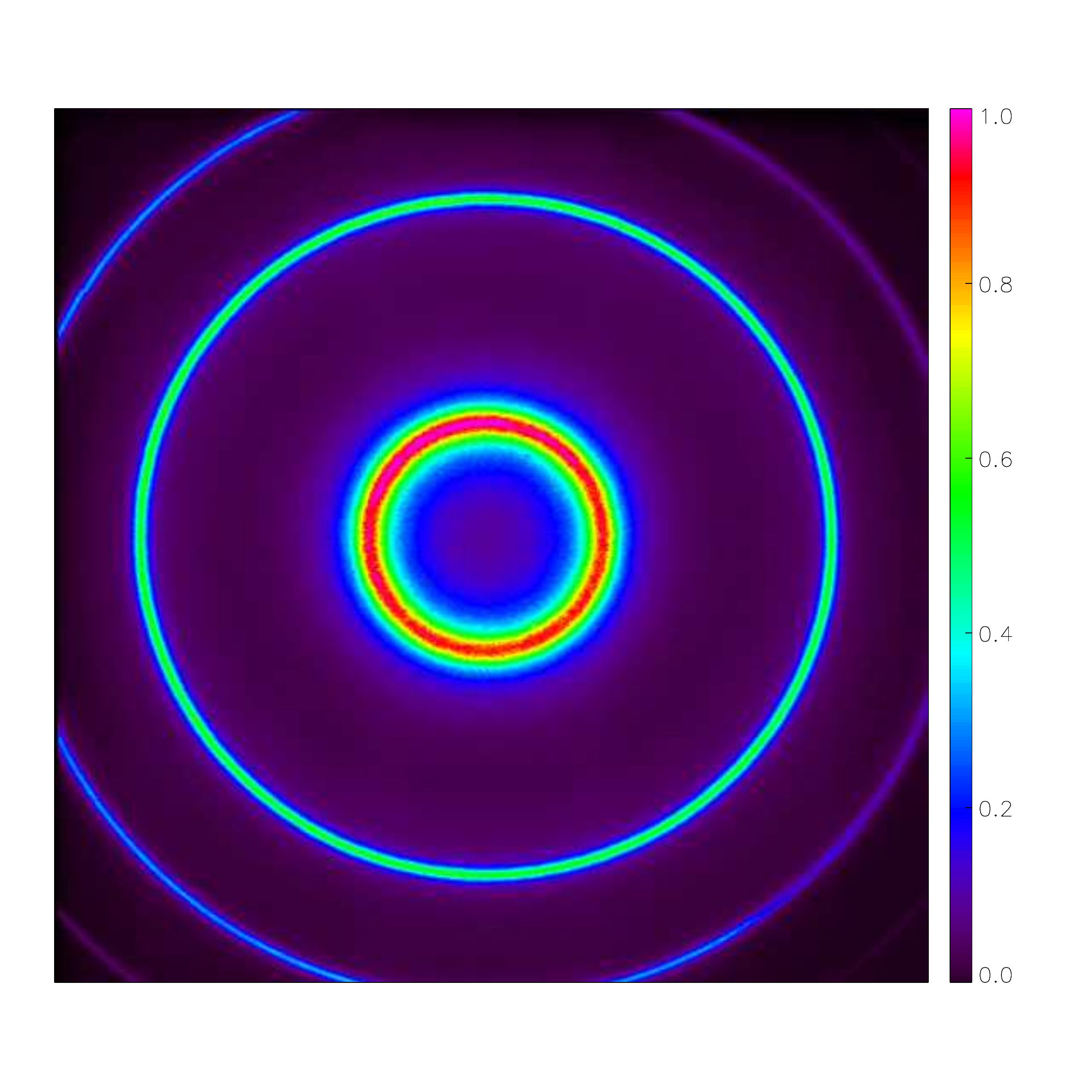}
\includegraphics[width=8cm]{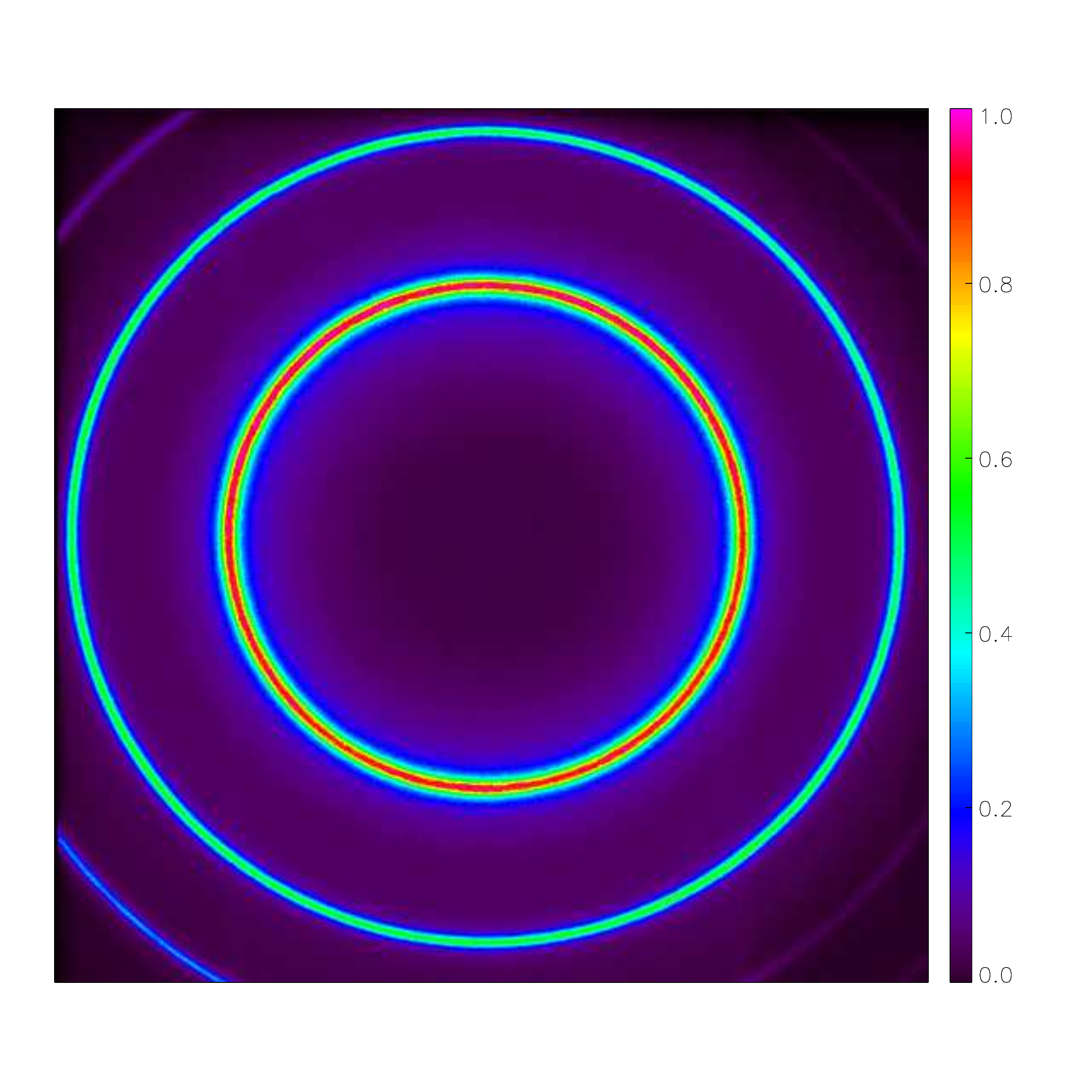}
\end{tabular}
\caption{Anneaux d'interférences obtenus lors d'une calibration utilisant la raie du Néon à $659.9~nm$ isolée grâce à un filtre interférentiel à bande étroite. Les différents anneaux correspondent à la même raie observée dans des ordres d'interférences distincts. Les deux images correspondent à deux espacements différents des lames (deux canaux spectraux): la raie n'est plus transmise à la même position, les longueurs d'onde transmises ont changé avec l'espacement des lames. Le \FP~utilisé fonctionne à l'ordre $798$ pour la raie \Ha~et donc à l'ordre $793$ pour cette raie du Néon (équation \ref{fp_fondamental}).}
\label{anneaux_interference}
\end{center}
\end{figure}

\begin{figure}[h]
\begin{center}
\includegraphics[width=8cm]{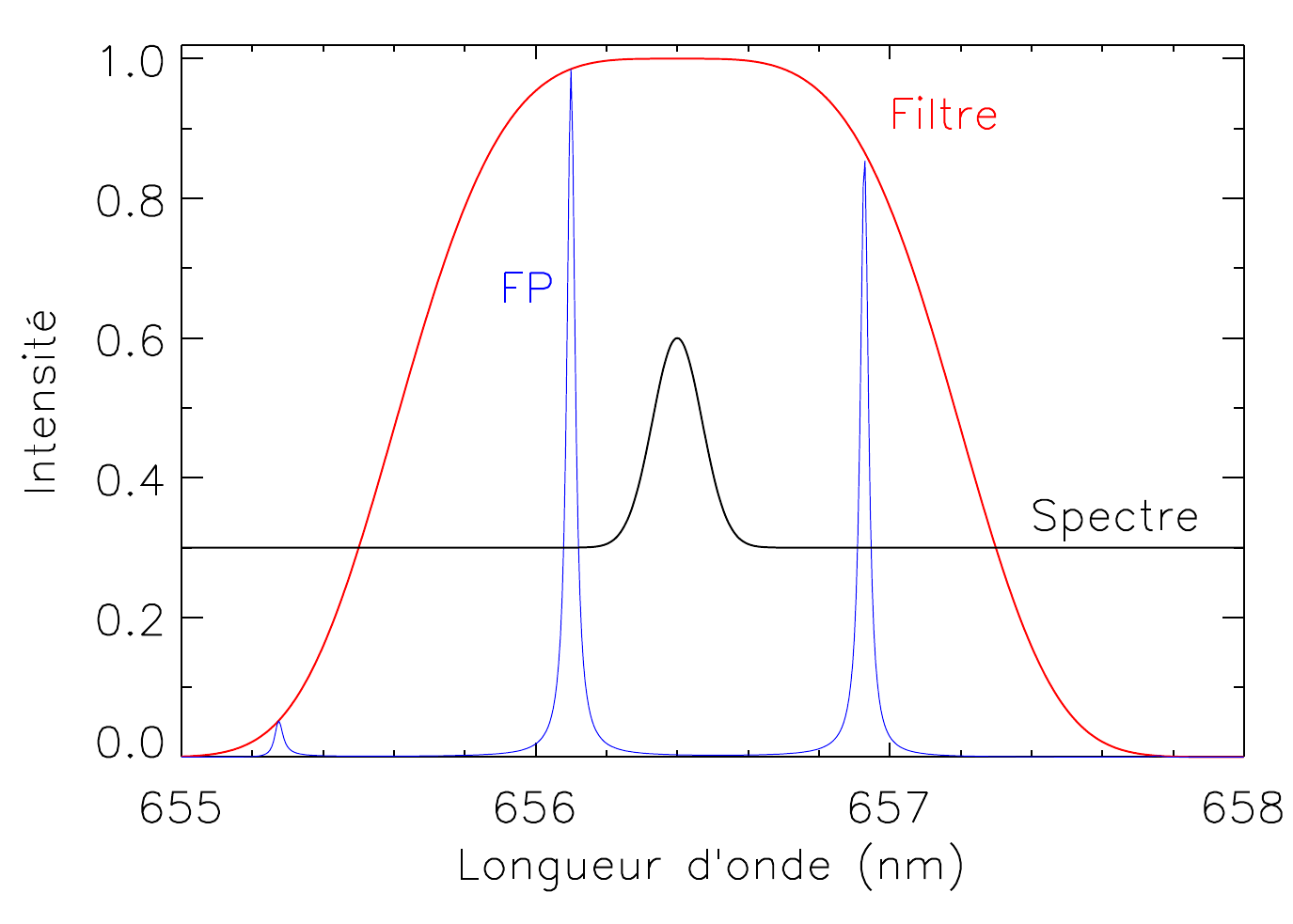}
\caption{La fonction d'Airy du Fabry-Perot est modulée par la réponse du filtre interférentiel. Ainsi, seuls les pics de transmission sur la région du spectre à étudier ont une bonne transmission. La raie est échantillonnée en balayant le spectre avec le Fabry-Perot.}
\label{filtre_interferentiel}
\end{center}
\end{figure}

L'équation \ref{fp_fondamental} est l'équation fondamentale du \FP. \`A partir de cette équation, on voit clairement que la longueur d'onde va varier dans le champ avec une symétrie circulaire, ce qui explique l'existence d'anneaux d'interférence (correspondant aux divers ordres) dans les interférogrammes lorsqu'on observe une source monochromatique (Figure \ref{anneaux_interference}).
On voit également que, pour une position donnée, plusieurs longueurs d'onde sont transmises. Afin d'éviter la superposition de ces longueurs d'onde, des filtres interférentiels sont généralement utilisés afin de sélectionner l'intervalle spectral contenant la raie d'intérêt (voir Figure \ref{filtre_interferentiel}). C'est le cas des instruments \FP~présentés dans cette thèse.
%

On appelle intervalle spectral libre ($ISL$) l'intervalle en longueur d'onde séparant deux pics de transmission (voir Figure \ref{FP}):\\
$$\lambda_p p=\lambda_{p+1} (p+1)\mathrm{~~~~et~~~~}2~ISL=\lambda_{p-1} - \lambda_{p+1}$$
Donc si l'ordre est grand, on a\footnote{En toute rigueur on a $ISL=\frac{\lambda_p}{p-1/p}$}:
\begin{equation}
ISL=\frac{\lambda}{p}
\label{isl}
\end{equation}
On peut montrer que la finesse d'un \FP~est égale à
\begin{equation}
F=\frac{ISL}{\delta\lambda}
\label{finesse_def}
\end{equation}
où $\delta\lambda$ est l'élément de résolution, typiquement la largeur à mi-hauteur de la PSF
(voir Figure \ref{FP}). L'acuité de la fonction d'Airy est donc déterminée par le choix de la réflectivité des lames (équations \ref{finesse} et \ref{finesse_def}).

Par définition, la résolution spectrale d'un instrument est
\begin{equation}
R=\frac{\lambda}{\delta\lambda}
\label{resolution_def}
\end{equation}
On en déduit l'expression de la résolution d'un \FP~à l'aide des équations \ref{isl} et \ref{finesse_def}:
\begin{equation}
R=pF
\label{resolution_fp}
\end{equation}

Les équations \ref{fp_fondamental}, \ref{isl}, \ref{finesse_def} et \ref{resolution_fp} sont les relations de base de la spectroscopie \FP.

\subsection{Utilisation classique du \FP: pupille ou foyer?}
\label{foyer_pupille}

La relation \ref{fp_fondamental} nous montre que les longueurs d'onde transmises par l'interféromètre de \FP~changent avec l'angle d'incidence du rayon lumineux.

Lorsque le \FP~est placé dans un faisceau collimaté, chaque angle d'incidence correspond à une position différente sur le champ et la longueur d'onde transmise varie dans le champ. Les défauts de la surface de l'interféromètre interceptée par le faisceau sont moyennés, ce qui provoque généralement un élargissement du pic de transmission, soit une diminution de la finesse, dépendant des défauts moyens.
Dans cette configuration, afin de minimiser la surface du \FP, ce dernier est généralement placé dans la pupille, d'où le nom d'utilisation en pupille.
\par
\begin{figure}[h]
\begin{center}
\includegraphics[width=8cm]{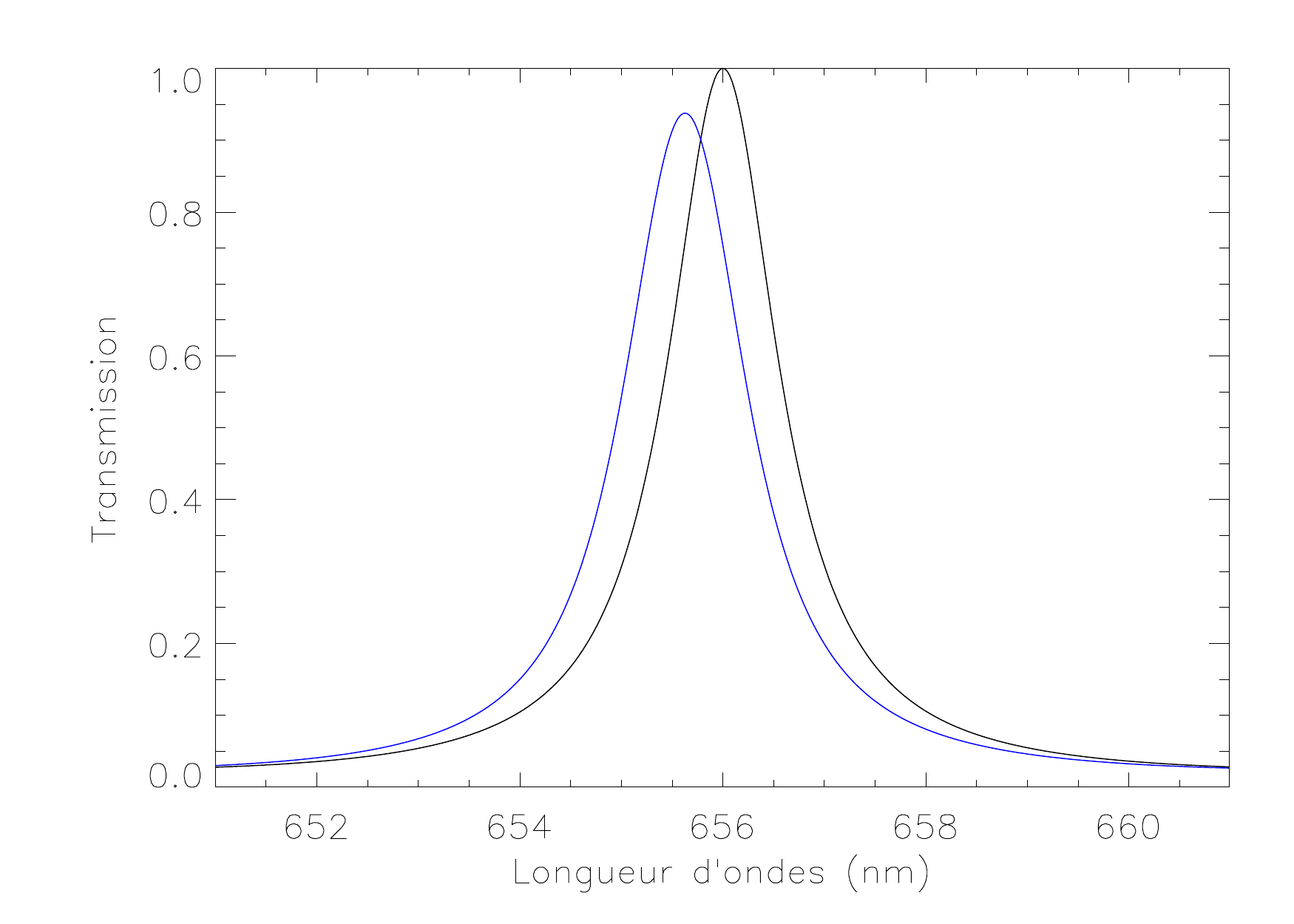}
\caption{Effet de l'utilisation d'un \FP~théorique au foyer du NTT ($F/11$ avec miroir primaire de diamètre $3.58~m$ et obstruction centrale de $1.16~m$), fonctionnant à l'ordre $50$ pour la raie \Ha~($656.3~nm$) et possédant un facteur de réflexion de $0.73$. Trait noir: réponse avec une ouverture nulle. Trait bleu: réponse au foyer Nasmyth du NTT.}
\label{effet_foyer}
\end{center}
\end{figure}
\begin{figure}[h]
\begin{center}
\includegraphics[width=10cm]{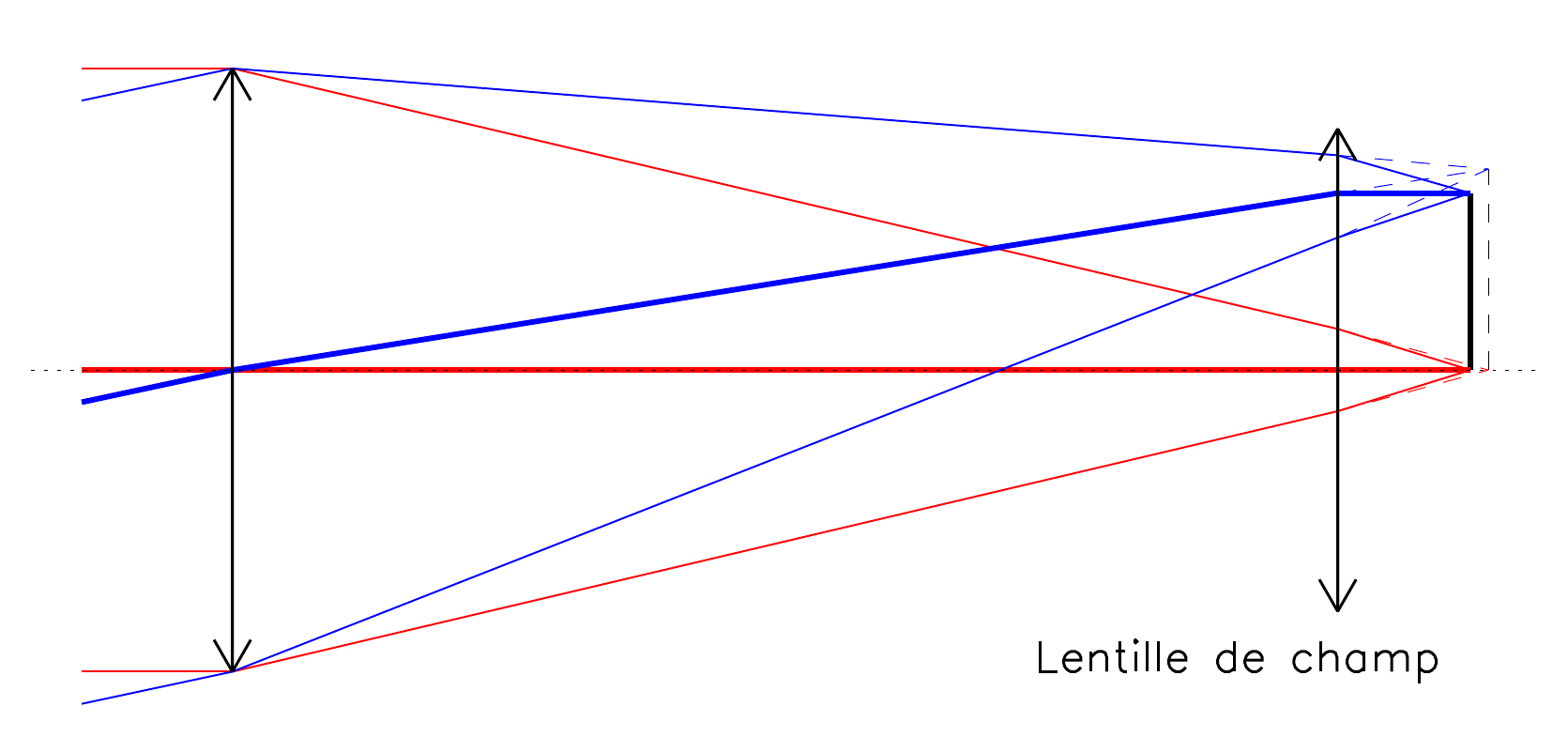}
\caption{La lentille de champ permet de rabattre les rayons afin d'avoir une incidence normale sur le Fabry-Perot, ou sur le filtre interférentiel, placé au plan focal quelle que soit la position dans le champ.}
\label{effet_lentille_champ}
\end{center}
\end{figure}

Lorsque le \FP~est placé en faisceau convergent, la longueur d'onde transmise pour chacun des rayons composant le faisceau varie selon leur angle d'incidence. Le pic de transmission est alors élargi en chaque point du champ et centré sur une longueur d'onde plus faible qu'à ouverture nulle car l'inclinaison moyenne des rayons est non nulle (Figure \ref{effet_foyer}). De plus, la longueur d'onde centrale peut changer selon la position dans le champ. En général, pour que la longueur d'onde ne varie pas, on utilise une lentille de champ afin d'avoir le même angle d'attaque sur le \FP~en tout point du champ (Figure \ref{effet_lentille_champ}). Dans cette configuration, afin de maximiser le champ de vue, le \FP~est placé dans le plan focal, d'où la dénomination d'utilisation au foyer. Les filtres interférentiels sont habituellement utilisés dans cette configuration.


Les observations présentées dans cette thèse utilisent l'interféromètre en pupille.

\subsection{Réalisation d'un sujet de travaux pratiques pour illustrer les possibilités du Fabry-Perot}

Les parties précédentes montrent que l'interféromètre de \FP~peut être utilisé de plusieurs manières.
J'ai repris les concepts des instruments CIGALE et PYHEAS afin de mettre en évidence différentes méthodes permettant de faire de la spectroscopie 3D à partir des propriétés de l'interféromètre de \FP~dans le cadre des travaux pratiques d'optique des Masters d'Astrophysique et d'Instrumentations Optique et Lasers de l'Université de Provence.

Dans l'instrument CIGALE (ou encore GHASP, FaNTOmM, GH$\alpha$Fas ou 3D-NTT dont il est question dans la partie \ref{ghasp_fantomm_ghafas}), le \FP~est placé en pupille, entre un collimateur et un objectif de caméra. Ainsi, pour chaque position, plusieurs longueurs d'onde sont transmises. Afin d'éviter la superposition de ces longueurs d'onde, un filtre interférentiel est placé au foyer du collimateur. L'échantillonnage spatial est réalisé par la caméra, et l'échantillonnage spatial est à la fois réalisé par les canaux de balayage du \FP~et par la caméra puisque la longueur d'onde dépend de la position.

Dans l'instrument PYTHEAS
\citep{Le-Coarer:1995}, la lumière traverse un \FP~placé dans un faisceau convergent très ouvert avant de former une image sur une trame de micro-lentilles. Cette trame réalise une inversion champ-pupille: l'échantillonnage spatial et réalisé par la trame qui crée des micro-pupilles en sortie. Celles-ci sont imagées sur la caméra par un système afocal constitué d'un collimateur et d'un objectif. Leur lumière est alors dispersée dans le faisceau collimaté par un réseau, ce qui permet de séparer les différents ordres du \FP, et ainsi d'utiliser un spectre plus étendu que CIGALE. PYTHEAS est d'une certaine manière l'instrument idéal pour faire de la spectro-imagerie puisqu'il permet de totalement découpler l'information spatiale de l'information spectrale.

Un intérêt pédagogique de PYTHEAS est qu'il emploie plusieurs éléments permettant habituellement de faire de la spectroscopie indépendamment (un \FP~et un réseau). Par ailleurs, l'utilisation d'une trame de micro-lentilles illustre le fait que l'imagerie et l'échantillonnage spatial ne sont pas forcément réalisés par la caméra.
De même l'échantillonnage spectral n'est pas non plus réalisé par la caméra, puisque dans le cas de PYTHEAS, ce sont les différents ordres du \FP~et ses pas de balayage qui réalisent l'échantillonnage spatial.

La comparaison entre PYTHEAS et CIGALE , quant à elle, permet de mettre en évidence qu'un instrument doit être utilisé et pensé en fonctions des besoins scientifiques:
\begin{itemize}
\item CIGALE possède un grand champ de vue pour un intervalle spectral réduit (un seul ordre du \FP) et est ainsi adapté à l'étude d'objets très étendus pour lesquels une seule raie est nécessaire, ce qui est le cas de l'étude de la cinématique de galaxies proches.
\item PYTHEAS possède un champ de vue réduit mais un intervalle spectral étendu correspondant à quelques centaines d'ordres du \FP. Il est ainsi adapté à l'étude d'objets faiblement étendus pour lesquels plusieurs raies sont utiles tels que les noyaux de galaxies ou les amas globulaires.
\end{itemize}
Cette comparaison permet également de souligner le compromis qui est fait en pratique: il peut être préférable d'utiliser un instrument comme CIGALE permettant d'utiliser pleinement la capacité du récepteur plutôt qu'un instrument idéal comme PYTHEAS qui contiendra moins d'information.

\subsection{Calibration en longueur d'onde}
\label{calibration_lambda}

En utilisation en pupille, nous avons montré que la longueur d'onde varie selon la position sur le champ de vue. Il est donc nécessaire de calibrer les observations en longueur d'onde afin de connaître avec précision la longueur d'onde transmise en chaque position du champ et pour chaque espacement des lames.
\par
En toute rigueur, si on veut une calibration précise, il est nécessaire de calibrer l'instrument à la longueur d'onde d'utilisation car les revêtements de surface des lames du \FP~peuvent induire un changement de phase à la réflexion dépendant de la longueur d'onde. En d'autre terme, l'épaisseur de la cavité vue par deux ondes planes monochromatiques de longueurs d'onde différentes n'est pas exactement la même. Cela va créer un décalage entre les longueurs d'onde théorique et réellement transmise.
Cet effet peut être mesuré et calibré \citep{Garrido:2005}.
Toutefois, l'effet est généralement suffisamment faible (de l'ordre de quelques pour cent) pour qu'une calibration à une longueur d'onde $\lambda_c$ voisine de la longueur d'onde de l'objet observé $\lambda$ soit correcte. De plus, les études cinématiques s'intéressent aux vitesses relatives plutôt qu'aux vitesses absolues.
\par
Pour faire la calibration en longueur d'onde, on utilise une source monochromatique, habituellement une lampe spectrale (au Néon) dont on isole une raie ($\lambda_c=6592$ \AA) en utilisant un filtre interférentiel à bande très étroite ($\sim 1.5~nm$) centré sur cette longueur d'onde (Figure \ref{filtre_interferentiel}). On réalise l'acquisition d'un cube de calibration en faisant balayer le \FP, généralement juste avant et juste après l'observation, pour être dans des conditions de flexion et de température similaires à l'observation. Deux dimensions sont spatiales, la troisième est spectrale. Chaque canal spectral correspond donc à un espacement donné des lames et, spatialement, on observe des anneaux (Figure \ref{anneaux_interference}).

\subsubsection{La phase parabolique}
\label{phase_parabolique}

\begin{figure}[h]
\begin{center}
\begin{tabular}{c c}
\includegraphics[width=8cm]{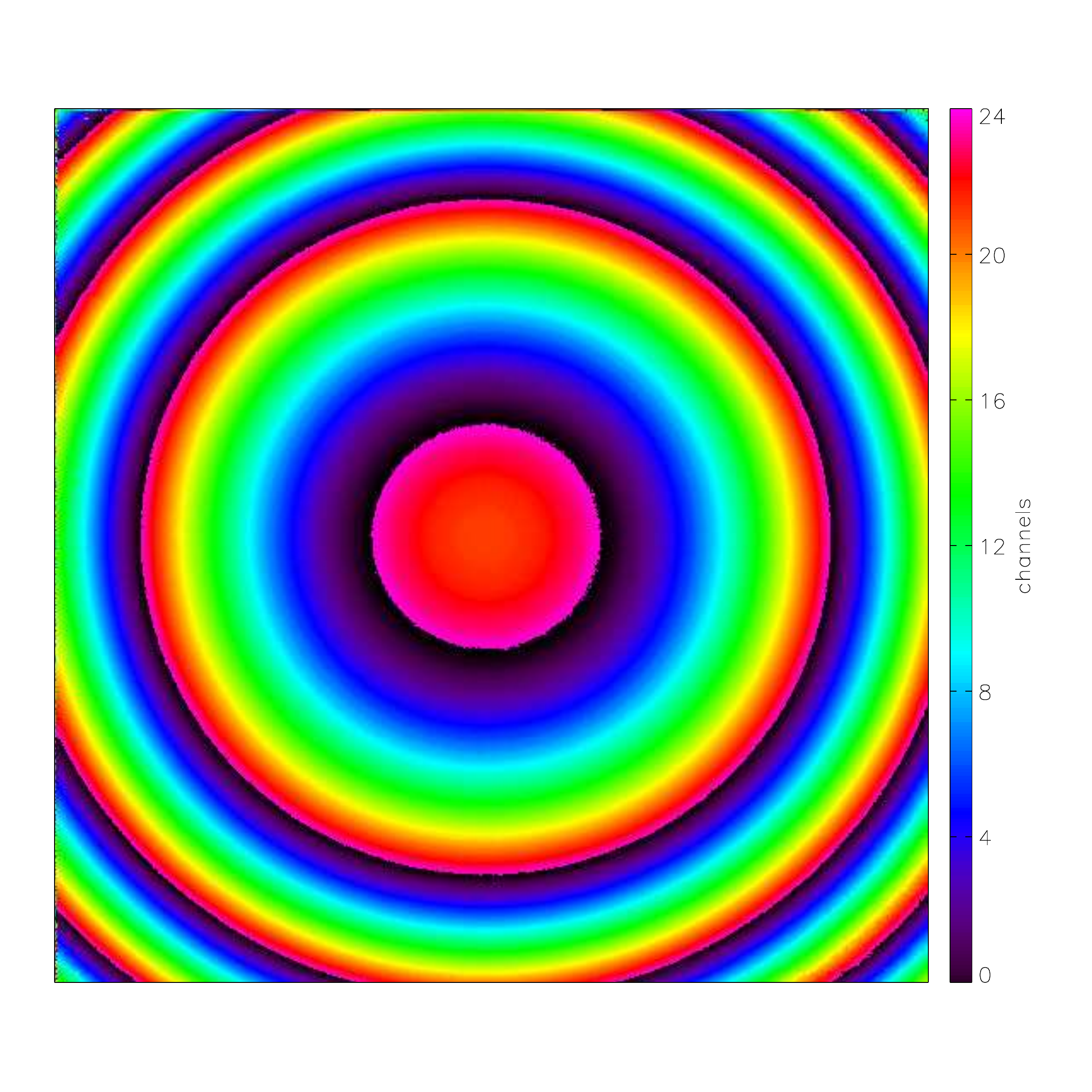}
\includegraphics[width=8cm]{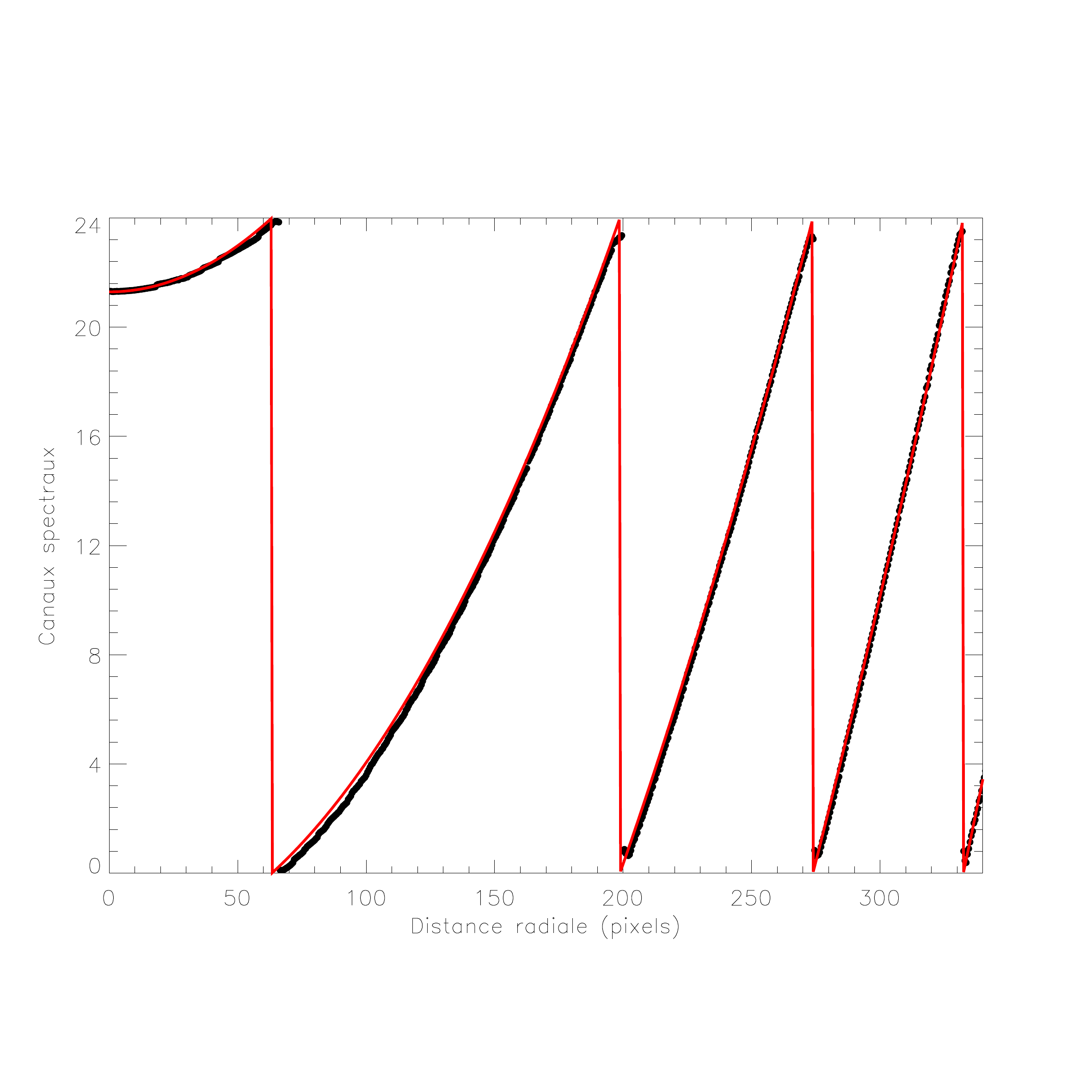}
\end{tabular}
\caption{Carte de phase brute (gauche) et variation de la phase brute avec le décalage en pixels par rapport à l'axe optique (droite). La phase est exprimée en canaux spectraux. La courbe en rouge correspond à l'ajustement d'une fonction parabolique repliée.}
\label{brut_phase}
\end{center}
\end{figure}

\begin{figure}[h]
\begin{center}
\begin{tabular}{c c}
\includegraphics[width=8cm]{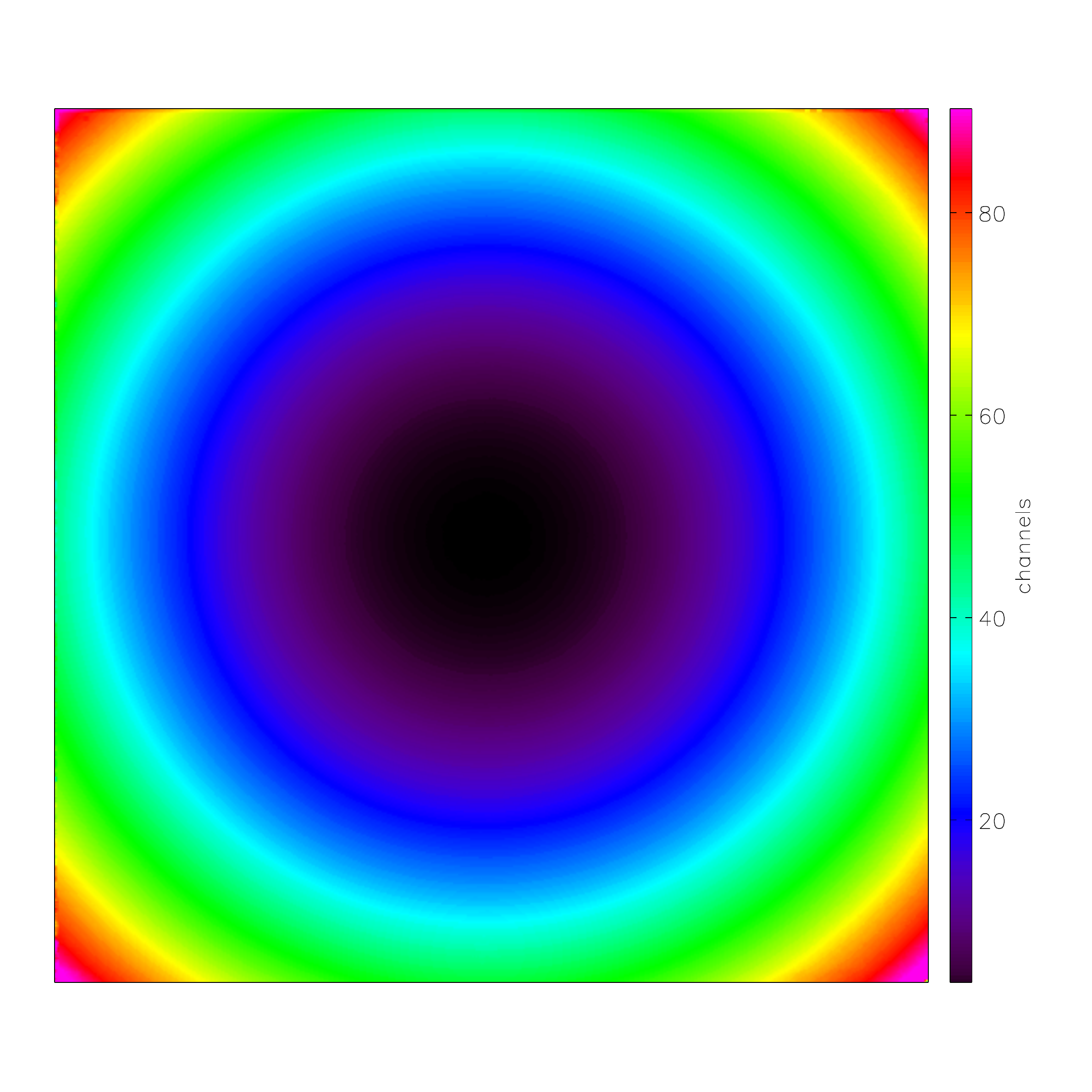}
\includegraphics[width=8cm]{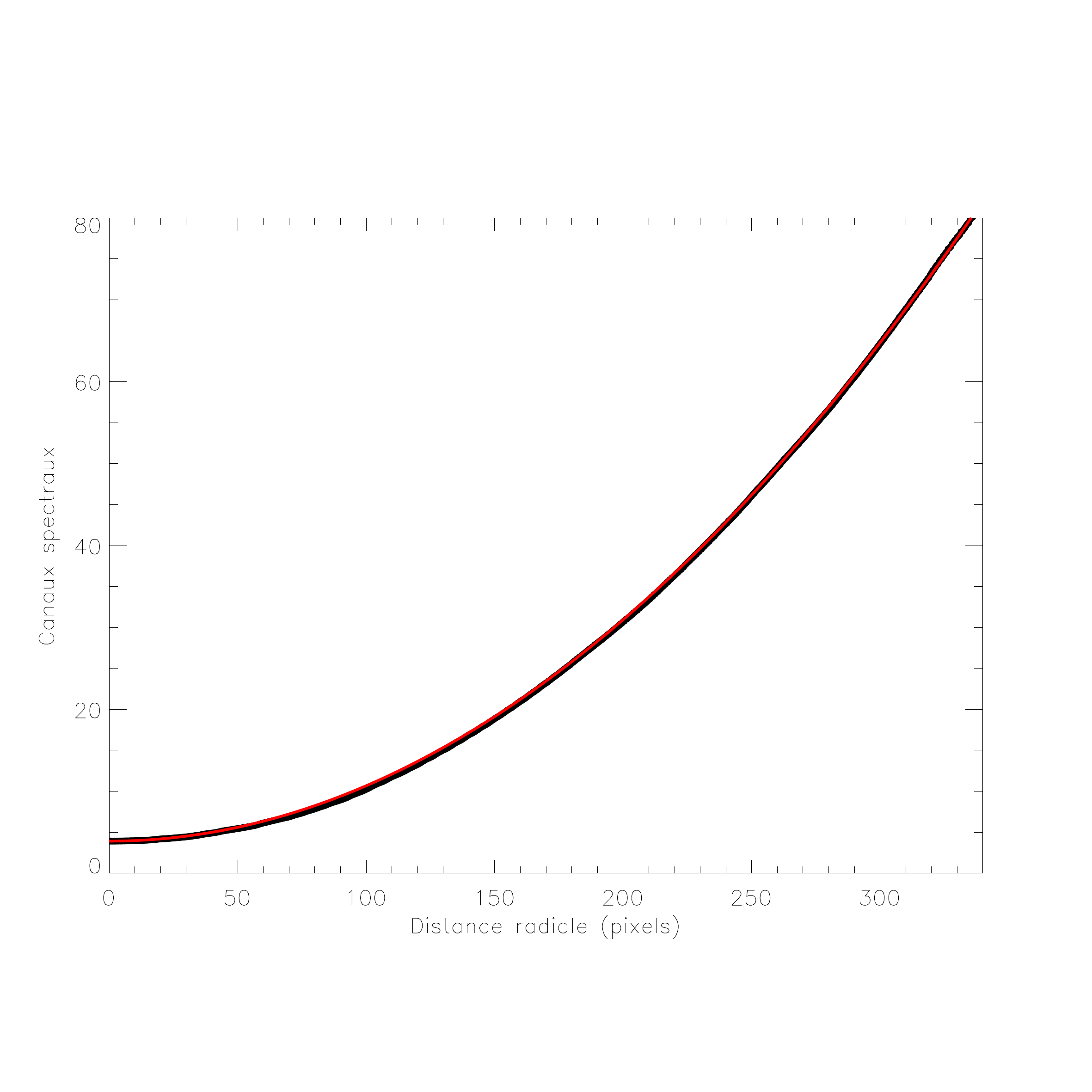}
\end{tabular}
\caption{Carte de phase après déroulement de la phase, dite carte de phase parabolique (gauche) et variation de la phase parabolique avec le décalage en pixels par rapport à l'axe optique (droite). La phase est exprimée en canaux spectraux. La courbe en rouge correspond à l'ajustement d'une fonction parabolique.}
\label{parabolic_phase}
\end{center}
\end{figure}

Pour un ordre d'interférence donné, la longueur d'onde $\lambda_c$ d'une source monochromatique est transmise pour des espacements $e$ différents selon la position dans le champ $\theta$ (d'après l'équation \ref{fp_fondamental}).
En notant $e_{\theta}$ l'épaisseur pour laquelle $\lambda_c$ est transmise à l'incidence $\theta$ par l'ordre $p$ on obtient
$$e_{\theta}-e_0=e_0\times\left(\frac{1}{\cos{\theta}}-1\right)$$
Le développement limité de cette expression donne
$$e_{\theta}-e_0=e_0\frac{\theta^2}{2}$$
D'où la dénomination de phase parabolique. La carte de phase parabolique est donc une carte où la valeur en chaque point correspond au canal spectral contenant une longueur d'onde de référence. On l'obtient en cherchant pour chaque position spatiale la position de la raie dans la dimension spectrale (Figure \ref{brut_phase}). Il est nécessaire de dérouler la phase pour obtenir la phase parabolique (Figure \ref{parabolic_phase}).

\subsubsection{Lien entre longueurs d'onde de calibration et d'observation}

Du fait que les longueurs d'onde de calibration $\lambda_c$ et d'observation $\lambda_s$ sont différentes, les ordres d'interférences utilisés sont différents pour la calibration ($p_c$) et l'observation ($p_s$). Il est donc nécessaire que la calibration soit transposée pour la longueur d'onde d'observation. Le raisonnement présenté dans ce paragraphe est celui que j'ai utilisé afin de programmer la calibration en longueur d'onde du logiciel de réduction de données présenté en Annexe \ref{help_computeeverything}.
\par
Considérons le canal qui transmet la longueur d'onde de référence $\lambda_r$ à l'ordre (entier) de référence $p_r$ au centre optique ($\lambda_r$ et $p_r$ sont fournis par le constructeur du \FP, pour une position moyenne des lames).
On a alors $$\lambda_r p_r =\lambda_c (p_c+\epsilon_c) =\lambda_s (p_s+\epsilon_s)$$ avec $p_c$ et $p_s$ entiers et $|\epsilon_c|$ et $|\epsilon_s|$ inférieurs à $0.5$ dans un premier temps (les longueurs d'ondes $\lambda_c$ et $\lambda_s$ ne sont pas nécessairement transmises pour ce canal au centre optique).
A priori, nous voulons qu'après calibration le canal numéroté $n/2$ (si les canaux sont numérotés de $0$ à $n-1$) contienne la longueur d'onde $\lambda_s$.
La calibration nous permet de savoir pour quel canal $\phi$ la longueur d'onde $\lambda_c$ est transmise. Nous pouvons donc déterminer quelle longueur d'onde se trouve au canal $\phi$ pour l'observation : $\lambda_{s\phi} = \lambda_c p_c/p_s$. On peut alors déterminer le canal $\psi$ où se trouve $\lambda_s$ :
$$\psi=\phi+n\frac{\lambda_s - \lambda_{s\phi}}{ISL_s}$$
Pour mettre $\lambda_s$ au canal $n/2$, il faut donc appliquer un décalage $\Delta$ en canaux de
$$\Delta=\psi-n/2=\phi+n\left(\frac{\lambda_s - \lambda_{s\phi}}{ISL_s}-\frac{1}{2}\right)$$
On généralise sur le champ entier en identifiant la phase à $\phi$.



\subsection{Réflexions parasites}
\label{reflexions_parasites}

Du fait de l'utilisation en pupille de l'interféromètre de \FP, de nombreuses réflexions sont générées. Certaines d'entre elles peuvent être gênantes. Il est donc nécessaire de comprendre leur origine pour pouvoir corriger numériquement ces réflexions parasites.
\par
\begin{figure}[h]
\begin{center}
\includegraphics[width=8cm]{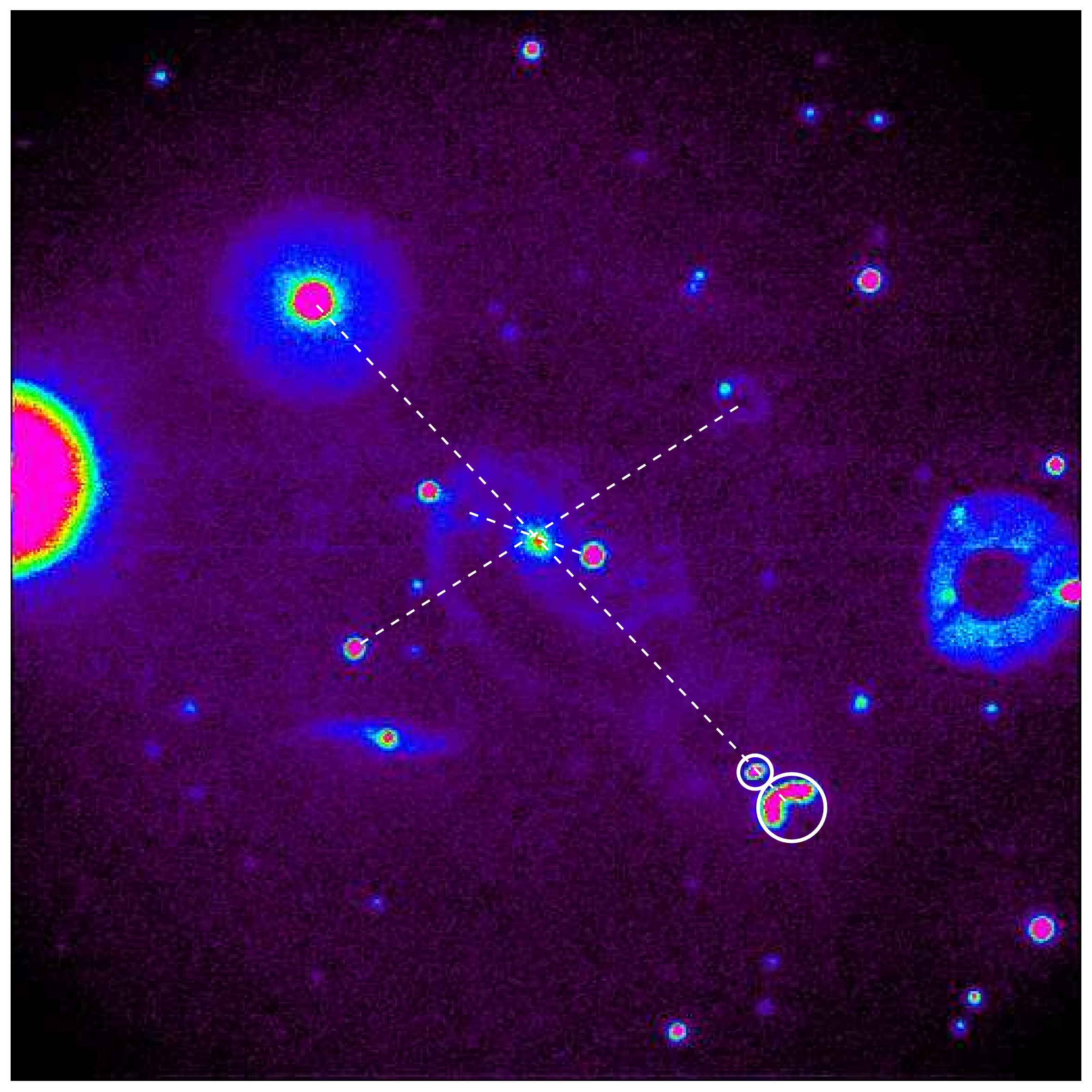}
\caption{Image (champ de UGC 1810) contenant plusieurs réflexions parasites. Les deux réflexions entourées correspondent aux deux types de réflexions corrigeables: focalisée et défocalisée. La réflexion défocalisée est vignettée. Les traits pointillés relient les étoiles et leur réflexion. On observe également sur cette image une réflexion de troisième ordre d'une étoile hors du champ. On remarque sur cet exemple un halo autour des étoiles très brillantes (voir partie \ref{ipcs}).}
\label{ghost_exemple}
\end{center}
\end{figure}
\begin{figure}[h]
\begin{center}
\includegraphics[width=12cm]{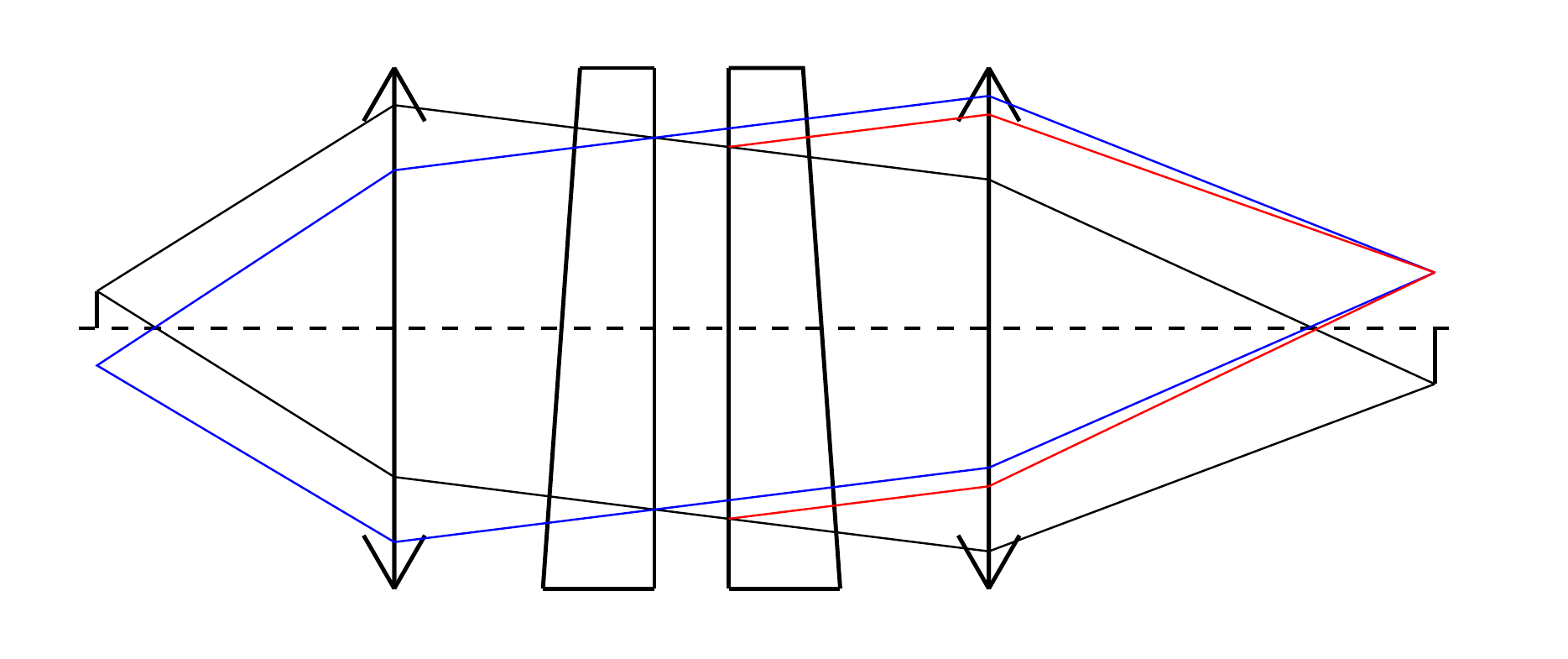}
\caption{Schéma de l'ensemble filtre interférentiel, collimateur, \FP, objectif, récepteur. Les rayons bleus correspondent à la réflexion générée par la première lame et le filtre interférentiel. Les rayons rouges correspondent à la réflexion générée par la seconde lame et la fenêtre du détecteur.}
\label{schema_ghosts}
\end{center}
\end{figure}

Lorsqu'elle entre dans la cavité, la lumière est réfléchie sur les deux interfaces du \FP. Pour éviter les reflets, les lames de \FP~sont trapézoïdales. Ainsi, les faces externes ne peuvent pas induire de réflexion parasite. En revanche, les faces internes ne peuvent pas être inclinées puisque ce sont ces faces qui forment la cavité et vont induire les réflexions parasites, appelées \textit{ghosts} en anglais. Le problème des réflexions parasites dans les étalons de \FP~a été étudié par \citet{Georgelin:1970}.
En utilisation en pupille, les reflets qui nous gênent principalement sont ceux qui ont lieu sur des surfaces placées dans les plans focaux: les filtres interférentiels ainsi que la fenêtre du détecteur (voir Figure \ref{schema_ghosts}). Ces deux reflets sont visibles dans nos données et différenciables par le fait que la défocalisation de ces deux reflets n'est pas nécessairement identique (Figure \ref{ghost_exemple}). En particulier, la position du filtre interférentiel n'a pas besoin d'être strictement au foyer, même si c'est la position qui induit le moins de vignettage, alors que la qualité de l'image nécessite de placer la caméra exactement au foyer. Il faut tout de même noter que si le foyer du collimateur ne coïncide pas exactement avec le foyer du télescope, ce défaut de réglage sera corrigé en ajustant la position de la caméra. Dans ce cas, les deux reflets sont défocalisés, mais en général la défocalisation du reflet ayant lieu sur la caméra est négligeable.
Dans les deux cas, le spectre de la réflexion est le même que celui de l'objet réfléchi une fois le cube calibré en longueur d'onde. En effet, dans le cas d'une réflexion sur le filtre interférentiel, le reflet passe à nouveau à travers le \FP~qui le filtre en longueur d'onde; dans le cas d'une réflexion sur la fenêtre du détecteur, la longueur d'onde a été filtrée au préalable et la réflexion est exactement symétrique.



\subsubsection{Méthode pour soustraire les réflexions parasites}

J'ai mis au point une méthode afin de soustraire le mieux possible les reflets parasites dans le cube de données (voir l'Annexe \ref{help_computeeverything} pour l'utilisation pratique de cette méthode). Pour faire ce travail rigoureusement, il faudrait connaître les réglages exacts de l'instrument lors de l'observation tels que la position et l'orientation du filtre interférentiel, de la caméra et du \FP. Il faudrait également disposer d'au moins une calibration de ces réflexions parasites afin de connaître les coefficients de réflexion du filtre et de la caméra.
Cependant, dans la pratique, aucune calibration autre que celle en longueur d'onde n'est faite. En revanche, dans certaines observations, les étoiles brillantes du champ peuvent servir de référence pour calibrer la soustraction des réflexions parasites.
Grâce à la calibration en longueur d'onde, l'axe optique du \FP, qui devrait être le centre de réflexion théorique, est déterminé avec précision.

La défocalisation est modélisée par un anneau dont le rayon varie. L'anneau correspond à la pupille du télescope: le miroir primaire et l'obstruction du miroir secondaire (donc le rapport des rayons est connu \textit{a priori}). Le flux réfléchi est réparti sur la surface de l'image de la pupille. On modifie donc les paramètres de réflexion, d'homothétie voire de décalage jusqu'à être satisfait de la correction sur l'étoile de calibration. L'idéal est d'avoir plusieurs étoiles afin d'ajuster au mieux la calibration, car il y a trois paramètres à déterminer pour la réflexion focalisée et quatre pour la réflexion défocalisée:
\begin{itemize}
\item le facteur de réflexion,
\item l'orientation de la surface réfléchissante par rapport à l'axe optique (deux paramètres),
\item la défocalisation de la réflexion défocalisée.
\end{itemize}
et seulement trois observables:
\begin{itemize}
\item le flux réfléchi (réflexion);
\item la taille de la pupille (défocalisation);
\item la position de la réflexion.
\end{itemize}

La présence de plusieurs objets induisant des réflexions parasites permet de lever la dégénérescence et aussi d'avoir une meilleure précision. Il est toutefois à noter que pour des observations utilisant une caméra à comptage de photons, la réponse n'est pas linéaire à forts flux. Ainsi, une calibration des reflets utilisant des étoiles ou des régions HII trop brillantes peut être biaisée puisque le flux de la réflexion devrait être proportionnel au flux réel de l'objet et non proportionnel au flux mesuré (voir partie \ref{ipcs}).

\subsection{Quel type de caméra pour des observations \FP?}
\label{ipcs}

Les observations \FP~sont principalement utilisées pour des applications cinématiques sur des objets possédant des raies d'émission (nébuleuses, galaxies, ...). Les temps de pose nécessaires pour obtenir un rapport signal sur bruit suffisant afin de déterminer des cartes de vitesses sont généralement supérieurs à une heure. Selon la position de l'objet sur le ciel ou bien selon qu'on observe en tout début ou en toute fin de nuit, les conditions d'observation peuvent changer durant l'observation: le fond de ciel peut fluctuer, la masse d'air peut évoluer rapidement pour des objets bas sur l'horizon induisant des raies du ciel nocturne d'intensité variable avec le temps. \'Etant donné que, pour un écartement donné des lames du \FP, la longueur d'onde transmise n'est pas partout la même dans le champ et puisqu'il est nécessaire de balayer le spectre (en faisant varier l'espacement des lames) pour obtenir un spectre complet sur tout le champ, les variations des conditions d'observations sont particulièrement gênantes.
Afin de moyenner les variations, on procède généralement à l'acquisition de plusieurs spectres complets durant l'observation, en partant du premier canal spectral au dernier puis en repartant du dernier pour aller jusqu'au premier et ainsi de suite.
\par
Les caméras CCD
possèdent un bruit de lecture non nul. Une telle séquence d'observation implique donc une perte en terme de rapport signal sur bruit. De plus, le temps de lecture peut durer quelques minutes, ce qui diminue le temps de pose réel sur le ciel. Ces caméras sont tout de même utilisées pour des observations à basse résolution spectrale car le flux reçu est plus important.
\par
Les caméras à comptage de photons (IPCS),
en revanche, possèdent un bruit de lecture nul, au détriment d'une efficacité légèrement plus faible \citep{Gach:2002}.
C'est habituellement ce type de caméra qui est utilisé avec les instruments de type \FP, avec l'acquisition d'une trentaine de cycles courts (environ dix secondes par canal spectral). Ces caméras sont composées d'une photocathode (tube intensificateur) soumise à une très haute tension (supérieure à $1000~V$) qui génère $10^6$ photo-électrons en moyenne pour un photon incident. Par phosphorescence sur un écran, ces électrons créent un faisceau important de photons. Ces photons sont détectés par un détecteur CCD à lecture rapide (de l'ordre de la dizaine de millisecondes) (voir la Figure 4 de l'article présenté en Annexe \ref{ghafas_pasp}, notée Figure \ref{fig4_ghafas} dans la table des figures). Ainsi, un photon incident est détecté si le flux généré mesuré est supérieur à un certain seuil, sa position est alors attribuée au barycentre de la tache générée. Cela induit que les rayons cosmiques, habituellement très énergétiques ne sont comptés que comme un unique événement, ce qui rend ce type de caméra insensible au rayonnement cosmique. En revanche, lorsque le flux est important, la réponse devient non linéaire car plusieurs photons peuvent arriver pendant le même pas de lecture, alors qu'un seul photon est compté. Un autre artefact peut survenir lorsque le flux est trop important. Il se forme alors un halo autour de la source (exemple de halo autour d'une étoile visible sur la Figure \ref{ghost_exemple}). Ce halo peut devenir gênant lorsqu'on utilise un \FP~à balayage en pupille. Cela est dû au fait que la longueur d'onde transmise change avec le champ et que le spectre du halo n'est pas filtré par le \FP. Imaginons qu'une région HII ait un flux induisant un halo, ce halo n'apparaîtra que dans un nombre restreint de canaux spectraux (ceux pour lesquels la transmission du \FP~sur la région HII correspond à la raie \Ha). Après calibration, ce halo sera donc vu comme une région HII étendue dont le spectre varie avec la position. Il est nécessaire de connaître l'existence de ce phénomène afin d'interpréter correctement le champ de vitesses autour de régions HII de flux très intense. Plus la caméra a une lecture rapide, moins ses artefacts sont importants.
\par
De nouvelles générations de caméras CCD à très faible bruit de lecture et à lecture rapide sont en cours de développement et seront donc utilisables avec des instruments de type \FP.

\section{Développement instrumental pour le 3D-NTT}
\label{3dntt}

L'utilisation d'instruments \FP\ sur des télescopes de classe $4~m$ de diamètre dans des sites dont la qualité du ciel est bonne n'est pas fréquente. Pourtant, ces instruments peuvent permettre de répondre à de nombreuses questions scientifiques. D'un côté, lorsque la résolution spectrale des \FP\ est grande, la cinématique des galaxies proches, des nébuleuses planétaires ainsi que des régions HII de la Galaxie peut être étudiée. D'un autre côté, lorsque la résolution spectrale des \FP\ est faible, l'étude de la physique des régions de formation d'étoiles peut être examinée et il est également envisageable de détecter des galaxies lointaines à fort taux de formation d'étoiles afin d'étudier les phénomènes d'accrétion du gaz et de ``feed-back'' radiatif.

Afin de permettre de telles études, le projet 3D-NTT (PI: M. Marcelin) a vu le jour, regroupant des scientifiques impliqués dans ces diverses thématiques. Ce projet qui proposait la construction d'un instrument visiteur pour le télescope de l'ESO NTT (La Silla, CHILI) a été accepté.
L'instrument 3D-NTT est actuellement en cours de développement et de construction au LAM
en collaboration avec le laboratoire du G\'EPI
(Observatoire de Paris) et le LAE
(Université de Montréal) et sera opérationnel au cours du second semestre de l'année 2009.
Le 3D-NTT sera installé au foyer Nasmyth ($F/11$) du NTT, télescope de $3.58~m$ de type Ritchey-Chrétien dont le miroir secondaire qui mesure $0.875~m$ crée une obstruction de $1.16~m$ de diamètre.
Les caractéristiques générales du 3D-NTT ainsi que ses objectifs scientifiques sont présentés dans l'article de l'Annexe \ref{3dntt_spie}.

\subsection{Le 3D-NTT, successeur de CIGALE, GHASP, FaNTOmM et GH$\alpha$FaS}
\label{ghasp_fantomm_ghafas}

Le projet 3D-NTT est une évolution en termes de technologies utilisées et de méthodes observationnelles qui s'inscrit dans la lignée de CIGALE (PI: J. Boulesteix), GHASP (PI: P. Amram),
FaNTOmM (PI: C. Carignan)
et GH$\alpha$FaS (PI: J. Beckman et C. Carignan),
précédents instruments développés par le LAM (Observatoire de Marseille) dont les plus récents en collaboration étroite avec le LAE de Montréal. L'IAC
a participé à la fabrication de GH$\alpha$FaS.
Comme ses prédécesseurs, le 3D-NTT consiste en un réducteur focal permettant de réduire le nombre d'ouverture du télescope afin d'adapter l'échantillonnage spatial en agrandissant le champ de vue et de diminuer la taille de la pupille pour y placer un \FP~à balayage. Son dessin optique (Figure 1 de l'article présenté en Annexe \ref{3dntt_spie}, notée Figure \ref{fig1_3dntt} dans la table des figures) a été réalisé par la société Immervision (QU\'EBEC).
Je ne discute ici que des instruments avec lesquels j'ai eu l'occasion de travailler, CIGALE ayant été utilisé au télescope de $3.6~m$ de l'ESO (CHILI) de 1990 à 2003.
\par
GHASP est le nom du programme initié en 1998 pour lequel cet instrument a été fabriqué et sur lequel la majeure partie de ma thèse a porté. Cet instrument a été utilisé au foyer Cassegrain du télescope de $1.93~m$ de diamètre de l’Observatoire de Haute Provence (FRANCE).
Deux caméras à comptage de photons ont été utilisées pour GHASP. La première, de $256\times256$ pixels couvre un champ de vue de $4'\times 4'$ avec un pixel de $0.96''$. Une nouvelle caméra à comptage de photons de $512\times512$ pixels, avec un pixel de $0.68''$ et un champ couvert de $5.8'\times 5.8'$ est utilisée depuis octobre 2000. Il s'agit d'un tube AsGa (photocathode) qui possède un meilleur rendement quantique que la précédente. Le refroidissement de la caméra est géré par un élément Pelletier dont la température est régulée par un circuit d'eau.
\par
Comme son nom l'indique, FaNTOmM est utilisé au sommet du Mont Mégantic (QU\'EBEC), au foyer du télescope de type Ritchey-Chrétien de $1.6~m$ de diamètre. Il est opérationnel depuis 2002. Il a également déjà été utilisé au CFHT
ainsi qu'au foyer du télescope de $3.6~m$ de l'ESO
à La Silla (CHILI). C'est avec cet instrument que j'ai appris à observer et à réduire des données \FP, dans le cadre d'un programme d'observations dont l'objectif est de réaliser une mosaïque de la galaxie d'Andromède (M31).
La différence principale de cet instrument avec GHASP est le système de refroidissement de la caméra. En effet, cet instrument utilise la cryogénie de Ranque-Hilsh qui consiste à diminuer la température grâce à un système de circulation d'air utilisant de l'air comprimé, un échangeur d'air et un tube de Ranque-Hilsh.
La caméra utilisée est une caméra à comptage de photons de $1024\times1024$ pixels de $0.8''$ sur le ciel, permettant d'obtenir un champ de $19.4'\times19.4'$.
\par
GH$\alpha$FaS est utilisé au foyer Nasmyth du WHT
($4.2~m$ de diamètre) à l'observatoire de la Roque de Los Muchachos sur l'île de La Palma (ESPAGNE) depuis l'été 2007 (voir l'article présenté en Annexe \ref{ghafas_pasp}). Ce foyer se trouve sur l'axe de rotation horizontal de la monture alt-azimuthale du télescope, dans une plate-forme qui tourne avec le télescope autour de l'axe de rotation vertical.
Afin de ne pas limiter le champ de vue à $2.5'\times2.5'$, GH$\alpha$FaS n'utilise pas de dérotateur d'image.
Ce type d'observation est rendu possible par l'utilisation d'une caméra à comptage de photons, car le temps d'intégration typique de $10$ secondes est suffisamment court pour pouvoir négliger la rotation du champ. Cette caméra possède $1024\times1024$ pixels couvrant un champ de $3.5'\times3.5'$ avec un pixel de $0.2''$. Elle utilise le même système de refroidissement que la caméra de GHASP. En revanche, le traitement des données issues de cet instrument doit inclure une routine de dérotation des cubes (Figure 7 de l'article de l'Annexe \ref{ghafas_pasp}, notée Figure \ref{fig7_ghafas} dans la table des figures).
L'avantage du foyer Nasmyth est que l'instrument est fixe sur table optique. Il n'y a donc pas de flexion et il est aisé de modifier des éléments du montage.
Suite à mes observations avec cet instrument, j'ai participé à la rédaction d'un manuel expliquant les étapes du montage de l'instrument qui est présenté en Annexe \ref{ghafas_setup}.
C'est l'utilisation de cet instrument sur table optique qui a motivé le choix d'utiliser un montage sur marbre plutôt qu'un montage en ligne pour le 3D-NTT afin de faciliter les interventions dans l'instrument, en particulier sur le positionnement des filtres et des \FP.
\par
Le 3D-NTT pourra utiliser une caméra CCD de $4096 \times 4096$ pixels et une caméra L3CCD de \hbox{$1600 \times 1600$} pixels pouvant fonctionner en mode comptage de photons qui permettront respectivement d'obtenir un champ de vue circulaire de $23'$ et de $12'$ pour un pixel respectif de $0.24''$ et $0.32''$. La quantité d'information sera donc bien supérieure aux précédents instruments.
\par
GHASP, FaNTOmM et GH$\alpha$FaS utilisent les mêmes \FP~de grande résolution spectrale fixe mesurant $50~mm$ de diamètre. Ces \FP~sont contrôlés par un appareil dénommé CS100 fabriqué par ICOS.
Des filtres interférentiels à bande étroite ($\sim1.5~nm$) sont nécessaires afin de sélectionner la raie d'intérêt. L'utilisation de ces filtres constitue une limitation pour couvrir tout le domaine spectral de par leur coût.
Le 3D-NTT se démarque par le fait qu'il propose deux modes d'utilisation détaillés dans l'article présenté en Annexe \ref{3dntt_spie}:
\begin{itemize}
\item un mode haute résolution, comme ses prédécesseurs, utilisant la caméra L3CCD en mode comptage de photons, dont l'objectif scientifique principal est l'étude de la cinématique des galaxies proches et des nébuleuses planétaires.
\item un mode basse résolution, dit filtre accordable, inspiré des instruments TTF
et MMTF,
utilisant l'une ou l'autre des caméras et dont les objectifs scientifiques principaux sont l'étude des régions de formation d'étoiles dans les galaxies proches (PI: M. Marcelin) ainsi que la détection de galaxies à fort taux de formation d'étoiles autour de quasars lointains afin d’en étudier les phénomènes d’accrétion de gaz et de ``feed-back'' radiatif (PI : J. Bland-Hawthorn).
\end{itemize}
Pour pouvoir répondre à ces objectifs, le 3D-NTT possède deux \FP~de nouvelle génération. Ces nouveaux \FP~permettent de couvrir un large domaine spectral ($350~nm$ à $800~nm$) grâce à des revêtements de surface optimisés et ont une résolution modulable ($100$ à $30000$ au total pour les deux \FP~du 3D-NTT) grâce à la grande amplitude de l'espacement entre les lames ($202.5~\mu m$) permis par des actuateurs piezo-électriques de nouvelle génération. Ces actuateurs sont asservis par un nouveau contrôleur en développement au LAM. Les deux \FP~possèdent une surface utile de $100~mm$. Le premier \FP~possède une finesse théorique de $50$ et un espacement des lames pouvant varier entre $140~\mu m$ et $340~\mu m$. Il fournira donc une haute résolution. Le second \FP~a une finesse théorique de $10$ pour un espacement minimum des lames de $2.5~\mu m$ contraint par l'espacement physique minimum et un espacement maximum de $202.5~\mu m$ fixé par la course des actuateurs piezo-électriques. Ce \FP~est également dénommé filtre accordable car il travaille à basse résolution. Le polissage des lames et le dessin de la mécanique des \FP~ont été réalisé par SESO
et les couches minces (revêtements de surface) sont développées et déposées à l'Institut Fresnel.
\par
Les deux \FP~pourront être utilisés simultanément dans le mode haute résolution: le \FP~de haute résolution est utilisé en pupille, mais les filtres interférentiels sont alors remplacés par le filtre accordable placé au foyer. Il est tout de même nécessaire d'utiliser des filtres dits de blocage afin de ne sélectionner que l'ordre utile du filtre accordable. Cependant, la largeur des filtres de blocage ($\sim15~nm$) est supérieure d'un ordre de grandeur à celle des filtres interférentiels ($\sim1.5~nm$). Ce dispositif présente l'avantage de permettre l'observation d'objets sur tout le spectre alors que jusqu'à présent, on était limité aux domaines de longueur d'onde pour lesquels on avait des filtres interférentiels. Le champ circulaire dans ce mode est limité à $9'$ de diamètre par la taille physique du \FP~placé au foyer, ce qui est adapté au champ de la caméra L3CCD.
\par
Le mode basse résolution qui emploie uniquement le filtre accordable et les filtres de blocage possède deux configurations possibles.
\begin{itemize}
\item La première est une utilisation classique similaire aux instruments TTF et MMTF, avec le \FP~placé en pupille. Il est alors nécessaire d'observer plusieurs canaux spectraux et d'effectuer une correction en longueur d'onde car la longueur d'onde varie avec la position sur le champ.
\item La seconde, plus inhabituelle, consiste à utiliser le \FP~au foyer $F/11$ du NTT. Dans ce cas, le champ est le même que pour le mode haute résolution et la longueur d'onde transmise est la même sur tout le champ sans qu'il ne soit utile d'utiliser de lentille de champ étant donné que le champ de vue est restreint. Ce mode d'observation est rendu possible car l'ouverture est faible au foyer utilisé. En revanche, il est limité à des ordres d'utilisation inférieurs à $80$ faute de quoi la transmission diminue de plus de $80$\% et la résolution est inférieure à la valeur théorique (voir discussion dans la partie \ref{foyer_pupille}).
\end{itemize}
Toutes ces caractéristiques sont récapitulées dans la Table \ref{instruments_fp}. On voit donc que le 3D-NTT présente un très bon compromis entre l'échantillonnage et le champ de vue. Il constitue véritablement un saut technologique par rapport à ses prédécesseurs de par sa résolution modulable et son domaine d'utilisation couvrant tout le visible.

\begin{table}[h]
\begin{center}
\begin{tabular}{| l | cc | c | c | cc |}
\hline
\multicolumn{1}{| r |}{Instrument} & \multicolumn{2}{c | }{GHASP} & FaNTOmM & GH$\alpha$FaS & \multicolumn{2}{c |}{3D-NTT} \\
\multicolumn{1}{| l |}{Caractéristique} & A & B & & & HR & BR \\
\hline
Mise en service & $1998$ & $2000$ & $2002$ & $2007$ & $2009$ & $2010$ \\
Télescope ($m$) & $1.96$ & $1.96$ & $1.6$ & $4.2$ & \multicolumn{2}{c |}{$3.6$} \\
Nombre d'ouverture & \multicolumn{2}{c |}{$15$} & $8$ & $11$ & \multicolumn{2}{c |}{$11$} \\
Nombre de pixels & $256^{2}$ & $512^{2}$ & $1024^{2}$ & $1024^{2}$ & $1600^{2}$ & $4096^{2}$ \\
Taille des pixels ($''$) & $0.96$ & $0.68$ & $0.8$ & $0.2$ & $0.32$ & $0.24$ \\
Champ circulaire ($'$) & $5.6$ & $8.2$ & $27.4$ & $4.9$ & $9$ - $12$ & $23$ \\
Diamètre FP ($mm$) & \multicolumn{2}{c |}{$50$} & $50$ & $50$ & $100$ & $100$ \\
Finesse des FP & \multicolumn{2}{c |}{$12$} & $12$ & $12$ & $10$ & $50$ \\
Domaine de $\lambda$ ($nm$) & \multicolumn{2}{c |}{$500$ - $750$} & $500$ - $750$ & $500$ - $750$ & \multicolumn{2}{c |}{$350$ - $800$} \\
Résolution spectrale & \multicolumn{2}{c |}{$\sim10000$} & $\sim10000$ & $\sim10000$ & \multicolumn{2}{c |}{$30000$ - $5000$ - $100$} \\
\hline
\multicolumn{7}{l}{{\small A: Caméra de première génération de GHASP}}\\
\multicolumn{7}{l}{{\small B: Caméra de seconde génération de GHASP}}\\
\multicolumn{7}{l}{{\small HR: Mode haute résolution du 3D-NTT}}\\
\multicolumn{7}{l}{{\small BR: Mode basse résolution du 3D-NTT}}\\
\end{tabular}
\caption{Comparatif des instruments utilisant un \FP~en pupille}
\label{instruments_fp}
\end{center}
\end{table}

Enfin, le logiciel d'acquisition des données ainsi que le logiciel de réduction des données seront basés sur les logiciels utilisés actuellement pour GHASP, FaNTOmM et GH$\alpha$FaS.

\subsection{Utilisation des deux \FP~en cascade}
\label{fpcascadeurs}

En mode haute résolution, le filtre accordable pourra être utilisé pour sélectionner l'ordre du \FP~de haut ordre (ou haute résolution spectrale) au lieu d'utiliser une série de filtres interférentiels. Cette utilisation simultanée de deux \FP~est une des originalités majeures du concept de l'instrument 3D-NTT.
Je présente ici une étude que j'ai réalisée montrant les critères qui doivent être respectés afin qu'une telle utilisation soit possible.


En mode haute résolution, on veut couvrir des ordres compris entre $200$ et $800$, afin d'atteindre des résolutions théoriques entre $10000$ et $40000$.


Il est alors nécessaire d'ajuster la résolution du \FP~de bas ordre afin de l'adapter à la dynamique du \FP~de haut ordre, typiquement son intervalle spectral libre.
L'indice $1$ réfère au \FP~de bas ordre, l'indice $2$ à celui de haut ordre, soit celui qui fait le balayage spectral.
Le \FP~de haut ordre va balayer le domaine de longueurs d'onde défini par le pic de transmission du \FP~de bas ordre. Si on veut une transmission supérieure à $50\%$ sur au moins un intervalle spectral libre du haut ordre, il faut alors satisfaire la relation suivante :
$$\delta \lambda_1 \ge ISL_2$$
Vues les relations \ref{isl} et \ref{resolution_def}, on peut écrire, après simplifications:
$$R_1 \le p_2$$
Cette relation doit être vérifiée pour toutes les longueurs d'onde, ainsi que pour tous les ordres $p_2$, en particulier pour l'ordre minimum d'utilisation du \FP~de haut ordre:
\begin{equation}
R_{1min} \le p_{2min}
\label{contrainte}
\end{equation}
L'ordre minimum du filtre accordable est directement relié à l'espacement minimum que l'on peut atteindre entre les deux lames (de l'ordre de quelques microns). Cet espacement va directement dépendre des revêtements de surface à cause de l'effet de phase qui induit un changement de la cavité vue par l'onde en fonction de la longueur d'onde.
Il est à noter que la finesse ne dépend en aucun cas de l'ordre d'utilisation. En revanche, selon les conditions d'utilisation (foyer ou pupille) et selon la qualité de planéité des lames, la finesse expérimentale ne peut qu'être plus faible que la finesse théorique, généralement d'un facteur deux.


Ce sont ces considérations, couplées aux objectifs scientifiques du mode filtre accordable, qui ont permis d'aboutir à la définition d'un \FP~de bas ordre avec une finesse de $10$. Cette finesse assure la possibilité d'atteindre une résolution minimale de $100$ sur tout le spectre.

\subsection{Effets de la non-uniformité des surfaces des \FP}

La non-uniformité de surface des lames induit des variations d'épaisseur de la cavité \FP. Ces variations d'épaisseur induisent donc un décalage de la longueur d'onde transmise selon la position sur la surface du \FP. J'ai participé à l'étude de la non-uniformité des surfaces afin de vérifier que les non-uniformités attendues sur les lames ne sont pas critiques pour le projet.

\subsubsection{Effet de la non-uniformité dans la pupille}
\begin{figure}[htbp]
\begin{center}
\includegraphics[width=8cm]{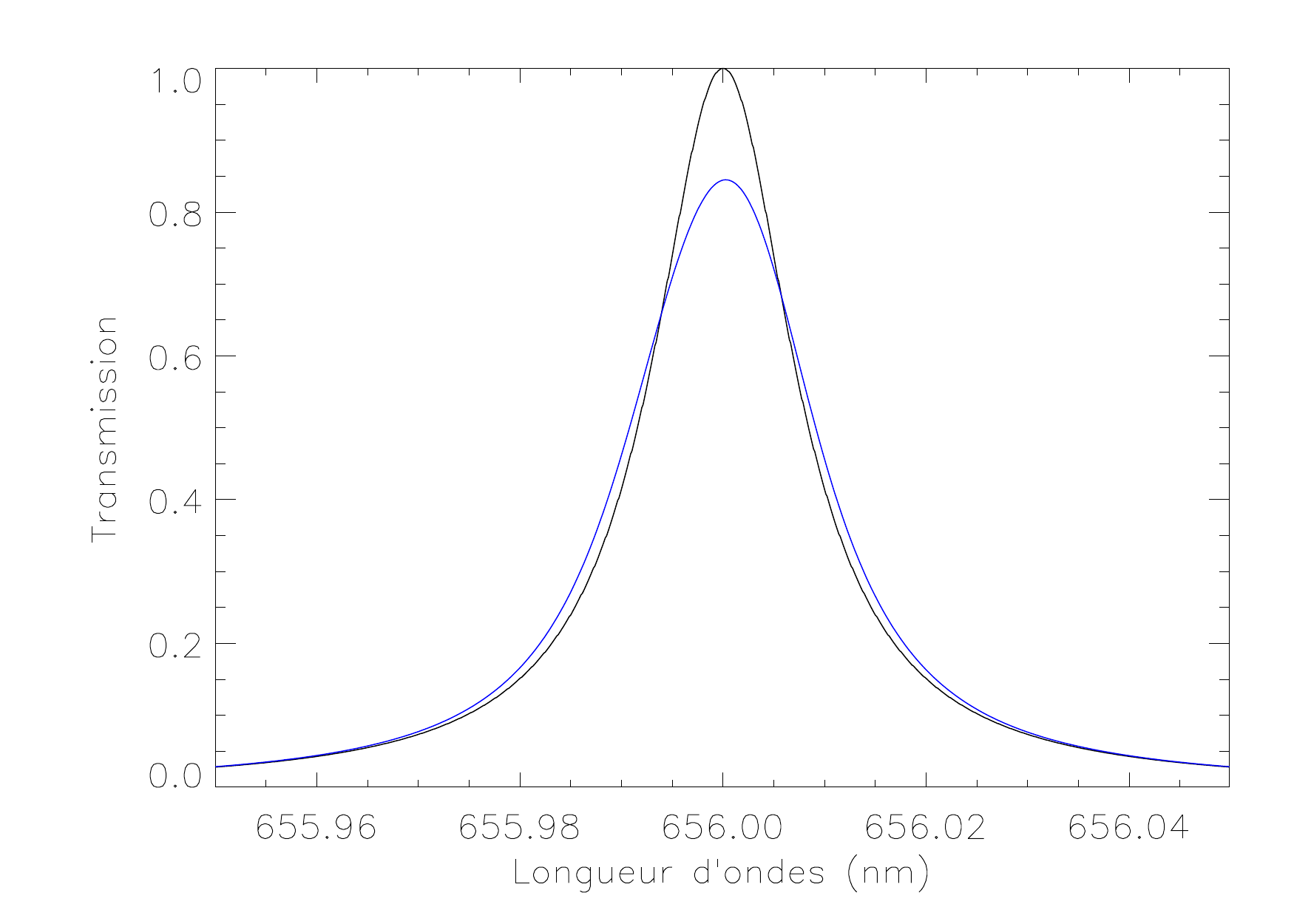}
\caption{Effet de la non-uniformité pour un \FP~théorique en pupille, fonctionnant à l'ordre $798$ pour la raie \Ha~($656.3~nm$) et possédant un facteur de réflexion de $0.94$. Trait noir: réponse sans défaut. Trait bleu: réponse avec les défauts mesurés sur les lames de la SESO.}
\label{effet_pupille}
\end{center}
\end{figure}

La totalité de l'intersection de la pupille et de la surface du \FP~est éclairée par chaque point du champ, par définition de la pupille. Les défauts de surface vont donc être moyennés et la transmission du \FP~sera constante sur tout le champ (même profil, même longueur d'onde transmise). L'effet va être uniquement un élargissement du profil théorique, soit une diminution de la finesse, et une diminution de la transmission (Figure \ref{effet_pupille}). L'ordre de grandeur de cet effet peut être relié à la valeur des défauts moyens.

\subsubsection{Effet de la non-uniformité au foyer}
Lorsqu'on place le \FP~au foyer, la largeur du profil ne dépend que de l'ouverture du faisceau (voir partie \ref{foyer_pupille}). J'ai travaillé sur ce point particulier afin de montrer que le filtre accordable devra être utilisé directement au foyer $F/11$ afin d'éviter la chute de la transmission et l'élargissement du pic.
En revanche, la longueur d'onde centrale transmise en chaque point du champ dépend de l'épaisseur locale de la cavité.
Le plus grand écart en longueur d'onde transmise dans le champ $d\lambda$ est celui qui va correspondre à la variation maximale d'épaisseur de cavité $\delta e$, soit l'erreur de polissage crête à crête:
$$d\lambda=2\delta e/p$$
On voit donc que l'effet sera d'autant plus important que l'ordre est petit. D'un autre côté, plus l'ordre est faible, plus la largeur théorique du pic de transmission est grande. On veut donc calculer la variation relative de longueur d'onde $d\lambda$ par rapport à la largeur à mi-hauteur du pic de transmission $\delta \lambda$.

\begin{figure}
\begin{center}
\includegraphics[width=8cm]{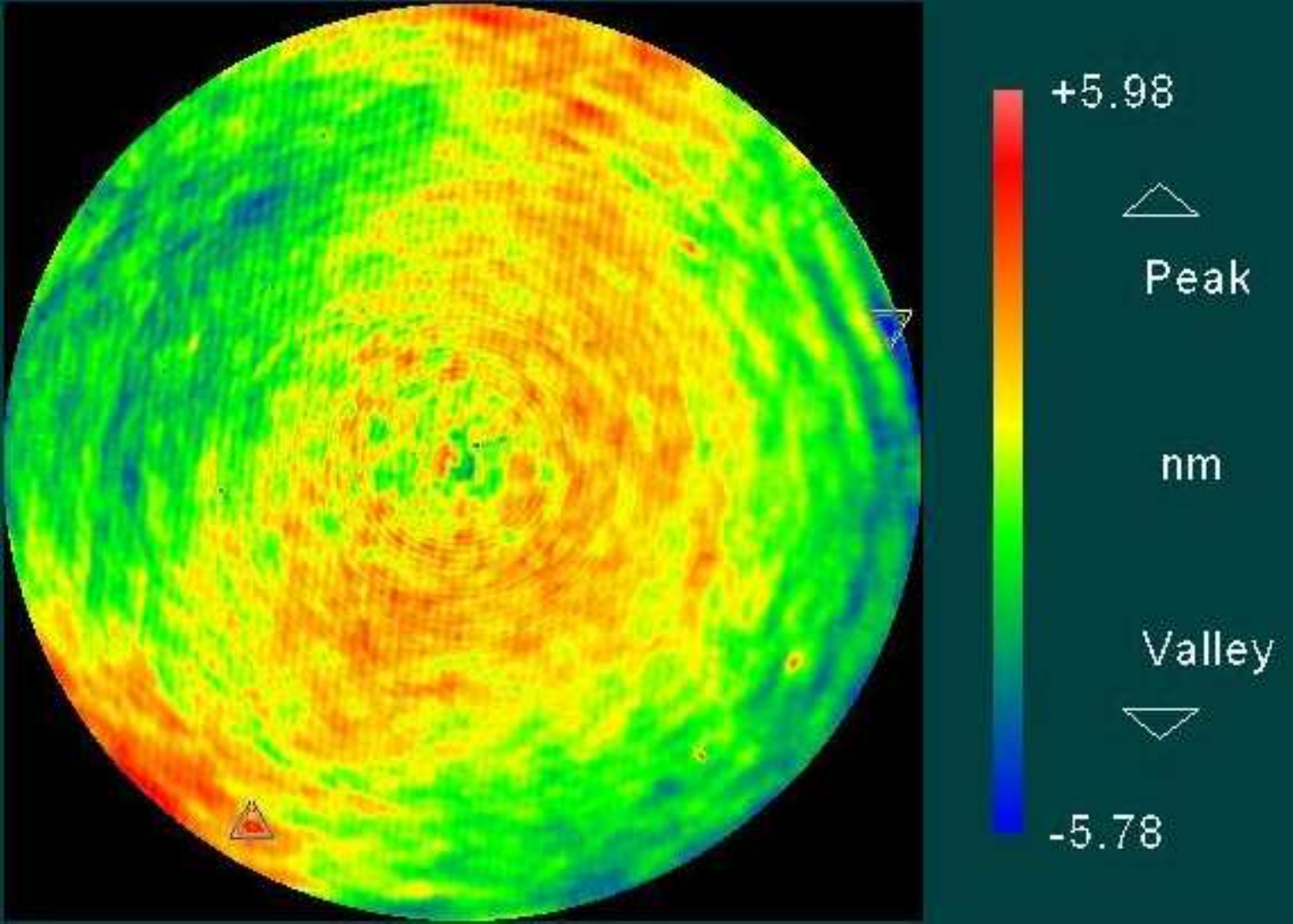}
\caption{Interférogramme d'une paire de lames, crédits SESO. L'échelle correspond aux variations d'épaisseur exprimées en nanomètres.}
\label{seso}
\end{center}
\end{figure}

D'après les équations \ref{isl} et \ref{finesse_def}, on déduit que $\delta \lambda=\lambda/(Fp)$.
La finesse étant fixe pour un \FP~donné, $\delta \lambda$ ne dépend que de l'ordre et de la longueur d'onde, on en déduit:
$$d\lambda/\delta \lambda=2F\delta e/\lambda$$
Cette expression ne dépend plus de l'ordre. Le décalage sera d'autant plus important que l'on travaillera à faibles longueurs d'onde. La finesse théorique du filtre accordable étant de $10$ et la variation d'épaisseur de la cavité crête à crête atteignant localement $11~nm$ selon l'interférogramme fourni par SESO (Figure \ref{seso}) on obtient pour le 3D-NTT:
\begin{itemize}
\item à $\lambda=350~nm$, l'erreur relative est proche de $0.5$;
\item à $\lambda=656~nm$, l'erreur relative est proche de $0.25$.
\end{itemize}
Si on tient compte de l'ouverture du faisceau ($F/11$), on doit utiliser la finesse réelle à la place de la finesse théorique. Toutefois, on montre que la finesse réelle n'est pas significativement dégradée pour une telle ouverture.
Lorsqu'on utilise le filtre accordable comme filtre pour le \FP~de haut ordre, la largeur à mi-hauteur du filtre accordable est du même ordre de grandeur que l'intervalle spectral libre du \FP~de haut ordre.
La configuration est donc acceptable mais pas optimale dans le bleu.

\begin{figure}
\begin{center}
\includegraphics[width=8cm]{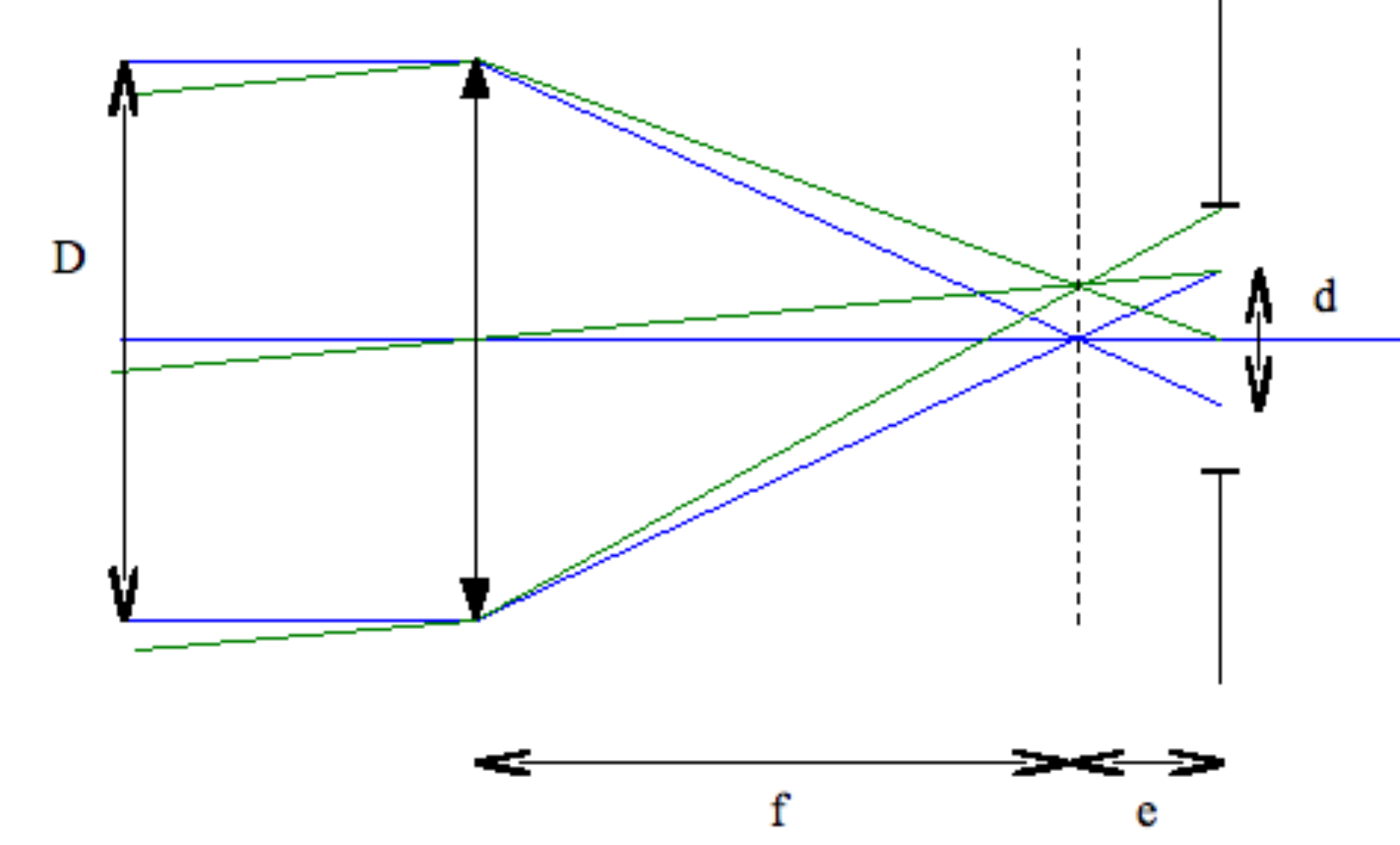}
\caption{Schéma du foyer du télescope et illustration du positionnement défocalisé de l'interféromètre: $f/D=e/d$.}
\label{defocus_fp}
\end{center}
\end{figure}

\'Etant donné qu'on ne peut pas diminuer la finesse du \FP~de bas ordre en deçà de $10$ (voir partie \ref{fpcascadeurs}), une solution pour palier à ce problème est de décaler légèrement le \FP~du foyer. Ainsi, les défauts de surface de hautes fréquences sont moyennés sur une surface égale à la défocalisation de l'image d'un point situé à l'infini.
La taille de la tache en fonction de la défocalisation est aisément calculable à partir des lois de l'optique géométrique (Figure \ref{defocus_fp}).
Cependant, la taille de la tache doit correspondre à une longueur de cohérence des défauts de surface des lames.

\begin{figure}
\begin{center}
\includegraphics[width=8cm]{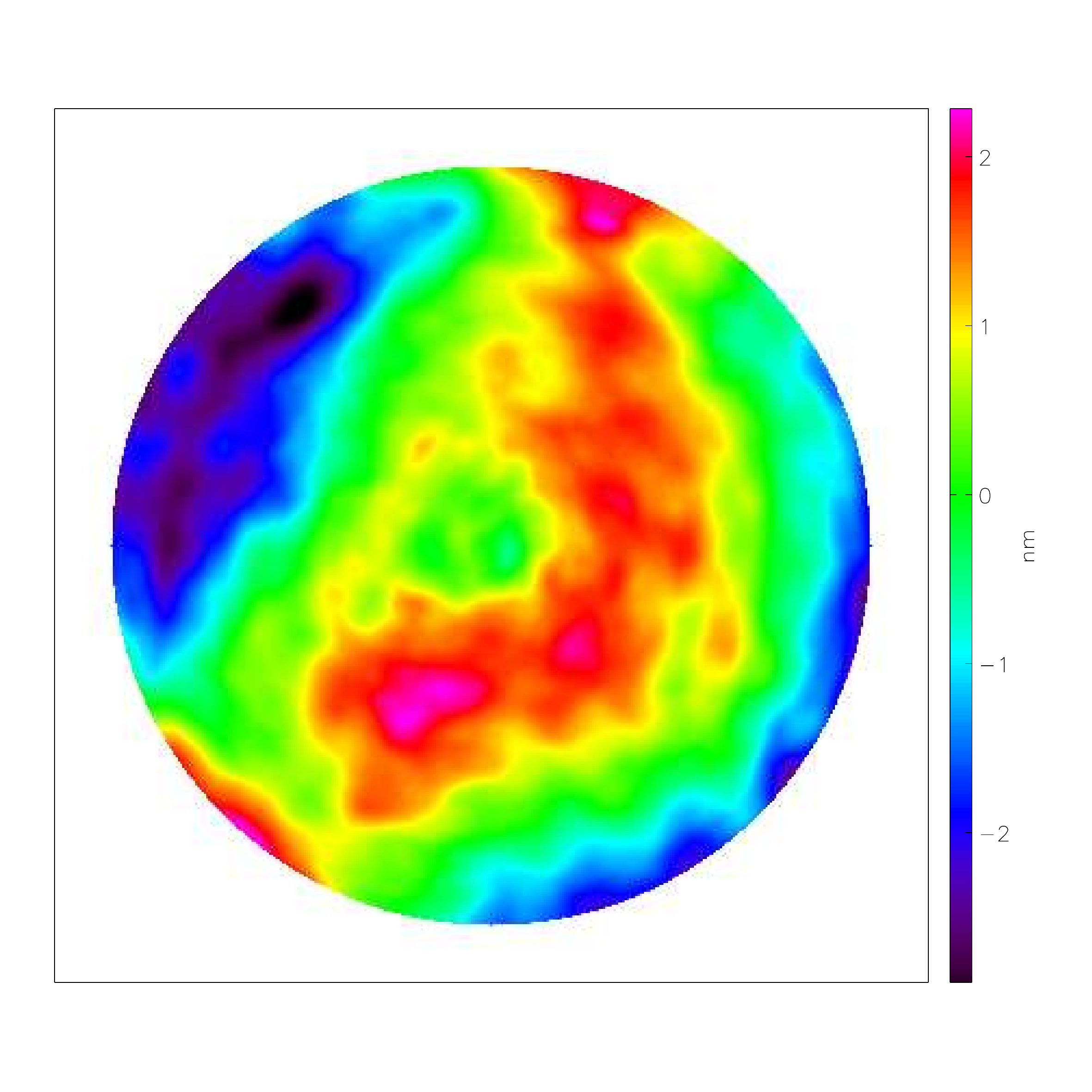}
\caption{Effet du positionnement du \FP~à $75~mm$ du foyer sur l'interférogramme. L'amplitude des défauts d'uniformité des surfaces des lames vus par le montage est atténuée: $min=-3~nm$, $max=+2.4~nm$.}
\label{seso_defocus}
\end{center}
\end{figure}

Pour diminuer l'effet par deux, l'interférogramme des lames montre qu'il est nécessaire d'avoir une tache de défocalisation d'au moins $7~mm$ (Figure \ref{seso_defocus}). Cela implique de décaler le \FP~d'environ $77~mm$ du plan focal. D'après les plans mécaniques (Figure 2 de l'article de l'Annexe \ref{3dntt_spie}, nommée Figure \ref{fig2_3dntt} dans la table des figures), le \FP~peut être décalé au maximum de $75~mm$ du foyer. Cet espacement est suffisant pour réduire l'effet de la non-uniformité des surfaces du \FP~d'un facteur proche de $2$. Cette solution présente l'avantage de libérer le plan focal, permettant par la même occasion d'y placer des masques pour la calibration.
De plus, les réflexions parasites éventuelles seront défocalisées et par conséquent moins intenses.
D'après le schéma optique, on déduit qu'un tel décalage de $75~mm$ du foyer va engendrer une perte de champ non vignetté par le \FP. Le rayon du champ non vignetté est supérieur à $92.5\%$ du rayon du champ qu'on aurait en plaçant le \FP~dans le plan focal.\\

En revanche, cette solution va induire une baisse de la transmission ainsi qu'une faible détérioration de la finesse qui ne devrait pas pénaliser le mode \FP~de bas ordre en pupille seul. Cette détérioration reste à quantifier.\\
\\
\underline{Remarque sur les revêtements de surface:} Il est probable que la non-uniformité des revêtements de surface soit plutôt à grande échelle, ce qui peut aggraver le problème présenté dans cette étude.

\subsection{Calibration en longueur d'onde}


Le système de calibration du 3D-NTT utilisera une source accordable (Photon etc) composée d'une lampe super-continuum couplée à une fibre optique afin de réaliser la sélection spectrale. Cette source permettra d'effectuer la calibration à la longueur d'onde d'observation et ainsi d'atténuer les effets de phase. En effet, il n'est pas possible de trouver des raies naturelles dans des sources spectrales à chaque longueur d'onde (environ une par $nm$). La résolution de cette source sera de $0.1~nm$.
Cependant, cette résolution sera trop grande pour pouvoir mesurer la réponse impulsionnelle de l'instrument. Il est donc prévu d'introduire dans cette source une lampe spectrale qui pourra être utilisée à la place de la source accordable ou bien en complément de celle-ci.
Le système optique adopté pour insérer cette source dans le trajet optique de l'instrument est un écran diffuseur (distribué par Schott). Cet écran sera placé à quelques centimètres du foyer du NTT. La source de calibration doit éclairer un faisceau de fibres optiques multimodes qui permet un éclairage uniforme au niveau de l'écran. Cette solution permet un encombrement minimum. Nous avons testé la faisabilité de cette configuration en laboratoire avec le réducteur focal de l'instrument GHASP, une caméra à comptage de photons, un écran diffuseur de surface quatre fois plus petite que celle de l'écran qui sera utilisé pour le 3D-NTT et avec une source accordable de résolution $0.3~nm$. Des anneaux d'interférence ont été obtenus.
J'ai vérifié que les flux seraient du même ordre de grandeur avec les composants qui seront utilisés pour le 3D-NTT en tenant compte de la surface de l'écran diffuseur, de l'ouverture de l'instrument, de l'efficacité de la caméra et de la taille du pixel.
\par
\'Etant donné que l'écran est positionné au foyer du NTT, il sera nécessaire de simuler la pupille du télescope afin de prendre en compte les effets de déplacement et de déformation du pic de transmission de la réponse instrumentale.
\par
La calibration en longueur d'onde du 3D-NTT en mode haute résolution utilisera la même procédure que celle décrite dans la partie \ref{calibration_lambda}.
La calibration en mode basse résolution, quant à elle, est différente car la résolution spectrale ne permet pas de mesurer avec précision des anneaux d'interférence. La méthode sera la même que celle des instruments TTF et MMTF.

\section{\'Etude du concept iBTF et d'une fonction de mérite dans le cadre du projet WFSpec}
\label{wfspec_ibtf}

Après les télescopes de classe $10~m$, la communauté scientifique se tourne vers des projets de télescopes de nouvelle génération de classe $40~m$. Ces télescopes auront tout particulièrement besoin des dernières technologies d'optique adaptative afin de pouvoir corriger sur un champ de l'ordre de $10'$ les effets de la turbulence atmosphérique en utilisant plusieurs étoiles lasers (LTAO,
MOAO,
...) afin d'atteindre des résolutions spatiales limitées par l'ouverture du télescope (tache d'Airy) dans le but d'observer des objets de petites tailles qui ne peuvent pas être résolus avec les télescopes actuels.
Les objectifs principaux de ces télescopes sont donc l'étude de populations stellaires résolues extragalactiques (dans d'autres galaxies que la Voie Lactée), la détection de planètes extra-solaires ainsi que l'étude de l'évolution de l'Univers, en particulier par l'étude à haute résolution de champs profonds.


Dans le cadre de cette dernière thématique, un groupe de travail s'est formé afin de réfléchir à diverses solutions pour réaliser la spectroscopie à champ intégral sur un grand champ pour le futur ELT européen: WFSpec (PI: J.-G. Cuby).
J'ai participé à ce groupe de travail pour étudier le concept iBTF, un des trois concepts instrumentaux qui ont émergé du projet et pour lesquels j'ai également défini et étudié la fonction de mérite afin de déterminer le plus performant qui fera l'objet d'une étude plus poussée. Les laboratoires ayant participé à ce projet sont le LAM (Marseille), le CRAL,
le G\'EPI (Paris) en collaboration avec les entreprises ON\'ERA et Photon etc.

\subsection{Les concepts de WFSpec}

Trois études ont été menées sur trois concepts différents avec des spécifications identiques: $48$ détecteurs de $4096\times 4096$ pixels, un échantillonnage spatial de $58~mas/pixel$ adapté à la résolution que l'optique adaptative souhaite atteindre sur ce type de télescopes, une résolution spectrale de $5000$ et une couverture spectrale dans l'infrarouge de $1.46~\mu m$ à $2.21~\mu m$. Le champ corrigé par l'optique adaptative est supposé atteindre $10' \times10'$. L'article présentant ces trois concepts est en Annexe \ref{wfspec_spie}. Toutefois, une description sommaire est présentée afin de mettre en évidence les caractéristiques utilisées pour la comparaison de leurs performances.

\subsubsection{Concept de champ monolithique: ``mono-IFU''}
Ce concept utilise un champ d'un seul bloc de $0.4'\times0.4'$ et une couverture spectrale continue.
Le champ est découpé en $48$ sous-champs par un diviseur de champ (``image splitter''). Chaque sous-champ est découpé et réordonné le long d'une fente virtuelle par un découpeur d'image (``image slicer''). Le spectre de chacune des fentes virtuelles est dispersé par un réseau sur un des $48$ récepteurs. Les Figures 1 et 2 (référencées Figures \ref{fig1_wfspec} et \ref{fig2_wfspec} dans la table des figures) de l'article présenté dans l'Annexe \ref{wfspec_spie} illustrent ce concept.

\subsubsection{Concept de champ subdivisé: ``multi-IFU''}
Ce concept peut utiliser jusqu'à $40$ IFU
indépendamment positionnés dans le champ corrigé de $10'\times10'$. Un IFU est composé d'une mosaïque de fibres permettant de couvrir un champ de vue modeste. Les fibres sont réordonnées le long d'une (ou plusieurs) fente et le spectre est dispersé par un réseau. Chaque IFU est imagé sur un (ou plusieurs) des $48$ détecteurs, pour un champ total de $0.4'\times0.4'$. Ce concept est illustré sur la Figure 3 (Figure \ref{fig2_wfspec} dans la table des figures) de l'article présenté dans l'Annexe \ref{wfspec_spie}. Dans la suite nous considérons que cet instrument dispose de $24$ IFU.


\subsubsection{Concept de filtre accordable: iBTF}

\begin{figure}[h]
\begin{center}
\begin{tabular}{cc}
\includegraphics[height=6.cm]{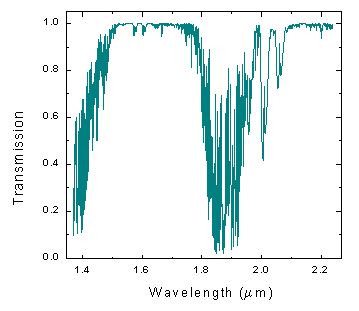} &
\includegraphics[height=6.cm]{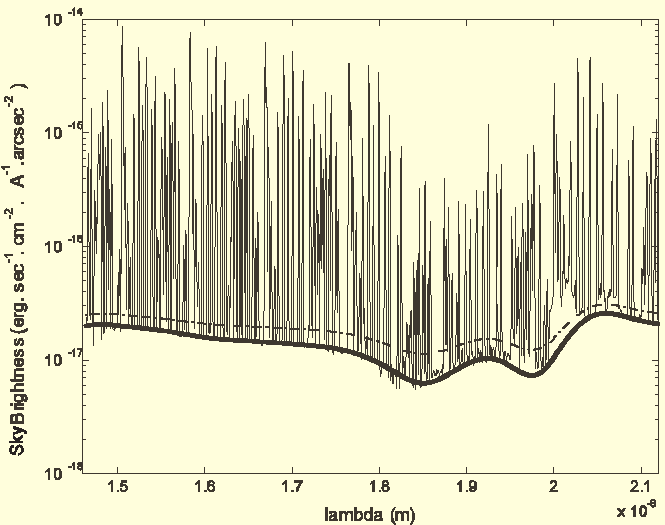}
\end{tabular}
\caption{\textbf{\`A gauche:} Transmission atmosphérique dans l'infrarouge proche ($1.4~\mu m$ à $2.2~\mu m$). La transmission est supérieure à $0.8$ sur des domaines de longueurs d'onde disjoints, couvrant au total moins de $0.5~\mu m$ sur les $0.8~\mu m$ définissant le domaine spectral des instruments. \textbf{\`A droite:} Spectre en émission du ciel nocturne dans l'infrarouge proche ($1.4~\mu m$ à $2.2~\mu m$). Des raies d'émission sont présentes sur tout le spectre.}
\label{sky_spectre}
\end{center}
\end{figure}

\begin{figure}[h]
\begin{center}
\includegraphics[width=13cm]{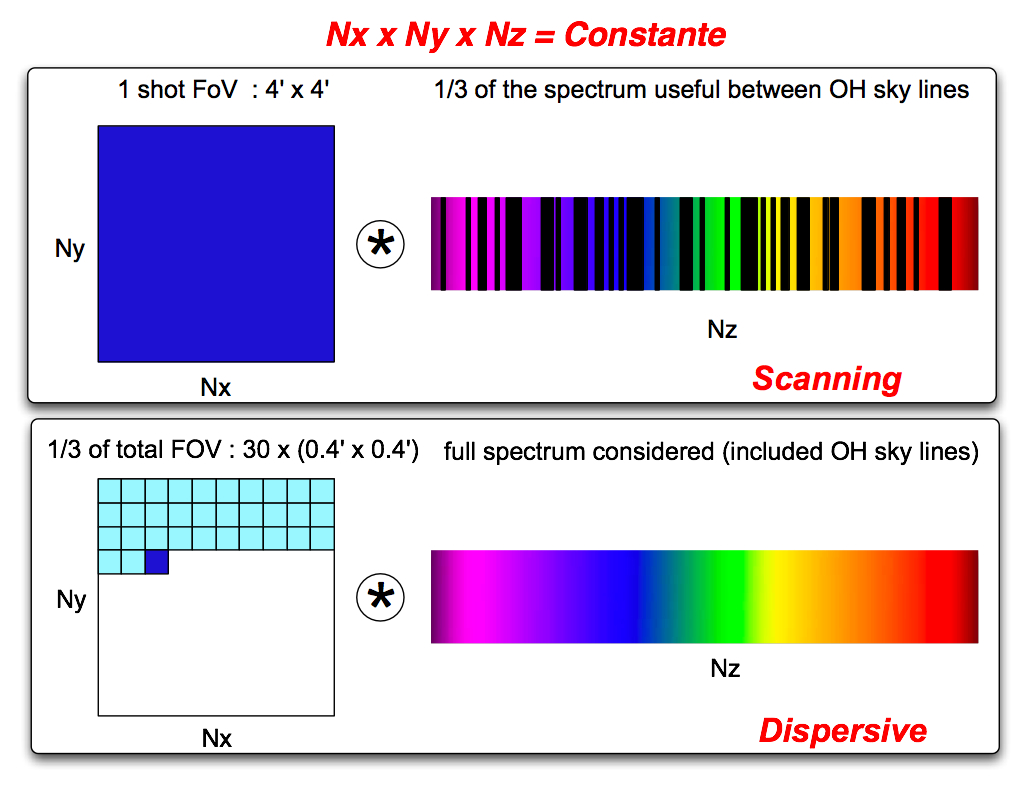}
\caption{Vue schématique comparant le fonctionnement des concepts dispersifs ``multi-IFU'' et ``mono-IFU'' (en bas) avec le concept à balayage iBTF (en haut). Pour un même temps de pose, tout le spectre est observé avec les systèmes dispersifs sur $30$ champs élémentaires (en bas) tandis qu'un seul grand champ est observé en effectuant $30$ pas de balayage pour obtenir le tiers utile du spectre (en haut). Au final, le nombre de pixels utilisés est le même: c'est le produit du nombre d'éléments d'échantillonnage spectral par le nombre d'éléments d'échantillonnage spatial.}
\label{information_spatiale_spectrale}
\end{center}
\end{figure}

Au lieu de découper le champ de vue comme dans les concepts précédents, nous avons réfléchi à un concept qui pourrait découper le spectre. En effet, dans le spectre infrarouge, certaines bandes sont très atténuées par l'absorption de l'atmosphère (voir Figure \ref{sky_spectre}, gauche). De plus, de nombreuses raies très intenses du ciel nocturne (voir Figure \ref{sky_spectre}, droite) font que toute étude spectrale est impossible. En découpant le spectre, les domaines non exploitables vont pouvoir être exclus afin de diminuer le nombre d'éléments spectraux sans perdre d'information utile. Ainsi, tout l'espace détecteur non utilisé par le spectre peut être utilisé pour observer un champ plus grand (voir Figure \ref{information_spatiale_spectrale}).
Mon travail sur ce projet a consisté à participer aux réflexions qui ont abouti aux caractéristiques techniques du concept iBTF décrites dans l'article \citet{Moretto:2006} (voir l'Annexe \ref{wfspec_spie}) et reprises ci-après. Je me suis plus particulièrement investi dans la définition des domaines spectraux utiles et dans les calculs de temps de pose et de rapports signal sur bruit.

Le concept iBTF,
comme son nom l'indique, utilise des filtres de Bragg. Ces filtres holographiques permettent d'extraire une composante spectrale d'un faisceau lumineux en la renvoyant dans une direction différente de celle du faisceau incident, idéalement perpendiculaire. La composante non déviée du faisceau (que l'on appelle l'ordre 0) contient tout le spectre hormis le domaine qui a été extrait. Ainsi en plaçant $48$ filtres de Bragg en cascade on va pouvoir extraire $48$ bandes spectrales du faisceau principal chacune imagée sur un des $48$ récepteurs. La Figure 4 (Figure \ref{fig4_wfspec} dans la table des figures) de l'article présenté dans l'Annexe \ref{wfspec_spie} illustre ce concept.
Cependant, chacune des bandes spectrales extraites de l'ordre 0 ne peut être suffisamment étroite pour atteindre la résolution souhaitée. En utilisant un \FP~à balayage en amont du faisceau (avant qu'il ne traverse la cascade de filtres de Bragg) et en le faisant balayer, on va ainsi atteindre la résolution souhaitée.
En pratique, ce concept ira explorer environ le tiers du domaine spectral total ($1400$ éléments de résolution sur $4000$ au total) en $30$ pas de balayage pour obtenir la résolution spectrale de $5000$ sur un champ total de $3.8'\times3.8'$.
En revanche, pour obtenir le même rapport signal sur bruit que les autres concepts, le temps de pose doit être $30$ fois plus élevé. Cependant, à temps de pose égal, la surface du champ est tout de même trois fois plus grande puisque l'étendue du spectre est réduite d'un facteur trois.
Le champ est tellement grand qu'il n'est pas nécessaire de réfléchir à une solution de type ``multi-IFU''.
\par
Ce concept purement théorique doit prouver sa faisabilité. En particulier, les premiers filtres de Bragg holographiques seront utilisés dans l'instrument BTFI (PI: C. Mendes de Oliveira)
pour le télescope SOAR,
un projet mené par l'USP/IAG (São Paulo) en collaboration avec l'INPE (São José dos Campos) et le LAM (Marseille).


\subsection{\'Etude comparative d'une fonction de mérite pour les trois concepts}

Afin de déterminer le concept le plus performant, j'ai été chargé de réaliser une étude des facteurs de mérite de ces instruments.
Traditionnellement, une fonction de mérite doit prendre en compte tous les raffinements, tels que la transmission des optiques, le rendement quantique des détecteurs, le coût de l'instrument, le poids, etc.
Pour le projet WFSpec, nous avons élaboré une fonction de mérite simple puisqu'il ne s'agit que de concepts:
$$M=\frac{FOV \times n}{m}$$
$FOV$ est le champ de l'instrument, $n$ est la densité d'objet dans le champ et $m$ est le nombre de poses élémentaires pour obtenir le spectre total. Cette fonction de mérite est simplement le nombre d'objets par exposition élémentaire.
Pour le concept iBTF, $m=30$ car il est nécessaire de faire $30$ pas d'échantillonnage spectral alors que pour les autres concepts, $m=1$ puisqu'une seule exposition permet d'observer l'ensemble du spectre.

Cette étude a nécessité une compréhension précise des spécifications et des contraintes liées aux objectifs scientifiques des instruments. En particulier, WFSpec cherche à répondre à deux objectifs majeurs de l'étude de l'évolution de l'Univers qui sont la détection de galaxies primordiales et l'étude de l'assemblage de masse des galaxies (voir Annexe \ref{wfspec_spie}).
L'interprétation de la fonction de mérite diffère selon le cas scientifique étudié.

\subsubsection{Détection de galaxies primordiales}

Ce type d'étude consiste à observer un champ dans une direction de l'espace donnée pendant une longue durée (pouvant aller jusqu'à plusieurs jours) de sorte que des sources faibles émergent du bruit.
L'obtention d'un spectre pour chaque élément de résolution spatiale permet d'améliorer la détection des sources par leurs raies d'émission. En effet, si une source possède des raies d'émission, alors le flux est bien plus intense aux longueurs d'onde où sont émises les raies qu'aux autres longueurs d'onde. Le rapport signal sur bruit est donc meilleur à cette longueur d'onde puisque le signal est celui de la raie et le bruit (poissonnien) est la racine carrée de l'ensemble du signal reçu à cette longueur d'onde, soit le signal de la raie ainsi que le flux provenant du fond de ciel et le flux continu de la source. Lorsqu'on utilise une imagerie classique, le flux reste celui de la raie (si on suppose que le continu de la source est faible) alors que le bruit est la racine carrée du flux détecté sur tout le spectre.
Outre la détection, un des objectifs est de mesurer le décalage spectral spectroscopique des sources détectées en mesurant la position de plusieurs raies.

\begin{figure}[h]
\begin{center}
\includegraphics[width=7.5cm]{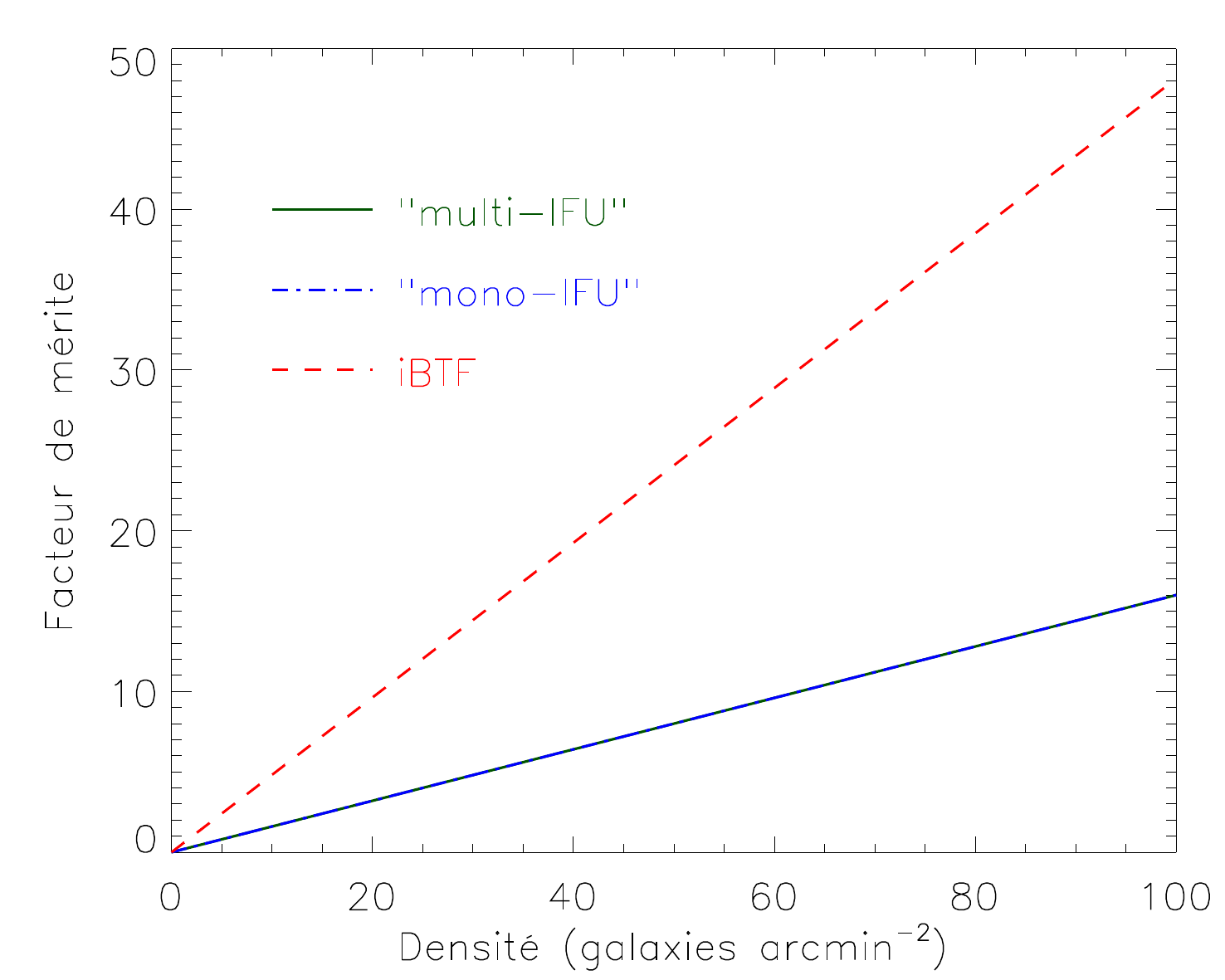}
\caption{Facteur de mérite dans le cadre de la détection de galaxies primordiales en fonction de la densité d'objets. La légende est indiquée sur la figure, cependant les courbes correspondant aux cas ``mono-IFU'' et ``multi-IFU'' sont confondues. Le facteur de mérite est supérieur pour le concept iBTF car le champ est plus grand et la couverture spectrale est moins importante.}
\label{merit_blind_detection}
\end{center}
\end{figure}

Un ELT devrait permettre de détecter des galaxies dont la magnitude en bande AB est inférieure à $28$ avec un rapport signal sur bruit de cinq en $15$ heures d'observation. \`A partir de l'extrapolation des observations actuelles ainsi qu'avec les simulations DM
et SAM,
la densité de galaxies devrait avoisiner les $100$ galaxies par minute d'arc au carré pour des décalages spectraux variant de $6$ à $15$. La fonction de mérite est tracée pour les trois concepts en fonction de la densité de galaxies sur la Figure \ref{merit_blind_detection}.

Le concept le plus performant est iBTF car le champ de vue est plus grand puisqu'on n'observe pas la partie inexploitable du spectre. En revanche, les temps de pose nécessaires sont très longs car on doit balayer le spectre. \'Etant donné qu'il faudrait typiquement $15$ heures par pose élémentaire, cet instrument nécessiterait $450$ heures pour observer tout le spectre. Un tel temps de pose, même s'il permet un champ très grand, est assez rédhibitoire puisqu'il sera impossible de faire un programme court montrant la faisabilité du projet. Puisque les sources ne sont pas connues \textit{a priori}, le concept ``multi-IFU'' ne présente aucun avantage en terme de champ de vue comparé au concept ``mono-IFU'' et leur fonction de mérite est donc identique.

\subsubsection{Assemblage de masse dans les galaxies à travers les âges}

L'objectif est d'étudier en détail la physique de galaxies déjà détectées et dont le décalage spectral spectroscopique est connu.
L'étude du spectre (en particulier les raies d'émission) de ces galaxies pour chaque élément de résolution spatiale, va ainsi permettre de cartographier leur taux de formation d'étoiles, leur métallicité mais aussi leur cinématique afin de contraindre les modèles d'évolution des galaxies.

\begin{figure}[h]
\begin{center}
\begin{tabular}{cc}
\includegraphics[width=7.5cm]{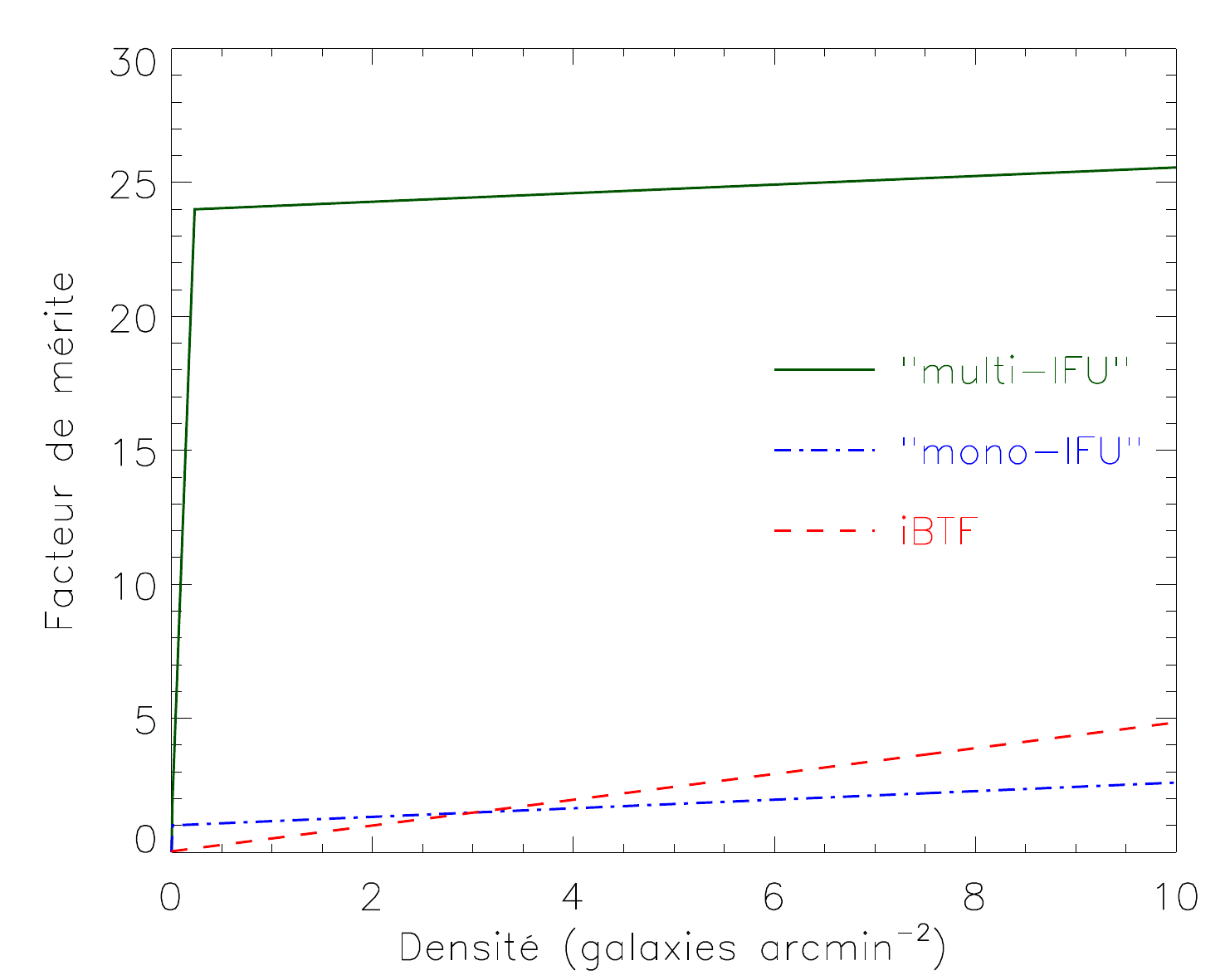} &
\includegraphics[width=7.5cm]{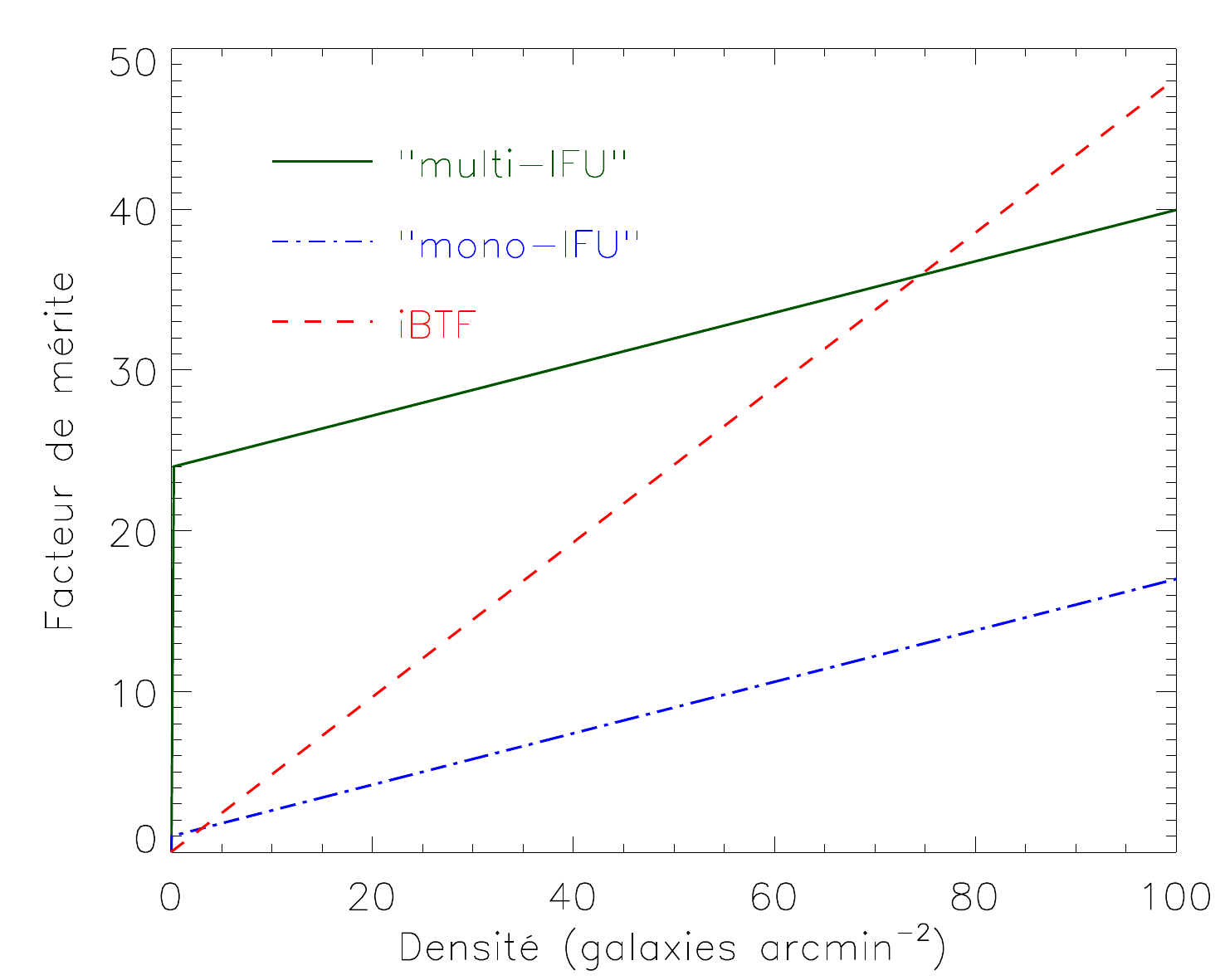}
\end{tabular}
\caption{Facteur de mérite dans le cadre de l'assemblage de masse des galaxies en fonction de la densité d'objets. La légende est indiquée sur la figure. \textbf{\`A gauche:} Pour des densités inférieures à dix galaxies par minute d'arc au carré, le concept ``multi-IFU'' est largement supérieur aux deux autres concepts. Le facteur de mérite de concept croît très vite tant qu'il y a moins d'objets dans le champ total corrigé par l'optique adaptative que d'IUF. Une fois que chaque IFU observe un objet, la fonction croît comme s'il n'y avait pas d'IFU. Le facteur de mérite du concept ``mono-IFU'' démarre à un puisqu'on va nécessairement choisir un champ avec au minimum un objet. Celui de l'iBTF démarre plus bas car même si on choisit le champ pour avoir un objet, il est nécessaire de faire $30$ expositions pour avoir le spectre complet. \textbf{\`A droite:} Dès lors que la densité dépasse $80$ galaxies par minute d'arc au carré, le concept iBTF devient plus intéressant que le concept ``multi-IFU''}
\label{merit_mass_assembly}
\end{center}
\end{figure}

Un ELT devrait permettre de détecter des galaxies dont la magnitude en bande AB est inférieure à $25$ avec un rapport signal sur bruit de cinq en huit heures d'observation.
\`A partir d'observations de galaxies à cassure de Lyman\footnote{en anglais: Lyman break galaxies} \citep{Steidel:2004,Lehnert:2003}, on s'attend, dans cette limite de magnitude, à pouvoir observer environ une dizaine de galaxies par minute d'arc au carré avec un décalage spectral compris entre $1$ et $7$, et ainsi observer plusieurs galaxies sur un seul champ. La fonction de mérite est tracée pour les trois concepts en fonction de la densité de galaxies sur la Figure \ref{merit_mass_assembly} (gauche).

Contrairement au cas scientifique précédent, le concept ``multi-IFU'' présente un avantage en terme de champ de vue du fait que les sources sont connues \textit{a priori}: les IFU vont pouvoir être placés n'importe où dans le champ de $10'\times10'$ corrigé par l'optique adaptative.
Le concept ``multi-IFU' est donc largement supérieur aux deux autres concepts du fait que le champ total est très large. Le nombre d'IFU a été déterminé pour être adapté aux densités attendues. Notons quand même que plus le décalage spectral est important, plus la densité d'objets devient faible. Cependant, dès lors que la densité est supérieure à $0.2$ galaxies par minute d'arc au carré, le concept ``multi-IFU'' est extrêmement performant puisqu'il y a assez d'objets dans le champ corrigé par l'optique adaptative pour utiliser tous les IFU. Dès lors que tous les IFU sont utilisés, le facteur de mérite croît au même rythme que pour le concept ``mono-IFU''.
\par
On observe tout de même que le facteur de mérite du concept iBTF croît plus rapidement que pour les deux autres concepts. Pour des densités très élevées, ce concept devient donc plus intéressant que le concept ``multi-IFU'' (Figure \ref{merit_mass_assembly}, droite).

\subsubsection{Choix du concept de type multi-IFU: EAGLE}

La fonction de mérite utilisée permet donc de différencier les différents concepts. En particulier, on voit bien que le projet iBTF constituerait un bon compromis entre les deux thématiques. Cependant, du fait des temps de pose induits par le mode d'observation (balayage du spectre dans le temps) et du fait du faible avancement technologique de ce projet, il n'est pas raisonnable d'envisager l'utilisation de ce concept pour une première génération d'instruments pour les ELT.
\par
\'Etant donnée l'équivalence en terme de fonction de mérite des deux concepts ``mono-IFU'' et ``multi-IFU'' pour la détection de galaxies primordiales et vu l'avantage incomparable que représente la solution ``multi-IFU'' pour l'étude de l'assemblage de masse dans les galaxies à travers les âges vu la faible densité d'objets attendue, ce concept a été retenu par le groupe de travail WFSpec, même s'il représente une difficulté technologique plus importante que le concept ``mono-IFU''. Ce projet a été baptisé EAGLE (PI: J.-G. Cuby), acronyme de Elt Ao for GaLaxy Evolution.

\subsection{Quel avenir pour iBTF?}
Le concept iBTF est un concept innovant tant au point de vue technologique qu'au point de vue observationnel. En effet, il permet de séparer plusieurs composantes spectrales d'une image en une unique exposition. Le couplage avec un \FP~permet de doper la résolution spectrale.
De plus, l'ordre zéro contenant tous les domaines non extraits peut être imagé. Ainsi, aucune lumière n'est perdue et on peut faire de l'imagerie profonde simultanément à l'acquisition du spectre. Il est dans ce cas intéressant d'utiliser un filtre suppresseur des raies d'émission du ciel nocturne (OH) afin de maximiser le rapport signal sur bruit dans l'ordre zéro (continuum).
Par ailleurs, ce concept peut être amélioré en utilisant des filtres de Bragg accordables. Pour le projet WFSpec, la majorité des filtres de Bragg utilisés extraient un domaine spectral fixe. Il est possible de faire varier la bande spectrale moyennant un encombrement plus important afin d'ajouter un système de contrôle. L'accordabilité des filtres de Bragg peut se faire en modifiant leur inclinaison. La difficulté principale vient du fait que pour être capable de faire de l'imagerie avec les filtres de Bragg, il est nécessaire d'en utiliser deux. Le premier sélectionne la bande spectrale mais disperse la lumière comme le ferait un réseau. Le second est utilisé pour compenser cette dispersion. L'accordabilité nécessite de contrôler l'inclinaison des deux filtres simultanément et d'ajuster la position de l'image sur le détecteur. Une autre solution est d'utiliser les propriétés acousto-optiques de cristaux biréfringents: en appliquant une onde ultra sonore au cristal biréfringent (filtre de Bragg) et en faisant varier sa longueur d'onde, la biréfringence du cristal va changer induisant une variation de la bande spectrale.

L'étude de la fonction de mérite a montré que cet instrument est particulièrement adapté pour des thématiques où la densité d'objets est très élevée et pour des objets s'accommodant de temps d'exposition courts ou d'une basse résolution spectrale afin de réduire le nombre de pas de balayage.
L'étude de populations stellaires résolues dans les galaxies proches semble donc être la thématique la plus appropriée pour l'utilisation de ce nouveau concept.

\chapter{GHASP, échantillon cinématique de 203 galaxies locales isolées}
\label{ghasp_donnees}
\minitoc
\textit{
Ce chapitre présente le projet GHASP et ses objectifs. La production de masse des données cinématiques pour l'échantillon GHASP est décrite dans deux publications. Les nouvelles méthodes utilisées pour obtenir les cartes cinématiques et en extraire les paramètres de projection cinématiques ainsi que les courbes de rotation y sont également détaillées, donnant lieu à une comparaison des paramètres de projection cinématiques et morphologiques et à l'étude de la relation de Tully-Fisher.
Ces articles représentent le c\oe ur du travail effectué durant cette thèse. Enfin, une ébauche de l'analyse des propriétés cinématiques de cet échantillon local est présentée sous forme d'un article en préparation.}
\hl

\section{Le projet GHASP}
\label{ghasp_projet}

\subsection{Définition et objectifs de l'échantillon GHASP}

\begin{figure*}[htbp]
\begin{center}
\begin{turn}{270}
\includegraphics[width=22cm]{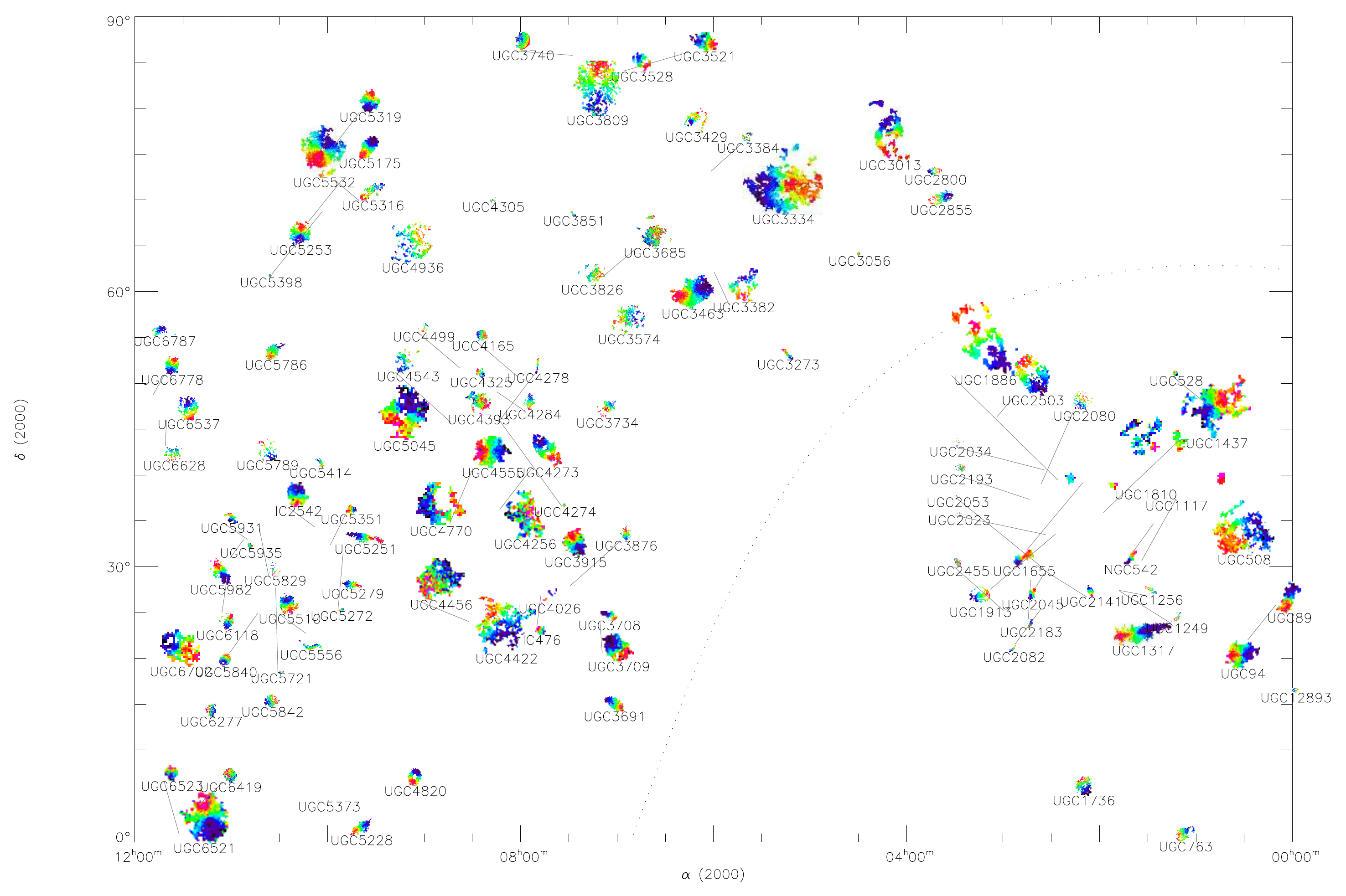}
\end{turn}
\caption{Distribution des galaxies GHASP sur la voûte céleste pour des ascensions droites comprises entre $0$h et $12$h. Les tailles physiques relatives des galaxies (et non les dimensions angulaires) sont respectées. La même échelle est utilisée pour dimensionner les galaxies de la Figure \ref{distribution_ghasp2}. Le trait pointillé correspond au plan galactique. On observe une répartition uniforme de galaxies GHASP, hormis dans le plan galactique.}
\label{distribution_ghasp1}
\end{center}
\end{figure*}

\begin{figure}[htbp]
\begin{center}
\begin{turn}{270}
\includegraphics[width=22cm]{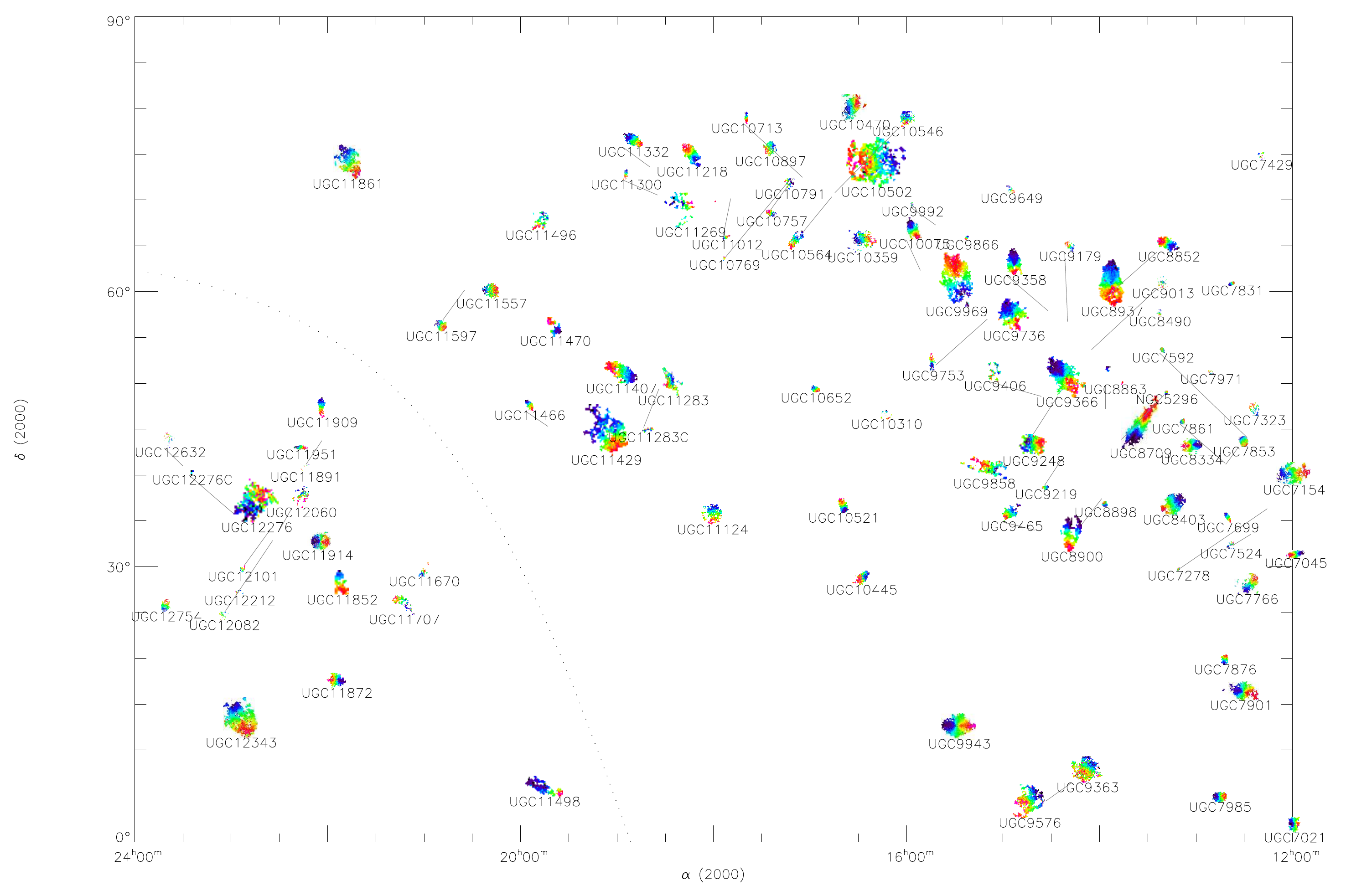}
\end{turn}
\caption{Distribution des galaxies GHASP sur la voûte céleste pour des ascensions droites comprises entre $12$h et $24$h. Les tailles physiques relatives des galaxies (et non les dimensions angulaires) sont respectées. La même échelle est utilisée pour dimensionner les galaxies de la Figure \ref{distribution_ghasp1}. Le trait pointillé correspond au plan galactique. On observe une répartition uniforme de galaxies GHASP, hormis dans le plan galactique.}
\label{distribution_ghasp2}
\end{center}
\end{figure}

L'échantillon GHASP est composé de $203$ galaxies spirales et irrégulières isolées, observées autour de la raie H$\alpha$ avec l'instrumentation GHASP (partie \ref{ghasp_fantomm_ghafas}) entre 1998 et 2004 sur $14$ périodes d'observation. C'est l'échantillon contenant le plus grand nombre de galaxies observées avec les techniques de \FP~à ce jour. Les galaxies sont situées dans l'hémisphère Nord puisqu'elles ont été observées depuis l'Observatoire de Haute Provence dont la latitude est légèrement inférieure à $+44$\degr~(voir Figures \ref{distribution_ghasp1} et \ref{distribution_ghasp2}).


Les détails concernant l'élaboration de l'échantillon sont détaillés dans la thèse d'Olivia Garrido \citep{Garrido:thesis} qui a effectué une grande partie des observations de ce programme.
Les objectifs scientifiques qui ont abouti à la définition de cet échantillon sont multiples:
\begin{itemize}
\item Calculer la relation de Tully-Fisher locale;
\item \'Etudier la distribution de matière de la composante lumineuse et du halo de matière sombre, à travers la séquence de Hubble, pour des galaxies à faible et forte brillance de surface et pour un grand intervalle de magnitude. Cette étude se fait habituellement par l'ajustement de modèles de masse sur des courbes de rotation hybrides combinant les données optiques et radio \citep{Barnes:2004,Spano:2008}. Une des ambitions de ce programme est de chercher la distribution de masse non plus sur une dimension spatiale (le rayon), mais sur les deux dimensions spatiales en utilisant les champs de vitesses 2D, de l'imagerie à bande large pour déterminer le potentiel 2D des étoiles et des modèles d'évolution spectrophotométrique des étoiles afin de contraindre le rapport masse sur luminosité des galaxies;
\item Comparer la cinématique dans des environnements variés (champ, paires, groupes compacts, galaxies d'amas) afin de dissocier l'évolution propre des galaxies dite séculaire de l'évolution due à une origine extérieure \citep{Garrido:2005};
\item Produire un échantillon de référence de galaxies locales afin de comparer leur cinématique avec la cinématique de galaxies à grand décalage spectral \citep{Epinat:2007,Puech:2008}. Il est primordial de pouvoir dissocier les effets de résolution spatiale et spectrale des effets d'évolution des galaxies;
\item Modéliser les effets de structures non-axisymétriques comme les bras spiraux, les barres, les distorsions ovales en utilisant des simulations numériques $N$-corps et hydrodynamiques, des études kinémétriques \citep{Krajnovic:2006} et la méthode Tremaine-Weinberg pour mesurer les motifs de vitesse des barres, des spirales et des structures internes \citep{Hernandez:2005a};
\item Analyser la dispersion de vitesses du gaz ionisé et chercher des liens avec la dispersion de vitesses des étoiles;
\item Créer des courbes de rotation et des champs de vitesses types \citep{Persic:1991,Persic:1996,Catinella:2006};
\item Comprendre les liens entre la cinématique (forme de la courbe de rotation, moment angulaire, ...) et les propriétés physiques des galaxies telles que le taux de formation stellaire en comparant, par exemple, à ce qui se passe pour des galaxies à fort taux de formation d'étoiles comme les galaxies bleues compactes.
\end{itemize}
Certains de ces objectifs ont été abordés pendant la thèse et sont exposés dans les sections \ref{etude_dynamique} et \ref{ghasp_highz}.
Ces données servent de point de départ à l'élaboration d'une nouvelle base de données \FP~présentée en Annexe \ref{annexe_bddfp}.

\subsection{\'Etude du gaz ionisé à partir de la raie \Ha}
\label{regions_h2}

La raie H$\alpha$ est le nom que les astronomes utilisent pour faire référence à la raie d'émission produite par les atomes d'hydrogène neutre correspondant à la transition de Balmer $\alpha$ observable dans le rouge ($656.3~nm$)\footnote{Historiquement, la série de Balmer à été découverte en premier car c'est la seule à émettre des photons dans le domaine visible.}. Elle est générée par la désexcitation des électrons passant du niveau d'énergie $n=3$ au niveau $n=2$.
Pourtant, la raie H$\alpha$ trace les régions contenant du gaz ionisé.
En effet, les régions de gaz ionisé sont des zones de formation stellaire: le rayonnement ultra-violet d'énergie supérieure à $13.6~eV$ émis par les étoiles jeunes ionise le gaz d'hydrogène neutre à partir duquel elles ont été formées. Dans ce milieu de gaz ionisé (atomes d'hydrogène neutres, protons, électrons, photons ionisants) protons et électrons sont en recombinaison et ionisation permanentes. Lors de la recombinaison d'un électron avec un atome d'hydrogène, la probabilité qu'il utilise la transition de Balmer $\alpha$ pour revenir au repos est environ de $0.5$, ce qui explique que le milieu de gaz ionisé émette intensément dans le raie H$\alpha$. Ces régions de gaz ionisé, également appelées régions HII, contiennent également des ions He$^+$, O$^+$, O$^{++}$, N$^+$, S$^+$ et S$^{++}$ qui possèdent également des raies d'émission dans le domaine visible. Elles sont principalement situées dans les zones de surdensité de matière que sont les bras spiraux. Cette surdensité compresse le gaz moléculaire provoquant son effondrement et la formation de proto-étoiles. Les étoiles excitatrices étant des étoiles de type OB dont la température effective est comprise entre $25000$ et $50000~K$, dont la masse est supérieure à $10~M_{\odot}$ et dont la durée de vie est inférieure à $10$ millions d'années, les régions HII sont donc de bons traceurs de la formation stellaire récente. Les régions HII observables ont une faible densité électronique (quelques électrons par $cm^3$), une masse pouvant atteindre $10^4~M_{\odot}$ pour une dimension de l'ordre de la centaine de parsecs et une température électronique de $10000~K$. Au delà de la sphère Strömgren à l'intérieur duquel tout le rayonnement ionisant est absorbé, le milieu n'est pas ionisé puisque l'ensemble du rayonnement ionisant a été absorbé en chemin. Ce milieu est donc froid ($\sim 100~K$), ce qui induit l'expansion de ces nébuleuses à une vitesse de l'ordre de la dizaine de $km~s^{-1}$.
\par
Il existe également dans les galaxies des régions de gaz diffus ionisé. Ce gaz diffus a été mis en évidence dès 1971 \citep{Monnet:1971} dans les régions inter-bras de nombreuses galaxies Sc. Il se présente sous forme de filaments, boucles ou coquilles, notamment au sein du halo et est probablement ionisé par les étoiles du disque. L'émission de ce milieu représente entre $25$ et $50$\%~de l'émission \Ha~totale \citep{Ferguson:1996}.
\par
Outre l'information stellaire que les raies d'émission apportent, l'observation de ces raies est particulièrement adaptée à l'étude de la cinématique des galaxies car elles permettent d'obtenir un rapport signal sur bruit bien plus élevé que l'observation de raies d'absorption. Le moment d'ordre $0$ de la raie nous donne accès au flux, ce qui est une information quant au taux de formation stellaire et les moments d'ordres supérieurs sont utilisés pour les études cinématiques. L'analyse des moments de raies est détaillée en Annexe \ref{annexe_moment}. La raie \Ha, qui a une intensité élevée par rapport aux autres raies d'émission du domaine visible, est d'autant plus adaptée aux études cinématiques puisqu'une mesure précise des moments de raies nécessite un rapport signal sur bruit qui augmente avec l'ordre du moment. Toutefois, par rapport aux raies d'absorption stellaires, les raies d'émission présentent l'inconvénient de tracer principalement la cinématique du gaz, donc des structures telles que les spirales, dont les mouvements ne sont pas purement circulaires.
\par
Pendant longtemps, les observations cinématiques ont été réalisées avec des spectrographes à fente placée le long du grand axe déterminé par la morphologie afin d'étudier la raie \Ha. La plupart des études sur la forme des courbes de rotation ont ainsi utilisé de très grands échantillons observés à partir de spectroscopie à fente \citep{Mathewson:1992,Persic:1996,Catinella:2006}.
Cependant, depuis une vingtaine d'années, les observations spectroscopiques à champ intégral se multiplient: les techniques de \FP~permettent d'observer tout le disque optique et les IFU permettent d'observer soit les régions internes de galaxies proches (e.g. \citealp{Falcon-Barroso:2006}), soit les galaxies distante (e.g. \citealp{Flores:2006,Genzel:2006,Law:2007}).
L'utilisation d'observations spectroscopiques à champ intégral permet de sonder les mouvements non circulaires dans les galaxies et également de déterminer des courbes de rotation moins biaisées par la présence de structures telles que des bras spiraux ou des barres. Elle permet également d'étudier la cinématique indépendamment de l'étude morphologique.
Des cartes cinématiques sont également déduites des observations de la raie HI réalisées avec des radiotélescopes à synthèse d'ouverture. L'avantage des techniques radio est que le gaz neutre peut être détecté sur des échelles supérieures à plusieurs diamètres optiques. La résolution spatiale de ces instruments dépend de la longueur des lignes de base $l$ séparant les diverses antennes ($\lambda/l$ est la résolution angulaire qui peut être atteinte), de la direction de ces lignes de base et de la latitude de la source. La meilleure résolution pourrait ainsi être atteinte sur Terre en plaçant deux antennes ou plus en des lieux diamétralement opposés de la planète pour avoir les lignes de base les plus longues possibles.
Pour des raisons de flux et donc de durée d'exposition, la résolution spatiale typique en radio est inférieure d'un facteur $10$ à celle que le domaine optique atteint usuellement, ce qui fait que les données optiques, en particulier la raie \Ha, sont bien mieux adaptées à l'étude de la courbe de rotation dans les régions internes. Cela est important notamment pour contraindre la forme des halos de matière sombre.

\section{Présentation des données scientifiques}

Les données \FP~nécessitent une réduction adaptée. Un des premiers logiciels développés pour réduire les données \FP~est le logiciel CIGALE développé par Jacques Boulesteix au début des années 1980 sur des machines utilisant le système d'exploitation VMS pour réduire les données de l'instrument CIGALE alors installé au CFHT. Par la suite, Jacques Boulesteix développa ADHOC
en 1993 puis l'adapta pour Windows (ADHOCw) en 1999. Ce logiciel très éprouvé et très didactique bénéficie de très bons outils de visualisation et de manipulation des données, ainsi que de programmes connexes extrêmement utiles lors des observations.
Parallèlement, un autre outil de réduction a été développé sous UNIX en langage C à partir de 1988, puis C$++$ à partir de 2000 par \'Etienne Le Coarer pour le dépouillement des données CIGALE observées avec le télescope de $36~cm$ de la Silla (ESO) correspondant au sondage du plan Galactique et aux nuages de Magellan \citep{Le-Coarer:1992}. Ce programme dispose d'outils spécifiques pour la visualisation et la décomposition des profils. Il cherche a minimiser l'occupation de l'espace disque par les produits de réduction en utilisant un algorithme de compression et en n'enregistrant que le produit final.

Afin de pouvoir étudier l'échantillon complet de façon homogène, de nouvelles procédures de réduction et d'analyse des données ont été développées \citep{Daigle:2006b} sous IDL.
Ces nouvelles procédures tendent à être le plus automatisé possible afin de rendre les résultats indépendants de la personne qui analyse les données. J'ai participé activement au développement de certaines routines de ce logiciel de réduction.
L'utilisation de l'environnement IDL a été motivé par le fait (i) qu'il fonctionne quel que soit le système d'exploitation utilisé, (ii) qu'IDL possède un environnement de base très riche ainsi que de librairies astronomiques très utiles et (iii) que ce logiciel peut se prêter à modification ou à des ajouts car le code est disponible librement.

\subsection{Des données d'observation aux cartes cinématiques}

La réduction des données a pour objectif de créer un cube de données calibré à partir des interférogrammes d'observation et d'en extraire les cartes cinématiques.
Le manuel d'utilisation du programme de réduction présentant les étapes de réduction ainsi que toutes les options de réduction est disponible en Annexe \ref{help_computeeverything}. Ce manuel d'utilisation a initialement été écrit par Olivier Daigle en français. Je l'ai traduit en anglais et mis à jour avec les modifications que j'ai apportées au programme. Les étapes de réduction appliquées pour réduire l'échantillon GHASP et leurs options sont les suivantes:
\begin{itemize}
\item Intégration des données en corrigeant la dérive de suivi du télescope lorsque le champ contient suffisamment d'étoiles. Cette dérive atteint jusqu'à $8$ pixels dans certains cas. Les poses élémentaires ayant un flux incompatible avec le flux médian sont supprimées ou remplacées par les poses des cycles d'observation adjacents de manière automatique.
\item Création d'un cube calibré en longueur d'onde.
\item Soustraction des raies d'émission du ciel nocturne en utilisant un ajustement polynomial d'ordre $2$. Dans les quelques cas où la galaxie est trop grande, un spectre médian a été soustrait. Ces raies sont majoritairement produites par les radicaux OH présents dans les couches atmosphériques d'une dizaine de kilomètres d'épaisseur situées à une altitude d'environ $100~km$ \citep{Rousselot:2000}. Ces radicaux sont créés par la réaction entre l'hydrogène et l'ozone ($H+O_3\rightarrow OH+O_2$). Le problème principal de ces raies est leur forte et rapide variation spatiale et spectrale.
\item \'Eventuelle suppression des réflexions parasites.
\item \'Eventuelle addition d'observations multiples d'un même objet en utilisant l'astrométrie.
\item Lissage spectral de Hanning.
\item Lissage spatial adaptatif (inspiré de \citealp{Cappellari:2003}) avec un rapport signal sur bruit défini comme étant la racine carrée du flux. L'objectif est d'atteindre un rapport signal sur bruit de $7$.
\item Calcul de l'astrométrie.
\item Correction de l'incertitude sur l'intervalle spectral libre.
\item Création des cartes et filtrage.
\end{itemize}

Un cube de données possède deux dimensions spatiales et une dimension spectrale. Ainsi, un spectre est mesuré pour chaque élément de résolution spatiale (pixel). Pour chaque pixel, on va utiliser le spectre pour mesurer le flux émis par la raie, sa position en longueur d'onde et sa largeur, et ainsi obtenir des cartes pour chacune de ces grandeurs. La méthode utilisée sur l'échantillon GHASP pour extraire ces grandeurs est la méthode des moments: le flux est l'ordre $0$ du spectre, la position est l'ordre $1$ et la dispersion est l'ordre $2$  (voir l'Annexe \ref{info_spectre} pour les détails). On déduit la vitesse et la dispersion de vitesses à partir des mesures en longueur d'onde grâce à l'effet Doppler-Fizeau. L'Annexe \ref{doppler_fizeau} détaille le calcul utilisé pour déterminer la vitesse et la dispersion de vitesses dans le cadre relativiste de l'effet Doppler-Fizeau. La technique de lissage adaptatif est utilisée pour avoir un rapport signal sur bruit correct même dans les régions où le flux est faible, en particulier pour le gaz diffus ionisé qui n'est pas aisé à détecter. Il s'agit d'une accrétion de pixels: tant que le rapport signal sur bruit recherché dans le spectre n'est pas atteint, les pixels sont accrétés. Ainsi, le spectre des régions de faible flux n'est pas corrélé avec les régions à fort flux.

\subsection{Analyse des cartes de vitesses}

L'analyse dont il est question ici consiste à extraire les paramètres cinématiques déduits des champs de vitesses et leurs erreurs associées. Ces paramètres sont les paramètres de projection de la galaxie (inclinaison, angle de position du grand axe, vitesse globale de la galaxies dite vitesse systémique) ainsi que la courbe de rotation de la galaxie.
Le c\oe ur des méthodes utilisées est décrit dans les articles, je me contente donc ici de mentionner les grandes lignes.
\par
Un modèle de disque fin avec une courbe de rotation paramétrique tirée de \citet{Kravtsov:1998} est ajusté aux données par une méthode des moindres carrés afin de déterminer ses paramètres. L'algorithme utilisé est l'algorithme de Levenberg-Marquardt \citep{numericalrecipes}, qui permet d'effectuer l'ajustement de modèles non linéaires aux données. La librairie IDL de Markwardt MPFIT a été utilisée. Cet algorithme utilise les dérivées partielles par rapport aux variables pour converger vers une vallée de moindres carrés. Des cas d'école de champs de vitesses obtenus avec quatre formes de courbes de rotation sont présentés dans la Figure 6 (Figure \ref{fig6_p4} dans la table des figures) de l'article présenté dans la partie \ref{ghasp_highz}.
\par
Cette méthode a été développée pour deux raisons. Jusqu'à présent, deux méthodes étaient couramment utilisées. La première est une détermination manuelle des paramètres afin de symétriser au mieux la courbe de rotation. Un premier inconvénient de cette méthode est que la détermination des paramètres est dépendante de l'utilisateur, mais le principal problème est que les erreurs ne peuvent pas vraiment être mesurées.
Une solution serait la détermination des paramètres indépendamment par plusieurs utilisateurs, ce qui, en pratique, demande beaucoup de temps et de ressources humaines.
La seconde méthode, utilisée par le logiciel GIPSY,
est une méthode dite de ``tilted ring model''. La galaxie est découpée en anneaux. Pour chaque anneau, un jeu de paramètres est alors ajusté par une méthode de moindres carrés, avec une erreur statistique pour chaque paramètre et pour chaque couronne. Cette méthode est particulièrement adaptée aux données possédant beaucoup de points afin d'avoir une statistique suffisante pour chaque anneau, ce qui est le cas des données HI pour lesquelles un gauchissement des disques est potentiellement observable. De tels gauchissements sont bien plus rares dans le disque optique et donc en \Ha. La décomposition en anneaux engendre donc une difficulté quant à une définition unique des paramètres de projection pour chaque galaxie. Il était donc nécessaire de faire l'ajustement d'un modèle possédant une seule inclinaison, un seul grand axe, une seule vitesse systémique et un seul et unique centre pour chaque galaxie. Par ailleurs, une détermination robuste et réaliste des erreurs a été mise au point en utilisant la carte des résidus du modèle. Ces méthodes sont décrites en détail dans les articles qui suivent.

Grâce à ces nouvelles méthodes, j'ai pu réduire et analyser l'ensemble de l'échantillon.
Deux articles acceptés dans une revue à comité de lecture\footnote{Monthly Notices of the Royal Astronomical Society} ont résulté de ces travaux \citep{Epinat:2008a,Epinat:2008b}. Ces articles présentent les cartes déduites de la réduction des données ainsi que la comparaison des paramètres de projection déduits de la cinématique et de la morphologie. Une relation de Tully-Fisher est également dérivée à partir de l'échantillon GHASP et des commentaires galaxie par galaxie sont présentés. Ils exposent la description de la cinématique déduite de nos données et la mettent principalement en regard avec les études cinématiques de la littérature.


\subsection{\underline{Article I:}~GHASP: an \Ha~kinematic survey of spiral and irregular galaxies - VI. New \Ha~data cubes for 108 galaxies}
\label{articleI}
Cet article présente les données correspondant aux sept dernières périodes d'observation du programme GHASP. La méthode d'ajustement des modèles y est présentée et chacune des galaxies a été commentée individuellement.
Les cartes ainsi que les courbes de rotation sont respectivement présentées en Annexe \ref{annexe_maps} et \ref{annexe_rcs} de la thèse afin de regrouper l'intégralité des images des deux articles.
\par
Nous présentons les observations \FP~pour un nouvel ensemble de $108$ galaxies dans le cadre du programme GHASP.
Ce relevé est composé de cubes de données pour $203$ galaxies spirales et irrégulières autour de la raie \Ha~afin d'en étudier la cinématique et couvre un grand intervalle de types morphologiques et de magnitudes. Les nouvelles données présentées dans cet article complètent le relevé. L'échantillon GHASP est désormais le plus grand échantillon de données \FP~jamais publié. L'analyse de l'échantillon complet sera présenté dans de prochains articles. En utilisant les techniques de lissage adaptatif basées sur un maillage de Voronoï, nous avons créé des cubes de données \Ha~à partir desquels les cartes de flux \Ha, les champs de vitesses ainsi que les champs de vitesses résiduels, les diagrammes position-vitesse, les courbes de rotation et les paramètres cinématiques ont été extraits pour presque toutes les galaxies. La détermination des paramètres cinématiques, des courbes de rotation et des incertitudes associées résulte de nouvelles méthodes implémentées dans la procédure de traitement des données. Cette nouvelle détermination repose sur l'utilisation de la totalité du champ de vitesses 2D et du spectre de puissance du champ de vitesses résiduel alors que les techniques classiques utilisent une décomposition du champ de vitesses en anneaux. Parmi les résultats présentés, nous montrons que les angles de position du grand axe morphologique ont des barres d'erreurs systématiquement plus importantes que pour les déterminations cinématiques, en particulier pour les systèmes faiblement inclinés. Les inclinaisons morphologiques des galaxies pour lesquelles la détermination de l'angle de position morphologique est très incertaine ne sont pas très fiables. Les galaxies qui ont une forte inclinaison présentent un meilleur accord entre l'inclinaison cinématique et l'inclinaison morphologique calculée en supposant un disque fin. Nous avons utilisé la relation de Tully-Fisher pour vérifier la consistance de nos mesures de vitesses maximales à partir de nos courbes de rotation. Nos données sont en accord avec les précédentes déterminations de la relation de Tully-Fisher provenant de la littérature. Néanmoins, les galaxies de faible inclinaison présentent statistiquement des vitesses supérieures à ce qui est attendu et les galaxies en rotation rapide sont moins lumineuses qu'attendu.

\includepdf[pagecommand={\pagestyle{headings}},scale=1.,offset=0 -5,pages={1-35},
addtotoc={
1,subsubsection,3, Introduction,intropaper1,
2,subsubsection,3, Observations et réduction des données,obspaper1,
4,subsubsection,3, Analyse des données, analysispaper1,
7,subsubsection,3, Relation de Tully-Fisher, tfpaper1,
9,subsubsection,3, Résumé et perspectives, conclpaper1,
11,subsubsection,3, Annexe A: Construction des courbes de rotation, methodepaper1,
12,subsubsection,3, Annexe B: Notes individuelles sur les galaxies, notespaper1,
24,subsubsection,3, Annexe C: Tableaux, tablespaper1,
31,subsubsection,3, Annexe D: Cartes et diagrammes position-vitesse, mappaper1,
35,subsubsection,3, Annexe E: Courbes de rotation, crmaper1
},
addtolist={
3,figure,{\underline{Article I}, Figure 1: Top panel: distribution of morphological type for almost all of the GHASP sample (201 out of 203 galaxies). Middle panel: distribution of the absolute $B$-band magnitude for almost all of the GHASP sample (198 out of 203 galaxies). For both the top and middle panels: the blue hash, red hash and residual white represent, respectively, the strongly barred, the moderately barred and the non-barred galaxies. Bottom panel: distribution for almost all of the GHASP sample (198 out of 203 galaxies) in the `magnitude-morphological type' plane distinguishing strongly barred (blue squares), moderately barred (red triangles) and unbarred galaxies (black circles).},fig1_p1,
5,figure,{\underline{Article I}, Figure 2:  Dispersion in residual velocity field versus maximum velocity, sorted by Hubble morphological type: black circles $0 \le t < 2$, red triangles $2 \le t < 4$, blue squares $4 \le t < 6$, green rhombuses $6 \le t < 8$ and pink stars $8 \le t < 10$. The dashed line represents the linear regression on the data. The points above the dotted line are discussed in Section 3.3. UGC 3334 labelled with an arrow has actually a huge residual velocity dispersion of $54~km~s^{-1}$ (see Table C2).},fig2_p2,
6,figure,{\underline{Article I}, Figure 3: Top panel: kinematical versus morphological (HyperLeda) PAs of the major axis. Galaxies for which no accurate morphological PA has been computed are shown by red open circles; galaxies having an inclination lower than 25\degr~are displayed by the blue squares; the other galaxies are represented by black circles. Bottom panel: histogram of the variation between kinematical and morphological PAs. The red hash, blue hash and residual white represent, respectively, the galaxies for which no accurate PA has been measured, for which inclination is lower than 25\degr~and the other galaxies of the sample.},fig3_p1,
7,figure,{\underline{Article I}, Figure 4: Top panel: kinematical versus thick disc morphological inclinations. Middle panel: kinematical versus thin disc morphological inclinations. Top and middle panels: galaxies for which no accurate morphological PA has been computed are shown by red open circles; galaxies with a difference between the kinematical and morphological PAs larger than 20\degr~are displayed with blue squares; the other galaxies are represented by black circles. Bottom panel: histogram of the variation between kinematical and morphological inclinations. The red hash, blue hash and residual white represent, respectively, the galaxies for which no accurate PA has been measured, for which the difference between the kinematical and morphological PAs is larger than 20\degr~and the other galaxies of the sample.},fig4_p1,
8,figure,{\underline{Article I}, Figure 5: Tully-Fisher relation for our sample of galaxies. The solid line
represents the $B$ magnitude Tully-Fisher relation determined by Tully \& Pierce (2000) from nearby galaxies in clusters (Ursa Major, Pisces filament, Coma). Top panel: sorted by inclination -- low-inclination galaxies ($i < 25$\degr): blue squares; other galaxies ($i \ge 25$\degr): black circles. Middle panel: sorted by $V_{max}$ flags -- $V_{max}$ reached: black dots, large size; $V_{max}$ probably reached: blue squares, medium size; $V_{max}$ probably not reached: red triangles, small size. Bottom panel: sorted by morphological type -- black circles from 0 to 2; red triangles from 2 to 4; blue squares from 4 to 6; green rhombuses from 6 to 8; pink stars from 8 to 10 and the dashed line represents the best linear fit to the data.},fig5_p1,
24,table,{\underline{Article I}, Table C1: Log of the observations.},logpaper1,
26,table, {\underline{Article I}, Table C2: Model parameters.},modparpaper1,
28,table, {\underline{Article I}, Table C3: Galaxy parameters.},galparpaper1,
31,figure,{\underline{Article I}, Figure D19: UGC 3740. Top left-hand panel: XDSS blue-band image. Top right-hand panel: \Ha~velocity field. Middle left-hand panel: \Ha~monochromatic image. Middle right-hand panel: \Ha~residual velocity field. The white and black cross is the kinematical centre. The black line is the major-axis, its length represents $D_{25}$ . Bottom panel: position-velocity diagram along the major-axis (full width of 7 pixels), arbitrary flux units. The red line plots the rotation curve computed from the model velocity field along the major-axis (full width of 7 pixel).},figd19,
32,figure,{\underline{Article I}, Figure D31: UGC 4820. Top left-hand panel: XDSS blue-band image. Top right-hand panel: \Ha~velocity field. Middle left-hand panel: \Ha~monochromatic image. Middle right-hand panel: \Ha~residual velocity field. The white and black cross is the kinematical centre. The black line is the major-axis, its length represents $D_{25}$ . Bottom panel: position-velocity diagram along the major-axis (full width of 7 pixels), arbitrary flux units. The red line plots the rotation curve computed from the model velocity field along the major-axis (full width of 7 pixel).}, figd31,
33,figure,{\underline{Article I}, Figure D45: UGC 5786. Top left-hand panel: XDSS blue-band image. Top right-hand panel: \Ha~velocity field. Middle left-hand panel: \Ha~monochromatic image. Middle right-hand panel: \Ha~residual velocity field. The white and black cross is the kinematical centre. The black line is the major-axis, its length represents $D_{25}$ . Bottom panel: position-velocity diagram along the major-axis (full width of 7 pixels), arbitrary flux units. The red line plots the rotation curve computed from the model velocity field along the major-axis (full width of 7 pixel).}, figd45,
34,figure,{\underline{Article I}, Figure D56: UGC 7154. Top left-hand panel: XDSS blue-band image. Top right-hand panel: \Ha~velocity field. Middle left-hand panel: \Ha~monochromatic image. Middle right-hand panel: \Ha~residual velocity field. The white and black cross is the kinematical centre. The black line is the major-axis, its length represents $D_{25}$ . Bottom panel: position-velocity diagram along the major-axis (full width of 7 pixels), arbitrary flux units. The red line plots the rotation curve computed from the model velocity field along the major-axis (full width of 7 pixel).}, figd56,
35,figure,{\underline{Article I}, Figure E1: From the top left-hand panel to the bottom right-hand panel: \Ha~rotation curve of UGC 12893, UGC 89, UGC 94, UGC 1317, UGC 1437 and UGC 1655. },fige1_p1
}]{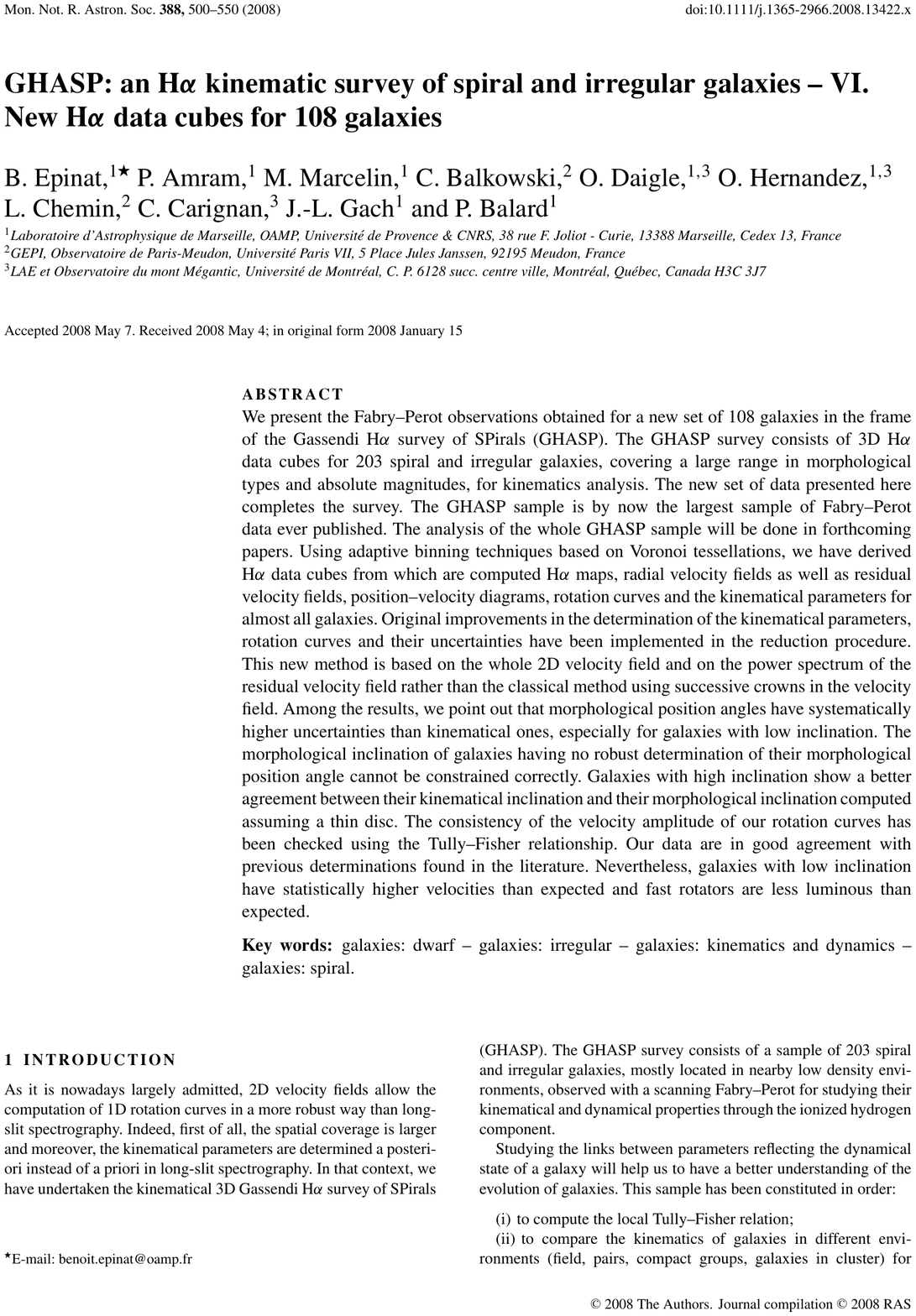}

\subsection{\underline{Article II:}~GHASP: An \Ha~kinematic survey of 203 spiral and irregular galaxies - VII. Revisiting the analysis of \Ha~data cubes for 97 galaxies.}
\label{articleII}

De même que pour l'article précédent, les cartes ainsi que les courbes de rotation sont respectivement présentées en Annexe \ref{annexe_maps} et \ref{annexe_rcs} de la thèse.
\par
Afin de disposer d'un échantillon homogène, réduit et analysé en utilisant les mêmes procédures, nous présentons dans cet article une nouvelle réduction et une nouvelle analyse pour un ensemble de $97$ galaxies déjà publiées dans des articles antérieurs \citep{Garrido:2002,Garrido:2003,Garrido:2004,Garrido:2005}, mais en utilisant désormais les nouvelles procédures adoptées pour l'ensemble de l'échantillon (voir partie \ref{articleI}). Le programme d'observations GHASP est à présent terminé. Cet article présente les paramètres cinématiques pour l'ensemble de l'échantillon. Pour la première fois, les profils intégrés de la raie \Ha~ont été calculés et sont présentés pour tout l'échantillon. Le flux \Ha~total déduit de ces profils a été utilisé afin de déterminer une calibration en flux pour les $203$ galaxies de GHASP. Cet article confirme les conclusions déjà suggérées par l'étude de la moitié de l'échantillon concernant (i) la précision de mesure des angles de position du grand axe qui est accrue en utilisant des données cinématiques, (ii) la difficulté d'obtenir des estimations robustes des inclinaisons morphologiques et cinématiques, en particulier pour les galaxies de faible inclinaison et (iii) l'accord excellent entre la relation de Tully-Fisher déterminée à partir de notre échantillon et ces précédentes déterminations provenant de la littérature.

\includepdf[pagecommand={\pagestyle{headings}},scale=1.,offset=0 -5,pages={1-25},
addtotoc={
1,subsubsection,3, Introduction,intropaper2,
2,subsubsection,3, Calibration et  profils \Ha,obspaper2,
3,subsubsection,3, Analyse des données, analysispaper2,
4,subsubsection,3, Relation de Tully-Fisher, tfpaper2,
6,subsubsection,3, Résumé et conclusions, conclpaper2,
7,subsubsection,3, Annexe A: Notes individuelles sur les galaxies, notespaper2,
9,subsubsection,3, Annexe B: Tableaux, tablespaper2,
20,subsubsection,3, Annexe C:  Profils \Ha, profilespaper2,
24,subsubsection,3, Annexe D: Cartes et diagrammes position-vitesse, mappaper2,
25,subsubsection,3, Annexe E: Courbes de rotation, crmaper2
},
addtolist={
2,figure,{\underline{Article II}, Figure 1: \Ha~flux measured by GHASP versus \Ha~flux from James et al. (2004). The dashed line represents the linear regression on the data from which results our calibration. Top panel: calibration for the IPCS $512\times 512$. Bottom panel: calibration for the IPCS $256\times 256$.},fig1_p2,
3,figure,{\underline{Article II}, Figure 2: Dispersion in residual velocity field versus maximum velocity, subdivided by Hubble morphological type: black circles $0 \le t < 2$, red triangles $2 \le t < 4$, blue squares $4 \le t < 6$, green rhombuses $6 \le t < 8$ and pink stars $8 \le t < 10$. The dashed line represents the linear regression on the data. The points above the dotted line are discussed in Section 3. UGC 3334 labelled with an arrow has actually a huge residual velocity dispersion of $54~km~s^{-1}$ (see Table B2).},fig2_p2,
4,figure,{\underline{Article II}, Figure 3: Top panel: kinematical versus morphological (HyperLeda) position angles of the major axis. Galaxies for which no accurate morphological position angle has been computed are shown by red open circles; galaxies having an inclination lower than 25\degr~are displayed by blue squares; the other galaxies are represented by black circles. Bottom panel: histogram of the variation between kinematical and morphological position angles. The red hash, blue hash and residual white represent, respectively, the galaxies for which no accurate position angle has been measured, for which inclination is lower than 25\degr~and the other galaxies of the sample.},fig3_p2,
5,figure,{\underline{Article II}, Figure 4: Top panel: kinematical versus thick disc morphological inclinations. Middle panel: kinematical versus thin disc morphological inclinations. Top and middle panels: galaxies for which no accurate morphological position angle has been computed are shown by red open circles; galaxies with a difference between the kinematical and morphological position angles larger than 20\degr~are displayed with blue squares; the other galaxies are represented by black circles. Bottom panel: histogram of the variation between kinematical and morphological inclinations. The red hash, blue hash and residual white represent, respectively, the galaxies for which no accurate position angle has been measured, for which the difference between the kinematical and morphological position angles is larger than 20\degr~and the other galaxies of the sample.},fig4_p2,
5,figure,{\underline{Article II}, Figure 5: Tully-Fisher relation for our sample of galaxies. The solid line
represents the $B$ magnitude Tully-Fisher relation determined by Tully \& Pierce (2000) from nearby galaxies in clusters (Ursa Major, Pisces filament, Coma). Top panel: subdivided by inclination -- low-inclination galaxies ($i < 25$\degr): blue squares; other galaxies ($i \ge 25$\degr): black circles. Middle panel: subdivided by $V_{max}$ flags -- $V_{max}$ reached: black dots, large size; $V_{max}$ probably reached: blue squares, medium size; $V_{max}$ probably not reached: red triangles, small size. Bottom panel: subdivided by morphological type -- black circles from 0 to 2; red triangles from 2 to 4; blue squares from 4 to 6; green rhombuses from 6 to 8; pink stars from 8 to 10 and the dashed line represents the best linear fit to the data.},fig5_p2,
9,table,{\underline{Article II}, Table B1: Calibration parameters.},logpaper2,
12,table, {\underline{Article II}, Table B2: Model parameters.},modparpaper2,
16,table, {\underline{Article II}, Table B3: Galaxy parameters.},galparpaper2,
20,figure,{\underline{Article II}, Figure C1: Integrated \Ha~profiles. The profiles have been displayed over three times the spectral range ($\sim25~A$, top label or $\sim1100~km~s^{-1}$, bottom label). The instrumental intensity in photoelectron per second and per channel is given on the left-hand side Y-axis. The calibrated intensity is displayed on the right-hand side Y-axis. The dashed vertical line indicates the systemic velocity provided by our kinematical models (see Table B2).},figc1_p2,
24,figure,{\underline{Article II}, Figure D1: UGC 508. Top left-hand panel: XDSS blue band image. Top right-hand panel: \Ha~velocity field. Middle left-hand panel: \Ha~monochromatic image. Middle right-hand panel: \Ha~residual velocity field. The white and black cross is the kinematical centre. The black line is the major axis, its length represents the $D_{25}$. Bottom panel: position-velocity diagram along the major axis (full width of 7 pixels), arbitrary flux units. The red line plots the rotation curve computed from the model velocity field along the major axis (full width of 7 pixel).},figd1_p2,
25,figure,{\underline{Article II}, Figure E1: From top left-hand panel to bottom right-hand panel: \Ha~rotation curve of UGC 508, UGC 528, UGC 763, UGC 1117, UGC 1256 and UGC 1736.},fige1_p2
}]{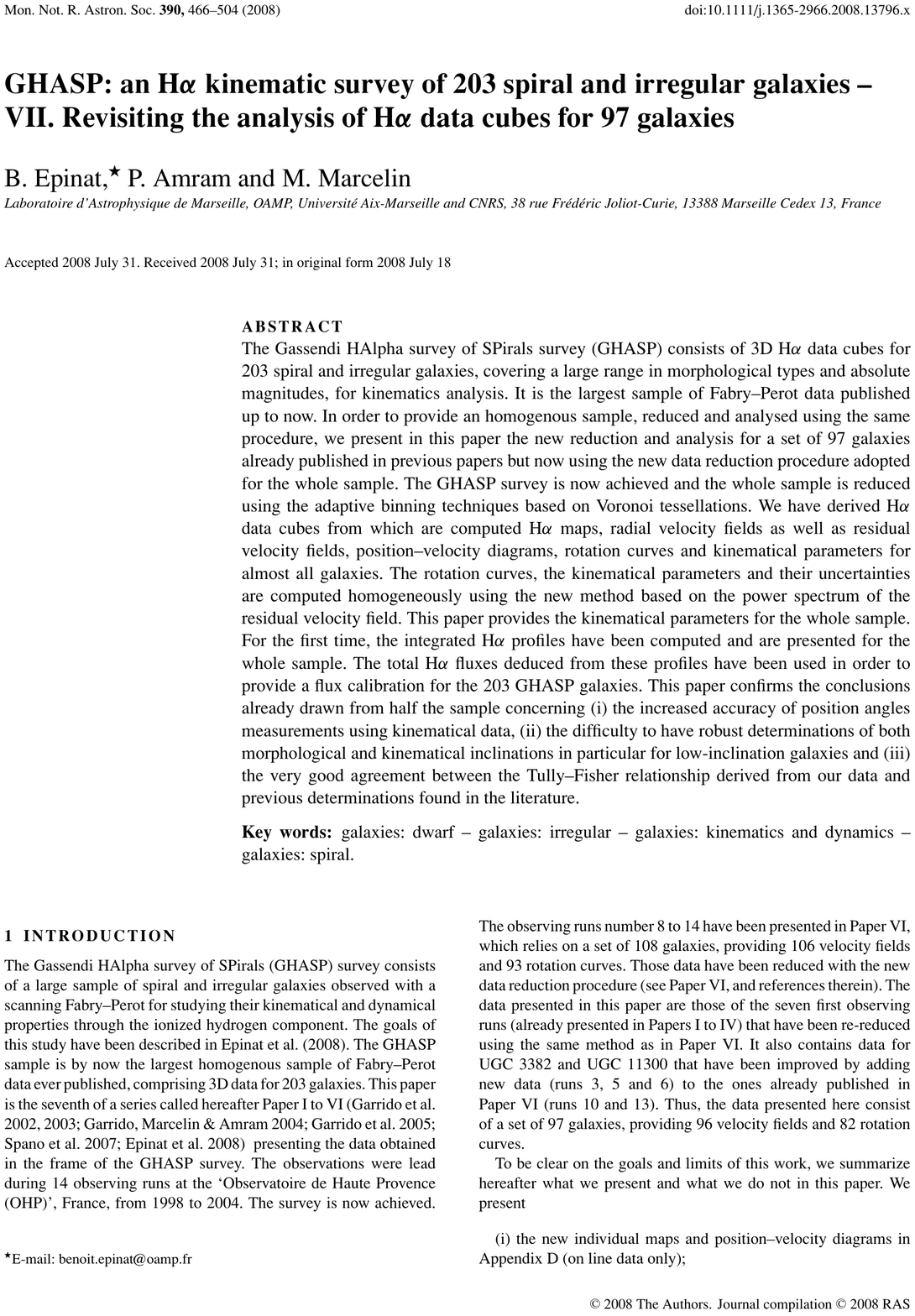}

\section{Exploitation des données cinématiques}
\label{etude_dynamique}

Les données de l'échantillon GHASP sont désormais réduites de manière homogène et les paramètres de projection cinématiques sont déterminés pour l'ensemble de l'échantillon. L'étude cinématique et dynamique de l'échantillon complet peut désormais être effectuée.
Des études ont déjà été réalisées par \citet{Garrido:2005,Spano:2008} sur des parties de l'échantillon. Des études similaires en cours sur l'ensemble de l'échantillon sont présentées dans cette partie.

\subsection{\'Etude de la forme des halos de matière sombre}

J'ai participé à une étude présentée dans l'Annexe \ref{analyse_ghasp} sur les modèles de masse de $36$ galaxies de l'échantillon \citep{Spano:2008}. Les courbes de rotation des galaxies GHASP sont combinées à des données radio lorsque celles-ci existent dans la littérature. Cela permet d'accroître l'extension des courbes afin de mieux contraindre les modèles. Ces modèles décomposent les courbes de rotation en une composante lumineuse déduite d'observations photométriques, une composante gazeuse déduite des observation du gaz HI et d'un halo de matière sombre. Les données photométriques sont elles-mêmes décomposées en un disque confiné dans un plan et en un bulbe sphérique.
Deux types de halos de matière sombre ont été utilisés: un modèle de sphère isotherme et un modèle de Navarro-Frenk-White suggéré par les simulations cosmologiques.
Les observations photométriques des galaxies GHASP auxquelles j'ai participé sont actuellement en cours à l'Observatoire de Haute Provence.
Les données photométriques ne sont donc pas disponibles pour tout l'échantillon GHASP. C'est pourquoi cette étude a été restreinte à $36$ galaxies.
Les principaux résultats sur ce sous-échantillon sont (i) que le modèle isotherme ajuste mieux les données que le modèle Navaro-Frenk-White et (ii) que quel que soit le modèle, le rayon de c\oe ur des halos est corrélé à leur densité centrale: les galaxies de faible luminosité ont un petit rayon de c\oe ur et une grande densité centrale alors que les halos de galaxies plus lumineuses ont un rayon de c\oe ur plus grand et une densité centrale plus faible. Il en résulte que la densité de surface des halos est à peu près constante quel que soit le modèle et quelle que soit la magnitude absolue, et donc le type de la galaxie.
\par
Suite à cette étude, j'ai travaillé sur de nouveaux programmes permettant d'ajuster d'autres types de halos de matière sombre. En particulier, j'ai co-encadré deux stages de Master première année (\'Elodie Giovannoli, Loïc Guennou) afin de tester respectivement un modèle de type Einasto \citep{Merritt:2006} et un modèle avec un potentiel de Yukawa \citep{Piazza:2003}. Ces deux modèles possèdent des paramètres libres additionnels par rapport aux modèles utilisés par \citet{Spano:2008} et sont donc moins contraints. Une investigation plus poussée de ces modèles est encore nécessaire.
\par
Des routines permettant l'ajustement d'un modèle de disque exponentiel afin de palier au manque de données photométriques ont été écrites dans le but de réaliser une étude préliminaire des modèles de masse de l'ensemble de l'échantillon GHASP.

\subsection{\underline{Article III:}~\textit{En préparation} - GHASP IX: Kinematical analysis of the whole sample}

J'ai initié des études cinématiques à partir de l'échantillon GHASP complet durant ma thèse. Je présente ces études sous forme d'article en préparation \citep{Epinat:prep} dont la forme finale sera probablement très différente de la forme actuelle puisque chaque domaine d'investigation sera développé pour donner éventuellement lieu, au final, à plusieurs articles disjoints.
\par
Ce travail est en cours. L'échantillon GHASP complet compte $203$ galaxies spirales et irrégulières principalement dans des environnements peu denses et est désormais réduit de manière homogène. L'analyse cinématique de l'ensemble de l'échantillon GHASP est présentée. Nous confirmons les résultats obtenus par \citet{Garrido:2005} à partir de la moitié de l'échantillon: (i) le gaz ionisé est présent sur l'ensemble du disque optique; (ii) l'asymétrie cinématique croît pour les galaxies plus bleues, plus faiblement lumineuses et de type plus tardif ce qui met en évidence que les galaxies en rotation lente ne sont pas capable d'homogénéiser leur distribution de masses; (iii) en moyenne, la pente interne des courbes de rotation décroît lorsque la luminosité diminue, lorsque le type morphologique devient tardif ou lorsque l'asymétrie cinématique croît; (iv) la pente externe des courbes de rotation tend à croître lorsque l'on considère des galaxies plus faiblement lumineuses, de type plus tardif ou présentant une plus forte asymétrie cinématique. Ces résultats sont préliminaires en vue d'étudier la possibilité d'ajuster un Courbe de Rotation Universelle \citep{Persic:1996,Catinella:2006,Noordermeer:2007}. En plus de ces analyses déjà ébauchées par le passé, une comparaison des profiles \Ha~et HI intégrés est menée: l'accord est correct pour plus de 75\% de l'échantillon. Par ailleurs, les signatures de barres sur les champs de vitesses sont étudiées à partir de simulations numériques \citep{Hernandez:thesis} en vue de futures études sur les échantillons GHASP et BH$\alpha$Bar (échantillon de galaxies barrées). Nous montrons que la présence d'une barre devrait être détectable à partir du champ de vitesses \Ha. Les cartes de dispersion de vitesses sont également présentées ainsi que les profils de dispersion de vitesses qui en ont été déduits (voir l'Annexe \ref{annexe_disp} de la thèse). \`A partir de ces données, nous montrons que la dispersion de vitesses du gaz est assez homogène pour l'échantillon local GHASP et qu'elle est plus faible que la dispersion de vitesses stellaire. Seule une minorité des galaxies présente un pic de dispersion central ou bien d'autres caractéristique marquées. L'effet du seeing (``beam smearing effect'') sur les cartes de dispersion de vitesses est également présenté.

\includepdf[pagecommand={\pagestyle{headings}},scale=1.,offset=0 0,pages=-,
addtotoc={
1,subsubsection,3, Introduction,intro,
2,subsubsection,3, Critère d'isolement,isol,
3,subsubsection,3, Profils \Ha~intégrés, prof,
3,subsubsection,3, Analyse des courbes de rotation, shape,
7,subsubsection,3, Signature cinématique des barres, bar,
8,subsubsection,3, Dispersion de vitesses du gaz, disp,
8,subsubsection,3, Conclusions, concl,
10,subsubsection,3, Annexe A: Dispersion de vitesses du disque projetée dans le plan, appendix_disp,
10,subsubsection,3, Annexe B: Cartes et profils radiaux de dispersion de vitesses, appendix_dispmaps
},
addtolist={
2,figure,{\underline{Article III}, Figure 1: Histogram of the isolation criterion for 121 out of 203
GHASP galaxies for which the systemic velocity is lower than $800~km~s^{-1}$.},fig1_prep,
3,figure,{\underline{Article III}, Figure 2: Velocity field asymmetry versus rotation curve asymmetry.},fig2_prep,
4,figure,{\underline{Article III}, Figure 3: Asymmetry as a function of (top) magnitude, (middle) morphological type and (bottom) $B-V$ color. Black squares are barred galaxies (B), green triangles are softly barred galaxies (AB) and red stars are unbarred galaxies (A).},fig3_prep,
4,figure,{\underline{Article III}, Figure 4: Comparison of the optical radius and the last radius of the rotation curve with a subdivision by Hubble morphological type: black circles $t\le2$, red triangles $2<t\le4$, blue squares $4<t\le6$, green rhombuses $6<t\le8$ and pink stars $t >8$. Opened symbols are galaxies that are larger than the GHASP field-of-view. The $y=x$ line is plotted as a reference. },fig4_prep,
4,figure,{\underline{Article III}, Figure 5: Tully-Fisher like relation with radius. Only galaxies for which the maximum velocity is reached or probably reached are displayed. Star symbols are used for unbarred, triangles for softly barred and squares for barred galaxies.},fig5_prep,
5,figure,{\underline{Article III}, Figure 6:  Inner slope as a function of (top) magnitude, (middle) morphological type, and (bottom) velocity field asymmetry. The left column shows non normalized inner slope whereas the right column shows the inner slope normalized with the maximum velocity and the optical radius. Black open squares are for barred galaxies (B), green triangles are for softly barred galaxies (AB) and red stars are for unbarred galaxies (A).},fig6_prep,
6,figure,{\underline{Article III}, Figure 7: Outer slope as a function of (top) magnitude, (middle) morphological type, and (bottom) velocity field asymmetry. The left column shows non normalized inner slope whereas the right column shows the inner slope normalized with the maximum velocity and the optical radius. Black open squares are for barred galaxies (B), green triangles are for softly barred galaxies (AB) and red stars are for unbarred galaxies (A).},fig7_prep,
7,figure,{\underline{Article III}, Figure 8: Modeled bar galaxy (Hernandez 2005c) with an inclination of 45\degr. Left: continuum. Right: gas velocity field. The scale is $0.1~kpc$ per pixel.},fig8_prep,
7,figure,{\underline{Article III}, Figure 9: Bar evolution. Top: position angle of the kinematical minor axis as a function of the position angle of the bar in the plane of the galaxy and bar strength. Bottom: position angle of the kinematical minor axis as a function of the position angle of the bar in the plane of the sky for only 180\degr. Green dotted line: inclination is 60\degr. Red dashed line: inclination is 30\degr.},fig9_prep,
9,figure,{\underline{Article III}, Figure 10: From left to right: UGC 5786 monochromatic map, velocity field, velocity dispersion and velocity dispersion modeled from the velocity field and the seeing.},fig10_prep
}]{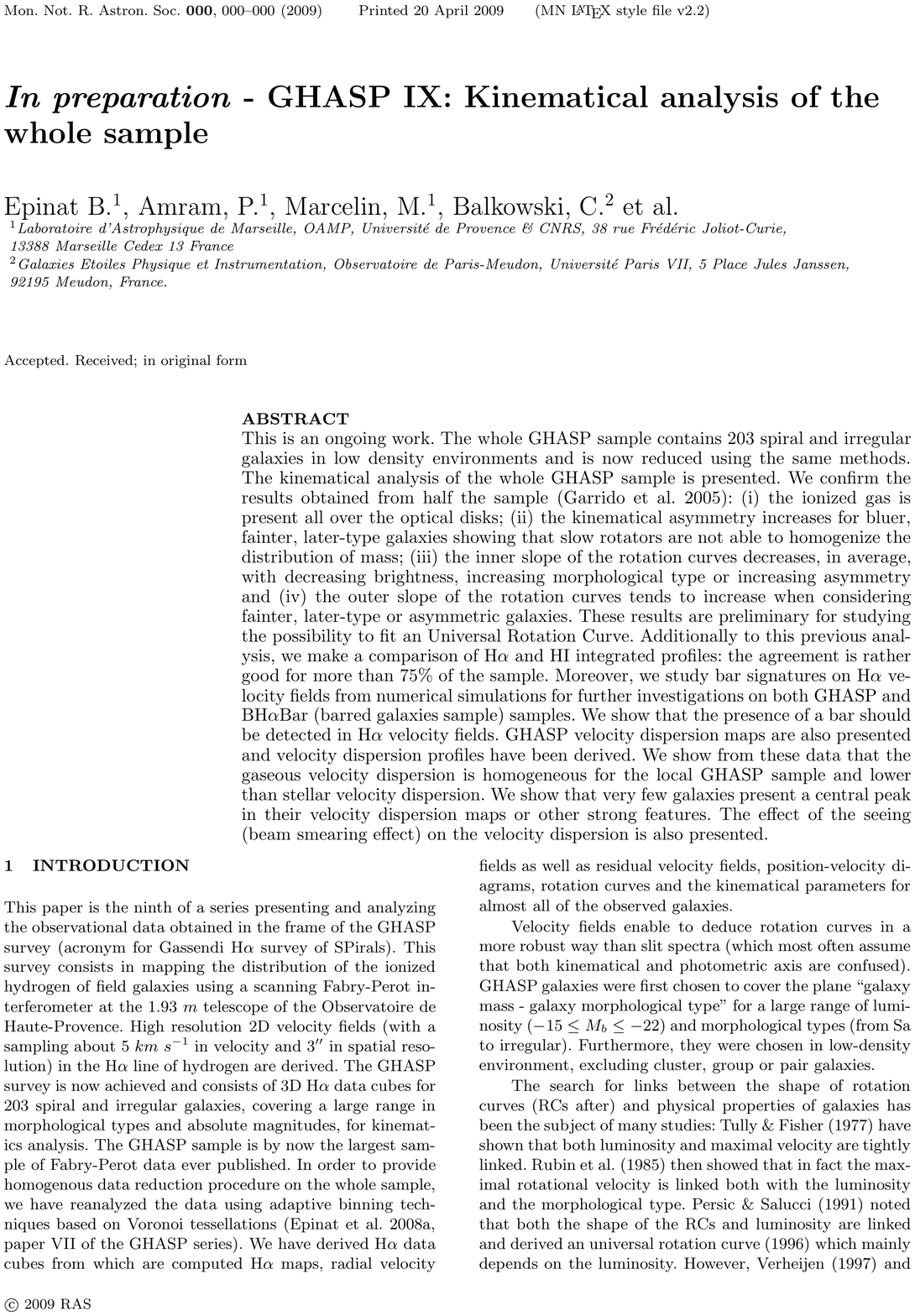}

\chapter{Cinématique des galaxies à grand décalage spectral}
\label{etudes_highz}
\minitoc
\textit{Ce chapitre expose les objectifs de l'étude de la cinématique résolue des galaxies à grand décalage spectral. De nouvelles données cinématiques obtenues par SINFONI sur le VLT sont présentées et interprétées. Enfin, un article présente l'utilisation de l'échantillon cinématique de galaxies proches GHASP afin de simuler des galaxies observées à grand décalage spectral. Les biais d'observations sont ainsi mis en évidence et la comparaison des propriétés dynamiques des galaxies locales et des galaxies à grand décalage spectral est faite en s'appuyant sur les observations de galaxies à grand décalage spectral de la littérature. {\bf Cet article est un des points clés de cette thèse}.}
\hl
\section{Problématique des galaxies à grand décalage spectral}

\subsection{Cosmologie moderne}
Le développement de la cosmologie moderne a été rendu possible grâce à des avancées théoriques, avec le développement de la théorie de la relativité générale par Einstein en 1915, et observationnelles, lorsque Edwin Hubble mit fin en 1924 au débat quant à la nature extragalactique des nébuleuses spirales grâce à l'observation d'étoiles variables dans la galaxie NGC 6822.
\par
\`A partir des équations de la relativité générale, le cadre théorique prend forme. Le principe cosmologique énoncé par Einstein repose sur le principe Copernicien (l'homme n'a pas une position privilégiée et l'Univers doit donc être partout le même) et stipule que l'Univers est uniforme et isotrope et que le temps est universel. Cela permet de définir des métriques et ainsi, à partir des équations d'Einstein, d'établir des relations théoriques entre la courbure de l'Univers et son contenu en matière-énergie.
La seule métrique générale à respecter le principe cosmologique est la métrique dite de Friedmann-Robertson-Walker:
$$ds^2=c^2dt^2-R(t)^2\left[\frac{dr^2}{1-kr^2}+r^2(d\theta^2+\sin^2{\theta}d\phi^2) \right]$$
Les distances sont décomposées en un produit entre un facteur d'échelle $R(t)$ dépendant du temps et une coordonnée comobile indépendante du temps. La géométrie de l'Univers dépend de la valeur du paramètre $k$.
\par
\begin{figure}[h]
\begin{center}
\includegraphics[width=10cm]{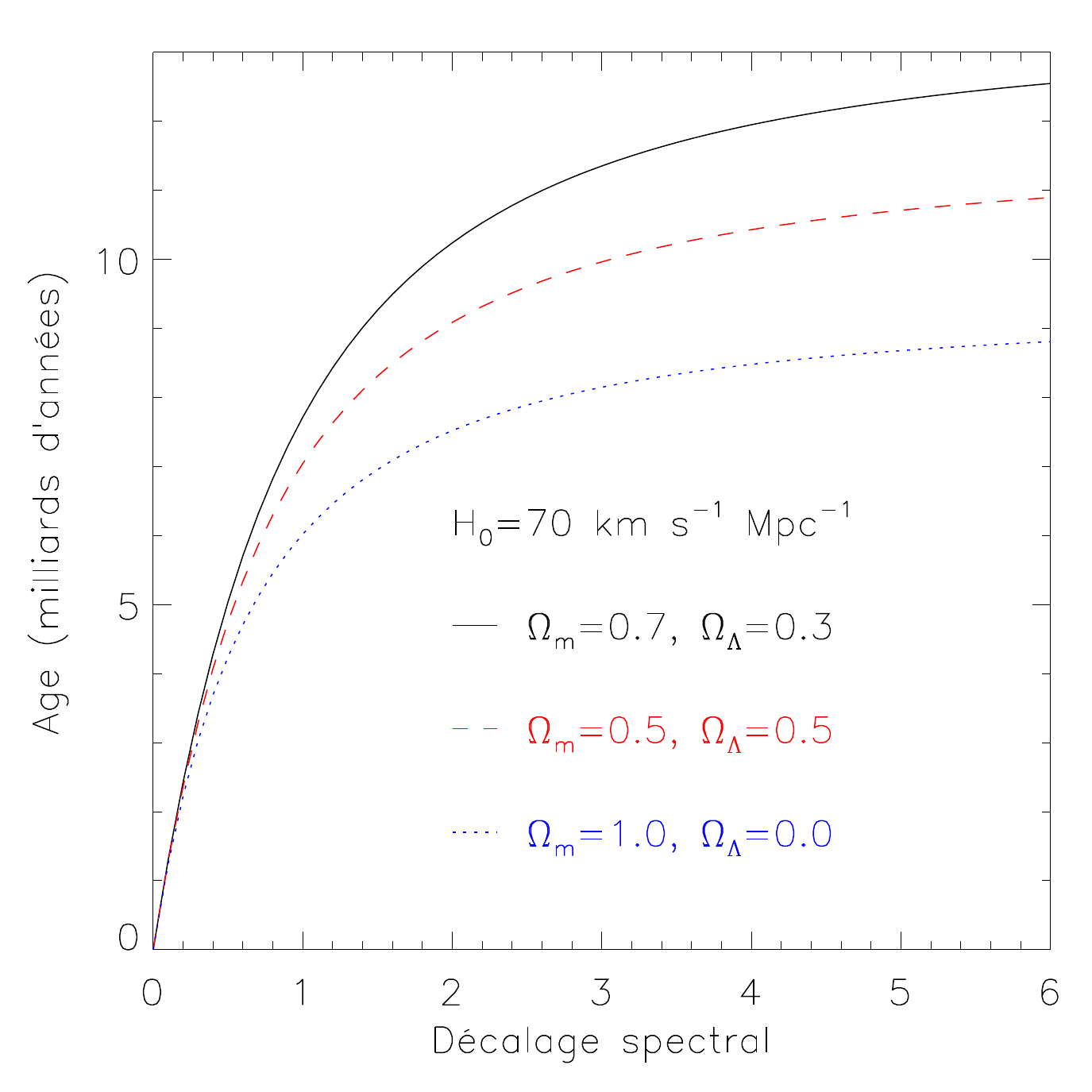}
\caption{Correspondance entre le décalage spectral et l'âge des galaxies (par rapport à l'époque actuelle)}
\label{z_t}
\end{center}
\end{figure}

La découverte de Hubble, quant à elle, est la première d'une série de découvertes observationnelles. Elle permit à Hubble de montrer que l'Univers est en expansion et d'estimer celle-ci par la constante de Hubble, alors estimée à $H_0=540~km~s^{-1}~Mpc^{-1}$. Cette découverte est explicable dans le cadre de la cosmologie, le paramètre de Hubble est alors égal à:
$$H(t)\equiv\frac{\dot{R}(t)}{R(t)}$$
La constante de Hubble est la valeur actuelle de la vitesse d'expansion de l'Univers (à $t=0$).
Dès lors les observations de galaxies distantes se sont multipliées afin de déterminer les paramètres cosmologiques avec la précision la meilleure possible.
Ces paramètres cosmologiques ont été introduits par les divers modèles d'Univers. Ils consistent en:
\begin{itemize}
\item le facteur de décélération $q_0=-\frac{\ddot{R}}{\dot{R}^2}(t=0)=-\frac{\ddot{R}(t=0)}{H_0^2}$
\item le paramètre de densité $\Omega_{m}$, défini comme le rapport de la densité de matière-énergie actuelle de l'Univers $\rho_0$ sur la densité critique $\rho_c=3 H_0^2/8\pi G$, $G$ étant la constante universelle de gravitation
\item la constante cosmologique $\Lambda$ qui intervient dans le calcul du paramètre de densité du vide $\Omega_{\Lambda}=\Lambda/3H_0^2$.
\end{itemize}
Pour un Univers plat, on a $\Omega_{m}+\Omega_{\Lambda}=1$.
Actuellement, la détermination des paramètres cosmologiques à partir des observations de WMAP
donne $H_0=72\pm5~km~s^{-1}~Mpc^{-1}$ et $\Omega_{m}=0.27\pm0.04$. De plus, les modèles privilégiés actuellement sont des modèles d'Univers plat.
Ces valeurs ont été utilisées pour les études présentées dans cette thèse. Elles permettent de dater les galaxies observées à un certain décalage spectral en calculant le temps de parcours des photons dans le modèle cosmologique adopté. La Figure \ref{z_t} présente la relation entre le décalage spectral et l'âge des galaxies pour trois cosmologies différentes.
On parle de grand décalage spectral à partir de $1$ et de décalage spectral intermédiaire entre $0.1$ et $1$.

\subsection{Histoire de l'Univers}
\begin{figure}[h]
\begin{center}
\includegraphics[width=12cm]{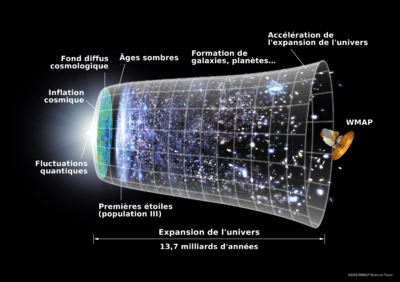}
\caption{Vue d'artiste de l'évolution de l'Univers. Crédits: NASA/WMAP science team.}
\label{evolution_univers}
\end{center}
\end{figure}

D'après ces modèles cosmologiques d'évolution de l'Univers, celui-ci est né il y a $13.7\pm0.13$ milliards d'années à partir d'une singularité lors du Big Bang.
Après le Big Bang, l'Univers s'est refroidi et neutrons et protons se sont combinés pour former de l'hydrogène neutre. Plus tard, les premières étoiles se sont formées par effondrement des nuages d'hydrogène. Le flux ultraviolet alors émis par ces étoiles ionisa à nouveau l'Univers le rendant ainsi transparent. La datation exacte de la ré-ionisation de l'Univers est encore très incertaine ($6<z<14$), c'est pourquoi les preuves observationnelles de cette ré-ionisation sont aujourd'hui recherchées \ref{evolution_univers}).

\subsection{Galaxies lointaines et spectrographie}

Jusqu'à présent, les observations de galaxies à moyen et grand décalage spectral utilisaient des spectrographes à longue fente. Ces observations ont permis d'obtenir des échantillons de galaxies conséquents \citep{Erb:2003,Erb:2004,Weiner:2006}. Cependant, la largeur de la fente est bien souvent trop importante pour pouvoir réaliser un échantillonnage spatial fin: les points hors de l'axe de la fente contribuent fortement au spectre mesuré et la courbe de rotation obtenue est biaisée \citep{Weiner:2006}. De plus, les erreurs d'alignement de la fente avec le grand axe cinématique sont fréquentes puisqu'il n'est pas aisé de mesurer ce paramètre à partir d'observations photométriques peu résolues.

L'utilisation de spectrographes à champ intégral tels que SINFONI ou FLAMES/GIRAFFE
sur le VLT et OSIRIS
sur un des télescopes de $10~m$ de l'observatoire du Keck a été une avancée significative puisqu'elle a permis l'observation résolue de la cinématique de galaxies jusqu'à un décalage spectral de 3 \citep{Forster-Schreiber:2006,Genzel:2006,Law:2007,Yang:2008}. Grâce à ces observations, l'information cinématique est disponible sur l'ensemble du champ. Il devient donc envisageable d'utiliser des méthodes sophistiquées afin de corriger la courbe de rotation de ces galaxies de la contribution des points hors du grand axe. Cependant, étant donné que la taille des dégradations dues à la turbulence atmosphérique est du même ordre de grandeur que les galaxies observées, ces corrections doivent être validées et les biais estimés.

Un des objectifs de ces observations est de pouvoir détecter des proto-galaxies dont les caractéristiques seraient bien différentes des galaxies connues à ce jour.
L'utilisation combinée d'imagerie à haute résolution (observation à partir de l'espace ou bien utilisant l'optique adaptative) avec des observations cinématiques est actuellement le meilleur moyen pour répondre à cet objectif. Cependant, le faible signal en provenance des objets les plus lointains constitue la principale limitation.
Toutefois, on peut d'ores et déjà sonder l'évolution des structures à travers les âges et, en particulier, essayer de comprendre si les galaxies actuelles se sont formées par accrétion lente de matière, par fusion de galaxies plus petites ou par d'autres phénomènes. L'évolution du support gravitationnel des structures peut également être étudiée afin de déterminer l'époque à laquelle ce support est devenu organisé.

\section{Nouvelles observations de galaxies à décalage spectral supérieur à un avec SINFONI}
\label{sinfoni_donnees}

\subsection{Les caractéristiques du spectrographe SINFONI}

SINFONI est un spectrographe à champ intégral fonctionnant dans le proche infrarouge ($1.1-2.45~\mu m$) installé au foyer Cassegrain du VLT UT4 ($8~m$, au CHILI) qui peut utiliser un module d'optique adaptative. Cet instrument utilise un découpeur d'image (``image slicer'') afin de placer tout le champ le long d'une fente virtuelle et de disperser la lumière avec un réseau. Quatre réseaux sont employés afin de couvrir quatre bandes spectrales (J, H, K, H+K) avec quatre résolutions spectrales différentes (respectivement $2000$, $3000$, $4000$ et $1500$). Selon le mode d'utilisation de l'instrument plusieurs tailles de pixels peuvent être utilisées. Généralement, lors d'observations sans optique adaptative, un pixel de $0.25''$ pour un champ de $8''\times 8''$ est utilisé alors que les observations avec optique adaptative utilisent un pixel de $0.1''$ pour un champ de $3''\times 3''$. \'Etant donné que la fente doit être échantillonnée sur au moins deux pixels afin de respecter le critère de Shannon pour l'échantillonnage spectral, l'échantillonnage spatial dans la direction de largeur de la fente est deux fois plus fin. Ainsi généralement les données sont interpolées suivant l'autre direction (la longueur) pour fournir un échantillonnage spatial de $0.125''$ ou de $0.05''$ selon le mode d'observation.
\par
\'Etant donnée l'amplitude spectrale, il est fréquent que plusieurs raies d'émission soient présentes. En particulier, lorsque la raie \Ha~est étudiée, les doublets d'azote [NII] ($6548$ et $6584$ \AA) et du soufre [SII] ($6716$ et $6731$ \AA) sont également observés dans le domaine spectral.

\subsection{MASSIV: observation de galaxies à grand décalage spectral}
Durant ma thèse, j'ai pris part à la collaboration internationale MASSIV (PI: T. Contini).
Comme son nom l'indique, le programme MASSIV a pour objectif l'étude de l'assemblage de masse des galaxies en utilisant des données spectroscopiques obtenues avec SINFONI
sur le VLT.
Pour cela, environ $140$ galaxies sont choisies parmi les galaxies du catalogue VVDS
pour des décalages spectraux compris entre $1$ et $1.8$. L'échantillon de base du VVDS présente l'avantage d'être complet et représentatif des populations de galaxies lointaines. C'est le sondage spectroscopique le plus profond et couvrant le plus grand champ. Ainsi, pour chaque galaxie, un décalage spectral spectroscopique est déterminé avec précision \citep{Le-Fevre:2005}. Cela permet de sélectionner des galaxies dont les raies d'émission sont intercalées entre les plus fortes raies du ciel nocturne dans le proche infrarouge, ce qui est impossible à réaliser lorsqu'on ne dispose que du décalage spectral photométrique à cause de son incertitude .
\`A partir des observations du programme MASSIV, il sera possible de réaliser une description détaillée de la proportion des différents types dynamiques (disques en rotation, sphéroïdes, galaxies en fusion, ...) dans cette gamme de décalage spectral, et de suivre l'évolution des lois d'échelle comme la relation Masse-Métallicité et la relation de Tully-Fisher afin de contraindre les scénarios d'évolution des galaxies.

\subsection{Observations SINFONI et réduction des données}

Les données SINFONI présentées ici sont des données préliminaires au programme MASSIV.
De même que pour le programme MASSIV, ces sept galaxies ont été choisies dans l'échantillon VVDS. Elles ont été observées en bande H durant des périodes d'observations de quatre nuits, du $5$ au $8$ septembre 2005 et du $12$ au $15$ novembre 2006 sans optique adaptative avec un pixel de $0.250''\times0.125''$ et un champ carré de $8''$ de côté (voir Table \ref{sinfoni_table}).

\begin{table}[h]
\begin{center}
\begin{tabular}{cccccc}
\hline
VVDS ID & $\alpha$ & $\delta$ & z$_{VVDS}$ & Exposition & Seeing \\
& (J2000) & (J2000) & & (hours) & ($''$) \\
\hline
$020182331$ & $02:26:44.260$ & $-04:35:51.89$ & $1.2286$ & $3$ & $0.93$ \\
$020261328$ & $02:27:11.049$ & $-04:25:31.60$ & $1.5291$ & $1$ & $0.61$ \\
$220596913$ & $22:14:29.184$ & $+00:22:18.89$ & $1.2667$ & $1.75$ & $0.47$ \\
$220584167$ & $22:15:23.038$ & $+00:18:47.01$ & $1.4637$ & $1.75$ & $0.77$ \\
$220544103$ & $22:15:25.708$ & $+00:06:39.53$ & $1.3970$ & $1$ & $0.69$ \\
$220015726$ & $22:15:42.455$ & $+00:29:03.59$ & $1.3091$ & $2$ & $0.55$ \\
$220014252$ & $22:17:45.690$ & $+00:28:39.47$ & $1.3097$ & $2$ & $0.61$ \\
\hline
\end{tabular}
\end{center}
\caption{Paramètres des observations.}
\label{sinfoni_table}
\end{table}

La réduction des données SINFONI a pour but de créer un cube de données calibré en longueur d'onde, corrigé du flat instrumental et du biais de la caméra, et pour lequel les raies d'émission du ciel nocturne ont été soustraites. La réjection des rayons cosmiques doit également être effectuée puisque ces données sont obtenues par une caméra CCD. La réduction doit également permettre de combiner les différentes observations nécessaires pour pouvoir éviter d'avoir un nombre trop important de rayons cosmiques.
L'ESO
met à la disposition des utilisateurs de SINFONI les logiciels de réduction ESOREX et GASGANO qui utilisent tous deux les mêmes procédures de réduction des données. ESOREX utilise la ligne de commande alors que GASGANO utilise une interface graphique.
Les étapes les plus importantes de la réduction des données SINFONI sont la soustraction des raies du ciel nocturne et la combinaison des différentes expositions en un cube final. La soustraction du ciel est effectuée en utilisant les expositions adjacentes. Toutefois, d'autres méthodes sont possibles, comme la soustraction d'un spectre médian. La recombinaison des cubes est faite en utilisant les décalages relatifs des pointés du télescope par rapport à une étoile brillante de référence.
Par ailleurs, pour chaque observation de galaxies, une étoile standard tellurique est observée afin de permettre la calibration en flux. Pour l'étude cinématique qui est présentée ici, cette calibration en flux n'est pas primordiale. En revanche, les étoiles de référence sont utiles pour mesurer le seeing des observations afin de l'inclure dans les modèles.

\begin{figure}
\begin{center}
\includegraphics[width=12.1cm]{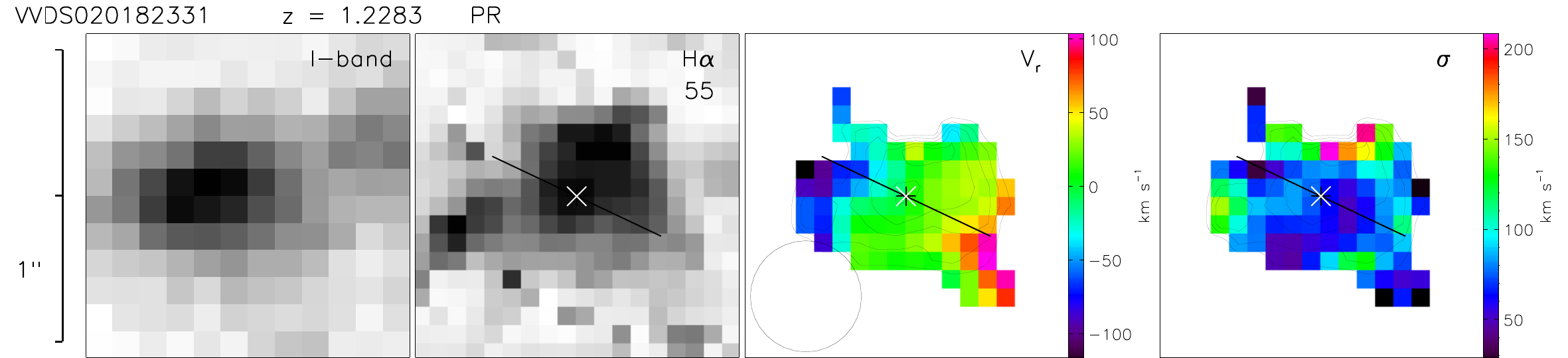}
\includegraphics[width=12.1cm]{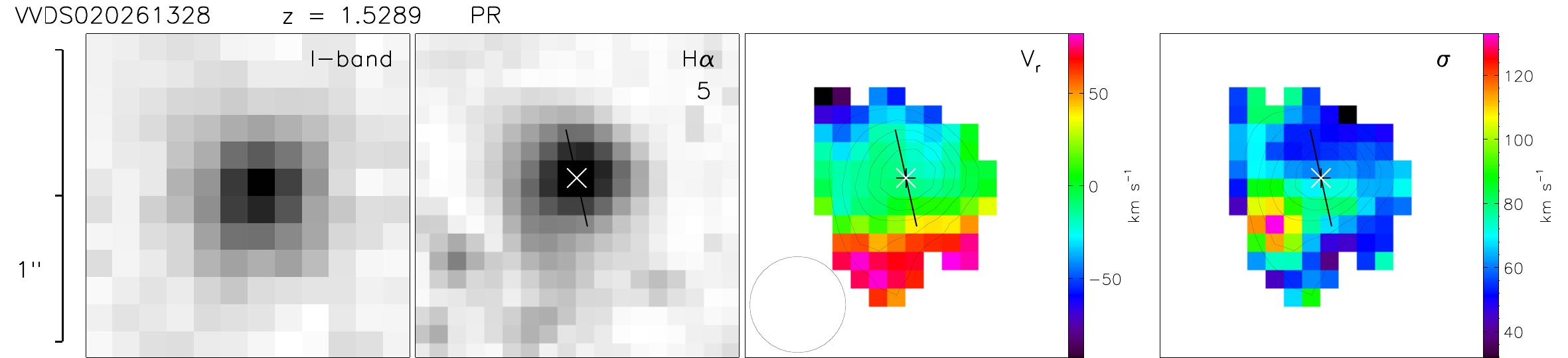}
\includegraphics[width=12.1cm]{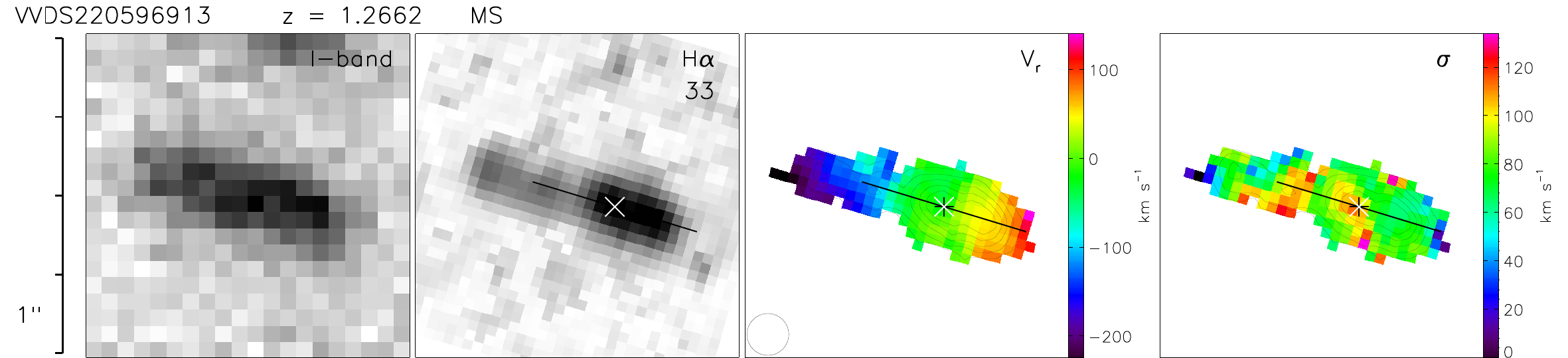}
\includegraphics[width=12.1cm]{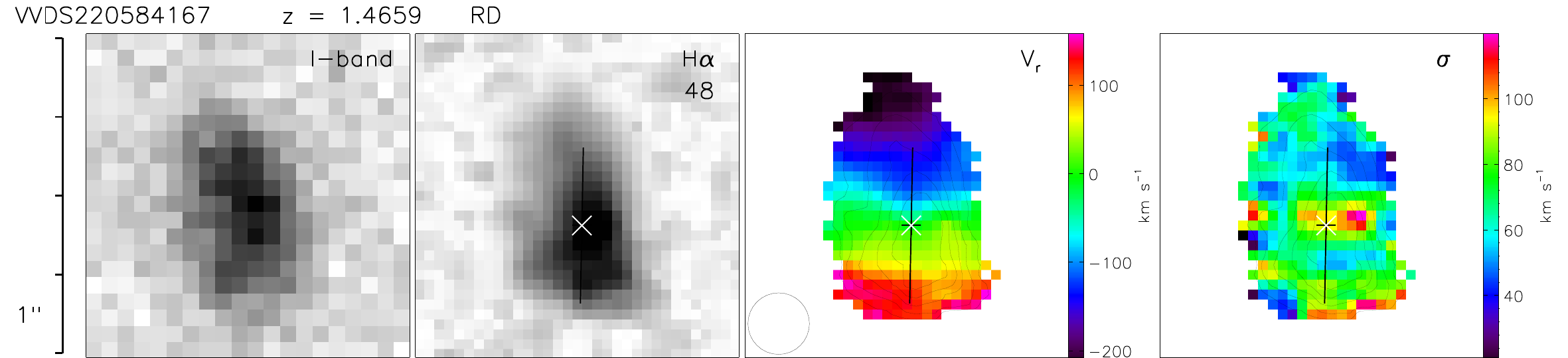}
\includegraphics[width=12.1cm]{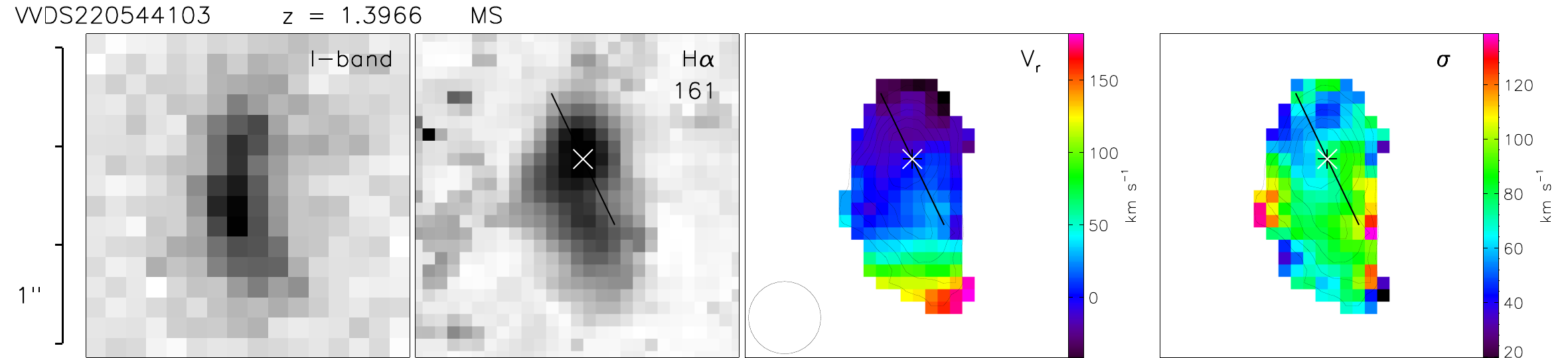}
\includegraphics[width=12.1cm]{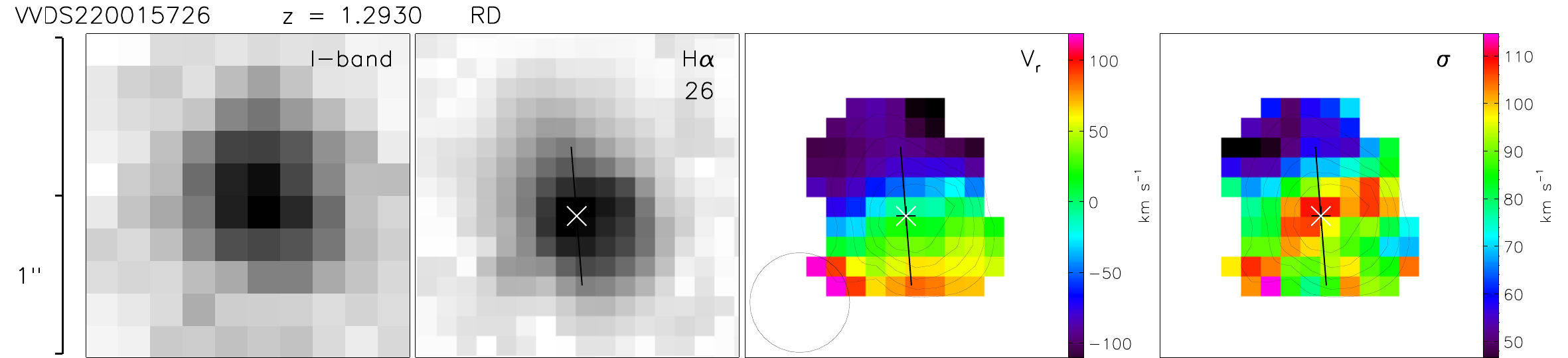}
\includegraphics[width=12.1cm]{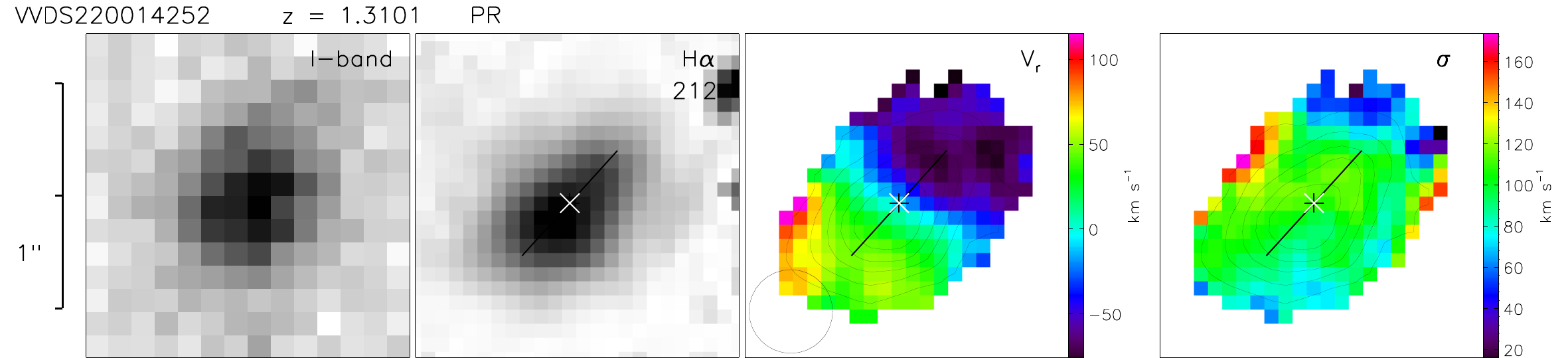}
\end{center}
\caption{De gauche à droite: image du CFHT en bande I, flux \Ha, champ de vitesses et carte de dispersion de vitesses. Le flux \Ha\ intégré est indiqué sur chaque carte ($10^{-17}~ergs~s^{-1}~cm^{-1}$). L'échelle est indiquée à gauche et la résolution spatiale (FWHM du seeing plus lissage spatial) est représentée par un cercle sur le champ de vitesses. Le Nord est en haut et l'Est est à gauche. Le centre utilisé pour la cinématique est indiqué par la double croix, et trait indique la position du grand axe du modèle. Les contours de la carte de flux sont superposés sur les cartes cinématiques. Le redshift déduit des données SINFONI est indiqué ainsi que la classification dynamique (RD: disque en rotation, PR: rotation perturbée, MS: système en fusion).}
\label{cartes_sinfoni}
\end{figure}

Les cartes cinématiques présentées en Figure \ref{cartes_sinfoni} ont été obtenues en utilisant un programme que j'ai développé sous IDL à partir des routines utilisées par le programme de réduction des données Fabry-Perot. Ce programme détermine les paramètres de la raie en utilisant un ajustement gaussien (voir Annexe \ref{spectre_vitesse}) et détermine également des cartes d'erreurs. Il permet également l'ajustement simultané de plusieurs raies. Un lissage spatial gaussien (largeur à mi-hauteur de $2$ pixels) a été effectué sur le cube avant le calcul des cartes cinématiques. Les cartes de dispersion de vitesses ont été corrigées de la résolution spectrale de SINFONI déterminée à partir du spectre de raies du ciel nocturne. Ce programme sera utilisé pour le programme MASSIV conjointement à GIPSY.
Les images de gauche ont été obtenues en bande I au CFHT. Ce sont les images avec le meilleur seeing ($<0.65''$) du ``CFHT legacy survey'' pour les galaxies du champs à deux heures et celle du ``CFH12K/CFHT'' \citep{McCracken:2003} pour les galaxies du champ à 22 heures.


\subsection{Analyse cinématique des galaxies observées par SINFONI}

\subsubsection{Méthode d'analyse}

J'ai contribué de manière significative à l'analyse cinématique de ces galaxies.
Pour chaque galaxie, l'inclinaison et l'angle de position du grand axe morphologique ont été estimés à partir des images en bande I avec le logiciel GALFIT \citep{Peng:2002}. Ce programme prend en compte la valeur du seeing afin d'ajuster un modèle. Les inclinaisons ont également été mesurées à partir de la mesure des rapports d'axe en utilisant la méthode décrite dans la section \ref{analysispaper4} de l'article de la partie \ref{ghasp_highz} et présentent un très bon accord avec les valeurs estimées par GALFIT. Pour les VVDS020261328 et VVDS220015726 pour lesquelles GALFIT détermine un inclinaison nulle, les inclinaisons déterminées à partir du rapport d'axe ont été utilisées.

\begin{table}[h]
\begin{center}
\begin{tabular}{cccccccc}
\hline
    VVDS ID & z$_{H\alpha}$ & Inclination & Position Angle & $r_t$ & $V_t$ & $\sigma_0$ & $V_{max}$ \\
            &               & (\degr)     & (\degr)        & ($''$)& ($km~s^{-1}$) & ($km~s^{-1}$) & ($km~s^{-1}$) \\
\hline
$020182331$ & $1.22832$ & $52$  & $244$ & $0.55$ & $134$ &  $71$ & $134$ \\
$020261328$ & $1.52891$ & $47$  & $192$ & $0.77$ & $195$ &  $55$ & $194$ \\
$220596913$ & $1.26619$ & $66$  & $253$ & $2.80$ & $325$ &  $76$ & $177$ \\
$220584167$ & $1.46588$ & $49$  & $178$ & $1.24$ & $280$ &  $47$ & $280$ \\
$220544103$ & $1.39659$ & $61$  & $205$ & $6.73$ & $762$ &  $70$ & $146$ \\
$220015726$ & $1.29300$ & $25$  & $184$ & $0.12$ & $323$ &  $38$ & $323$ \\
$220014252$ & $1.31014$ & $48$  & $137$ & $0.12$ & $103$ &  $92$ & $103$ \\
\hline
\end{tabular}
\end{center}
\caption{Paramètres des modèles cinématiques.}
\label{sinfoni_params}
\end{table}

Ces paramètres morphologiques ont servi de paramètres de départ à l'ajustement d'un modèle cinématique sur les champs de vitesses 2D. Ce modèle consiste en un disque fin en rotation dont la courbe de rotation est croissante dans les parties internes et plate à grand rayon. Il est similaire à celui utilisé par \citet{Wright:2007}:
$$V(r) = V_t \frac{r}{r_t}$$
lorsque $r < r_t$ et pour $r \ge r_t$:
$$V(r) = V_t$$
La méthode utilisée pour prendre en compte le seeing de l'observation est décrite dans l'article présenté dans la partie \ref{ghasp_highz}. Les paramètres du modèle sont l'inclinaison, l'angle de position du grand axe, la vitesse d'éloignement global de la galaxie (convertie en terme de redshift dans la Table \ref{sinfoni_params}), le centre cinématique, $V_t$ et $r_t$. \'Etant donné le faible nombre de pixels, l'inclinaison est difficilement contrainte. Elle a donc été fixée à la valeur déterminée par la morphologie. Cependant, lorsque cette inclinaison est nulle, une inclinaison de $10$\degr~a été utilisée. De plus, le centre a également été contraint à être situé au centre de l'émission \Ha\ (soit le pic, soit le centre des isophotes externes, voir Figure \ref{cartes_sinfoni}). \`A partir des meilleurs ajustements, une carte de dispersion de vitesses modèle a été calculée. Celle-ci ne contient que les effets liés au gradient de vitesse non résolu dû au manque de résolution spatiale (voir l'Annexe \ref{modelpaper4} de l'article présenté dans la partie \ref{ghasp_highz}). Cette carte de dispersion est soustraite quadratiquement aux cartes de dispersion de vitesses observées afin d'examiner la dispersion de vitesses intrinsèque de la galaxies.
La dispersion moyenne (pondérée par l'inverse de la carte d'erreur associée) $\sigma_0$ mesurée sur les cartes ainsi corrigées est présentée dans la Table \ref{sinfoni_params} avec les autres paramètres du modèle. La vitesse maximale est déterminée comme étant la vitesse atteinte par le modèle au point le plus éloigné du centre.
Cette modélisation a pour but d'aider dans l'interprétation cinématique des galaxies, afin de voir si leur cinématique peut être expliquée par un disque en rotation.

\subsubsection{Commentaires galaxie par galaxie}
Une interprétation de la cinématique individuelle de chaque galaxie à laquelle j'ai contribué est présentée.
\\
\par
\textbf{VVDS020182331}
\par
Le flux \Ha\ de cette galaxie est principalement émis au centre. Le champ de vitesses de cette galaxie est perturbé. Les résidus important d'une raie du ciel ($14605$\AA) perturbent la measure des paramètres des raies et sont à l'origine de la forte dispersion de vitesses ainsi que du maximum local du flux, tous deux au Nord. L'incertitude sur la mesure des vitesses n'est inférieure à $20$\kms\ que sur un diamètre d'une seconde d'arc au centre. Sur cette région, un modèle de disque en rotation est acceptable. La dispersion de vitesses corrigée est élevée ($71$\kms) et le support dynamique de cet objet n'est pas clairement dominé par la rotation ($V_{max}/\sigma_0 =1.9$). L'objet proche, visible sur l'image en bande I, n'est pas détecté en \Ha, ce qui indique que cet objet est soit une source d'avant ou d'arrière plan, soit un compagnon qui forme peu ou pas d'étoiles et avec lequel VVDS020182331 pourrait interagir. Cette dernière hypothèse pourrait expliquer que VVDS020182331 ait une rotation perturbée.
%
\\
\par
\textbf{VVDS020261328}
\par
La carte de flux est piquée au centre et présente une extension faiblement lumineuse au Sud. La carte de dispersion de vitesses présente un pic à $\sim120$\kms\ au Sud-Est. Le champ de vitesses montre clairement un gradient de vitesses d'environ $100$\kms\ sur un rayon de $0.8''$ qui est toutefois irrégulier. L'ajustement d'un modèle de disque en rotation suggère que seul le centre de la galaxie est détecté puisque le plateau n'est pas atteint. Il montre également que le pic de dispersion de vitesses ne peut pas être attribué aux effets du seeing. Cette galaxie est considérée comme un disque avec une rotation perturbée et dont le support dynamique est principalement dû à la rotation ($V_{max}/\sigma_0 =3.5$).
%
\\
\par
\textbf{VVDS220596913}
\par
La carte de flux \Ha\ présente deux composantes principales séparées de $12.5~kpc$ ($1.5''$) qu'on ne distingue pas sur l'image en bande I. La composante la plus lumineuse est située du côté Ouest et est elle-même constituée de deux pics d'intensité égale. L'autre composante située à l'Est est bien moins lumineuse. Elle présente un faible gradient de vitesses et deux pics de dispersion de vitesses ($\sim100$\kms). L'un correspond à la région la plus intense, l'autre est plus diffus et est situé sur la zone de transition entre les deux composantes principales où l'émission est la plus diffuse (rapport signal sur bruit d'environ quatre). Cette forte dispersion de vitesses pourrait être une signature de fusion comme c'est le cas pour le groupe compact de Hickson H31 \citep{Amram:2007}. Le gradient de vitesses de la composante faiblement lumineuse fait un angle de $20$\degr\ avec celui de la composante la plus brillante, ce qui suggère, de même que la distribution de flux irrégulière, que ce système est composé d'au moins deux galaxies en cours de fusion.
Bien qu'un unique disque en rotation composé de grands blocs (comme observés par \citealp{Bournaud:2008}) ne puisse être exclus (induisant $V_{max}/\sigma_0=2.3$), l'ajustement de la composante la plus brillante seule donne de meilleurs résultats. Cela suggère que cette composante est en rotation avec une vitesse de roation maximale de $\sim200$\kms. Le centre a été fixé entre les deux pics de flux de la composante brillante et l'inclinaison a été estimée à $55$\degr. Cet objet est donc classé comme étant un système en cours de fusion.
%
\\
\par
\textbf{VVDS220584167}
\par
VVDS220584167 est l'objet qui possède l'émission \Ha\ la plus étendue de l'échantillon. Il présente un pic de dispersion de vitesses proche du centre allongé. L'observation est affectée par de forts résidus dus à une raie du ciel ($16195$\AA) qui induisent de fortes incertitudes sur la mesure des paramètres des raies, plus particulièrement au Sud. Le champ de vitesses est asymétrique et la galaxie possède une morphologie irrégulière en bande I, ce qui suggère l'existence d'une forte barre. De plus, le décalage observé entre les morphologies en bande I et \Ha\ suggère qu'un violent épisode de formation stellaire est en cours. Le champ de vitesses est ajusté raisonnablement par un modèle de disque en rotation et le pic de dispersion de vitesses n'est pas expliqué par ce modèle puisque ce dernier ne simule pas l'effet de la barre sur le champ de vitesses. Le support dynamique de cet objet est dominé par la rotation ($V_{max}/\sigma_0 = 5.6$). Cette glalaxie est donc classée comme étant un disque en rotation possédant une barre.
%
\\
\par
\textbf{VVDS220544103}
\par
La carte de flux \Ha\ montre deux pics séparés d'environ $6~kpc$ ($0.75''$). Cette galaxie possède une morphologie irrégulière aussi bien en \Ha\ qu'en bande I. La variation de vitesse le long de la composante principale (Nord) est faible ($\sim 30$\kms) alors que celle de la composante la plus faible (Sud) est élevée ($\sim 120$\kms). Les deux gradients ont globalement la même direction (la vitesse augmente du Nord au Sud). Les lignes d'isovitesses sont perpendiculaires à la morphologie tordue et allongée de l'objet. Les pics les plus intenses de dispersion de vitesses sont situés sur les bords où le rapport signal sur bruit est le plus faible. Hormis ces pics, la carte de dispersion de vitesses présente un pic allongé proche du pic de flux \Ha. L'ajustement d'un modèle de disque en rotation au système entier induit $V_{max}/\sigma_0=2.1$, mais ne semble cependant pas optimum et l'hypothèse d'un objet constitué de deux galaxies (ou plus) en cours de fusion semble plus appropriée. C'est objet est donc classé comme un système en fusion.
%
\\
\par
\textbf{VVDS220015726}
\par
Les régions centrales ont un signal important alors que les régions avec un faible signal sont affectées par des résidus dus à la proximité de la raie \Ha\ avec deux raies de ciel très intenses ($15053$\AA et $15056$\AA).
La carte de flux est piquée au centre, au même endroit où la dispersion de vitesses est maximale ($\sim 100$\kms). Les cartes cinématiques de cette galaxie sont bien reproduites par un modèle de disque en rotation. En particulier, la dispersion de vitesses centrale est très bien expliquée par la faible résolution spatiale. VVDS220015726 est clairement dominée par la rotation ($V_{max}/\sigma_0=8.5$) puisque c'est le disque en rotation le plus rapide de l'échantillon ($V_{max}=323$\kms) mais aussi l'objet avec la plus faible dispersion de vitesses ($\sigma_0=38$\kms). Cette galaxie est classée comme étant un disque en rotation.
%
\\
\par
\textbf{VVDS220014252}
\par
La carte de flux \Ha\ présente un pic allongé ainsi qu'une émission diffuse dans les régions externes. Le pic ne correspond pas exactement au centre des isophotes externes. La carte de flux suggère également un bras à l'Ouest. Le champ de vitesses ressemble à celui d'un disque en rotation sauf à l'extrême Ouest, où les vitesses sont plus grandes qu'attendues. Un modèle de disque en rotation s'ajuste toutefois bien au champ de vitesses, en particulier lorsque le centre cinématique est fixé au centre des isophotes externes. La vitesse de rotation maximale est relativement basse ($103$\kms) mais le modèle montre qu'elle est atteinte proche du centre.
Le côté Nord-Ouest du champ de vitesses présente un maximum local suivi d'une faible chute de vitesses. Autours de ce maximum, les profils sont larges et asymétriques. Des profils larges (plus de $150$\kms) sont également observés au Nord-Est avec un rapport signal sur bruit plus grand que cinq.
La dispersion de vitesses n'est pas piquée au centre et est élevée partout. De plus, la dispersion de vitesses moyenne de cette galaxies est la plus grande de tout l'échantillon ($92$\kms). Le modèle montre clairement que la dispersion de vitesses n'est pas due à la résolution. Les profils larges et asymétriques observés laissent supposer qu'ils sont en fait des profils doubles non résolus, ce qui pourrait être une signature d'interactions \citep{Amram:2007}. De plus, le support dynamique n'est pas dominé par la rotation puisque $V_{max}/\sigma_0=1.1$. Cette galaxie est donc classée comme ayant une rotation perturbée, et connaissant potentiellement un épisode de fusions mineures.
%
\\

Finalement, sur l'ensemble de ces observations, seule une minorité est bien reproduite par un modèle de disque en rotation. En effet, des écarts au modèle et des pics de dispersion de vitesses sont observés pour presque toutes les galaxies présentées, sans qu'il soit pour autant possible de les attribuer au gradient de vitesse non résolu. Par ailleurs, la dispersion de vitesses est de l'ordre de $80~km~s^{-1}$, ce qui n'est pas habituel dans l'Univers local. Cela suggère que les disques sont de nature différente.
L'interprétation de ces observations pourrait être améliorée en utilisant des images observée à partir de l'espace ou avec optique adaptative.

\section{Observations cinématiques de galaxies à un décalage spectral voisin de 0.6 avec FLA\-MES/GIRAFFE}
\label{giraffe_donnees}

\subsection{Les caractéristiques du spectrographe FLAMES/GIRAFFE}

FLAMES/GIRAFFE (PI : F. Hammer) est un spectrographe à champ intégral fonctionnant dans le visible ($370-900~nm$) installé au foyer Nasmyth du VLT UT2 ($8~m$, au CHILI). FLAMES est le nom du système permettant de positionner les fibres dans un champ de vue de $25$ minutes d'arc de diamètre. GIRAFFE est le nom du spectrographe qui permet d'obtenir des résolutions spectrales allant de $7500$ à $30000$ grâce à deux réseaux. Trois modes d'utilisation sont possibles. Le premier permet de positionner $132$ fibres indépendantes dans le champ de vue (MEDUSA) afin d'observer des objets ponctuels. Le second consiste en un unique champ rectangulaire de $22\times14$ micro-lentilles (ARGUS). Enfin, le dernier mode (IFU) possède $15$ sous-champs de $20$ micro-lentilles qui peuvent être placés simultanément mais indépendamment dans le champ de vue de FLAMES. Pour ce dernier mode, l'échantillonnage spatial est de $0.52''$ par micro-lentille, ce qui induit un champ de $\sim3.12''\times2.08''$. C'est le mode utilisé pour observer la cinématique de galaxies lointaines, grâce au doublet [OII] ($3727$ et $3728.5$ \AA).

\subsection{\'Etude de la relation de Tully-Fisher avec le programme IMAGES}

L'objectif du programme IMAGES (PI : F. Hammer) est d’étudier l’assemblage de masses des galaxies depuis $z = 1$. Pour cela, un échantillon de galaxies avec un décalage spectral $0.4<z<0.75$ est en cours d'observation par l'instrument FLAMES/GIRAFFE. Cet échantillon compte actuellement $68$ galaxies \citep{Flores:2006,Puech:2006,Yang:2008,Neichel:2008,Puech:2008}.
\par
\`A partir de ces observations, l'évolution de relation de Tully-Fisher a été étudiée par \citep{Puech:2008}. J'ai contribué à cette étude en utilisant un partie réduite de l'échantillon GHASP afin d'évaluer la qualité de la détermination des vitesses de rotation de l'échantillon IMAGES. Pour cela, j'ai projeté les galaxies GHASP (la méthode est décrite dans la partie \ref{ghasp_highz}) à un décalage spectral de $0.6$ dans les conditions d'observation de l'échantillon IMAGES: un pixel de $0.52''$ et un seeing de $0.8''$. L'utilisation des données GHASP a permis de vérifier sur des données réelles les barres d'erreurs estimées à partir de simulations numériques. Le résultat principal de cette étude présentée en Annexe \ref{annexe_highz} est que, d'après la comparaison entre la relation de Tully-Fisher locale et celle obtenue à partir de l'échantillon IMAGES, les galaxies auraient doublé leur masse stellaire entre $z\sim0.6$ et $z\sim0$.

\section{\underline{Article IV:}~Evidence for strong dynamical evolution in disk galaxies through the last 11 Gyr. \emph{GHASP VIII: A local reference sample of rotating disk galaxies for high redshift studies}}
\label{ghasp_highz}
Un des objectifs de l'échantillon GHASP annoncés dans le chapitre \ref{ghasp_donnees} est la création d'un échantillon de référence afin de mieux comprendre l'étude des galaxies à grand décalage spectral et de les comparer aux galaxies locales.
L'échantillon GHASP étant désormais complet, j'ai
utilisé l'ensemble des données de l'échantillon afin de simuler un échantillon de référence de $153$ galaxies locales observées dans les mêmes conditions de résolution spatiale que les galaxies dont le décalage spectral est $z\sim1.7$. Ce travail est présenté sous forme d'un article\footnote{Monthly Notices of the Royal Astronomical Society} \citep{Epinat:2008c}.
\par
Les galaxies à grand décalage spectral sont observées avec une faible résolution spatiale à cause de leur éloignement. Nous avons utilisé les cubes de données \Ha~observés avec un \FP~de $153$ galaxies proches isolées sélectionnées dans le programme GHASP (Gassendi \Ha~survey of SPirals) afin de dissocier les effets de résolution spatiale de l'évolution cinématique des galaxies. Nous avons simulé des cubes de données de galaxies à un décalage spectral $z=1.7$ en utilisant un pixel de $0.125''$ et un seeing de $0.5''$ à partir desquels ont été déterminées des cartes de flux \Ha, de vitesses et de dispersion de vitesses. Nous avons montré que le gradient interne est affaibli et qu'il est responsable d'un pic de dispersion de vitesses. Des modèles simples de disques en rotation ont été ajustés à ces données possédant une faible résolution spatiale afin de déterminer les paramètres cinématiques des galaxies et leur courbe de rotation. La détermination de l'inclinaison n'est pas fiable et la position du centre est délicate. L'angle de position du grand axe est retrouvé avec une précision meilleure que $5$\degr\ pour $70$\%\ de l'échantillon. Ces modèles permettent également de retrouver statistiquement la vitesse de rotation maximale ainsi de la dispersion de vitesses intrinsèque. Ceci valide l'utilisation de la relation de Tully-Fisher pour les galaxies à grand décalage spectral. Nous avons cependant noté que le manque de résolution induit une pente plus faible à grand décalage spectral. Nous avons également conclu que les principaux paramètres cinématiques sont mieux contraints pour les galaxies dont le rayon est supérieur à trois fois la largeur à mi-hauteur du seeing. Les données simulées ont été comparées aux données de galaxies observées avec VLT/SINFONI, Keck/OSIRIS et VLT/GIRAFFE dans la gamme de décalage spectral $3>z>0.4$ permettant de suivre l'évolution de onze à quatre Gyr. Nous avons montré que, pour le galaxies dominées par la rotation, la classification basée sur l'utilisation du pic de dispersion de vitesses comme signature de diques en rotation peut s'avérer erronée pour les rotateurs lents ou présentant une courbe de rotation en corps solide. C'est le cas pour $\sim 30$\%\ de notre échantillon.
Nous avons mis en évidence que la grande dispersion de vitesses observée dans les galaxies à grand décalage spectral n'est pas reproduite avec l'échantillon local projeté, à moins d'utiliser la dispersion de vitesses non corrigée des effets de résolution. Cela signifie sans ambiguïté que, contrairement aux galaxies locales évoluées, il existe à grand décalage spectral au moins une famille de galaxies pour laquelle une fraction significative du support dynamique est due à des mouvements aléatoires.
Il faudra néanmoins s'assurer que ces propriétés ne sont pas dues à d'importants biais de sélection avant de pouvoir conclure que la formation d'un disque gazeux instable et transitoire est un processus commun dans la formation des galaxies.

\includepdf[pagecommand={\pagestyle{headings}},scale=1.,offset=0 0,pages={1-38},
addtotoc={
2,subsection,2, Introduction,intropaper4,
3,subsection,2, Des galaxies locales pour simuler des galaxies distantes,simgalpaper4,
4,subsection,2, \'Echantillon,datasetpaper4,
10,subsection,2, Signatures cinématiques de disques en rotation à grand décalage spectral,classpaper4,
11,subsection,2, Ajustement de modèles,fitpaper4,
13,subsection,2, Analyse, analysispaper4,
26,subsection,2,Discussion, discussionpaper4,
30,subsection,2,Conclusion, conclusionpaper4,
33,subsection,2, Annexe A: Modélisation, modelpaper4,
35,subsection,2, Annexe B: Tables des paramètres, tablepaper4
},
addtolist={
4,figure,{\underline{Article IV}, Figure 1: Evolution of the scale with the redshift using the canonical cosmological parameters $H_0= 71~km~s^{-1}~Mpc^{-1}$ , $\Omega_m = 0.27$, and $\Omega_{\Lambda}= 0.73$.},fig1_p4,
6,figure,{\underline{Article IV}, Figure 2: Relative distribution of galaxy properties. Top: optical radius; Middle: maximum rotation velocity; Bottom: masses. The black stairs indicates the GHASP local sub-sample, the red hatchings the IMAGES sample and the blue hatchings the SINS sample. In order to show the respective size of the samples (153, 63 and 26 galaxies respectively for GHASP, IMAGES and SINS), arrows and letters with the same colors indicate five galaxies for each sample (G for GHASP, I for IMAGES and S for SINS).},fig2_p4,
8,figure,{\underline{Article IV}, Figure 3: Spatial resolution effects illustrated on eight galaxies illustrating an unambiguous case and the cases described from (i) to (vii) in section 4. The following comments concern each galaxy. Top line: actual high resolution data at $z=0$. Bottom line: data projected at $z=1.7$. The spatial scale is labelled in arcsecond on the left side of both lines. From left to right: \Ha\ monochromatic maps, \vfs\ and \vdms. The rainbow scale on the right side of each image represents the flux for the first column and the line-of-sight velocities corrected from instrumental function for the two next columns. The black and white double crosses mark the kinematical center at low redshift, while the black line represents the major axis and ends at the optical radius. More projected galaxies are presented in Appendix C.},fig3_p4,
9,figure,{\underline{Article IV}, Figure 4: Beam smearing effects on a simulation (\vfs\ on the top line and \vdms\ on the bottom line) depending on increasing blurring parameter. From left to right, the seeing (represented by a dark disk on the six images) increases from 0.25$''$ to 1$''$. The pixel size is 0.125$''$. The disk scale length is set to 5 kpc (observed at $z=1.7$), the inclination is 45\degr\ and the maximum velocity in the plane of the disk is 200$~km.s^{-1}$.},fig4_p4,
9,figure,{\underline{Article IV}, Figure 5: Example of a \rc\ obtained for a redshifted galaxy. Both \rcs\ have been computed from both local and projected UGC 7901 \vfs\ presented in Figure 3: red-open triangles correspond to local full resolution data while black dots come from the data projected at $z=1.7$.},fig5_p4,
12,figure,{\underline{Article IV}, Figure 6: High resolution rotation curves (black curve) superimposed on velocity fields (color image) of the four models used. From left to right and from top to bottom: exponential disk, isothermal sphere, ``flat'' and arc-tangent models. The radius (x-axis) is common for the four velocity fields and the four rotation curves. The velocity fields scale is given by the rainbow scale on the right side of the images. The velocity amplitude of the rotation curves is given by the scale on the left side of the y-axis.},fig6_p4,
14,table,{\underline{Article IV}, Table 1: Successfulness of the four $z=1.7$ models to recover $z=0$ actual parameters.},tab1_p4,
14,figure,{\underline{Article IV}, Figure 7: Kinematical inclination computed at $z=1.7$ using a ``flat model'' vs actual kinematical inclination evaluated at $z=0$. Each circle represents a galaxy. The open ones are galaxies which are stacked to the boundaries allowed for any fit (10 and 80\degr). The line indicates $y=x$.},fig7_p4,
15,figure,{\underline{Article IV}, Figure 8: Difference in the kinematical inclination between actual $z=0$ galaxies and simulated galaxies $z=1.7$ using a ``flat model'' vs the beam smearing parameter $B$. Each circle represents a galaxy. The open ones are galaxies which are stacked to the boundaries allowed for any fit (10 and 80\degr). The two dashed lines represent the mean positive and negative errors.},fig8_p4,
17,figure,{\underline{Article IV}, Figure 9: Histogram of the difference in kinematical position angles computed at $z=0$ and at $z=1.7$ using a ``flat model''.},fig9_p4,
17,figure,{\underline{Article IV}, Figure 10: Difference in the position angle between actual $z=0$ galaxies and simulated galaxies $z=1.7$ using a ``flat model'' vs the beam smearing parameter $B$. Each circle represents a galaxy.},fig10_p4,
18,figure,{\underline{Article IV}, Figure 11: Mean rotation velocity difference $\Delta V_c^{mean}$ between actual $z=0$ and different $z=1.7$ rotation curves vs the actual $z=0$ rotation curve inner slope. Top: $\Delta V_c^{mean}$ is the difference between actual $z=0$ and non-corrected $z=1.7$ rotation curves. The $z=1.7$ rotation curves are computed along the major axis of the galaxies. Bottom: $\Delta V_c^{mean}$ is the mean difference between actual $z=0$ and model $z=1.7$ rotation curves. The $z=1.7$ rotation curves are computed using a ``flat model''.},fig11_p4,
19,figure,{\underline{Article IV}, Figure 12: Comparison between the maximum rotation velocities at $z=0$ (x-axis) and $z=1.7$ (y-axis). The blue squares (left column) and the black dots (right column) represent respectively the maximum velocities measured along the major axis of the blurred \vfs\ and the maximum deduced from the ``flat model'' fitting. (Top) \Vfs\ are truncated at diameters $D_{25}$ (along the major axis). (Bottom) \Vfs\ are truncated at diameters $D_{25}/2$.},fig12_p4,
20,figure,{\underline{Article IV}, Figure 13: Relative difference between the maximum rotation velocities at $z=1.7$ and $z=0$ as a function of the beam smearing parameter $B$. The symbols are the same than in Figure 12.},fig13_p4,
20,figure,{\underline{Article IV}, Figure 14: Example of comparison between high redshift simulated data (left column) and high redshift model mimicking the data (middle column) for the galaxy UGC 7901. A ``flat model'' has been used here. Top line: velocity field. Bottom line: velocity dispersion map. The difference between the simulated high redshift data (left columm) and the model (middle columm) is given for both the velocity field and the velocity dispersion map (quadratic difference) on the right column. The velocities are given by the rainbow scales on the right side of the images.},fig14_p4,
21,figure,{\underline{Article IV}, Figure 15: Velocity dispersion as a function of projected maximum velocity measured on $z=0$ \vfs. Each point represents a galaxy. Blue squares correspond to the seeing-induced central velocity dispersion measured on $z=1.7$ maps (without applying any corrections). Red-open triangles represent the mean velocity dispersions measured on $z=0$ galaxies. The black dots correspond to the mean velocity dispersion measured on corrected \vdms\ of $z=1.7$ galaxies using a ``flat model''. The grey dashed and dotted lines respectively indicate the mean velocity dispersion in the IMAGES $z\sim0.6$ sample and in a sample of forty-two $1<z<3$ objects observed with OSIRIS and SINFONI.},fig15_p4,
22,figure,{\underline{Article IV}, Figure 16: Minimum velocity dispersion of $z = 1.7$ galaxies as a function of the mean velocity dispersion at $z = 0$.},fig16_p4,
23,figure,{\underline{Article IV}, Figure 17: Ratio between the maximum rotation velocity and the mean velocity dispersion as a function of the maximum rotation velocity. Top. GHASP projected sub-sample. Circles: values uncorrected for the beam smearing. 85\%\ of these circles are within the grey horizontal hatchings zone. Dots: values corrected for the beam smearing using the rotating disk modeling. 85\%\ of these dots are within the purple vertical hatchings zone. The grey and purple hatchings are reported in the Bottom figure for reference. Bottom. Observed high redshift galaxies. Open symbols correspond to observations using AO. Red rhombuses: SINS rotating disks at $z\sim2$ (Cresci et al. 2009). Orange squares: MASSIV pilot run galaxies at $z\sim 1.5$ (Epinat et al. 2009c). Green triangles: $z\sim 3$ Law et al. (2009) OSIRIS observations. Blue upside down triangles: rotating disks in Wright et al. (2007, 2009) at $z\sim 1.5$ observed with OSIRIS.},fig17_p4,
25,table,{\underline{Article IV}, Table 2: Fits of local and distant Tully-Fisher relation.},tab2_p4,
26,figure,{\underline{Article IV}, Figure 18: Tully-Fisher relation at $z=0$ -- red line -- compared with Tully-Fisher relation computed at $z=1.7$ -- black dots and black dashed line --. Open symbols correspond to galaxies that would not be classified as rotating disks at redshift 1.7. The two linear regressions (red line and black dotted line) are computed using only galaxies classified as rotating disks.},fig18_p4,
36,table,{\underline{Article IV}, Table B1: Galaxy parameters at $z=0$.},tabB1_p4
}]{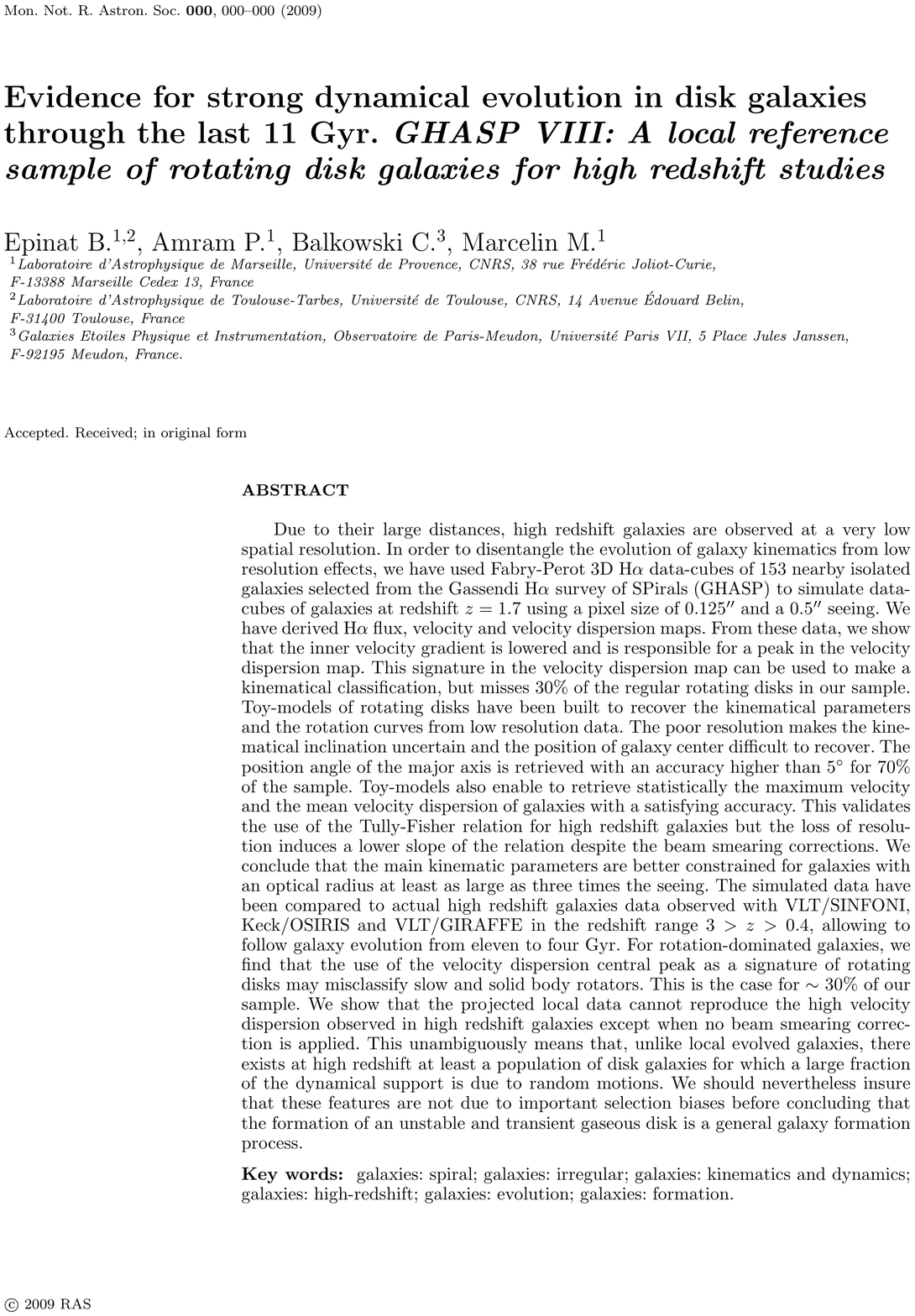}

\includepdf[pagecommand={\pagestyle{headings}},scale=1.,landscape=true,offset=0 0,pages={39-54},
addtolist={
39,table,{\underline{Article IV}, Table B1: Exponential disk model on the sample projected at $z=1.7$.},tabB2_p4,
43,table,{\underline{Article IV}, Table B1: Isothermal sphere model on the sample projected at $z=1.7$.},tabB3_p4,
47,table,{\underline{Article IV}, Table B1: ``Flat model'' on the sample projected at $z=1.7$.},tabB4_p4,
51,table,{\underline{Article IV}, Table B1: Arctangent model on the sample projected at $z=1.7$.},tabB5_p4
}]{articles/epinat_etal_ghz_vastroph.pdf}

\includepdf[pagecommand={\pagestyle{headings}},scale=1.,offset=0 0,pages={55-91},
addtotoc={
58,subsection,2, Annexe C: Cartes cinématiques, mapspaper4,
81,subsection,2, Annexe D: Courbes de rotation, rcspaper4
},
addtolist={
55,table,{\underline{Article IV}, Table B1: Parameters computed without beam smearing correction for the sample projected at $z=1.7$.},tabB6_p4,
59,figure,{\underline{Article IV}, Figures C: From left to rigth: XDSS image, \Ha\ monochromatic image, velocity field, velocity dispersion map. The white \& black crosses mark the kinematical center. The black line is the major axis, its length represents the $D_{25}$. These maps are not truncated.},figC_p4,
81,figure,{\underline{Article IV}, Figures D: Rotation curves.},figD_p4
}]{articles/epinat_etal_ghz_vastroph.pdf}

\cleardoublepage
\phantomsection
\addstarredchapter{Conclusions et perspectives}
\chaptermarkfree{Conclusions et perspectives}
\chapter*{Conclusions et perspectives}
L'étude de la cinématique et de la dynamique des galaxies proches et lointaines permet de con\-traindre les mécanismes physiques à l'origine de la formation des galaxies, de leur évolution et de leur stabilité actuelle.
Grâce à l'utilisation des données spectroscopiques à champ intégral obtenues par interférométrie \FP~sur l'échantillon de galaxies proches GHASP, j'ai pu, au cours de cette thèse, initier l'étude cinématique de ces galaxies et sonder les effets liés à la faible résolution spatiale des données cinématiques actuelles des galaxies lointaines afin de dissocier les effets de distance des effets d'évolution.

\subsection*{Instrumentation}

L'importance d'utiliser une instrumentation adaptée aux besoins spécifiques de mon travail de thèse m'a amené à m'impliquer dès le début de ma thèse dans le projet 3D-NTT,
instrument qui sera sur le ciel fin 2009.
J'ai ainsi participé à la définition des spécifications des deux \FP~contenus dans cet instrument et de leurs modes d'utilisation respectifs. Un de ces \FP~possédera une faible résolution spectrale et pourra être utilisé en pupille ou au foyer du télescope. Dans ce dernier cas, il pourra être utilisé conjointement au second \FP~de grande résolution afin de sélectionner la bande passante utile. Le choix des résolutions de chacun des deux \FP~est le résultat de cette utilisation combinée ainsi que des besoins scientifiques des différents modes d'observation du 3D-NTT: le mode basse résolution permettra principalement d'étudier les caractéristiques du milieu interstellaire dans les galaxies proches grâce à des cartes d'extinction et d'abondances,
ainsi que de détecter des galaxies lointaines grâce à leurs raies d'émission et d'étudier les phénomènes d'accrétion de gaz et de ``feed-back'' radiatif;
le mode haute résolution sera destiné à l'étude de la cinématique d'objets à raies d'émission tels que des nébuleuses planétaires, des régions HII Galactiques et des galaxies locales.
Je me suis également investi dans la calibration en longueur d'onde de cet instrument et, étant donnée l'expérience que j'ai acquise pour la réduction de données \FP~durant cette thèse, je prendrai activement part à la mise en place du logiciel de réduction des données issues du 3D-NTT. Par ailleurs, le VLT étant une réplique du NTT à plus grande échelle, le 3D-NTT a été conçu pour pouvoir devenir un 3D-VLT moyennant un nombre de modifications minimes. Je pourrais participer à une telle mise à niveau de l'instrument.

Par ailleurs, en vue des futures observations des galaxies lointaines et dans le cadre du développement de l'ELT européen, j'ai pris part à un projet concernant un spectrographe à grand champ dédié à l'étude de la formation des galaxies primordiales: WFSpec. Cela a été pour moi l'occasion de travailler sur un concept innovant, l'iBTF. Ce concept propose d'utiliser des filtres de Bragg afin de sélectionner plusieurs bandes spectrales tout en préservant l'imagerie, et un Fabry-Perot afin de doper la résolution spectrale. Dans le cadre de ce projet je me suis également penché sur les deux autres concepts proposés afin d'en étudier le facteur de mérite. Cette étude a permis de déterminer qu'un concept ``multi-IFU'', rebaptisé EAGLE,
est le plus adapté à l'étude des galaxies primordiales. Il en résulte aussi que le concept iBTF est adapté à des thématiques scientifiques possédant une très forte densité d'objets dans le champ, comme par exemple l'étude de populations stellaires résolues dans les galaxies proches. Je compte prolonger mon investissement dans les projets d'instruments spectroscopiques pour les ELT que sont l'iBTF et EAGLE.
Par ailleurs, les ELT observeront principalement dans l'infrarouge car cette bande spectrale permet d'effectuer une correction plus fine des effets de la turbulence atmosphérique. Il est donc primordial de réfléchir à des solutions permettant de réduire la contribution des nombreuses raies du ciel dans ce domaine de longueurs d'onde pour faire de l'imagerie en bande large. Dans cette perspective, j'envisage de m'impliquer dans le développement de filtres suppresseurs des raies du ciel nocturne.

Enfin, la réduction et l'utilisation de données \FP~nécessitant une bonne compréhension de cet interféromètre, je me suis penché durant ma thèse sur  diverses solutions de spectroscopie \FP. Je les ai reprises pour élaborer un sujet de travaux pratiques pour des étudiants de Master afin de les initier à la spectroscopie \FP.

\subsection*{Galaxies de l'Univers local: GHASP}

Le projet GHASP a pour objectif principal de constituer un échantillon de référence de galaxies spirales et irrégulières locales situées dans des environnements peu denses pour des études cinématiques et dynamiques et notamment pour la comparaison avec les galaxies distantes. Cet échantillon est composé de $203$ galaxies pour lesquelles des cubes de données ont été obtenus autour de la raie \Ha~avec un instrument utilisant un interféromètre de \FP~au télescope de $1.93~m$ de l'OHP.

Durant ma thèse, afin de disposer de données réduites de manière homogène pour effectuer l'analyse cinématique, j'ai, dans un premier temps, réalisé la réduction de l'ensemble des données cinématiques de l'échantillon GHASP en utilisant de nouvelles procédures de réduction. Ces procédures ont permis d'obtenir des cartes cinématiques de meilleure qualité notamment grâce à la correction des erreurs de guidage du télescope, à l'exclusion des poses élémentaires présentant des anomalies de flux, à la soustraction éventuelle de reflets parasites et surtout grâce à l'emploi d'un maillage adaptatif pour assurer un rapport signal sur bruit correct tout en préservant l'indépendance de l'information entre mailles adjacentes.
\`A partir des cartes cinématiques ainsi déduites, les paramètres de projection cinématiques et les courbes de rotation ont été déterminés en vue de leur exploitation scientifique.
Une extraction robuste des paramètres de projection cinématiques et des courbes de rotation mettant à profit l'intégralité de l'information à deux dimensions du champ de vitesses en lui ajustant un modèle de disque fin en rotation a été mise en place parallèlement à une détermination réaliste des erreurs associées utilisant le spectre de puissance du champ de vitesses résiduel.
Les principaux résultats de l'analyse cinématique de l'échantillon GHASP, initiée durant cette thèse, sont énumérés ci-après.

\begin{enumerate}
\item La relation de Tully-Fisher obtenue à partir de l'échantillon GHASP est en très bon accord avec les déterminations précédentes de la littérature. Cependant, l'échantillon GHASP suggère l'existence d'une famille de galaxies en rotation rapide pour lesquelles la luminosité est plus faible qu'attendue d'après la relation de Tully-Fisher. Par ailleurs, nous avons montré que les galaxies de faible inclinaison doivent être exclues de l'étude de cette relation car l'incertitude sur leur vitesse de rotation déprojetée est grande et cette vitesse est de ce fait souvent surestimée.
\item L'angle de position du grand axe est déterminé de manière plus précise à partir des données cinématiques, en particulier pour les galaxies de faible inclinaison. Pour les galaxies fortement barrées, la détermination de l'angle de position cinématique peut être biaisée et dépend de l'orientation de la barre.
\item Les inclinaisons cinématiques et morphologiques sont déterminées avec une précision équivalente ($\sim 7\pm1$\degr).
\item \`A partir des courbes de rotation de $36$ galaxies GHASP, il a été montré qu'un profil de sphère isotherme décrit mieux la densité des halos de matière sombre qu'un profil de Navarro-Frenk-White. De plus, les halos de matière sombre possèdent une densité de surface constante, indépendante du type morphologique ou de la magnitude. Les observations photométriques des galaxies GHASP sont en cours afin de pouvoir utiliser l'échantillon complet. Toutefois, des études préliminaires sur tout l'échantillon peuvent d'ores et déjà être menées en utilisant un modèle de disque exponentiel pour décrire le disque stellaire.
L'ajustement de modèles de masses sur l'ensemble des galaxies de l'échantillon permettra de mieux contraindre les caractéristiques des halos de matière sombre. Des modèles plus sophistiqués seront aussi utilisés. De plus, le développement de méthodes d'ajustement à partir d'imagerie et de champs de vitesses à deux dimensions, au lieu de profils de luminosité et de courbes de rotation à une dimension, devrait permettre de mieux contraindre la forme des halos de matière sombre.
\item L'utilisation de l'échantillon complet permet de confirmer que la pente interne des courbes de rotation des galaxies croît avec leur masse. Une légère corrélation entre la pente externe et la magnitude a également pu être mise en évidence. Ces indices suggèrent qu'une description de la courbe de rotation à partir d'un nombre réduit de paramètres doit être possible ``en moyenne''. Les propriétés et la validité d'une courbe de rotation universelle seront donc étudiées en détail. Des courbes de rotation et des champs de vitesses typiques (``templates'') seront également déduits à partir des données GHASP.
\item L'étude préliminaire des cartes et des profils de dispersion des vitesses du gaz dans les galaxies GHASP montre que la dispersion de vitesses du gaz varie assez faiblement avec la position. De même la valeur moyenne de la dispersion de vitesses varie très peu d'une galaxie à l'autre et est proche de $20~km~s^{-1}$. Nous avons également montré que la dispersion de vitesses centrale augmente lorsque la résolution spatiale est trop faible par rapport au gradient de vitesses central, ce qui n'est généralement pas le cas pour les galaxies proches étant donnée la grande résolution spatiale des observations. Une étude plus poussée de la dispersion de vitesses du gaz ionisé sera menée, afin \textit{(i)} de chercher des corrélations avec les paramètres physiques des galaxies comme le type morphologique ou la magnitude, \textit{(ii)} de la comparer à la dispersion de vitesses des étoiles et \textit{(iii)} de déterminer les liens qui existent entre les dispersions azimutale, radiale et perpendiculaire au plan du disque, en utilisant des modèles de déprojection des cartes de dispersion de vitesses.
\end{enumerate}

En plus des travaux présentés précédemment initiés à partir de l'échantillon GHASP, d'autres points clés seront étudiés. En particulier, l'effet des potentiels barrés sur les champs de vitesses sera approfondi. Une étude des effets de la barre sur la cinématique des galaxies à partir de simulations numériques de galaxies barrées a mis en évidence une corrélation entre l'angle de la barre et la position du petit axe cinématique.
La recherche de cette corrélation dans l'échantillon de galaxies barrées BH$\alpha$Bar et dans l'échantillon GHASP fait partie des perspectives de cette thèse. De plus, les résultats pourront être confrontés à l'étude kinémétrique de ces échantillons qui devrait permettre de mettre en évidence des potentiels non axi-symmétriques.
Enfin, l'influence du milieu dans lequel évoluent les galaxies sur leur cinématique sera étudiée. Pour cela, il sera nécessaire de confronter l'échantillon GHASP à des échantillons de galaxies dans des environnements différents (amas, groupes, groupes compacts, binaires, etc.). L'objectif de la base de données Fabry-Perot, projet auquel j'ai participé, est justement de mettre à la disposition de la communauté scientifique les observations de galaxies de ces divers échantillons déjà observés et à venir. En particulier, le 3D-NTT permettra d'observer de nouveaux échantillons.

\subsection*{Galaxies de l'Univers lointain}

La compréhension des mécanismes de formation des galaxies et leur évolution à travers les âges est un des défis de ce début de siècle. Quand et comment la matière s'est-elle condensée pour former les structures observées dans l'Univers proche? D'où leur stabilité résulte-t-elle? Quand les galaxies ont-elles formé leurs étoiles? Quand les disques stellaires se sont-ils formés? L'étude de la cinématique des galaxies permet de donner des indices pour répondre à ces questions. Cependant, l'observation des galaxies lointaines est fortement handicapée par la faible résolution spatiale due à leur faible dimension angulaire.

Afin d'aider à l'interprétation des données cinématiques de ces galaxies lointaines, l'échantillon GHASP a été projeté à grand décalage spectral ($\sim1.7$), dans les conditions d'observation obtenues sans optique adaptative. Les biais observationnels dus au manque de résolution spatiale ont ainsi pu être étudiés.
\begin{enumerate}
\item Le manque de résolution spatiale induit une diminution du gradient de vitesse, cependant, celui-ci se retrouve en partie dans les cartes de dispersion de vitesses sous forme d'un pic central.
\item La courbe de rotation mesurée le long du grand axe des galaxies lointaines présente une forme typique de rotation en corps solide, en particulier car la pente interne est fortement diminuée par les effets de résolution. De plus, la vitesse maximale déterminée à partir de cette courbe de rotation est fortement sous-estimée pour les galaxies dont la dimension est inférieure à trois fois la tache du seeing.
\item L'ajustement de modèles cinématiques aux champs de vitesses prenant en compte la faible résolution spatiale permet statistiquement de retrouver la vitesse maximale avec une précision meilleure que 25\% et l'angle de position du grand axe cinématique avec une précision meilleure que 5\degr~pour des galaxies aussi petites que la tache du seeing. Cependant, le centre et l'inclinaison sont bien plus délicats à déterminer, ce qui démontre l'intérêt d'utiliser des images du continuum en bande large obtenues par optique adaptative ou à partir de l'espace. En particulier, la valeur de l'inclinaison fixe l'amplitude de la courbe de rotation déprojetée.
\item La dispersion de vitesses intrinsèque peut être retrouvée à partir des modèles car ils permettent de soustraire la contribution due au gradient de vitesse.
\item Les courbes de rotation des modèles sont plus fidèles aux courbes de rotation réelles que les courbes de rotation mesurées le long du grand axe. Cependant, la forme de la courbe des modèles est simpliste puisque ces modèles ne possèdent que deux paramètres.
\end{enumerate}

L'utilisation conjointe des données GHASP projetées et des données cinématiques à champ intégral de galaxies à grand décalage spectral de la littérature a permis de mettre en évidence des effets d'évolution.
\begin{enumerate}
\item Le résultat majeur est que la dispersion de vitesses des galaxies possédant un décalage spectral $z>1$ est plus grand que pour les galaxies locales, impliquant que les disques observés à grand décalage spectral sont de nature différente. Il pourrait s'agir de disques épais créés par accrétion de nuages de gaz possédant une forte formation stellaire dont le support dynamique serait moins organisé que localement. Par contre, la dispersion de vitesses des galaxies possédant un décalage spectral $z\sim0.6$ est assez similaire à celle des galaxies locales, ce qui implique que les disques actuels étaient déjà formés il y a six milliards d'années.
\item Le programme IMAGES,
pour lequel $68$ galaxies ont actuellement été observées à un décalage spectral $z\sim0.6$ par FLAMES/GIRAFFE dans le visible, a été mis en place afin d'étudier l'assemblage de masses des galaxies depuis $z=1$. En particulier, l'évolution de la relation de Tully-Fisher a été étudiée.
L'étude comparative de cette relation à partir de l'échantillon IMAGES et à partir de l'échantillon GHASP (projeté et local) semble confirmer la conclusion avancée par \citet{Puech:2008} selon laquelle les galaxies auraient doublé leur masse stellaire depuis six milliards d'années. Par ailleurs, une étude est actuellement en cours avec l'utilisation conjointe des échantillons GHASP et IMAGES afin de sonder l'évolution du moment cinétique des galaxies. Un sous-échantillon de GHASP représentatif de l'échantillon IMAGES sera extrait et projeté dans les mêmes conditions d'observation.
\item Le projet MASSIV
a pour objectif d'observer un échantillon de plus de cent galaxies avec un décalage spectral $1<z<1.8$ dans un large intervalle de masses stellaires en vue d'étudier l'assemblage de masses dans ce domaine de décalage spectral. Ces observations sont obtenues par SINFONI, avec optique adaptative pour certaines d'entre elles. Les données SINFONI préliminaires à ce programme présentées dans cette thèse montrent que la cinématique des galaxies avec un décalage spectral $z>1$ est comparable à la cinématique de disques en rotation mais avec une forte dispersion de vitesses ($\sim100~km~s^{-1}$), souvent piquée à diverses positions et corrélée au flux de la raie observée. Ces observations sont en accord avec les autres observations de la littérature et avec les conclusions avancées au point 1. Dans le cadre du projet MASSIV, je mettrai à profit l'expérience acquise durant cette thèse afin de prendre part au dépouillement des données et à leur analyse. La constitution d'un échantillon complet et représentatif et sa comparaison avec des échantillons locaux et des simulations numériques permettra d'aboutir à des conclusions solides quant à la formation et l'évolution des structures.
\end{enumerate}

Afin de réaliser la comparaison des galaxies lointaines et locales, il est nécessaire de disposer d'échantillons de référence locaux de galaxies dans divers environnements. Les méthodes de projection et d'analyse appliquées à l'échantillon GHASP seront employées afin de projeter des galaxies binaires, des galaxies d'amas, des galaxies bleues compactes et des groupes compacts. L'utilisation du 3D-NTT pour observer de tels échantillons de galaxies sera déterminante.
Par ailleurs, la signature cinématique des barres mise en évidence sur les simulations numériques sera recherchée sur les échantillons GHASP et BH$\alpha$Bar projetés afin de déterminer quelle résolution est nécessaire pour mettre en évidence des structures barrées à grand décalage spectral sur les données cinématiques.
Enfin, de la même manière que les effets de la résolution spectrale ont été étudiés à partir de l'échantillon GHASP, j'utiliserai cet échantillon afin de déterminer si la relativement faible résolution spectrale des données SINFONI ($\sim80~km~s^{-1}$ contre $\sim20~km~s^{-1}$ pour GHASP) a une incidence majeure sur leur analyse cinématique.\\
\\
\par
L'arrivée future des ELT permettra d'observer les galaxies lointaines avec une résolution spatiale comparable à celle qui est aujourd'hui obtenue sur les galaxies proches. Cela ouvrira la voie à des études cinématiques poussées incluant l'évolution de la distribution de matière dans les galaxies, et de la forme des halos de matière sombre à travers les âges...

\cleardoublepage
\phantomsection
\addstarredchapter{Bibliographie du texte en Français}
\bibliographystyle{astron}
\bibliography{biblio_these}

\begin{thebibliography}{}

\bibitem[\protect\astroncite{{Amram}}{1991}]{Amram:1991}
{Amram}, P.: 1991,
\newblock {\em Ph.D. thesis}, Universit\'e de Provence (France)

\bibitem[\protect\astroncite{{Amram} et~al.}{2007}]{Amram:2007}
{Amram}, P., {Mendes de Oliveira}, C., {Plana}, H., {Balkowski}, C., et
  {Hernandez}, O.: 2007,
\newblock {\em \aap} {\bf 471}, 753

\bibitem[\protect\astroncite{{Barnes} et~al.}{2004}]{Barnes:2004}
{Barnes}, E.~I., {Sellwood}, J.~A., et {Kosowsky}, A.: 2004,
\newblock {\em \aj} {\bf 128}, 2724

\bibitem[\protect\astroncite{{Bland} et {Tully}}{1989}]{Bland:1989}
{Bland}, J. et {Tully}, R.~B.: 1989,
\newblock {\em \aj} {\bf 98}, 723

\bibitem[\protect\astroncite{{Bosma}}{1978}]{Bosma:1978}
{Bosma}, A.: 1978,
\newblock {\em Ph.D. thesis}, Groningen Universiteit

\bibitem[\protect\astroncite{{Boulesteix} et~al.}{1984}]{Boulesteix:1984}
{Boulesteix}, J., {Georgelin}, Y., {Marcelin}, M., et {Monnet}, G.: 1984,
\newblock in A. {Boksenberg} et D.~L. {Crawford} (eds.), {\em Society of
  Photo-Optical Instrumentation Engineers (SPIE) Conference Series}, Vol. 445
  of {\em Society of Photo-Optical Instrumentation Engineers (SPIE) Conference
  Series}, pp 37--41

\bibitem[\protect\astroncite{{Boulesteix} et~al.}{1987}]{Boulesteix:1987}
{Boulesteix}, J., {Georgelin}, Y.~P., {Le Coarer}, E., {Marcelin}, M., et
  {Monnet}, G.: 1987,
\newblock {\em \aap} {\bf 178}, 91

\bibitem[\protect\astroncite{{Bournaud} et~al.}{2008}]{Bournaud:2008}
{Bournaud}, F., {Daddi}, E., {Elmegreen}, B.~G., {Elmegreen}, D.~M.,
  {Nesvadba}, N., {Vanzella}, E., {di Matteo}, P., {Le Tiran}, L., {Lehnert},
  M., et {Elbaz}, D.: 2008,
\newblock {\em \aap} {\bf 486}, 741

\bibitem[\protect\astroncite{{Buisson} et~al.}{1914}]{Buisson:1914}
{Buisson}, H., {Fabry}, C., et {Bourget}, H.: 1914,
\newblock {\em \apj} {\bf 40}, 241

\bibitem[\protect\astroncite{{Burbidge} et~al.}{1960}]{Burbidge:1960}
{Burbidge}, E.~M., {Burbidge}, G.~R., et {Prendergast}, K.~H.: 1960,
\newblock {\em \apj} {\bf 131}, 282

\bibitem[\protect\astroncite{{Cappellari} et {Copin}}{2003}]{Cappellari:2003}
{Cappellari}, M. et {Copin}, Y.: 2003,
\newblock {\em \mnras} {\bf 342}, 345

\bibitem[\protect\astroncite{{Catinella} et~al.}{2006}]{Catinella:2006}
{Catinella}, B., {Giovanelli}, R., et {Haynes}, M.~P.: 2006,
\newblock {\em \apj} {\bf 640}, 751

\bibitem[\protect\astroncite{{Christodoulou} et~al.}{1993}]{Christodoulou:1993}
{Christodoulou}, D.~M., {Tohline}, J.~E., et {Steiman-Cameron}, T.~Y.: 1993,
\newblock {\em \apj} {\bf 416}, 74

\bibitem[\protect\astroncite{{Cole} et~al.}{2001}]{Cole:2001}
{Cole}, S., {Norberg}, P., {Baugh}, C.~M., {Frenk}, C.~S., {Bland-Hawthorn},
  J., {Bridges}, T., {Cannon}, R., {Colless}, M., {Collins}, C., {Couch}, W.,
  {Cross}, N., {Dalton}, G., {De Propris}, R., {Driver}, S.~P., {Efstathiou},
  G., {Ellis}, R.~S., {Glazebrook}, K., {Jackson}, C., {Lahav}, O., {Lewis},
  I., {Lumsden}, S., {Maddox}, S., {Madgwick}, D., {Peacock}, J.~A.,
  {Peterson}, B.~A., {Sutherland}, W., et {Taylor}, K.: 2001,
\newblock {\em \mnras} {\bf 326}, 255

\bibitem[\protect\astroncite{{Court{\`e}s}}{1972}]{Courtes:1972}
{Court{\`e}s}, G.: 1972,
\newblock {\em Vistas in Astronomy} {\bf 14}, 81

\bibitem[\protect\astroncite{{Daigle} et~al.}{2006}]{Daigle:2006b}
{Daigle}, O., {Carignan}, C., {Hernandez}, O., {Chemin}, L., et {Amram}, P.:
  2006,
\newblock {\em \mnras} {\bf 368}, 1016

\bibitem[\protect\astroncite{{de Vaucouleurs}
  et~al.}{1995}]{de-Vaucouleurs:1995}
{de Vaucouleurs}, G., {de Vaucouleurs}, A., {Corwin}, H.~G., {Buta}, R.~J.,
  {Paturel}, G., et {Fouque}, P.: 1995,
\newblock {\em VizieR Online Data Catalog} {\bf 7155}, 0

\bibitem[\protect\astroncite{{de Vaucouleurs}
  et~al.}{1974}]{de-Vaucouleurs:1974}
{de Vaucouleurs}, G., {de Vaucouleurs}, A., et {Pence}, W.: 1974,
\newblock {\em \apjl} {\bf 194}, L119

\bibitem[\protect\astroncite{{Epinat} et~al.}{2007}]{Epinat:2007}
{Epinat}, B., {Amram}, P., et {Balkowski}, C.: 2007,
\newblock in F. {Combes} et J. {Palous} (eds.), {\em IAU Symposium}, Vol. 235
  of {\em IAU Symposium}, pp 401--401

\bibitem[\protect\astroncite{{Epinat} et~al.}{2009a}]{Epinat:2008c}
{Epinat}, B., {Amram}, P., {Balkowski}, C., et {Marcelin}, M.: 2009a

\bibitem[\protect\astroncite{{Epinat} et~al.}{2008a}]{Epinat:2008b}
{Epinat}, B., {Amram}, P., et {Marcelin}, M.: 2008a,
\newblock {\em \mnras} {\bf 390}, 466

\bibitem[\protect\astroncite{{Epinat} et~al.}{2009b}]{Epinat:prep}
{Epinat}, B., {Amram}, P., {Marcelin}, M., et {al.}: 2009b,
\newblock {\em en pr\'eparation}

\bibitem[\protect\astroncite{{Epinat} et~al.}{2008b}]{Epinat:2008a}
{Epinat}, B., {Amram}, P., {Marcelin}, M., {Balkowski}, C., {Daigle}, O.,
  {Hernandez}, O., {Chemin}, L., {Carignan}, C., {Gach}, J.-L., et {Balard},
  P.: 2008b,
\newblock {\em \mnras} {\bf 388}, 500

\bibitem[\protect\astroncite{{Epinat} et~al.}{2009c}]{Epinat:2009}
{Epinat}, B., {Contini}, T., {Le Fevre}, O., {Vergani}, D., {Garilli}, B.,
  {Amram}, P., {Queyrel}, J., {Tasca}, L., et {Tresse}, L.: 2009c

\bibitem[\protect\astroncite{{Erb} et~al.}{2003}]{Erb:2003}
{Erb}, D.~K., {Shapley}, A.~E., {Steidel}, C.~C., {Pettini}, M., {Adelberger},
  K.~L., {Hunt}, M.~P., {Moorwood}, A.~F.~M., et {Cuby}, J.-G.: 2003,
\newblock {\em \apj} {\bf 591}, 101

\bibitem[\protect\astroncite{{Erb} et~al.}{2004}]{Erb:2004}
{Erb}, D.~K., {Steidel}, C.~C., {Shapley}, A.~E., {Pettini}, M., et
  {Adelberger}, K.~L.: 2004,
\newblock {\em \apj} {\bf 612}, 122

\bibitem[\protect\astroncite{{Falc{\'o}n-Barroso}
  et~al.}{2006}]{Falcon-Barroso:2006}
{Falc{\'o}n-Barroso}, J., {Bacon}, R., {Bureau}, M., {Cappellari}, M.,
  {Davies}, R.~L., {de Zeeuw}, P.~T., {Emsellem}, E., {Fathi}, K.,
  {Krajnovi{\'c}}, D., {Kuntschner}, H., {McDermid}, R.~M., {Peletier}, R.~F.,
  et {Sarzi}, M.: 2006,
\newblock {\em \mnras} {\bf 369}, 529

\bibitem[\protect\astroncite{{Fathi} et~al.}{2008}]{Fathi:2008}
{Fathi}, K., {Beckman}, J.~E., {Lundgren}, A.~A., {Carignan}, C., {Hernandez},
  O., {Amram}, P., {Balard}, P., {Boulesteix}, J., {Gach}, J.-L., {Knapen},
  J.~H., et {Rela{\~n}o}, M.: 2008,
\newblock {\em \apjl} {\bf 675}, L17

\bibitem[\protect\astroncite{{Ferguson} et~al.}{1996}]{Ferguson:1996}
{Ferguson}, A.~M.~N., {Wyse}, R.~F.~G., {Gallagher}, III, J.~S., et {Hunter},
  D.~A.: 1996,
\newblock {\em \aj} {\bf 111}, 2265

\bibitem[\protect\astroncite{{Flores} et~al.}{2006}]{Flores:2006}
{Flores}, H., {Hammer}, F., {Puech}, M., {Amram}, P., et {Balkowski}, C.: 2006,
\newblock {\em \aap} {\bf 455}, 107

\bibitem[\protect\astroncite{{F{\"o}rster Schreiber}
  et~al.}{2006}]{Forster-Schreiber:2006}
{F{\"o}rster Schreiber}, N.~M., {Genzel}, R., {Lehnert}, M.~D., {Bouch{\'e}},
  N., {Verma}, A., {Erb}, D.~K., {Shapley}, A.~E., {Steidel}, C.~C., {Davies},
  R., {Lutz}, D., {Nesvadba}, N., {Tacconi}, L.~J., {Eisenhauer}, F., {Abuter},
  R., {Gilbert}, A., {Gillessen}, S., et {Sternberg}, A.: 2006,
\newblock {\em \apj} {\bf 645}, 1062

\bibitem[\protect\astroncite{{Gach} et~al.}{2002}]{Gach:2002}
{Gach}, J.-L., {Hernandez}, O., {Boulesteix}, J., {Amram}, P., {Boissin}, O.,
  {Carignan}, C., {Garrido}, O., {Marcelin}, M., {{\"O}stlin}, G., {Plana}, H.,
  et {Rampazzo}, R.: 2002,
\newblock {\em \pasp} {\bf 114}, 1043

\bibitem[\protect\astroncite{{Garrido}}{2003a}]{Garrido:thesis}
{Garrido}, O.: 2003a,
\newblock {\em Ph.D. thesis}, Universit\'e de Provence (France)

\bibitem[\protect\astroncite{{Garrido} et~al.}{2004}]{Garrido:2004}
{Garrido}, O., {Marcelin}, M., et {Amram}, P.: 2004,
\newblock {\em \mnras} {\bf 349}, 225

\bibitem[\protect\astroncite{{Garrido} et~al.}{2005}]{Garrido:2005}
{Garrido}, O., {Marcelin}, M., {Amram}, P., {Balkowski}, C., {Gach}, J.~L., et
  {Boulesteix}, J.: 2005,
\newblock {\em \mnras} {\bf 362}, 127

\bibitem[\protect\astroncite{{Garrido} et~al.}{2003b}]{Garrido:2003}
{Garrido}, O., {Marcelin}, M., {Amram}, P., et {Boissin}, O.: 2003b,
\newblock {\em \aap} {\bf 399}, 51

\bibitem[\protect\astroncite{{Garrido} et~al.}{2002}]{Garrido:2002}
{Garrido}, O., {Marcelin}, M., {Amram}, P., et {Boulesteix}, J.: 2002,
\newblock {\em \aap} {\bf 387}, 821

\bibitem[\protect\astroncite{{Genzel} et~al.}{2006}]{Genzel:2006}
{Genzel}, R., {Tacconi}, L.~J., {Eisenhauer}, F., {F{\"o}rster Schreiber},
  N.~M., {Cimatti}, A., {Daddi}, E., {Bouch{\'e}}, N., {Davies}, R., {Lehnert},
  M.~D., {Lutz}, D., {Nesvadba}, N., {Verma}, A., {Abuter}, R., {Shapiro}, K.,
  {Sternberg}, A., {Renzini}, A., {Kong}, X., {Arimoto}, N., et {Mignoli}, M.:
  2006,
\newblock {\em \nat} {\bf 442}, 786

\bibitem[\protect\astroncite{{Georgelin} et~al.}{1987}]{Georgelin:1987}
{Georgelin}, Y.~M., {Boulesteix}, J., {Georgelin}, Y.~P., {Laval}, A., et
  {Marcelin}, M.: 1987,
\newblock {\em \aap} {\bf 174}, 257

\bibitem[\protect\astroncite{{Georgelin}}{1970}]{Georgelin:1970}
{Georgelin}, Y.~P.: 1970,
\newblock {\em \aap} {\bf 9}, 441

\bibitem[\protect\astroncite{{Georgelin} et {Amram}}{1995}]{Georgelin:1995}
{Georgelin}, Y.~P. et {Amram}, P.: 1995,
\newblock in G. {Comte} et M. {Marcelin} (eds.), {\em IAU Colloq. 149:
  Tridimensional Optical Spectroscopic Methods in Astrophysics}, Vol.~71 of
  {\em Astronomical Society of the Pacific Conference Series}, p. 382

\bibitem[\protect\astroncite{{Gottesman} et {Weliachew}}{1975}]{Gottesman:1975}
{Gottesman}, S.~T. et {Weliachew}, L.: 1975,
\newblock {\em \apj} {\bf 195}, 23

\bibitem[\protect\astroncite{{Gunn}}{1980}]{Gunn:1980}
{Gunn}, J.~E.: 1980,
\newblock {\em Royal Society of London Philosophical Transactions Series A}
  {\bf 296}, 313

\bibitem[\protect\astroncite{{Hernandez}}{2005a}]{Hernandez:thesis}
{Hernandez}, O.: 2005a,
\newblock {\em Ph.D. thesis}, Universit\'e de Montr\'eal (Canada)

\bibitem[\protect\astroncite{{Hernandez} et~al.}{2008b}]{Hernandez:2008}
{Hernandez}, O., {Fathi}, K., {Carignan}, C., {Beckman}, J., {Gach}, J.-L.,
  {Balard}, P., {Amram}, P., {Boulesteix}, J., {Corradi}, R.~L.~M., {de
  Denus-Baillargeon}, M.-M., {Epinat}, B., {Rela{\~n}o}, M., {Thibault}, S., et
  {Vall{\'e}e}, P.: 2008b,
\newblock {\em \pasp} {\bf 120}, 665

\bibitem[\protect\astroncite{{Hernandez} et~al.}{2005}]{Hernandez:2005a}
{Hernandez}, O., {Wozniak}, H., {Carignan}, C., {Amram}, P., {Chemin}, L., et
  {Daigle}, O.: 2005,
\newblock {\em \apj} {\bf 632}, 253

\bibitem[\protect\astroncite{{Hicks} et~al.}{1976}]{Hicks:1976}
{Hicks}, T.~R., {Reay}, N.~K., et {Stephens}, C.~L.: 1976,
\newblock {\em \aap} {\bf 51}, 367

\bibitem[\protect\astroncite{{Kormendy}}{1977}]{Kormendy:1977}
{Kormendy}, J.: 1977,
\newblock {\em \apj} {\bf 217}, 406

\bibitem[\protect\astroncite{{Krajnovi{\'c}} et~al.}{2006}]{Krajnovic:2006}
{Krajnovi{\'c}}, D., {Cappellari}, M., {de Zeeuw}, P.~T., et {Copin}, Y.: 2006,
\newblock {\em \mnras} {\bf 366}, 787

\bibitem[\protect\astroncite{{Kravtsov} et~al.}{1998}]{Kravtsov:1998}
{Kravtsov}, A.~V., {Klypin}, A.~A., {Bullock}, J.~S., et {Primack}, J.~R.:
  1998,
\newblock {\em \apj} {\bf 502}, 48

\bibitem[\protect\astroncite{{Laval} et~al.}{1987}]{Laval:1987}
{Laval}, A., {Boulesteix}, J., {Georgelin}, Y.~P., {Georgelin}, Y.~M., et
  {Marcelin}, M.: 1987,
\newblock {\em \aap} {\bf 175}, 199

\bibitem[\protect\astroncite{{Law} et~al.}{2007}]{Law:2007}
{Law}, D.~R., {Steidel}, C.~C., {Erb}, D.~K., {Larkin}, J.~E., {Pettini}, M.,
  {Shapley}, A.~E., et {Wright}, S.~A.: 2007,
\newblock {\em \apj} {\bf 669}, 929

\bibitem[\protect\astroncite{{Le Coarer} et~al.}{1992}]{Le-Coarer:1992}
{Le Coarer}, E., {Amram}, P., {Boulesteix}, J., {Georgelin}, Y.~M.,
  {Georgelin}, Y.~P., {Marcelin}, M., {Joulie}, P., et {Urios}, J.: 1992,
\newblock {\em \aap} {\bf 257}, 389

\bibitem[\protect\astroncite{{Le Coarer} et~al.}{1995}]{Le-Coarer:1995}
{Le Coarer}, E., {Bensammar}, S., {Comte}, G., {Gach}, J.~L., et {Georgelin},
  Y.: 1995,
\newblock {\em \aaps} {\bf 111}, 359

\bibitem[\protect\astroncite{{Le F{\`e}vre} et~al.}{2005}]{Le-Fevre:2005}
{Le F{\`e}vre}, O., {Vettolani}, G., {Garilli}, B., {Tresse}, L., {Bottini},
  D., {Le Brun}, V., {Maccagni}, D., {Picat}, J.~P., {Scaramella}, R.,
  {Scodeggio}, M., {Zanichelli}, A., {Adami}, C., {Arnaboldi}, M., {Arnouts},
  S., {Bardelli}, S., {Bolzonella}, M., {Cappi}, A., {Charlot}, S., {Ciliegi},
  P., {Contini}, T., {Foucaud}, S., {Franzetti}, P., {Gavignaud}, I., {Guzzo},
  L., {Ilbert}, O., {Iovino}, A., {McCracken}, H.~J., {Marano}, B., {Marinoni},
  C., {Mathez}, G., {Mazure}, A., {Meneux}, B., {Merighi}, R., {Paltani}, S.,
  {Pell{\`o}}, R., {Pollo}, A., {Pozzetti}, L., {Radovich}, M., {Zamorani}, G.,
  {Zucca}, E., {Bondi}, M., {Bongiorno}, A., {Busarello}, G., {Lamareille}, F.,
  {Mellier}, Y., {Merluzzi}, P., {Ripepi}, V., et {Rizzo}, D.: 2005,
\newblock {\em \aap} {\bf 439}, 845

\bibitem[\protect\astroncite{{Lehnert} et {Bremer}}{2003}]{Lehnert:2003}
{Lehnert}, M.~D. et {Bremer}, M.: 2003,
\newblock {\em \apj} {\bf 593}, 630

\bibitem[\protect\astroncite{{Maillard}}{1996}]{Maillard:1996}
{Maillard}, J.~P.: 1996,
\newblock {\em \ao} {\bf 35}, 2734

\bibitem[\protect\astroncite{{Marcelin} et~al.}{2008}]{Marcelin:2008}
{Marcelin}, M., {Amram}, P., {Balard}, P., {Balkowski}, C., {Boissin}, O.,
  {Boulesteix}, J., {Carignan}, C., {Daigle}, O., {de Denus-Baillargeon},
  M.-M., {Epinat}, B., {Gach}, J.-L., {Hernandez}, O., {Rigaud}, F., et
  {Vall{\'e}e}, P.: 2008,
\newblock in {\em Ground-based and Airborne Instrumentation for Astronomy II,
  Ian S. McLean; Mark M. Casali, Editors, 701455}, Vol. 7014 of {\em Presented
  at the Society of Photo-Optical Instrumentation Engineers (SPIE) Conference}

\bibitem[\protect\astroncite{{Marcelin} et~al.}{1987}]{Marcelin:1987}
{Marcelin}, M., {Le Coarer}, E., {Boulesteix}, J., {Georgelin}, Y., et
  {Monnet}, G.: 1987,
\newblock {\em \aap} {\bf 179}, 101

\bibitem[\protect\astroncite{{Mathewson} et~al.}{1992}]{Mathewson:1992}
{Mathewson}, D.~S., {Ford}, V.~L., et {Buchhorn}, M.: 1992,
\newblock {\em \apjs} {\bf 81}, 413

\bibitem[\protect\astroncite{{McCracken} et~al.}{2003}]{McCracken:2003}
{McCracken}, H.~J., {Radovich}, M., {Bertin}, E., {Mellier}, Y., {Dantel-Fort},
  M., {Le F{\`e}vre}, O., {Cuillandre}, J.~C., {Gwyn}, S., {Foucaud}, S., et
  {Zamorani}, G.: 2003,
\newblock {\em \aap} {\bf 410}, 17

\bibitem[\protect\astroncite{{Merritt} et~al.}{2006}]{Merritt:2006}
{Merritt}, D., {Graham}, A.~W., {Moore}, B., {Diemand}, J., et {Terzi{\'c}},
  B.: 2006,
\newblock {\em \aj} {\bf 132}, 2685

\bibitem[\protect\astroncite{{Milgrom}}{1983}]{Milgrom:1983}
{Milgrom}, M.: 1983,
\newblock {\em \apj} {\bf 270}, 371

\bibitem[\protect\astroncite{{Monnet}}{1971}]{Monnet:1971}
{Monnet}, G.: 1971,
\newblock {\em \aap} {\bf 12}, 379

\bibitem[\protect\astroncite{{Moretto} et~al.}{2006}]{Moretto:2006}
{Moretto}, G., {Bacon}, R., {Cuby}, J.-G., {Hammer}, F., {Amram}, P.,
  {Blais-Ouellette}, S., {Blanc}, P.-E., {Devriendt}, J., {Epinat}, B.,
  {Fusco}, T., {Jagourel}, P., {Hernandez}, O., {Kneib}, J.-P., {Montilla}, I.,
  {Neichel}, B., {P{\'e}contal}, E., {Prieto}, E., et {Puech}, M.: 2006,
\newblock in {\em Ground-based and Airborne Instrumentation for Astronomy.
  Edited by McLean, Ian S.; Iye, Masanori. Proceedings of the SPIE, Volume
  6269, pp. 62692G (2006).}, Vol. 6269 of {\em Presented at the Society of
  Photo-Optical Instrumentation Engineers (SPIE) Conference}

\bibitem[\protect\astroncite{{Neichel} et~al.}{2008}]{Neichel:2008}
{Neichel}, B., {Hammer}, F., {Puech}, M., {Flores}, H., {Lehnert}, M., {Rawat},
  A., {Yang}, Y., {Delgado}, R., {Amram}, P., {Balkowski}, C., {Cesarsky}, C.,
  {Dannerbauer}, H., {Fuentes-Carrera}, I., {Guiderdoni}, B., {Kembhavi}, A.,
  {Liang}, Y.~C., {Nesvadba}, N., {{\"O}stlin}, G., {Pozzetti}, L.,
  {Ravikumar}, C.~D., {di Serego Alighieri}, S., {Vergani}, D., {Vernet}, J.,
  et {Wozniak}, H.: 2008,
\newblock {\em \aap} {\bf 484}, 159

\bibitem[\protect\astroncite{{Noordermeer} et~al.}{2007}]{Noordermeer:2007}
{Noordermeer}, E., {van der Hulst}, J.~M., {Sancisi}, R., {Swaters}, R.~S., et
  {van Albada}, T.~S.: 2007,
\newblock {\em \mnras} {\bf 376}, 1513

\bibitem[\protect\astroncite{{Peng} et~al.}{2002}]{Peng:2002}
{Peng}, C.~Y., {Ho}, L.~C., {Impey}, C.~D., et {Rix}, H.-W.: 2002,
\newblock {\em \aj} {\bf 124}, 266

\bibitem[\protect\astroncite{{Persic} et {Salucci}}{1991}]{Persic:1991}
{Persic}, M. et {Salucci}, P.: 1991,
\newblock {\em \apj} {\bf 368}, 60

\bibitem[\protect\astroncite{{Persic} et~al.}{1996}]{Persic:1996}
{Persic}, M., {Salucci}, P., et {Stel}, F.: 1996,
\newblock {\em \mnras} {\bf 281}, 27

\bibitem[\protect\astroncite{{Piazza} et {Marinoni}}{2003}]{Piazza:2003}
{Piazza}, F. et {Marinoni}, C.: 2003,
\newblock {\em Physical Review Letters} {\bf 91(14)}, 141301

\bibitem[\protect\astroncite{{Plana}}{1996}]{Plana:1996}
{Plana}, H.: 1996,
\newblock {\em Ph.D. thesis}, Universit\'e de Provence (France)

\bibitem[\protect\astroncite{{Pogge} et~al.}{1995}]{Pogge:1995}
{Pogge}, R.~W., {Atwood}, B., {Byard}, P.~L., {O'Brien}, T.~P., {Peterson},
  B.~M., {Lame}, N.~J., et {Baldwin}, J.~A.: 1995,
\newblock {\em \pasp} {\bf 107}, 1226

\bibitem[\protect\astroncite{{Puech} et~al.}{2008}]{Puech:2008}
{Puech}, M., {Flores}, H., {Hammer}, F., {Yang}, Y., {Neichel}, B., {Lehnert},
  M., {Chemin}, L., {Nesvadba}, N., {Epinat}, B., {Amram}, P., {Balkowski}, C.,
  {Cesarsky}, C., {Dannerbauer}, H., {di Serego Alighieri}, S.,
  {Fuentes-Carrera}, I., {Guiderdoni}, B., {Kembhavi}, A., {Liang}, Y.~C.,
  {{\"O}stlin}, G., {Pozzetti}, L., {Ravikumar}, C.~D., {Rawat}, A., {Vergani},
  D., {Vernet}, J., et {Wozniak}, H.: 2008,
\newblock {\em \aap} {\bf 484}, 173

\bibitem[\protect\astroncite{{Puech} et~al.}{2006}]{Puech:2006}
{Puech}, M., {Hammer}, F., {Flores}, H., {{\"O}stlin}, G., et {Marquart}, T.:
  2006,
\newblock {\em \aap} {\bf 455}, 119

\bibitem[\protect\astroncite{{Rousselot} et~al.}{2000}]{Rousselot:2000}
{Rousselot}, P., {Lidman}, C., {Cuby}, J.-G., {Moreels}, G., et {Monnet}, G.:
  2000,
\newblock {\em \aap} {\bf 354}, 1134

\bibitem[\protect\astroncite{{Rubin} et~al.}{1978}]{Rubin:1978}
{Rubin}, V.~C., {Thonnard}, N., et {Ford}, Jr., W.~K.: 1978,
\newblock {\em \apjl} {\bf 225}, L107

\bibitem[\protect\astroncite{{Russeil}}{1998}]{Russeil:1998}
{Russeil}, D.: 1998,
\newblock {\em Ph.D. thesis}, Universit\'e de Provence (France)

\bibitem[\protect\astroncite{{Russeil} et~al.}{2005}]{Russeil:2005}
{Russeil}, D., {Adami}, C., {Amram}, P., {Le Coarer}, E., {Georgelin}, Y.~M.,
  {Marcelin}, M., et {Parker}, Q.: 2005,
\newblock {\em \aap} {\bf 429}, 497

\bibitem[\protect\astroncite{{Spano} et~al.}{2008}]{Spano:2008}
{Spano}, M., {Marcelin}, M., {Amram}, P., {Carignan}, C., {Epinat}, B., et
  {Hernandez}, O.: 2008,
\newblock {\em \mnras} {\bf 383}, 297

\bibitem[\protect\astroncite{{Steidel} et~al.}{2004}]{Steidel:2004}
{Steidel}, C.~C., {Shapley}, A.~E., {Pettini}, M., {Adelberger}, K.~L., {Erb},
  D.~K., {Reddy}, N.~A., et {Hunt}, M.~P.: 2004,
\newblock {\em \apj} {\bf 604}, 534

\bibitem[\protect\astroncite{{Taylor} et {Atherton}}{1980}]{Taylor:1980}
{Taylor}, K. et {Atherton}, P.~D.: 1980,
\newblock {\em \mnras} {\bf 191}, 675

\bibitem[\protect\astroncite{Vetterling et~al.}{1989}]{numericalrecipes}
Vetterling, W.~T., Flannery, B.~P., Press, W.~H., et Teukolski, S.~A.: 1989,
\newblock {\em Numerical {R}ecipes in {FORTRAN} - {T}he {A}rt of {S}cientific
  {C}omputing - {S}econd {E}dition},
\newblock University Press, Cambridge

\bibitem[\protect\astroncite{{Vogt} et~al.}{1996}]{Vogt:1996}
{Vogt}, N.~P., {Forbes}, D.~A., {Phillips}, A.~C., {Gronwall}, C., {Faber},
  S.~M., {Illingworth}, G.~D., et {Koo}, D.~C.: 1996,
\newblock {\em \apjl} {\bf 465}, L15

\bibitem[\protect\astroncite{{Weiner} et~al.}{2006}]{Weiner:2006}
{Weiner}, B.~J., {Willmer}, C.~N.~A., {Faber}, S.~M., {Melbourne}, J.,
  {Kassin}, S.~A., {Phillips}, A.~C., {Harker}, J., {Metevier}, A.~J., {Vogt},
  N.~P., et {Koo}, D.~C.: 2006,
\newblock {\em \apj} {\bf 653}, 1027

\bibitem[\protect\astroncite{{Wright} et~al.}{2007}]{Wright:2007}
{Wright}, S.~A., {Larkin}, J.~E., {Barczys}, M., {Erb}, D.~K., {Iserlohe}, C.,
  {Krabbe}, A., {Law}, D.~R., {McElwain}, M.~W., {Quirrenbach}, A., {Steidel},
  C.~C., et {Weiss}, J.: 2007,
\newblock {\em \apj} {\bf 658}, 78

\bibitem[\protect\astroncite{{Yang} et~al.}{2008}]{Yang:2008}
{Yang}, Y., {Flores}, H., {Hammer}, F., {Neichel}, B., {Puech}, M., {Nesvadba},
  N., {Rawat}, A., {Cesarsky}, C., {Lehnert}, M., {Pozzetti}, L.,
  {Fuentes-Carrera}, I., {Amram}, P., {Balkowski}, C., {Dannerbauer}, H., {di
  Serego Alighieri}, S., {Guiderdoni}, B., {Kembhavi}, A., {Liang}, Y.~C.,
  {{\"O}stlin}, G., {Ravikumar}, C.~D., {Vergani}, D., {Vernet}, J., et
  {Wozniak}, H.: 2008,
\newblock {\em \aap} {\bf 477}, 789

\end{thebibliography}

\thispagestyle{empty}
\cleardoublepage
\appendix
\phantomsection
\addstarredchapter{\textsc{Annexes}}
\chaptermarkfree{\textsc{Annexes}}

\begin{centering}
{\large\textsc{Université de Provence - Aix-Marseille I}}\\
{\large\textsc{\'Ecole Doctorale de Physique et Sciences de la Matière}}\\
{\large\textit{Laboratoire d'Astrophysique de Marseille}}\\
\vspace{1.3cm}
\LARGE{\textsc{\textbf{Thèse de Doctorat}}}\\
\vspace{0.6cm}
\normalsize{
\textit{présentée pour obtenir le grade de}\\
\vspace{0.6cm}
{\large Docteur de l'Université de Provence}\\
{\large Spécialité: Astrophysique}}\\
\vspace{0.4cm}
{\large \textit{par}}\\
\vspace{0.5cm}
{\LARGE \textbf{Benoît \textsc{Epinat}}}\\
\vspace{1.cm}

\rule{16.cm}{0.5pt}
\huge{\textsc{\textbf{Des galaxies proches aux galaxies lointaines}}}\\
\LARGE{\textsc{\textbf{\'Etudes cinématique et dynamique}}}\\
\rule{16.cm}{0.5pt}

\vspace{3.5cm}
\normalsize{
\huge{\textsc{\textbf{Annexes}}}\\
}

\end{centering}

\thispagestyle{empty}
\clearpage

\clearpage
\thispagestyle{empty}

\chapter{Articles et documents concernant l'instrumentation présentée au chapitre \ref{instrumentation}}
\chaptermarkannexe{Articles et documents concernant l'instrumentation présentée au\\chapitre \ref{instrumentation}}
\label{annexe_instrumentation}
\minitoc
\textit{Cette annexe contient des documents concernant les instruments présentés dans le chapitre \ref{instrumentation}. En particulier, les plans, les schémas ainsi que les principaux objectifs scientifiques des instruments sont présentés dans des publications.}
\hl
\section{\underline{Article V:}~3D-NTT: a versatile integral field spectro-imager for the NTT}
\label{3dntt_spie}

Un premier article concernant le projet 3D-NTT (PI: M. Marcelin) (partie \ref{3dntt}) a été publié dans les comptes-rendus de la conférence SPIE qui s'est tenue à Marseille en juin 2008 \citep{Marcelin:2008} suite à la présentation d'un poster.
\par
Le 3D-NTT est un spectro-imageur à champ intégral pour le visible permettant deux modes d'observation: un mode basse résolution avec un filtre accordabe (R~$\sim$~300 à 6000) possédant un grand champ de vue ($17'\times17'$) et un mode haute résolution (R~$\sim$~10000 à 40000) avec un Fabry-Perot à balayage ($7'\times7'$). Il sera utilisé comme instrument visiteur sur le NTT à partir de 2009. Deux grands programmes seront menés: ``Caractérisation du milieu interstellaire de galaxies proches grâce à des cartes 2D d'extinction et d'abondances'' (PI: M. Marcelin) et ``Accrétion de gaz et feed-back radiatif dans les premiers âges de l'Univers'' (PI: J. Bland Hawthorn). Ils utiliseront tous deux principalement le mode filtre accordable. Cet instrument est en cours de construction au sein d'une collaboration entre le LAM (Marseille), le G\'EPI (Paris) et le LAE (Montréal). L'adresse du site internet de cet instrument est \url{http://www.astro.umontreal.ca/3DNTT}.

\includepdf[pagecommand={\pagestyle{headings}},scale=1.,offset=0 -50,pages=-,
addtotoc={
1,subsection,2, Introduction,intro3dnnt,
1,subsection,2, L'instrument,instrument3dntt,
7,subsection,2, Programmes scientifiques,programmes3dntt,
8,subsection,2, Comparaison du 3D-NTT avec d'autres instruments \FP,comparaison3dntt
}
,addtolist={
2,figure,{\underline{Article V}, Figure 1: 3D-NTT optical design (by Immervision, Quebec). The light beam is folded from the Nasmyth focus with 3 mirrors (M1 to M3). Fabry-Perot interferometers can be placed either in the focal plane or in the pupil. A filter wheel contains blocking filters for isolating the interference order of interest.},fig1_3dntt,
2,figure,{\underline{Article V}, Figure 2: Schematic drawing of the 3D-NTT attached at the Nasmyth derotator of the NTT (violet). The two interferometers are shown as red cylinders (one in the focal plane and one in the pupil plane). Design : F. Rigaud.},fig2_3dntt,
3,figure,{\underline{Article V}, Figure 3: Photograph of the mechanics of the Tunable Filter of the 3D-NTT. The overall diameter is 30 cm (the useful diameter is 10 cm). The height is 12 cm. The lower right corner insert shows a piezoactuator (not at scale). The three piezoactuators are located at 120\degr~around the central hole. The weight is 16 kg (without glass plates and accessories).},fig3_3dntt,
4,figure,{\underline{Article V}, Figure 4: Control Interferogram for the worst glass plate of the interferometers of the 3D-NTT. PTV value is 26 nm and rms value is 1.8 nm.},fig4_3dntt,
4,figure,{\underline{Article V}, Figure 5: Controller board of the interferometer. The "One Euro" coin gives the scale.},fig5_3dntt,
5,table,{\underline{Article V}, Table 1: Low resolution mode characteristics.},tab1_3dntt,
5,figure,{\underline{Article V}, Figure 6: Quantum efficiency of the 4k$\times$4k CCD used for the low resolution mode.},fig6_3dntt,
6,table,{\underline{Article V}, Table 2: High resolution mode characteristics.},tab2_3dntt,
6,figure,{\underline{Article V}, Figure 7: Quantum efficiency of the L3CCD used for the high resolution mode.},fig7_3dntt
}]{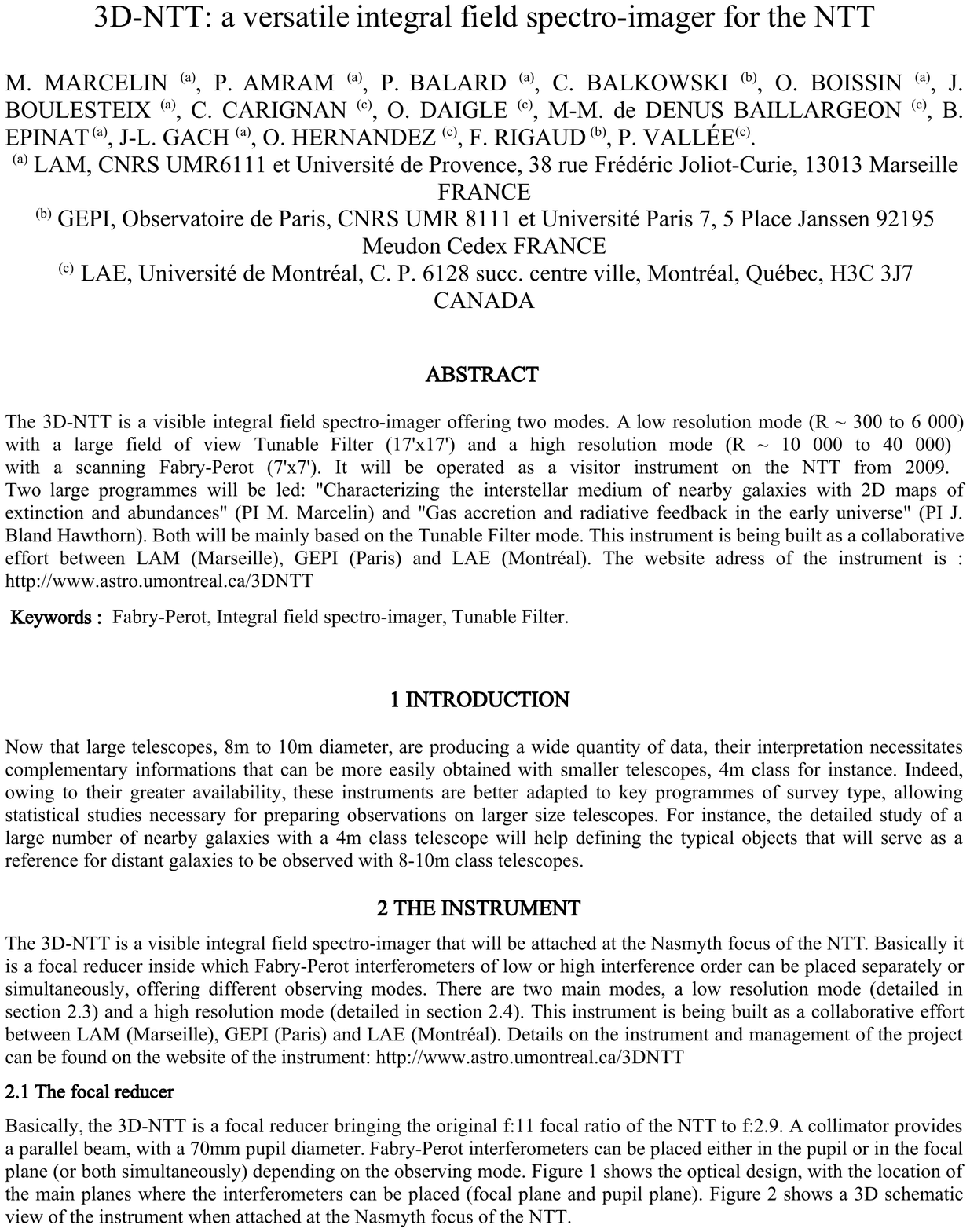}

\section{GH$\alpha$FaS}
\label{annexe_ghafas}

\subsection{\underline{Article VI:}~GH$\alpha$FaS: Galaxy H$\alpha$ Fabry-Perot System for the William Herschel Telescope}
\label{ghafas_pasp}
Un article sur la première lumière de l'instrument GH$\alpha$FaS (présenté dans la partie \ref{ghasp_fantomm_ghafas}) a été publié par \citet{Hernandez:2008} dans une revue à comité de lecture\footnote{Publications of the Astronomical Society of the Pacific}.
\par
GH$\alpha$FaS, un nouvel instrument à Fabry-Perot, est désormais opérationnel au William Herschel Telescope (WHT). Il a été monté pour la première fois au foyer Nasmyth du WHT ($4.2~m$) à La Palma en juillet 2007. Utilisant une technologie moderne, avec une résolution spectrale de l'ordre de R~$\sim15000$, une résolution spatiale limitée par la turbulence atmosphérique et un champ de vue circulaire de $4.8'$, GH$\alpha$FaS apportera un nouveau regard sur le gaz ionisé émettant dans la raie \Ha~dans l'univers proche. Beaucoup de programmes scientifiques peuvent être étudiés avec un Fabry-Perot à balayage sur un télescope de $4.2~m$ couplé à des conditions atmosphériques de qualité. Les galaxies, mais aussi les régions HII, les nébuleuses planétaires, les restes de supernova et le milieu interstellaire diffus sont autant de sujets pour lesquels des données uniques peuvent être obtenues rapidement. Des astronomes du Laboratoire d’Astrophysique Expérimentale (LAE) de Montréal, du Laboratoire d’Astrophysique de Marseille (LAM-OAMP) et de l'Instituto de Astrofísica de Canarias (IAC) ont inauguré GH$\alpha$FaS en étudiant en détail la dynamique de plusieurs galaxies spirales proches. Un ensemble d'outils robustes de réduction et d'analyse des cubes de données obtenus par GH$\alpha$FaS a également été développé.

\includepdf[pagecommand={\pagestyle{headings}},scale=1.,offset=0 0,pages={1-16},
addtotoc={
1,subsubsection,3, Introduction,introghafas,
2,subsubsection,3, Programmes scientifiques, programmesghafas,
4,subsubsection,3, Un instrument à champ intégral, ifughafas,
10,subsubsection,3, Réduction des données, reductionghafas,
14,subsubsection,3, Conclusions, conclusionghafas
}
,addtolist={
5,figure,{\underline{Article VI}, Figure 1: Two-dimensional optical layout of GH$\alpha$FaS. The optical design was done in collaboration with Immervision inc. Light beam from telescope comes from the left. The detector (IPCS) can been found at the right-hand side. Different shades of the optical rays represent different positions on the focal plane. See the electronic edition of the PASP for a color version of this figure.},fig1_ghafas,
5,figure,{\underline{Article VI}, Figure 2: Spot Diagram of GH$\alpha$FaS in the detector plane for five different positions of the optical rays at a wavelength of 6560 $A$. See the electronic edition of the PASP for a color version of this figure.},fig2_ghafas,
5,table,{\underline{Article VI}, Table 1: GH$\alpha$FaS optical characteristics.},tab1_ghafas,
6,table,{\underline{Article VI}, Table 2: GH$\alpha$FaS optical prescription.},tab2_ghafas,
7,table,{\underline{Article VI}, Table 3: GH$\alpha$FaS Fabry-Perot etalons characteristics at \Ha.},tab3_ghafas,
7,figure,{\underline{Article VI}, Figure 3: Top view and cut along the optical axis parallel to the optical breadboard at the Nasmyth focus. The optical and mechanical design of GH$\alpha$FaS has been especially customized for the GHRIL Nasmyth focus of the WHT. The filter wheel can fit four 75-mm filters and, like all the elements of the system, can be controlled remotely via ethernet cables.},fig3_ghafas,
8,table,{\underline{Article VI}, Table 4: Summary of the GH$\alpha$FaS automated parts.},tab4_ghafas,
8,figure,{\underline{Article VI}, Figure 4: Left: Principle of electron amplification using micro channel plates (MCP) in an image intensifier. Right: Details of high-voltage effect on a micro channel.},fig4_ghafas,
9,figure,{\underline{Article VI}, Figure 5: Signal-to-noise ratio comparison: IPCS and CCD at the WHT. The red line shows the S/N for an IPCS, for a 3 hr exposure of 40 cycles with 48 channels, on a 4.2-m telescope with a pixel size of 0.5$''$ assuming a trough-put of the telescope of 80\%. The blue line shows the S/N of a scientific grade CCD with a QE of 90\% and a readout noise ($\sigma$) of 3e$^{-}$ in the same exposure conditions. The faint detection limit is placed at a S/N of 3. See the electronic edition of the PASP for a color version of this figure.},fig5_ghafas,
10,figure,{\underline{Article VI}, Figure 6: Computed nonlinearity of an IPCS detector at different frame rates. See the electronic edition of the PASP for a color version of this figure.},fig6_ghafas,
11,figure,{\underline{Article VI}, Figure 7: Examples of derotation.},fig7_ghafas,
12,figure,{\underline{Article VI}, Figure 8: Kinematical maps of four objects. Top left: NGC 5427 with NGC 5426. Top right: NGC 4376. Bottom left: NGC 6384. Bottom right: NGC 5954 with NGC 5953.},fig8_ghafas,
13,figure,{\underline{Article VI}, Figure 9: \Ha~images of four objects. Top left: NGC 5427 with NGC 5426. Top right: NGC 4376. Bottom left: NGC 6384. Bottom right: NGC 5954 with NGC 5953.},fig9_ghafas,
14,figure,{\underline{Article VI}, Figure 10: Left: [N II]658.3 nm image of Planatary Nebula M1-75 with a pixel size of $0.2''$. Right: associated expansion velocity field. Range of velocity is from 76 km s$^{-1}$ to 231 km s$^{-1}$.},fig10_ghafas,
14,table,{\underline{Article VI}, Table 5: Observed objects with GH$\alpha$FaS during 6 nights.},tab5_ghafas
}]{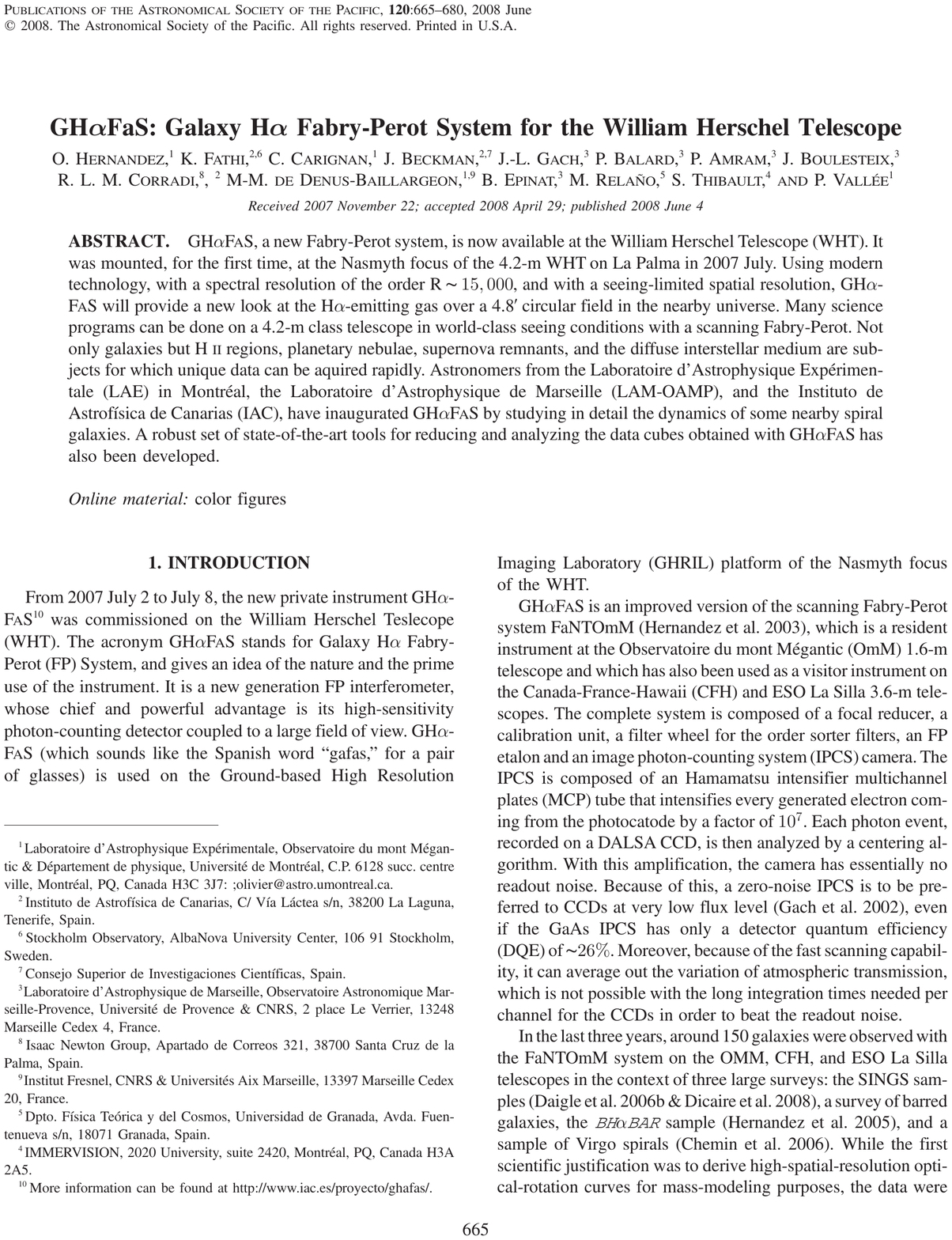}

\subsection{Comment monter GH$\alpha$FaS?}
\label{ghafas_setup}

Durant les deux périodes d'observation de l'année 2008 qui ont suivi les premières observations de juillet 2007, un manuel pour le montage de l'instrument a été rédigé par Marie-Maude de Denus Baillargeon et moi-même. Il décrit toutes les opérations à réaliser (assemblage des éléments optiques, câblage des éléments, réglages optiques, configuration réseau), étape par étape, images à l'appui afin que l'utilisateur de l'instrument ait tous les éléments pour que la mission d'observation soit un succès. Je ne présente ici que la première page de ce manuel.
\includepdf[pagecommand={\pagestyle{headings}},scale=0.9,pages={1}]{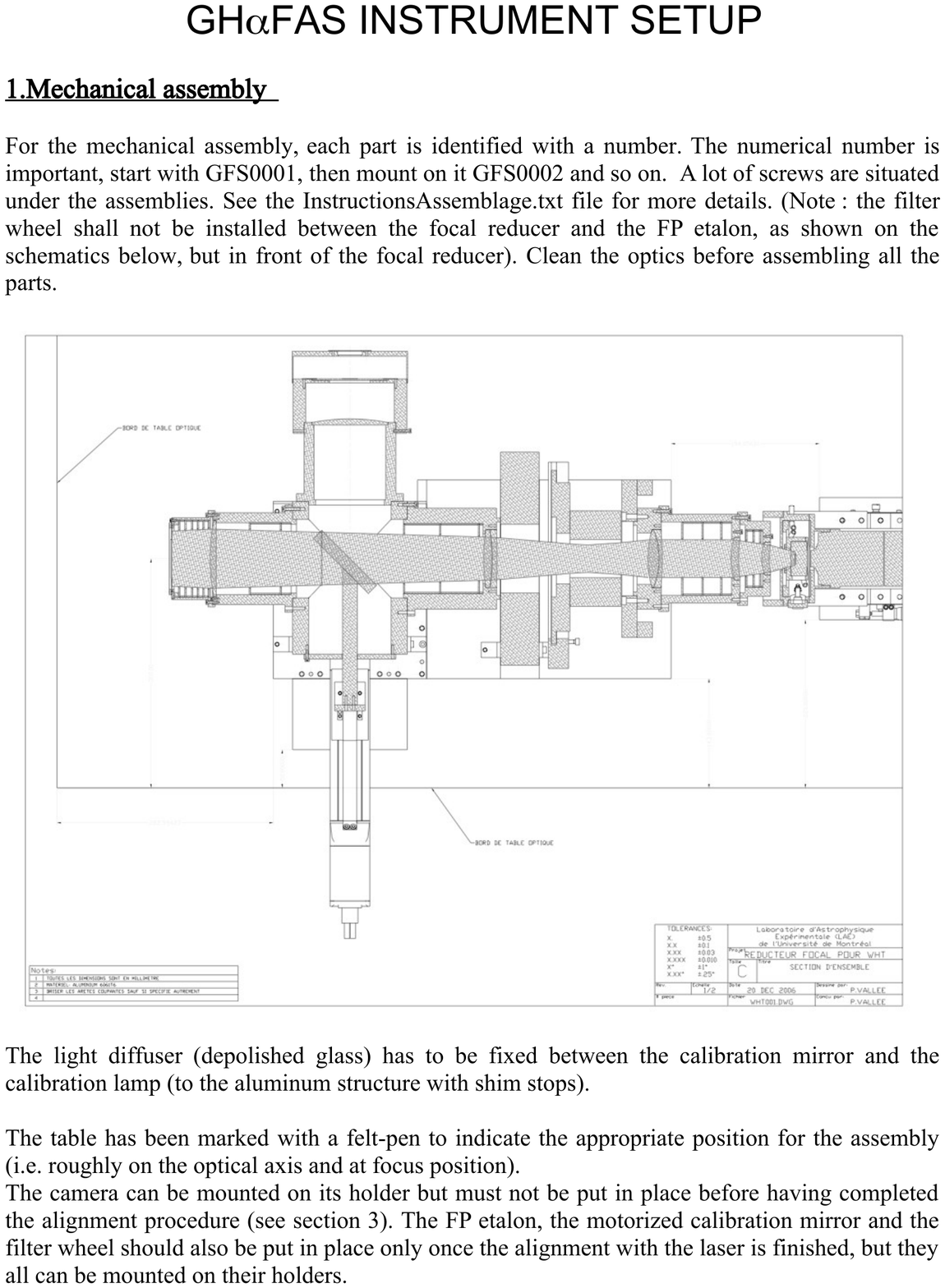}

\section{\underline{Article VII:}~Wide Field Spectrograph Concepts for the European Extremely Large Telescope}
\label{wfspec_spie}

Cet article a été publié dans les comptes-rendus de la conférence SPIE qui a eu lieu à Orlando en juin 2006 par \citet{Moretto:2006}. Il discute des objectifs scientifiques du projet WFSpec et présente les trois concepts présentés dans la section \ref{wfspec_ibtf}.
\par
Les spécifications de haut niveau liées aux programmes scientifiques pour un spectrographe à grand champ sur un Extremely Large Telescope (ELT) ainsi que les divers concepts envisagés sont présentés. Les dessins préliminaires correspondant aux différents concepts instrumentaux sont exposés: un champ intégral monolithique (IFU), un champ intégral subdivisé (multi-IFU) et un nouveau type de filtre accordable. Ce travail s'intègre dans les activités effectuées dans le cadre du groupe de travail ``Instrumentation'' du ``ELT Design Study'', un programme subventionné par la Communauté Européenne, Framework Programme 6.

\includepdf[pagecommand={\pagestyle{headings}},scale=1.,pages=-,offset=0 20,
addtotoc={
1,subsection,2, Introduction,introwfspec,
2,subsection,2, Programmes scientifiques et spécifications, programmeswfspec,
3,subsection,2, Concepts et performances, conceptswfspec,
10,subsection,2, Conclusions, conclusionwfspec
}
,addtolist={
2,table,{\underline{Article VII}, Table 1: High level specifications for science case\#1: the Early Universe, the first galaxies and the end of the re-ionization epoch. The objective is to observe statistical samples of z > 7 objects, up to redshifts $\sim$10-15 for a complete census of the formation of the galaxies and of the physical processes that led to the re-ionization of the Universe.},tab1_wfspec,
3,table,{\underline{Article VII}, Table 2: High level specifications for science case\#2: the mass assembly of galaxies across cosmic times (z=[1-5]). Moderate resolution spectroscopy of homogeneous samples of galaxies for chemical and dynamical analyzes (mapping of absorption and emission line velocities, metallicities, extinction and ionization, etc.) of the galaxies from redshift $\sim$1 to 5 for a complete census of baryonic and dark matter assembled and evolved in galaxies.},tab2_wfspec,
4,figure,{\underline{Article VII}, Figure 1: Monolithic wide-field mode. Left: schematic concept, similar to the MUSE instrument at the VLT. The entrance field of view is split over several modules, each consisting in an image slicer and a spectrograph. Right:  3D view of one of the modules of the instrument.},fig1_wfspec,
4,figure,{\underline{Article VII}, Figure 2: The 3D view of all 45 spectrographs constituting the monolithic wide-field mode.},fig2_wfspec,
5,table,{\underline{Article VII}, Table 3: The optical parameters for the monolithic IFU instrument concept. },tab3_wfspec,
6,table,{\underline{Article VII}, Table 4: The monolithic wide-field mode concept: expected performance at 2 $\mu$m.},tab4_wfspec,
6,figure,{\underline{Article VII}, Figure 3: Multi IFU Mode instrument: (a) the target acquisition system and (b) the 3D view of the 40 channels with the steering mirror, the WFS, and the central telescope focal plane.},fig3_wfspec,
7,table,{\underline{Article VII}, Table 5: The optical parameters for the multi-IFU mode instrument concept.},tab5_wfspec,
8,table,{\underline{Article VII}, Table 6: The inputs parameters for the Holographic-Tunable Filter concept.},tab6_wfspec,
9,figure,{\underline{Article VII}, Figure 4: The Holographic-Tunable Filter Instrument. Top:  fifty-two spectral bands (blue, green, yellow and red curves) are selected in between the main OH night sky lines (green lines) already lowered by an ad hoc OH-suppression filter. Bottom: imaging Bragg Tunable Filters (i-BTFs): The light is coming from the right (green arrow) and reaches successively: (1) the tunable filter (2) the 4 scanning i-BTFs (3) the first module containing 16 arms. Each arm contains one i-BTF and one 4k$\times$4k CCD. The second and third modules are identical to the first one, except that they select the bluer regions of the spectrum. Note: The central beam before the tunable filter contains the whole spectrum. The Airy function is transmitted (Finesse 20), during the scanning process, the whole spectrum is transmitted. Each i-BTF extracts a spectral band on the 0-order beam.},fig4_wfspec
}]{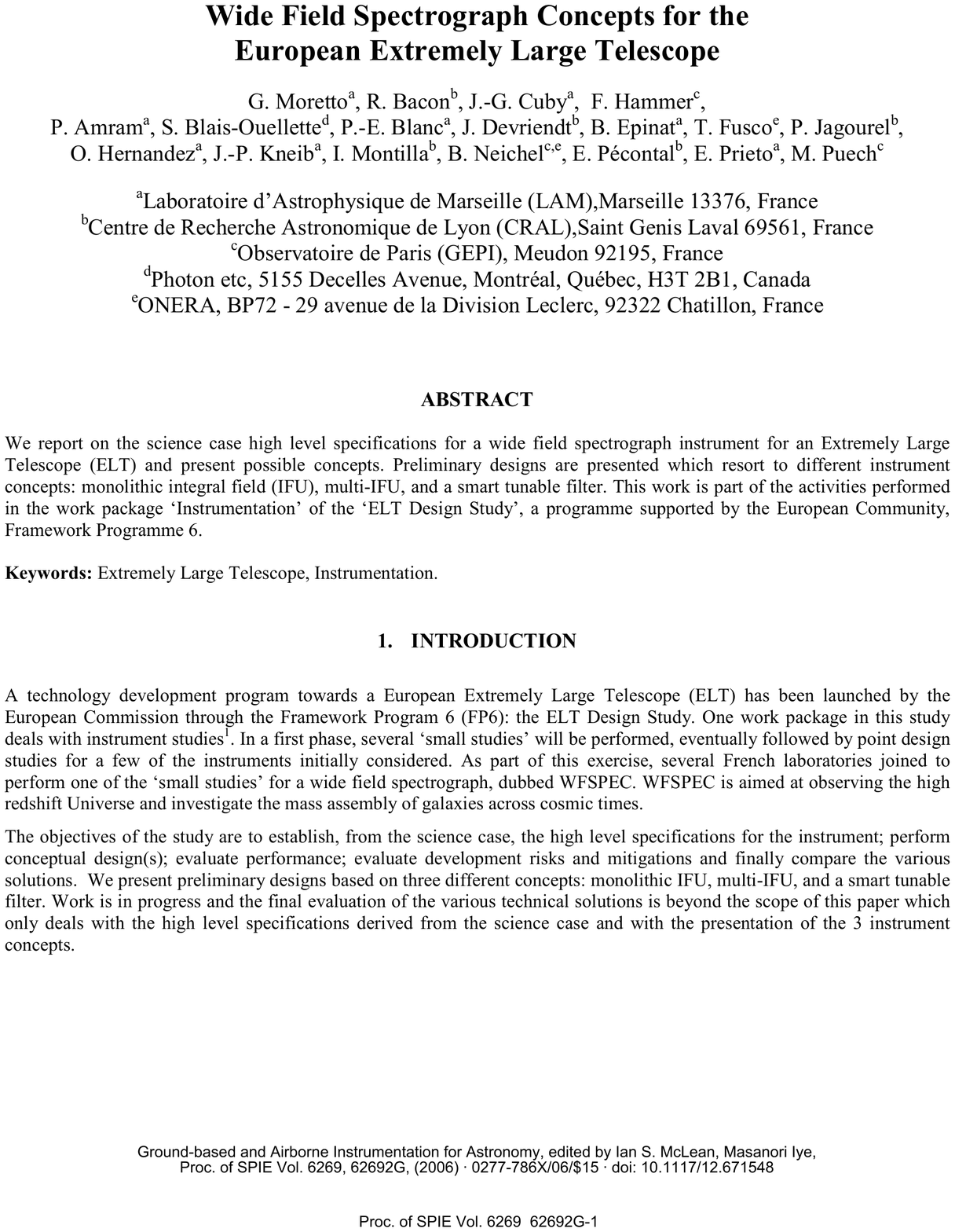}

\chapter{Aide du programme utilisé pour réduire les données Fabry-Perot présentées au chapitre~\ref{ghasp_donnees}}
\chaptermarkannexe{Aide du programme utilisé pour réduire les données Fabry-Perot présentées au chapitre~\ref{ghasp_donnees}}
\label{help_computeeverything}
\minitoc
\textit{Au cours de cette thèse, le programme de réduction écrit en langage IDL a été amélioré et utilisé pour la réduction des données GHASP présentées dans le chapitre \ref{ghasp_donnees}. J'ai donc mis le manuel d'utilisation à jour dans une version traduite en anglais présentée dans cette annexe.
Actuellement, ce programme de réduction ne gère que les données acquises par les caméras à comptage de photons. Pour les données obtenues avec une caméra CCD, il est donc nécessaire de créer préalablement un cube de données en utilisant par exemple le logiciel ``ADHOCw''.}
\hl

\section{Le programme principal: \textbf{computeeverything}}
The main procedure is called \textbf{computeeverything}. By launching this procedure, all the steps of the reduction will be managed:
\begin{itemize}
\item Integration of observation files (.ada);
\item Creation of a wavelength calibration phase map from calibration observations;
\item Wavelength calibration of the cube;
\item Spectral smoothing of the cube;
\item Spectral correction of the cube using night sky spectrum from the cube itself;
\item Night sky spectrum subtraction;
\item Adaptive binning and/or spatial smoothing of the data cube;
\item Construction of kinematical maps (continuum, line flux, radial velocity, velocity dispersion);
\item Computation of the astrometry for all the products using ``Karma/Koords'' software (semi-automatic);
\item Correction of the free spectral range uncertainty;
\item Cleaning of the maps;
\item Automatic adjustment of a data cube from a user modified velocity field.
\end{itemize}

\subsection{Description des options de \textbf{computeeverything}}

The syntax to call the procedure is:\\
\\
\textit{pro \textbf{computeeverything}, dir, cieldir, calibdir, adadir, conffile = conffile, name = name, astron = astron, plot = plot, fits = fits, correctguiding = correctguiding, guidingnobreak = guidingnobreak, ignorecycle = ignorecycle, spectral\_smooth = spectral\_smooth, snr = snr, targetsn = targetsn, fwhm = fwhm, remove\_sky = remove\_sky, ohmap\_remove\_sky = ohmap\_remove\_sky, ohmap\_degree = ohmap\_degree, maps = maps, clean = clean, manual\_clean = manual\_clean, fsr=fsr, align = align, thresholdalign = thresholdalign, maxalign = maxalign, passalign = passalign, coadd = coadd, continumflat = continumflat, noflat = noflat, autorvmonolevel = autorvmonolevel, expand = expand, adjustrv\_nosmooth = adjustrv\_nosmooth, tryhard\_voro = tryhard\_voro, calibadhoc = calibadhoc, startfromneb = startfromneb, adacieldir = adacieldir, scalefactorciel = scalefactorciel, exit\_at\_end = exit\_at\_end}\\
\\
The most important options of this procedure can be summarized in a configuration file (fits header format) produced by the user friendly Qt based program \textbf{reducWizard} available at \url{http://www.astro.umontreal.ca/~odaigle/reduction}. This program uses information given by the user and can also use information stored in observation files (.adt and .adp files).
Thus the syntax become lighter:\\
\\
\textit{pro \textbf{computeeverything}, conffile = conffile, align = align, thresholdalign = thresholdalign, maxalign = maxalign, passalign = passalign, coadd = coadd, continumflat = continumflat, noflat = noflat, autorvmonolevel = autorvmonolevel, expand = expand, adjustrv\_nosmooth = adjustrv\_nosmooth, tryhard\_voro = tryhard\_voro, calibadhoc = calibadhoc, startfromneb = startfromneb, adacieldir = adacieldir, scalefactorciel = scalefactorciel, exit\_at\_end = exit\_at\_end}\\
\\
For clarity, are first presented options that can be set using reducWizard facility. The location for setting the parameters is given in italic between brackets:
\begin{itemize}
\item \textbf{conffile:} String variable containing the name of the configuration file created by reducWizard.
\item \textbf{name:} String variable containing the name of the object (\textit{Params / Object name}). Note that it will be used for astrometry purpose only.
\item \textbf{dir:} String variable containing the directory name in which the products of the data reduction are stored (\textit{Data / Output directory}). This directory should contain an .adp file with the information about the observation when no configuration file created by \textbf{reducWizard} is not used.
\item \textbf{calibdir:} String vector variable containing the directory names where are stored the raw calibration files (\textit{Data / Calibration directory, directories separated by a coma}).
\item \textbf{adadir:} String variable containing the directory name in which the raw data is stored (\textit{Data / Observation directory}).
\item \textbf{cieldir:} String variable containing the directory name in which are located subdirectories with the calibration and the observation. This keyword is used for convenience, in order to avoid to have to specify the whole path twice for the two keywords calibdir and adadir. When using \textbf{reducWizard}, the whole path is automatically given in \textbf{adadir} and \textbf{calibdir}.
\item \textbf{astrom:} Binary value specifying that astrometry during the reduction is desired (\textit{Data / Misc / Semi-automatic astrometry (through koords)}). The astrometry is performed using the softaware ``Koords'' from ``Karma'' package. The user will have to interact manually with ``Koords''. However, reference XDSS data is automatically downloaded from DSS server. Be sure to have an internet connexion and to have provided a correct name for your object or to download reference images by yourself previously. Default is 0: no astrometry.
\item \textbf{plot:} Binary value indicating that maps and graphics have to be displayed (1) or not (0) during the reduction (\textit{Data / Misc / Plot while working}). Note that setting this option to 0 is problematic with options requiring user to interact. Default is 0.
\item \textbf{fits:} Binary value indicating that files must be save in fits format (1) or in ``ADHOCw'' format (0) (\textit{Data / Misc / Output to fits files (instead of AD? files)}). Default is 0, i.e. ``ADHOCw'' format.
\item \textbf{correctguiding:} Binary value setting the telescope guiding error correction option (\textit{Data / Integration / Guiding correction}). Set to one if guiding correction is required during integration of observation files (.ada). In that case, user is asked to choose at least three reference stars on the first cycle, the instructions are written in IDL terminal. If not enough stars are present, the guiding correction can be aborted by middle clicking and the configuration file (if used) will be updated accordingly. The position of the reference stars are stored in \textsf{stars.ad1} file. The file format is: line 0: x first star position, line 1: y first star position, line 2: x second star position, etc. Default is 0: no guiding correction.
\item \textbf{guidingnobreak:} Binary value specifying if a step of the guiding system has happened during the observation (\textit{also Data / Integration / Guiding correction}). This was frequently at Mont Megantic Observatory, Canada, during observations of SINGS sample. Set to 0 if a step happened. Default is 1: no step.
\item \textbf{ignorecycle:} Vector of integers indicating the cycles to ignore during the integration process of raw data (\textit{Data / Integration / Cycles to ignore separated by a coma}).
\item \textbf{spectral\_smooth:} Spectral smoothing options (2-elements vector). A post wavelength calibration spectral smoothing can be done (first element) (\textit{Data / Spectral / Neb spectral smoothing}). Spectral smoothing is usually done after the wavelength calibration (second element) (\textit{Data / Spectral / Lambda spectral smoothing}). Set to 0 for no spectral smoothing, to 1 for a hanning spectral smoothing, to 2 for a three channels gaussian smoothing. Default is [0,1]: hanning spectral smoothing after the calibration.
\item \textbf{spatial\_smooth:} Spatial smoothing options (2-elements vector). A post wavelength calibration spatial smoothing can be done (first element) (\textit{Data / Spatial / Neb spatial smoothing}). However spatial smoothing is usually done after the wavelength calibration (second element) (\textit{Data / Spatial / Lambda spatial smoothing}). Set to 0 if no spatial smoothing, to 1 if the voronoi adaptive binning technique have to be used (not available before wavelength calibration), to 2 for a $3\times3$ pixels binning or to 4 for a two dimension gaussian smoothing. Default is [0,0]: no spatial smoothing.
\item \textbf{snr:} Signal to noise ratio definition for voronoi adaptive binning (\textit{Data / Spatial / SN method}). Five definitions are available:\\
(1) $\frac{line\_flux}{\sqrt{line\_flux+line\_width\time \sigma\_{continuum}^2}}$, (2) $\frac{line\_flux}{line\_width\time \sigma\_{continuum}}$, (3) $\frac{barycenter\_heigth}{\sigma\_{continuum}}$,\\
(4) $\frac{line\_flux}{\sqrt{pixel\_flux}}$: the pixel flux is the sum of all the channels of the original data cube (without any correction), (5) $\sqrt{line\_flux}$: it is a poissonnian like signal to noise ratio definition. Default is 1.
\item \textbf{targetsn:} Target signal to noise ratio for voronoi adaptive binning (\textit{Data / Spatial / SN}). Default is 5.
\item \textbf{fwhm:} Full width at half maximum in pixels of the two dimension gaussian smoothing (\textit{Data / Spatial / FWHM}). This is a 2-elements vector, the first element used for the post wavelength calibration smoothing, and the second for the last smoothing. Default is [2,3].
\item \textbf{remove\_sky:} Binary value specifying if sky spectrum has to be removed from the data cube (\textit{Data / Sky / Sky subtraction}). 0: no. 1: yes. Default is 1.
\item \textbf{ohmap\_remove\_sky:} Method specifying the method that has to be used to subtract the sky (\textit{also Data / Sky / Sky subtraction}). Four methods are possibles. (1) Median spectrum; (2) Sky cube polynomial fitting; (3) Sky from the interferograms (non wavelength calibrated data cube), usual method for Tunable Filter data. Better if no smoothing before the wavelength calibration is done; (4) Sky cube interpolation. For photon counting camera data, (2) usually gives the best results. Default is 1.
\item \textbf{ohmap\_degree:} Degree of the polynomial sky cube fitting (\textit{Data / Sky / Polynomial degree}). Default is 4.
\item \textbf{maps:} 3-elements binary vector. First element: Total line width map (\textit{Data / Maps / Total width}). Second elements: Dispersion in the continuum map (\textit{Data / Maps / Continuum sigma}). Third element: Barycenter height (\textit{Data / Maps / Barycenter height}). Set to 1 for desired maps. Default is [0,0,0] (none of these maps).
\item \textbf{clean:} Binary value indicating if cleaning is desired (\textit{Data / Maps / Clean maps}). Set to 1 to try cleaning. Default is 0.
\item \textbf{manual\_clean:} Binary value indicating wether regions for cleaning are given by the user (1) or are computed automatically (0) (\textit{Data / Maps / Manually choose regions}). Default is 0.
\item \textbf{fsr:} Binary value indicating wether free spectral range jumps may be corrected (1) or not (0) (\textit{Data / Maps / Correct FSR jumps}). In the case the correction is requested, user is prompted to define a polygonal region by clicking on the map. When the region is selected, the user may indicate if the free spectral range has to be added (+) or subtracted (-). The map is then updated and the user can choose another region if necessary by answering the prompt. Default is 1.
\end{itemize}

The other options are only available using command line or using the option available in \textit{Data / Misc / Use modified launch command} with \textbf{reducWizard}. They are presented hereafter:
\begin{itemize}
\item \textbf{align:} Binary value that specifies that a spectral correction using sky spectrum has to be done. 0: no. 1: yes. Default is 0.
\item \textbf{thresholdalign:} Cross correlation factor value beyond which a spectrum is considered to be dominated by sky spectrum. Cross correlation is done with respect to the data cube median spectrum. The value must range from 0 to 1. Default is 0.9.
\item \textbf{maxalign:} Maximum spectral correction in channel that can be done by the spectral correction routine. Value must range from 2 and the number of spectral channels of the data cube. Default is the number of spectral channels.
\item \textbf{passalign:} Maximum number of iterations for spectral correction using sky cube. The value must be greater than 0. Usually, the algorithm should converge in two or three iterations. Default is 20.
\item \textbf{coadd:} Binary value that specifies if the directory \textsf{dir} contains a data cube that have been coadded (using the routine coadd\_lambda\_files). 0: no. 1: yes. When this option is set, only cube whose name contains the string \textsf{-coadd} are used. Default is 0.
\item \textbf{continumflat:} During spectral correction using sky cube, this option tells the routine that the \textit{flat} must be computed using the continuum level of each sky pixel (1) or that it must be computed using the data cube itself where it is not dominated by the object (0). Default is 0.
\item \textbf{noflat:} This option tells that no flat has to applied on the data cube (1) during the spectral correction using sky cube. Otherwise, the cube is corrected (0). Default is 1.
\item \textbf{autorvmonolevel:} Monochromatic flux beyond which a automatic search of emission lines correlated with the brightest neighbor lines has to be done.
\item \textbf{expand:} By default, the adaptive binning routine using only regions where calibration is considered as valid in order to eliminate vignetted areas. However, when using coadded cubes, this masking is no more valid. If the program \textsf{coaddlambdafiles} has been used with the option \textsf{expand = 1}, this keyword has to be set to 1.
\item \textbf{tryhard\_voro:} Adaptive binning algorithm accretes spectra in order to reach the requested signal to noise. However, it happens that the signal to noise diminishes when adding a new spectrum. This event makes the bin aborted and all the pixels of this bin are excluded for further bin accretion. Setting this parameter to 1 makes the previous pixels still usable (but not the one that made the signal to noise ration diminish in order to avoid an infinite loop on its spectrum) but increases the computing time. Default and recommended is 1.
\item \textbf{calib\_adhoc:} If one wants to use a calibration computed using ``ADHOCw'', this parameter must be set to one, and the file \textsf{\mbox{Phas\_prb.AD2}} must be present in the object directory. Otherwise, the calibration is computed by the data reduction software (0). Default is 0.
\item \textbf{startfromneb:} This option is particularly important since it enables to avoid the integration steps and to start from a non wavelength calibrated cube (has to be used for CCD observations instead of photon counting cameras). The cube must be called \textsf{neb.ad3}, be present in the object directory and the parameter must be set to one. Default is 0.
\item \textbf{adacieldir:} Specifies the directory in which are located observation files of the sky (when a sky has been observed, for instance because the object covers the whole field of view). By default, the sky is computed from the object cube.
\item \textbf{scalefactorciel:} This option must be used when the previous option is used. It tells the program how to compensate the integration time between object and sky observations. For instance, if the object has been observed twice the time for the sky, this parameter has to be set to 2.
\item \textbf{exit\_at\_end:} This option make the program quit IDL at the end of execution. Default is 0. When the reduction is launch from \textbf{reducWizard}, this keyword is set to 1.
\end{itemize}

\subsection{Exemple d'utilisation de \textbf{computeeverything}}

\textit{\textbf{computeeverything}, 'ngc5713', '/home/user/observations/ciel', ['M282', 'M284'], 'M283', targetsn = 7, spatial\_smooth = [0,1], align = 1, plot = 1, correctguiding = 1, spatial\_smooth = [0,0], ignorecycle = [5, 6, 8]}.\\
\par
This calling sequence of \textbf{computeeverything} proceeds to the data reduction of observations located in \textbf{/home/user/observations/ciel/M283} directory, using calibrations placed in \textbf{/home/user/obser-vations/ciel/M282} and \textbf{/home/user/observations/ciel/M284}. Cycles 5, 6 and 8 are discarded. A guiding correction will be done (\textbf{correctguiding = 1}) during the integration of observation data without searching for a guiding break (since \textbf{guidingnobreak} is set to 1 by default). The integrated data cube is then calibrated in wavelength before a hanning spectral smoothing is applied on the data (by default \textbf{spectral\_smooth} is set to [0,1]). The data cube wavelength calibration is calibrated by using sky lines (\textbf{align = 1}). The next step is the subtraction of a sky cube computed from a 4 degrees polynomial fit (default is \textbf{remove\_sky = 1}, \textbf{ohmap\_remove\_sky = 1} and \textbf{ohmap\_degree = 4}). Then an adaptive spatial binning with a target signal to noise ratio of 7 is done (\textbf{spatial\_smooth = [0,1]} and \textbf{targetsn = 5}), and the maps are created.
The following files are created:
\begin{itemize}
\item \textbf{calibration.ad3:} Calibration cube resulting from the addition of the two calibrations located in \textbf{M282} and \textbf{M284}.
\item \textbf{cal\_prb.ad2, cal\_bru.ad2, cal\_sum.ad2, cal\_rv.ad2, cal\_cen.ad1, cal\_finesse.ad1, cal\_valid.fits:} Files extracted from the calibration cube. \textbf{cal\_prb.ad2:} parabolic phase file used to convert an interferogram cube into a wavelength ordered cube. \textbf{cal\_bru.ad2:} raw phase file that contains the position in channel of the calibration line barycenter. \textbf{cal\_sum.ad2:} File containing the sum of the spectrum for each pixel of the calibration cube. \textbf{cal\_rv.ad2:} Position of the calibration line barycenter after applying the calibration to the calibration cube itself in order to check the validity of the parabolic phase. \textbf{cal\_cen.ad2:} Position (x,y) of the center of the calibration rings. \textbf{cal\_finesse.ad2:} Finesse computed in the whole calibration cube. \textbf{cal\_valid.fits:} Mask indicating where the calibration is valid in order to eliminated vignetted areas.
\item \textbf{neb.ad3:} Interferogram data cube obtained from the integration of the observation data located in \textbf{M283}.
\item \textbf{adasort.txt:} File containing the rejected channels and cycles during integration.
\item \textbf{valid\_reg.fits:} Mask indicating where the observation is valid in order to eliminate vignetted areas due to filters.
\item \textbf{stars.ad1:} File storing the position of stars used for guiding correction.
\item \textbf{lambda.ad3:} Wavelength calibrated cube after applying the parabolic phase correction to \textbf{neb.ad3}.
\item \textbf{lambda-SZ1.ad3:} Data cube after hanning spectral smoothing.
\item \textbf{lambda-SZ1-aligned.ad3:} Data cube after alinement using sky lines.
\item \textbf{lambda-SZ1-aligned-OHmap.ad3:} Sky subtracted data cube.
\item \textbf{iciel.ad3:} Sky cube.
\item \textbf{ohmap-SZ1.ad2:} Mask indicating the regions from which the sky cube as been computed.
\item \textbf{lambda-SZ1-aligned-OHmap-sn07-binned.ad3:} Cube on which have been applied the adaptive binning with a signal to noise ratio of 7. This cube contains the raw result of bin accretion.
\item \textbf{lambda-SZ1-aligned-OHmap-sn07.ad3:} Cube \textsf{lambda-SZ1-aligned-OHmap-sn07-binned.ad3} on which a Delaunay triangulation has been applied in order to smooth bin edges.
\item \textbf{lambda-SZ1-aligned-OHmap-sn07-binned-adjusted.ad3:} Cube \textsf{lambda-SZ1-aligned-OHmap-sn07-binned.ad3} on which an automatic search of emission lines in low line flux regions has been done using correlation with neighbor strong emission lines.
\item \textbf{lambda-SZ1-aligned-OHmap-sn07-adjusted.ad3:} Cube \textsf{lambda-SZ1-aligned-OHmap-sn07-binned-adjusted.ad3} on which a Delaunay triangulation has been applied in order to smooth bin edges.
\item \textbf{rv-SZ1-aligned-OHmap.ad2}, \textbf{mono-SZ1-aligned-OHmap.ad2}, \textbf{cont-SZ1-aligned-OHmap.ad2}, \textbf{disp-SZ1-aligned-OHmap.ad2:} Radial velocity (\textbf{rv}), line monochromatic flux (\textbf{mono}), continuum (\textbf{cont}) et velocity dispersion (\textbf{disp}) maps extracted from the cube \textsf{lambda-SZ1-aligned-OHmap.ad3}.
\item \textbf{rv-SZ1-aligned-OHmap-sn07-binned.ad2}, \textbf{mono-SZ1-aligned-OHmap-sn07-binned.ad2}, \textbf{cont-SZ1-aligned-OHmap-sn07-binned.ad2}, \textbf{disp-SZ1-aligned-OHmap-sn07-binned.ad2:} Radial velocity (\textbf{rv}), line monochromatic flux (\textbf{mono}), continuum (\textbf{cont}) et velocity dispersion (\textbf{disp}) maps extracted from the cube \textsf{lambda-SZ1-aligned-OHmap-sn07-binned.ad3}.
\item \textbf{rv-SZ1-aligned-OHmap-sn07.ad2}, \textbf{mono-SZ1-aligned-OHmap-sn07.ad2}, \textbf{cont-SZ1-aligned-OHmap-sn07.ad2}, \textbf{disp-SZ1-aligned-OHmap-sn07.ad2:} Radial velocity (\textbf{rv}), line monochromatic flux (\textbf{mono}), continuum (\textbf{cont}) et velocity dispersion (\textbf{disp}) maps extracted from the cube \textsf{lambda-SZ1-aligned-OHmap-sn07.ad3}.
\item \textbf{rv-SZ1-aligned-OHmap-sn07-binned-adjusted.ad2}, \textbf{mono-SZ1-aligned-OHmap-sn07-binned-adjusted.ad2}, \textbf{cont-SZ1-aligned-OHmap-sn07-binned-adjusted.ad2}, \textbf{disp-SZ1-aligned-OHmap-sn07-binned-adjusted.ad2:} Radial velocity (\textbf{rv}), line monochromatic flux (\textbf{mono}), continuum (\textbf{cont}) et velocity dispersion (\textbf{disp}) maps extracted from the cube \textsf{lambda-SZ1-aligned-OHmap-sn07-binned-adjusted.ad3}.
\item \textbf{rv-SZ1-aligned-OHmap-sn07-adjusted.ad2}, \textbf{mono-SZ1-aligned-OHmap-sn07-adjusted.ad2}, \textbf{cont-SZ1-aligned-OHmap-sn07-adjusted.ad2}, \textbf{disp-SZ1-aligned-OHmap-sn07-adjusted.ad2:} Radial velocity (\textbf{rv}), line monochromatic flux (\textbf{mono}), continuum (\textbf{cont}) et velocity dispersion (\textbf{disp}) maps extracted from the cube \textsf{lambda-SZ1-aligned-OHmap-sn07-adjusted.ad3}.
\item \textbf{binSize-SZ1-aligned-OHmap-sn07.ad2:} Map indicating for each pixel the size of the bin to which it belongs for an adaptive binning with a target signal to noise of 7.
\item \textbf{binnum-SZ1-aligned-OHmap-sn07.ad2:} Map indicating for each pixel the index of the bin to which it belongs for an adaptive binning with a target signal to noise of 7.
\item \textbf{binCentroid-SZ1-aligned-OHmap-sn07.ad2:} Mask indicating the centroids of the bins for an adaptive binning with a target signal to noise of 7. There is one point per bin.
\item \textbf{signal-SZ1-aligned-OHmap-sn07.ad2:} Signal map for an adaptive binning with a target signal to noise of 7.
\item \textbf{noise-SZ1-aligned-OHmap-sn07.ad2:} Noise map for an adaptive binning with a target signal to noise of 7.
\item \textbf{sn-SZ1-aligned-OHmap-sn07.ad2:} Signal to noise ratio map for an adaptive binning with a target signal to noise of 7.
\item \textbf{skyflat.ad2:} Amplitude correction applied on the cube when alining with sky lines (for creating \textsf{lambda-SZ1-aligned.ad3}).
\item \textbf{offset.ad2:} Spectral correction (expressed in spectral channels) applied on the cube when alining with sky lines (for creating \textsf{lambda-SZ1-aligned.ad3}).
\end{itemize}

\section{Quelques autres programmes}

\subsection{Addition de cubes de données}

Two routines are available to perform this job.

\subsubsection{Solution 1: \textbf{coaddlambdafiles}}
\label{coaddlambdafiles}

This procedure can be used in order to add two data cubes. Cubes must be observed with the same interferometer since no spectral adjustment is done (only a translation). The procedure can find the spatial and spectral translations to perform in order to add two cubes by itself. When the cubes have only few common information, a file called \textbf{alignzone.adz} can by created in the directories of both cubes to add. The zone defined in each files should correspond approximately to the same zone on the galaxy. The translation is then computed from this zone.
\par
The calling sequence of the procedure is \\
\\
\textit{\textbf{coaddlambdafiles}, dirs, align = align, remove\_sky = remove\_sky, spectral\_smooth = spectral\_smooth, ohmap\_remove\_sky = ohmap\_remove\_sky, expand = expand}.\\
\\
The options are the following:
\begin{itemize}
\item \textbf{dirs:} Directories in which is stored the data. At least two directories have to be specified. The result is created in the first directory.
\item \textbf{align, remove\_sky, ohmap\_remove\_sky, spectral\_smooth, :} These option must be identical to the one used to create the cube by \textbf{computeeverything}. They are used by \textbf{coaddlambdafiles} in order to find which cube to add.
\item \textbf{expand:} By default, \textbf{coaddlambdafiles} creates files with the same size as the file of the first directory. When addition is done in order to enlarge the field of view, this option must be set to one.
\end{itemize}

\subsubsection{Solution 2: \textbf{addcubesastrometry}}
\label{addcubesastrometry}

This procedure has been developed since the astrometry has been integrated to the reduction pipeline. It uses this information in order to recover the translations needed to add the cubes. It also uses information stored in cubes header in order to compute the translation needed in the spectral direction.
\par
The calling sequence of the procedure is \\
\\
\textit{\textbf{addcubesastrometry}, dir1, dir2, diradd, filename}.\\
\\
The options are the following:

\begin{itemize}
\item \textbf{dir1:} Directories in which is stored the first cube.
\item \textbf{dir2:} Directories in which is stored the second cube.
\item \textbf{diradd:} Directories in which is computed the added cube.
\item \textbf{filename:} Name of the files to be added that have to be the same for both cubes. It is also the name of the added cube but with the string \textsf{-coadd} before the extension.
\end{itemize}

\subsubsection{Protocole à suivre pour l'addition de cubes}

Before adding cubes, it is necessary to create them in an identical way. Here is an example for the reduction of galaxy \mbox{NGC 5033}. Since it has been observed using two different interference filters, the cubes have to be added before applying the last steps of reduction (adaptive binning, maps extraction, ...). \\
\\
\textit{\textbf{computeeverything}, 'ngc5033-1', '/home/user/observations/ciel', ['W021', 'W024'], 'W022', spetral\_smooth = [0,1], align = 1, plot = 1, correctguiding = 1, spatial\_smooth = [0,0]}.\\
\\
\textit{\textbf{computeeverything}, 'ngc5033-2', '/home/user/observations/ciel', ['W021', 'W024'], 'W023', spectral\_smooth = [0,1], align = 1, plot = 1, correctguiding = 1, spatial\_smooth = [0,0]}.\\
\\
\textit{\textbf{coaddlambdafiles}, ['ngc5033-1', 'ngc5033-2'], align = 1, spectral\_smooth = 1}\\
or\\
\textit{\textbf{addcubeastrometry}, 'ngc5033-1', 'ngc5033-2', 'ngc5033-1','lambda-SZ1-aligned-OHmap.ad3'}\\
\\
\textit{\textbf{computeeverything}, 'ngc5033-1', '/home/user/observations/ciel', ['W021', 'W024'], 'W022', spectral\_smooth = [0,1], align = 1, plot = 1, correctguiding = 1, spatial\_smooth = [0,1], tryhard\_voro = 1, targetsn = 7, coadd = 1}\\
\\
The two first calls to \textbf{computeeverything} make the basic data reduction for the two observations: integration with guiding correction, hanning spectral smoothing, sky line alinement, sky subtraction. Note that spatial adaptive binning is not done at this stage. The results are respectively stored in \textbf{ngc5033-1} and \textbf{ngc5033-2} directories.

The call of the two adding functions makes the same job. It uses files \textbf{lambda-SZ1-aligned-OHmap.ad3} previously created in directories \textbf{ngc5033-1} and \textbf{ngc5033-2} and make the addition in directory \textbf{ngc5033-1} with the name \textbf{lambda-SZ1-aligned-OHmap-coadd.ad3}.

Last call to \textbf{computeeverything} takes this cube and does the last reduction steps. All the created files will contain the substring \textbf{-coadd} in their names.

\subsection{Soustraction des reflets parasites: \textbf{ghost2d} et \textbf{ghost3d}}

It can happen that observations are affected by ghosts. A ghost subtraction routine is available in order to subtract two kinds of ghosts that are usually present on the data: a focused one and a defocused one. In order to do a clean ghost subtraction, a set of parameters needs to be adjusted by the user.
First of all, a full reduction has to be done in order to determine the calibration center (center of the reflections) but also to check the existence of such ghosts. Usually, they are present because of bright line emitting regions. In such a case, the ghost is visible on the velocity field as a symmetrical region with the same velocity. The use of adaptive binning makes the detection of ghost easier.
In order to do an accurate subtraction, it is better if the observation contains several bright stars that cause several ghosts.

The procedure is the following in order to adjust the parameters:
One must use the routine \textbf{ghost2d}. The calling sequence is:\\
\\
\textit{result = \textbf{ghost2d}(image = image, center = center, refl1 = refl1, refl2 = refl2, rin = rin, rout = rout, homot = homot, ex1 = ex1, ey1 = ey1, ex2 = ex2, ey2 = ey2)}\\
\\
Here are some explanations about the inputs:
\begin{itemize}
\item \textbf{image:} This image should be the sum over spectral dimension of the cube \textbf{neb.ad3} or \textbf{lambda.ad3}. The ghost subtraction will be calibrated using this image.
\item \textbf{center:} This is the center of reflection. If one is working in the reduction directory, by default the programs search for the center specified in \textbf{cal\_cen.ad1}, which should be the center for ghosts.
\item \textbf{refl1:} This is the reflection factor of the focused ghost. By default this parameters is set to 0.01.
\item \textbf{refl2:} This is the reflection factor of the defocused ghost. By default this parameters is set to 0.1.
\item \textbf{rin:} The defocused ghost is simply modeled as the pupil of the instrument, that is usually a circular mirror (primary) with a circular obstruction (secondary mirror). This parameter is the radius in pixel of the inner obstruction as observed on the ghost. Default is 10 pixels
\item \textbf{rout:} This is the radius in pixel of the primary mirror. Default is 38 pixels.
\item \textbf{homot:} Due to defocus, this reflection is usually not exactly symmetrical to the object. An homothetie factor is then needed to recover correctly the ghost position. Default is 1 (no homohetie).
\item \textbf{ex1:} It is possible that the center needs to be adjusted for the focused reflection. This is the shift in pixel in x direction.
\item \textbf{ey1:} This is the shift in pixel in y direction.
\item \textbf{ex2:} It is possible that the center needs to be adjusted for the defocused reflection. This is the shift in pixel in x direction.
\item \textbf{ey2:} This is the shift in pixel in y direction.
\end{itemize}
The function returns the corrected image. The user has to display it in order to check if the subtraction has been done correctly. Ghost parameters have to be adjusted separately for both reflections, by setting the reflection factor of the other reflection to 0. For the focused reflection, one should start to adjust the center (using parameters \textbf{ex1} and \textbf{ey1}) and then the reflection factor.
For the defocused reflection, one should start to adjust the center (using parameters \textbf{ex2} and \textbf{ey2}) and the homothetie factor. Once it is done, the pupil shape should be adjusted (parameters \textbf{rin} and \textbf{rout}. Then the reflection factor can be adjusted. Once these adjustments are done, a complete ghost subtraction has to be done, and eventually further adjustments can be made.

The ghost subtraction now needs to be performed on the data cube calibrated in wavelength or not by using the routine \textbf{ghost3d}. The calling sequence is:\\
\\
\textit{result = \textbf{ghost3d}(cube = cube, center = center, refl1 = refl1, refl2 = refl2, rin = rin, rout = rout, homot = homot, ex1 = ex1, ey1 = ey1, ex2 = ex2, ey2 = ey2)}\\
\\
The parameters must be set to the values previously determined. The function returns the ghost subtracted data cube that the user has to write using \textbf{writefits} or \textbf{writead3} functions (depending on the format chosen for reduction). It is better if the header/trailer of the non ghost subtracted cube is attached to the corrected cube. One can now restart data reduction by renaming the corrected cube as the non corrected one. A copy of the non corrected cube should be done for backup in case the subtraction is still not accurate enough. All the products of the reduction after cube creation have to be deleted in order that the reduction starts from the corrected cube.

\subsection{Autres routines utiles}

In this section, the most usefull routines for manipulating data are described. All the routines used for data reduction are available on the website \url{http://www.astro.umontreal.ca/fantomm/reduction}.

\begin{itemize}
\item \textbf{readad3(file\_name, trailer, xyz = re)}. This function returns the cube read from the \textsf{ad3} file \textsf{file\_name}. The cube trailer is put in the variable \textit{trailer} if specified. The trailer has the format of the structure \textsf{ad3\_trailer}, defined in the file \textsf{ad3\_trailer\_\_define.pro}. This routine can read gzipped files \textsf{.ad3.gz}. The cube is oriented as [x,y,z] when the variable \textsf{xyz = 1}. By default, the cube is oriented as [z,x,y] (z is the spectral dimension).
\item \textbf{writead3(file\_name, cube, trailer, data\_xyz = dx, write\_xyz = dw)}. This function writes the cube \textsf{cube} in the \textsf{.ad3} file \textsf{ad3}, with its trailer if specified in variable \textsf{trailer}. If no trailer is specified, a generic trailer is attached to the cube. The keyword \textsf{data\_xyz} specifies the orientation of the input cube: \textsf{data\_xyz = 0} means that the orientation is [z,x,y] (default) whereas \textsf{data\_xyz = 1} means that the orientation is [x,y,z]. The keyword \textsf{write\_xyz} tells how to write the file. When \textsf{write\_xyz = 0}, the cube is written as [z,x,y] (default). When \textsf{write\_xyz = 1}, the cube is written as [x,y,z].
The file is gzipped when the field \textsf{was\_compressed} of the \textsf{ad3\_trailer} structure is set to 1 or if the file name contains the extension \textsf{.gz}.
\item \textbf{readad2(file\_name, trailer):} This function returns the image read from the \textsf{ad2} file \textsf{file\_name}. The image trailer is put in the variable \textit{trailer} if specified. The trailer has the format of the structure \textsf{ad2\_trailer}, defined in the file \textsf{ad2\_trailer\_\_define.pro}
\item \textbf{writead2(file\_name, image, trailer):} This function writes the image \textsf{image} in the \textsf{.ad2} file \textsf{ad2}, with its trailer if specified in variable \textsf{trailer}. If no trailer is specified, a generic trailer is attached to the image.
\item \textbf{readad1(file\_name):} This function reads a \textsf{.ad1} file and returns its contents.
\item \textbf{writead1(file\_name, data):} This function creates a \textsf{.ad1} file and writes the contents of the variable \textsf{data}.
\item \textbf{ad3ToFits(ad3, fits, fits\_xyz = fx):} This function converts the \textsf{.ad3} file \textsf{ad3} into the \textsf{.fits} file \textsf{fits}. The \textsf{fits} header will be created from the \textsf{ad3} trailer. The option \textsf{fits\_xyz} tells how to write the file: \textsf{fits\_xyz = 0} writes the dimensions as [z,x,y] and \textsf{fits\_xyz = 1} writes the dimension as [x,y,z]. Default is \textsf{fits\_xyz = 1}.
\item \textbf{ad2ToFits(ad2, fits):} This function converts the \textsf{.ad2} file \textsf{ad2} into the \textsf{.fits} file \textsf{fits}. The \textsf{fits} header will be created from the \textsf{ad2} trailer.
\item \textbf{fitsToAd3(fits, ad3, trailer, ad3\_xyz = ax):} This function converts the \textsf{.fits} file \textsf{fits} into the \textsf{.ad3} file \textsf{ad3}. The variable \textsf{trailer} contains the trailer that will be happened to the file. The \textsf{ad3} trailer will be completed with the \textsf{fits} header information. The option \textsf{ad3\_xyz} tells how to write the file: \textsf{ad3\_xyz = 0} writes the dimensions as [z,x,y] and \textsf{ad3\_xyz = 1} writes the dimension as [x,y,z]. Default is \textsf{ad3\_xyz = 0}.
\item \textbf{fitsToAd2(fits, ad2, trailer):} This function converts the \textsf{.fits} file \textsf{fits} into the \textsf{.ad2} file \textsf{ad2}. The variable \textsf{trailer} contains the trailer that will be happened to the file. The \textsf{ad2} trailer will be completed with the \textsf{fits} header information.
\end{itemize}


\chapter{Détermination des vitesses et des dispersions de vitesses (chapitres \ref{ghasp_donnees} et \ref{etudes_highz})}
\label{spectre_vitesse}
\chaptermarkannexe{Détermination des vitesses et des dispersions de vitesses\\(chapitres \ref{ghasp_donnees} et \ref{etudes_highz})}
\minitoc
\textit{Cette annexe explique comment déterminer la vitesse et la dispersion de vitesses à partir d'un spectre d'une raie d'émission: les paramètres de la raie (flux, position, largeur) peuvent être déterminés par la méthode des moments (utilisée pour les données GHASP présentées au chapitre \ref{ghasp_donnees} et dans la partie \ref{ghasp_highz}) ainsi que par la méthode de l'ajustement d'une gaussienne à la raie d'émission (utilisée pour les données SINFONI présentées dans la partie \ref{sinfoni_donnees}); enfin l'effet Doppler-Fizeau dans le cadre relativiste est utilisé pour convertir les longueurs d'onde en vitesses.}
\hl
\section{Détermination des paramètres des spectres à raies d'émission}
\label{info_spectre}
\begin{figure}[h]
\begin{center}
\includegraphics[width=8cm]{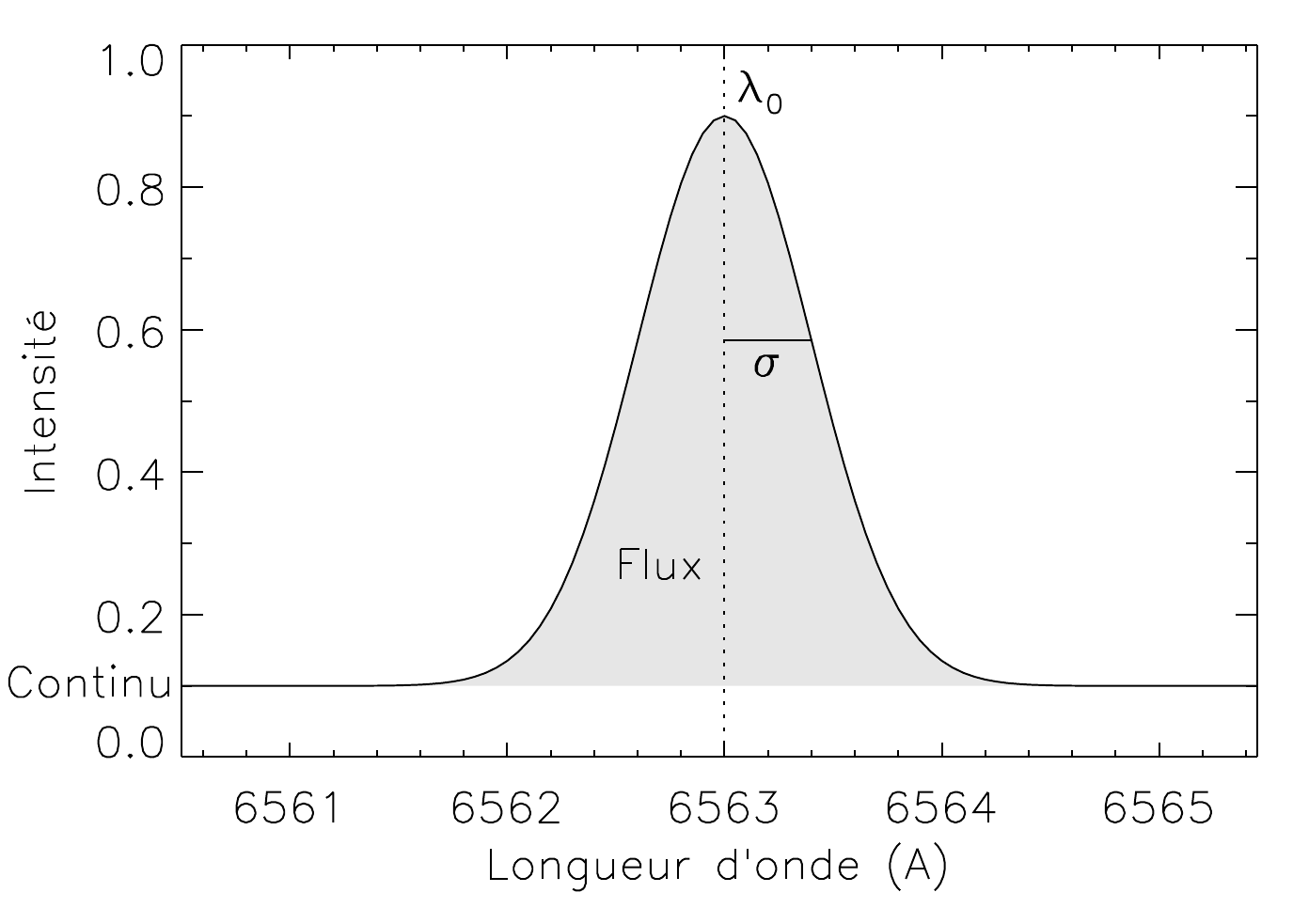}
\caption{Exemple d'une raie d'émission (gaussienne) centrée sur la longueur d'onde $\lambda_0=6563$ \AA, avec une dispersion $\sigma=0.4$ \AA~et un continuum de $0.1$. Le flux de la raie correspond à l'aire grisée.}
\label{spectre_info}
\end{center}
\end{figure}

Les paramètres d'intérêt des spectres à raies d'émission sont le niveau de continuum du spectre, le flux de la raie, sa position et sa largeur qui est généralement définie soit comme sa largeur à mi-hauteur, soit comme sa dispersion (voir Figure \ref{spectre_info}).
Il existe plusieurs méthodes pour déterminer ces paramètres. Certaines méthodes peuvent prendre en compte les effets d'asymétrie des raies dus à des phénomènes d'absorption liés à l'inclinaison des galaxies, à l'existence de profils doubles ou encore à des profils bruités. Cependant, les méthodes utilisées au cours de cette thèse n'ont pas pris en compte ces effets. En particulier, lorsque l'inclinaison des galaxies était supérieure à $75$\degr, plutôt que d'utiliser la courbe de rotation, nous avons préféré utiliser le diagramme position-vitesse à partir duquel il est possible de mesurer les vitesses en utilisant l'enveloppe externe des profils. Cette partie expose deux méthodes classiques: la méthode des moments, utilisée pour déterminer les cartes cinématiques des données GHASP (parties \ref{ghasp_projet} et \ref{ghasp_highz}), et la méthode d'ajustement d'une gaussienne, utilisée pour les données SINFONI (partie \ref{sinfoni_donnees}).

\subsection{La méthode des moments}
\label{annexe_moment}

Le spectre $S$ mesuré est discrétisé par l'échantillonnage en longueur d'onde. On peut le décomposer en une raie $R(\lambda_i)$ et un continuum $C$:
$$S(\lambda_i)=R(\lambda_i)+C$$
Le continuum est généralement estimé à partir des éléments du spectre qui sont hors de la raie: la méthode calcule la moyenne des éléments du spectre dont le flux est inférieur à un seuil. Celui-ci est déterminé à partir du profil: le maximum du profil est mesuré et le seuil correspond à la valeur du flux du premier élément autour de la raie pour lequel le flux remonte.
\par
Une fois le continuum déterminé, on peut calculer les moments de la raie d'émission.
Le flux $F$ est alors défini comme étant la somme du flux pour chaque longueur d'onde élémentaire:
$$F=\sum_i R(\lambda_i)=\sum_i \lambda_i^0 R(\lambda_i)$$
Cela correspond à la définition du moment d'ordre zéro de la raie.
La longueur d'onde $\Lambda$ de la raie va être définie comme étant son barycentre, soit sa longueur d'onde moyenne:
$$\Lambda=\frac{\sum_i \lambda_i^1 R(\lambda_i)}{F}$$
Ceci correspond au moment d'ordre un de la raie.
La dispersion $\sigma$ de la raie est l'écart moyen entre chaque longueur d'onde élémentaire du spectre de la raie et le barycentre $\Lambda$ précédemment défini:
$$\sigma=\sqrt{\frac{\sum_i (\lambda_i-\Lambda)^2 R(\lambda_i)}{F}}$$
La dispersion est ainsi déterminée à partir du moment centré d'ordre deux de la raie. En pratique, pour écourter le temps de calcul, l'algorithme utilisé emploie la relation suivante:
$$\sigma= \sqrt{\frac{\sum_i \lambda_i^2 R(\lambda_i)}{F}-\Lambda^2}$$

On pourrait également calculer les ordres successifs afin de déterminer des paramètres d'asymétrie de la raie. Cependant, plus l'ordre est élevé, plus il est nécessaire d'avoir un spectre de bonne qualité.
Cette méthode possède l'avantage de ne rien présupposer sur le spectre, en particulier sur sa symétrie.

\subsection{Ajustement d'une gaussienne}
\label{annexe_gaussienne}

Généralement, le spectre de raies d'émission est relativement bien décrit par une fonction gaussienne. Ainsi, on ajuste le spectre par une fonction gaussienne superposée à un continuum $C$:
$$S(\lambda)=I_0\exp{-\frac{(\lambda-\Lambda)^2}{2\sigma^2}}+C$$
où $I_0$ est le maximum de la raie, $\Lambda$ est la position de la raie et $\sigma$ est la dispersion en longueur d'onde de la raie. Ce sont les paramètres libres à ajuster par la méthode des moindres carrés.
Le flux $F$ est égal à l'intégrale de la fonction gaussienne:
$$F=I_0\sigma\sqrt{2\pi}$$
On relie également la largeur à mi-hauteur $FWHM$ à la dispersion de vitesses:
$$FWHM=2\sqrt{2 \ln{2}}\sigma$$
Cette méthode possède l'avantage d'être moins sensible au rapport signal sur bruit.
Par contre, cette méthode suppose que la raie est symétrique, ce qui n'est pas forcément le cas. Cependant, cette méthode a été principalement utilisée pour des données SINFONI (section \ref{sinfoni_donnees}) dont la faible résolution spectrale induit une symétrisation des profils.
De plus, il est plus aisé d'effectuer les corrections de la réponse instrumentale pour déterminer la dispersion de la raie. En effet, si on considère que la réponse instrumentale ainsi que le profil réel sont gaussiens, alors le profil observé est la convolution de deux gaussiennes. On montre alors que ce profil est également gaussien et que sa largeur est la somme quadratique des largeurs respectives de ces gaussiennes.
Cette correction quadratique est largement utilisée, même lorsque la dispersion de vitesses n'est pas déterminée à partir de cette méthode.

Un autre avantage de cette méthode est que le bruit du spectre peut être pris en compte pour l'ajustement. Il est également possible d'ajuster simultanément plusieurs raies d'émission en supposant que ces raies ont toutes les mêmes caractéristiques, ce qui peut accroître la précision de l'estimation.

\section{Des longueurs d'onde aux vitesses}
\label{doppler_fizeau}

Le passage des longueurs d'onde aux vitesses est possible grâce l'effet Doppler-Fizeau.
\'Etant donné que l'un des objets de cette thèse est l'étude de galaxies lointaines, les effets relativistes doivent être pris en compte pour déterminer les vitesses. L'effet Doppler-Fizeau doit alors être déterminé à partir des lois de la relativité restreinte.

\subsection{L'effet Doppler-Fizeau}

L'effet Doppler est un phénomène généré par le déplacement de la source d'une onde par rapport à un observateur. Il est dû à la vitesse de propagation finie de l'onde et est nommé effet Doppler-Fizeau dans le cas d'ondes lumineuses.
L'effet Doppler-Fizeau relie la fréquence émise par la source $\nu_0$ et la fréquence mesurée par le récepteur $\nu_m$ selon l'expression générale:
\begin{equation}
\nu_0=\nu_m\frac{\sqrt{1-\beta^2}}{1+\beta\cos{\theta}}
\end{equation}
$\beta=V/c$ où $V$ est la vitesse relative de la source par rapport au récepteur, $c$ est la célérité de la lumière et $\theta$ est l'angle formé entre la ligne de visée et la direction propre du mouvement.

Habituellement, les mouvements relativistes sont dus à l'expansion de l'Univers et se font principalement le long de la visée. On utilise donc la formule de l'effet Doppler-Fizeau longitudinal ($\theta=0$):
\begin{equation}
\nu_0=\nu_m\sqrt{\frac{1-\beta}{1+\beta}}
\label{nu_df}
\end{equation}

Pour nos applications, il est nécessaire d'exprimer la vitesse $V$ en fonction de la longueur d'onde mesurée $\lambda_m=c/\nu_m$ et de la longueur d'onde au repos $\lambda_0=c/\nu_0$. On obtient à partir de l'équation \ref{nu_df}:
\begin{equation}
V=c\frac{(z+1)^2-1}{(z+1)^2+1}
\label{v_df}
\end{equation}
où $z$ est le décalage spectral: $z=\frac{\lambda_m-\lambda_0}{\lambda_0}$.

\subsection{Loi de composition des vitesses}

Lorsqu'on observe la cinématique des galaxies, les vitesses mesurées correspondent à la composition d'un mouvement d'éloignement global de la galaxie et de mouvements internes à la galaxie. Il est nécessaire de prendre en compte le changement de référentiel afin d'avoir une détermination correcte des vitesses.

\subsubsection{Le décalage spectral est faible}

Lorsque le décalage spectral est faible ($z\ll1$), l'équation \ref{v_df} se simplifie:
\begin{equation}
V=c \times z
\label{v_dfz}
\end{equation}
Cela correspond donc au cas où la vitesse est faible. Dans ce cas, la composition des vitesses peut se faire dans l'approximation Newtonienne et la vitesse observée est simplement la somme de la vitesse d'entraînement et de la vitesse dans le référentiel de la galaxie.

\subsubsection{Le décalage spectral est important}

Des galaxies dont le décalage spectral est supérieur à $1$ sont présentées dans la partie \ref{sinfoni_donnees}. Pour ces galaxies, il est nécessaire de faire la composition des vitesses en utilisant les lois de la relativité restreinte. Notons $\lambda_0$ la longueur d'onde au repos de la raie observée. Notons $V$ les vitesses, $\lambda$ les longueurs d'onde et $z=\frac{\lambda-\lambda_0}{\lambda_0}$ les décalages spectraux. L'indice $e$ indique qu'il s'agit de valeurs correspondant à l'entraînement et l'indice $m$ est réservé aux valeurs mesurées.
Enfin, notons $V_r$ la vitesse dans le référentiel de la galaxie observée ($r$ pour relative). Toutes les vitesses sont des vitesses le long de la ligne de visée.

Les lois de composition des vitesses de la relativité restreinte donnent:
\begin{equation}
V_r=\frac{V_m-V_e}{1-\frac{V_m V_e}{c^2}}
\label{composition}
\end{equation}

On peut écrire l'équation \ref{v_df} pour $V_e$ et $V_m$ et ainsi obtenir en utilisant l'équation \ref{composition} après simplification:
\begin{equation}
V_r=c\frac{(z_m+z_e+2)(z_m-z_e)}{(z_m+1)^2+(z_e+1)^2}\sim c\frac{z_m-z_e}{1+z_e}
\end{equation}
Ce qui donne en termes de longueurs d'onde:
\begin{equation}
V_r\sim c\frac{\lambda_m-\lambda_e}{\lambda_e}
\label{v_l}
\end{equation}
L'équivalence finale provient du fait que les mouvements dans la galaxies sont d'amplitude négligeable par rapport à la vitesse d'éloignement.

Finalement, tout se passe comme si la longueur d'onde au repos était $\lambda_e$ au lieu de $\lambda_0$. Il n'y a qu'une dilatation du spectre.

\subsection{Dispersion de vitesses}

\'Etant donné que la largeur de raie naturelle est négligeable par rapport à la réponse impulsionnelle spectrale des spectrographes utilisés et que cette même réponse impulsionnelle est inférieure à la largeur de raie observée, cette dernière s'explique par une dispersion de vitesses le long de la ligne de visée. La dispersion de vitesses est donc directement reliée à la dispersion de la raie en longueur d'onde. Il est alors légitime d'exprimer la dispersion de la raie en unité de vitesses. En différenciant l'équation \ref{v_l} par rapport à la variable $\lambda_m$ on obtient:
\begin{equation}
dV_r\sim c\frac{d\lambda_m}{\lambda_e}
\label{dv_dl}
\end{equation}
Dans certains cas, plusieurs raies sont observées. Elles peuvent alors être utilisées simultanément pour obtenir la vitesse et la dispersion des vitesses. Il est donc utile d'exprimer la dispersion de vitesses en fonction du décalage spectral puisqu'il est le même quelle que soit la raie:
\begin{equation}
dV_r\sim c\frac{dz_m}{1+z_e}
\label{dv_dz}
\end{equation}

Ces équations ne sont rigoureusement valables que lorsque la dispersion de vitesse est négligeable par rapport à la célérité de la lumière car on a supposé que la vitesse varie linéairement avec la longueur d'onde (ou le décalage spectral) en utilisant la différenciation. Cette hypothèse est vérifiée en pratique.
\chapter{Données cinématiques de l'échantillon GHASP décrites au chapitre \ref{ghasp_donnees}}\chaptermarkannexe{Données cinématiques de l'échantillon GHASP décrites au chapitre \ref{ghasp_donnees}}
\label{ghasp_images}
\minitoc
\textit{Les deux articles présentés dans le chapitre \ref{ghasp_donnees} présentent les données GHASP. Les cartes cinématiques constituent deux annexes à chacun de ces articles qui sont uniquement disponibles en ligne. Les courbes de rotation sont incluses dans des annexes qui font partie du corps principal de l'article. Cependant, plutôt que de montrer les cartes et les courbes de rotation des deux sous-échantillons de manière disjointe, j'ai rassemblé dans cette annexe l'ensemble des cartes cinématiques et des courbes de rotation des articles pour toutes les galaxies en les ordonnant par ascension droite croissante. De plus, les cartes de dispersion de vitesses et les profils de dispersion de vitesses qui en sont déduits sont également présentés dans cette annexe. Une présentation succincte de la base de données \FP~est faite.}
\hl
\clearpage
\section{Cartes cin\'ematiques des galaxies GHASP}
\noindent
\begin{tabular}{p{8.cm}}
\centering
\includegraphics[width=8.cm,trim=0.cm 1.cm 0.cm 1.cm, clip=true]{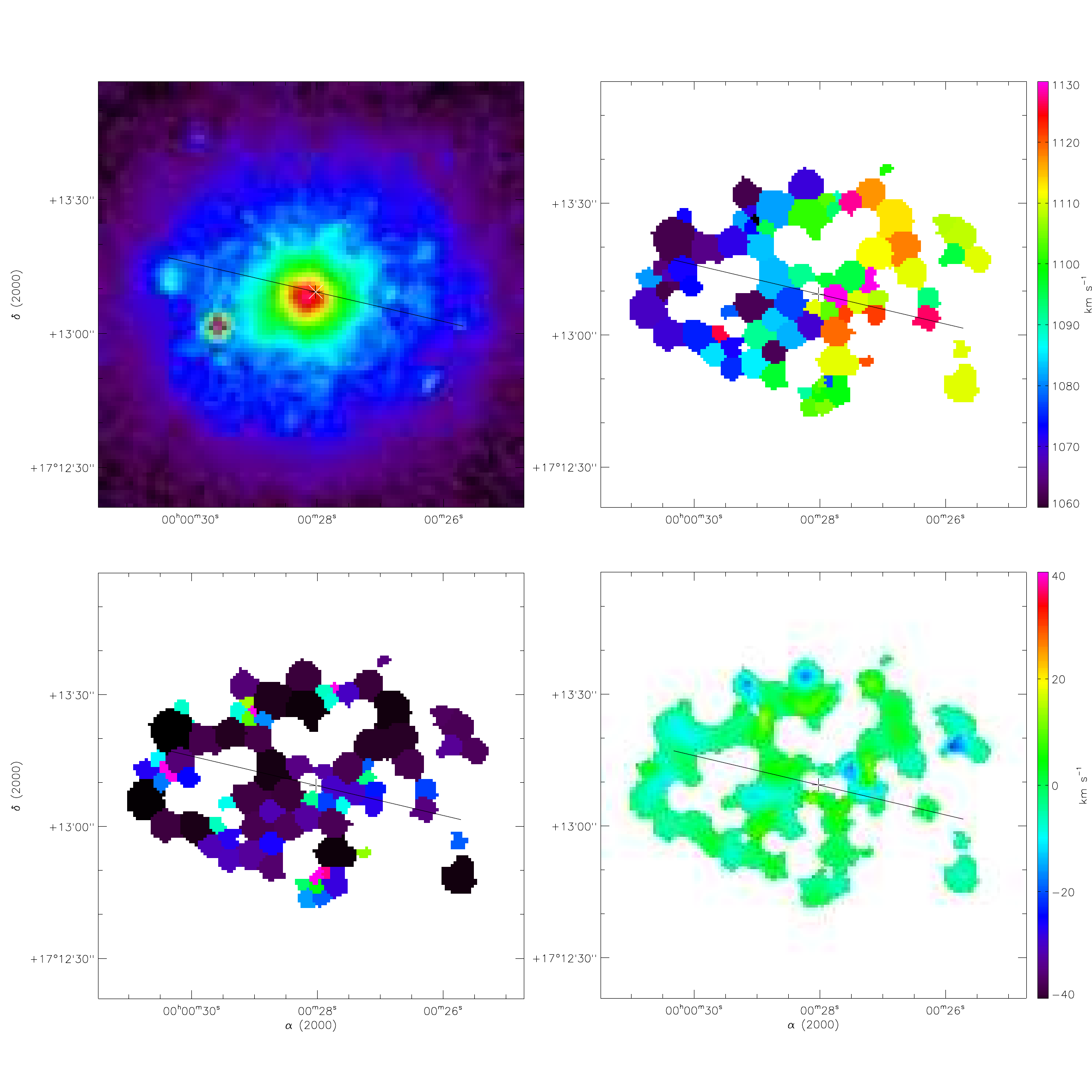}\\
\includegraphics[width=8.cm,trim=0.cm 0.cm 1.cm 0.cm, clip=true]{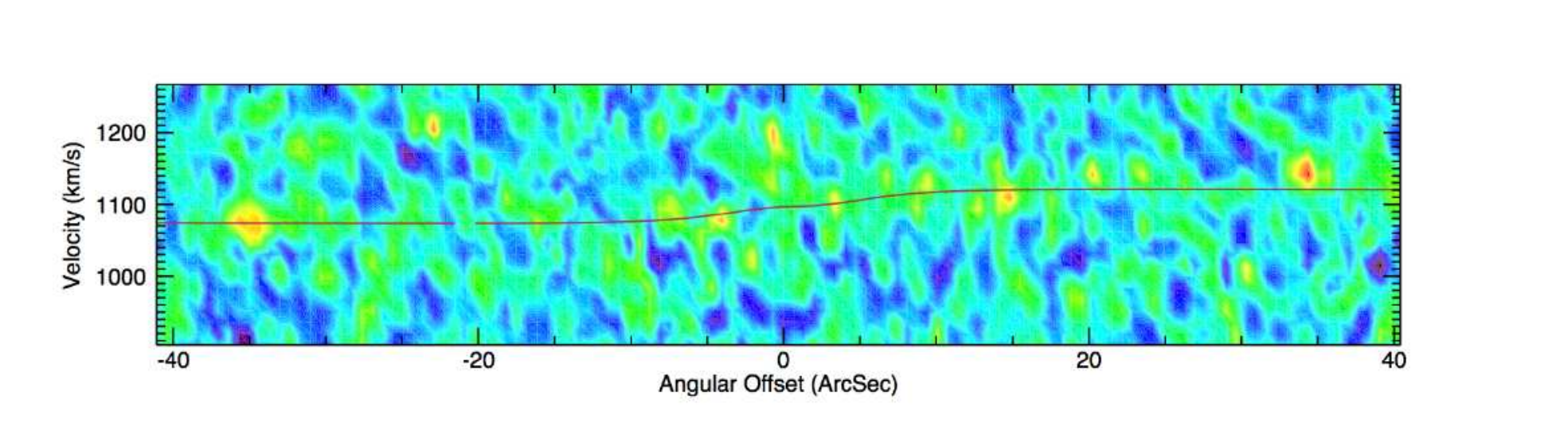}\\
(1) UGC 12893\\
\end{tabular}
\begin{tabular}{p{8.cm}}
\centering
\includegraphics[width=8.cm,trim=0.cm 1.cm 0.cm 1.cm, clip=true]{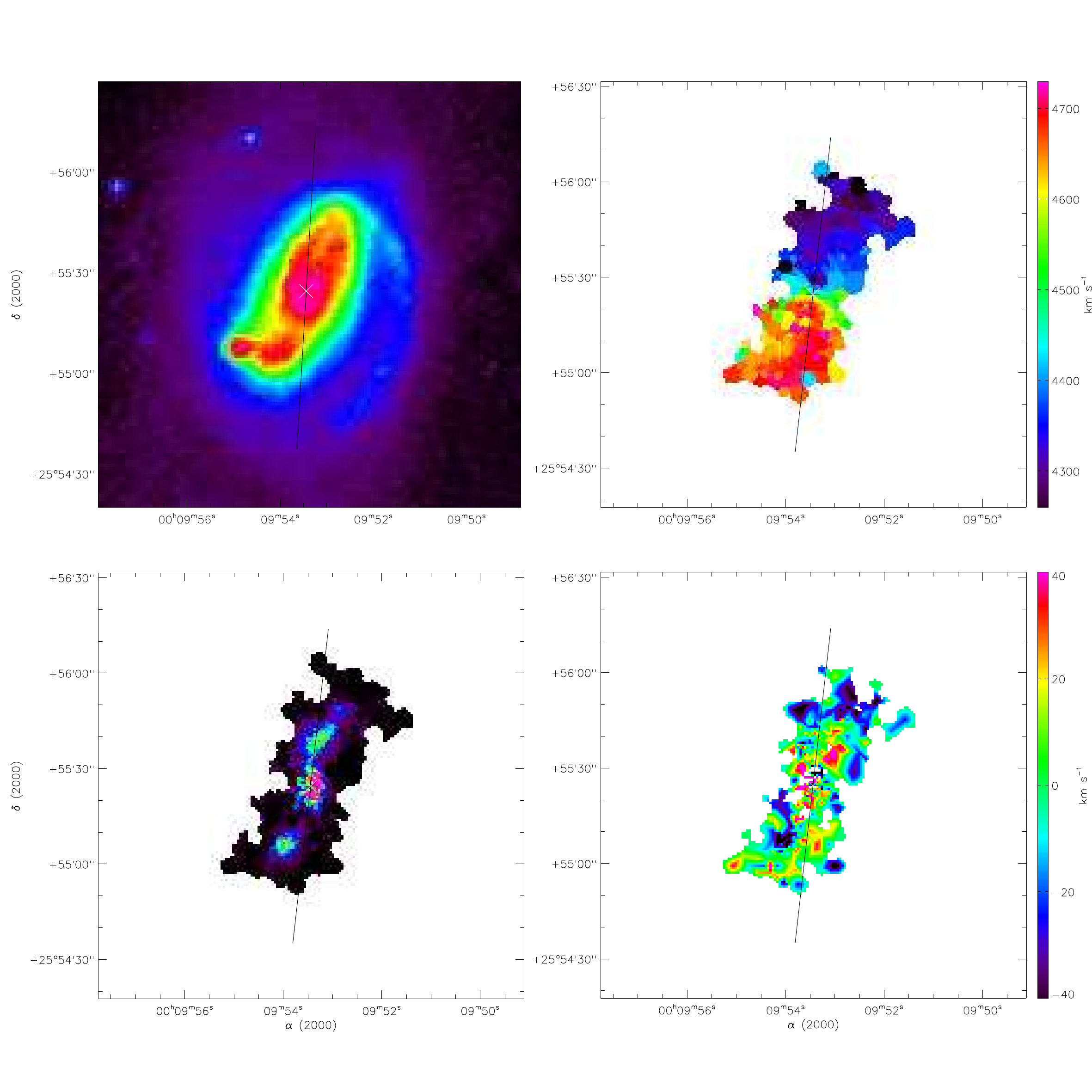}\\
\includegraphics[width=8.cm,trim=0.cm 0.cm 1.cm 0.cm, clip=true]{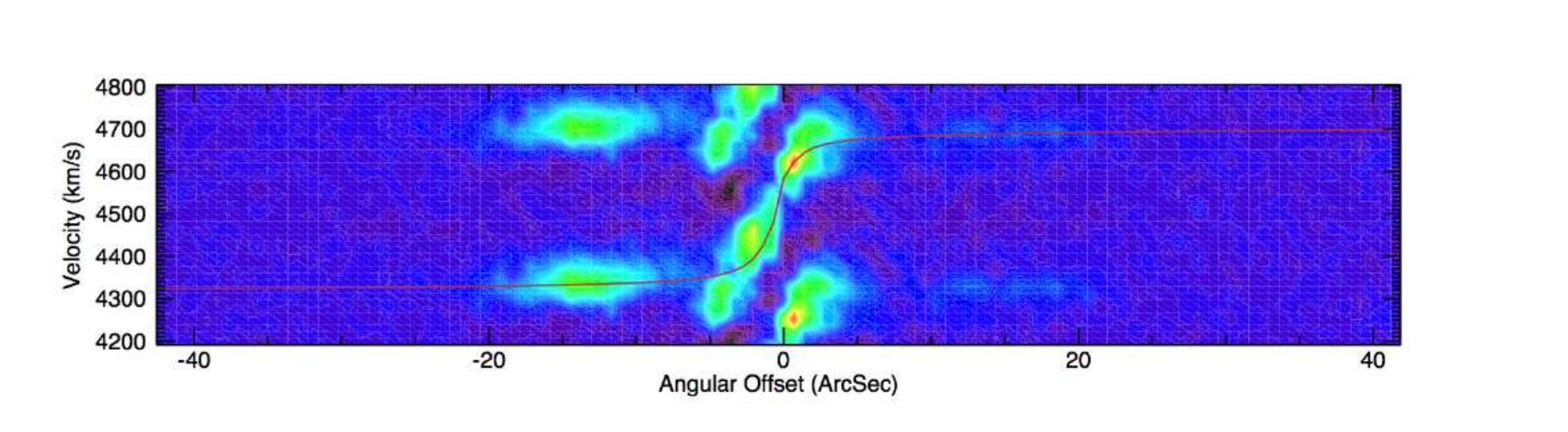}\\
(2) UGC 00089\\
\end{tabular}
\begin{figure}[h]
\caption{Cartes cin\'ematiques des galaxies GHASP. \textbf{En haut à gauche}: Image en bande B (ou R) du XDSS. \textbf{En haut à droite}: Champ de vitesse \Ha. \textbf{Au milieu à gauche}: Carte monochromatique de flux \Ha. \textbf{Au milieu à droite}: Champ de vitesse r\'esiduel lorsqu'il a été possible d'ajuster un modèle de champ de vitesses. La double croix blanche et noire repr\'esente le centre cin\'ematique. La ligne noire repr\'esente le grand axe et s'arr\^ete au rayon optique ($D_{25}/2$, \citealp{de-Vaucouleurs:1995}). \textbf{En bas}: Diagramme position-vitesse le long du grand axe (largeur totale de 7 pixels) avec une unit\'e de flux arbitraire. La ligne rouge correspond \`a la courbe de rotation cacul\'ee le long du grand axe du champ de vitesses mod\`ele.}
\label{annexe_maps}
\end{figure}
\clearpage
\noindent
\begin{center}
\textsc{Fig.} D.1: Disponible dans la version complète du manuscript \\ \url{http://tel.archives-ouvertes.fr/tel-00413769/fr/}.
\end{center}
\clearpage
\noindent
\begin{center}
\textsc{Fig.} D.1: Disponible dans la version complète du manuscript \\ \url{http://tel.archives-ouvertes.fr/tel-00413769/fr/}.
\end{center}
\clearpage
\noindent
\begin{center}
\textsc{Fig.} D.1: Disponible dans la version complète du manuscript \\ \url{http://tel.archives-ouvertes.fr/tel-00413769/fr/}.
\end{center}
\clearpage
\noindent
\begin{center}
\textsc{Fig.} D.1: Disponible dans la version complète du manuscript \\ \url{http://tel.archives-ouvertes.fr/tel-00413769/fr/}.
\end{center}
\clearpage
\noindent
\begin{center}
\textsc{Fig.} D.1: Disponible dans la version complète du manuscript \\ \url{http://tel.archives-ouvertes.fr/tel-00413769/fr/}.
\end{center}
\clearpage
\noindent
\begin{center}
\textsc{Fig.} D.1: Disponible dans la version complète du manuscript \\ \url{http://tel.archives-ouvertes.fr/tel-00413769/fr/}.
\end{center}
\clearpage
\noindent
\begin{center}
\textsc{Fig.} D.1: Disponible dans la version complète du manuscript \\ \url{http://tel.archives-ouvertes.fr/tel-00413769/fr/}.
\end{center}
\clearpage
\noindent
\begin{center}
\textsc{Fig.} D.1: Disponible dans la version complète du manuscript \\ \url{http://tel.archives-ouvertes.fr/tel-00413769/fr/}.
\end{center}
\clearpage
\noindent
\begin{center}
\textsc{Fig.} D.1: Disponible dans la version complète du manuscript \\ \url{http://tel.archives-ouvertes.fr/tel-00413769/fr/}.
\end{center}
\clearpage
\noindent
\begin{center}
\textsc{Fig.} D.1: Disponible dans la version complète du manuscript \\ \url{http://tel.archives-ouvertes.fr/tel-00413769/fr/}.
\end{center}
\clearpage
\noindent
\begin{center}
\textsc{Fig.} D.1: Disponible dans la version complète du manuscript \\ \url{http://tel.archives-ouvertes.fr/tel-00413769/fr/}.
\end{center}
\clearpage
\noindent
\begin{center}
\textsc{Fig.} D.1: Disponible dans la version complète du manuscript \\ \url{http://tel.archives-ouvertes.fr/tel-00413769/fr/}.
\end{center}
\clearpage
\noindent
\begin{center}
\textsc{Fig.} D.1: Disponible dans la version complète du manuscript \\ \url{http://tel.archives-ouvertes.fr/tel-00413769/fr/}.
\end{center}
\clearpage
\noindent
\begin{center}
\textsc{Fig.} D.1: Disponible dans la version complète du manuscript \\ \url{http://tel.archives-ouvertes.fr/tel-00413769/fr/}.
\end{center}
\clearpage
\noindent
\begin{center}
\textsc{Fig.} D.1: Disponible dans la version complète du manuscript \\ \url{http://tel.archives-ouvertes.fr/tel-00413769/fr/}.
\end{center}
\clearpage
\noindent
\begin{center}
\textsc{Fig.} D.1: Disponible dans la version complète du manuscript \\ \url{http://tel.archives-ouvertes.fr/tel-00413769/fr/}.
\end{center}
\clearpage
\noindent
\begin{center}
\textsc{Fig.} D.1: Disponible dans la version complète du manuscript \\ \url{http://tel.archives-ouvertes.fr/tel-00413769/fr/}.
\end{center}
\clearpage
\noindent
\begin{center}
\textsc{Fig.} D.1: Disponible dans la version complète du manuscript \\ \url{http://tel.archives-ouvertes.fr/tel-00413769/fr/}.
\end{center}
\clearpage
\noindent
\begin{center}
\textsc{Fig.} D.1: Disponible dans la version complète du manuscript \\ \url{http://tel.archives-ouvertes.fr/tel-00413769/fr/}.
\end{center}
\clearpage
\noindent
\begin{center}
\textsc{Fig.} D.1: Disponible dans la version complète du manuscript \\ \url{http://tel.archives-ouvertes.fr/tel-00413769/fr/}.
\end{center}
\clearpage
\noindent
\begin{center}
\textsc{Fig.} D.1: Disponible dans la version complète du manuscript \\ \url{http://tel.archives-ouvertes.fr/tel-00413769/fr/}.
\end{center}
\clearpage
\noindent
\begin{center}
\textsc{Fig.} D.1: Disponible dans la version complète du manuscript \\ \url{http://tel.archives-ouvertes.fr/tel-00413769/fr/}.
\end{center}
\clearpage
\noindent
\begin{center}
\textsc{Fig.} D.1: Disponible dans la version complète du manuscript \\ \url{http://tel.archives-ouvertes.fr/tel-00413769/fr/}.
\end{center}
\clearpage
\noindent
\begin{center}
\textsc{Fig.} D.1: Disponible dans la version complète du manuscript \\ \url{http://tel.archives-ouvertes.fr/tel-00413769/fr/}.
\end{center}
\clearpage
\noindent
\begin{center}
\textsc{Fig.} D.1: Disponible dans la version complète du manuscript \\ \url{http://tel.archives-ouvertes.fr/tel-00413769/fr/}.
\end{center}
\clearpage
\noindent
\begin{center}
\textsc{Fig.} D.1: Disponible dans la version complète du manuscript \\ \url{http://tel.archives-ouvertes.fr/tel-00413769/fr/}.
\end{center}
\clearpage
\noindent
\begin{center}
\textsc{Fig.} D.1: Disponible dans la version complète du manuscript \\ \url{http://tel.archives-ouvertes.fr/tel-00413769/fr/}.
\end{center}
\clearpage
\noindent
\begin{center}
\textsc{Fig.} D.1: Disponible dans la version complète du manuscript \\ \url{http://tel.archives-ouvertes.fr/tel-00413769/fr/}.
\end{center}
\clearpage
\noindent
\begin{center}
\textsc{Fig.} D.1: Disponible dans la version complète du manuscript \\ \url{http://tel.archives-ouvertes.fr/tel-00413769/fr/}.
\end{center}
\clearpage
\noindent
\begin{center}
\textsc{Fig.} D.1: Disponible dans la version complète du manuscript \\ \url{http://tel.archives-ouvertes.fr/tel-00413769/fr/}.
\end{center}
\clearpage
\noindent
\begin{center}
\textsc{Fig.} D.1: Disponible dans la version complète du manuscript \\ \url{http://tel.archives-ouvertes.fr/tel-00413769/fr/}.
\end{center}
\clearpage
\noindent
\begin{center}
\textsc{Fig.} D.1: Disponible dans la version complète du manuscript \\ \url{http://tel.archives-ouvertes.fr/tel-00413769/fr/}.
\end{center}
\clearpage
\noindent
\begin{center}
\textsc{Fig.} D.1: Disponible dans la version complète du manuscript \\ \url{http://tel.archives-ouvertes.fr/tel-00413769/fr/}.
\end{center}
\clearpage
\noindent
\begin{center}
\textsc{Fig.} D.1: Disponible dans la version complète du manuscript \\ \url{http://tel.archives-ouvertes.fr/tel-00413769/fr/}.
\end{center}
\clearpage
\noindent
\begin{center}
\textsc{Fig.} D.1: Disponible dans la version complète du manuscript \\ \url{http://tel.archives-ouvertes.fr/tel-00413769/fr/}.
\end{center}
\clearpage
\noindent
\begin{center}
\textsc{Fig.} D.1: Disponible dans la version complète du manuscript \\ \url{http://tel.archives-ouvertes.fr/tel-00413769/fr/}.
\end{center}
\clearpage
\noindent
\begin{center}
\textsc{Fig.} D.1: Disponible dans la version complète du manuscript \\ \url{http://tel.archives-ouvertes.fr/tel-00413769/fr/}.
\end{center}
\clearpage
\noindent
\begin{center}
\textsc{Fig.} D.1: Disponible dans la version complète du manuscript \\ \url{http://tel.archives-ouvertes.fr/tel-00413769/fr/}.
\end{center}
\clearpage
\noindent
\begin{center}
\textsc{Fig.} D.1: Disponible dans la version complète du manuscript \\ \url{http://tel.archives-ouvertes.fr/tel-00413769/fr/}.
\end{center}
\clearpage
\noindent
\begin{center}
\textsc{Fig.} D.1: Disponible dans la version complète du manuscript \\ \url{http://tel.archives-ouvertes.fr/tel-00413769/fr/}.
\end{center}
\clearpage
\noindent
\begin{center}
\textsc{Fig.} D.1: Disponible dans la version complète du manuscript \\ \url{http://tel.archives-ouvertes.fr/tel-00413769/fr/}.
\end{center}
\clearpage
\noindent
\begin{center}
\textsc{Fig.} D.1: Disponible dans la version complète du manuscript \\ \url{http://tel.archives-ouvertes.fr/tel-00413769/fr/}.
\end{center}
\clearpage
\noindent
\begin{center}
\textsc{Fig.} D.1: Disponible dans la version complète du manuscript \\ \url{http://tel.archives-ouvertes.fr/tel-00413769/fr/}.
\end{center}
\clearpage

\section{Courbes de rotation des galaxies GHASP}
\noindent
%
\begin{figure}[h]
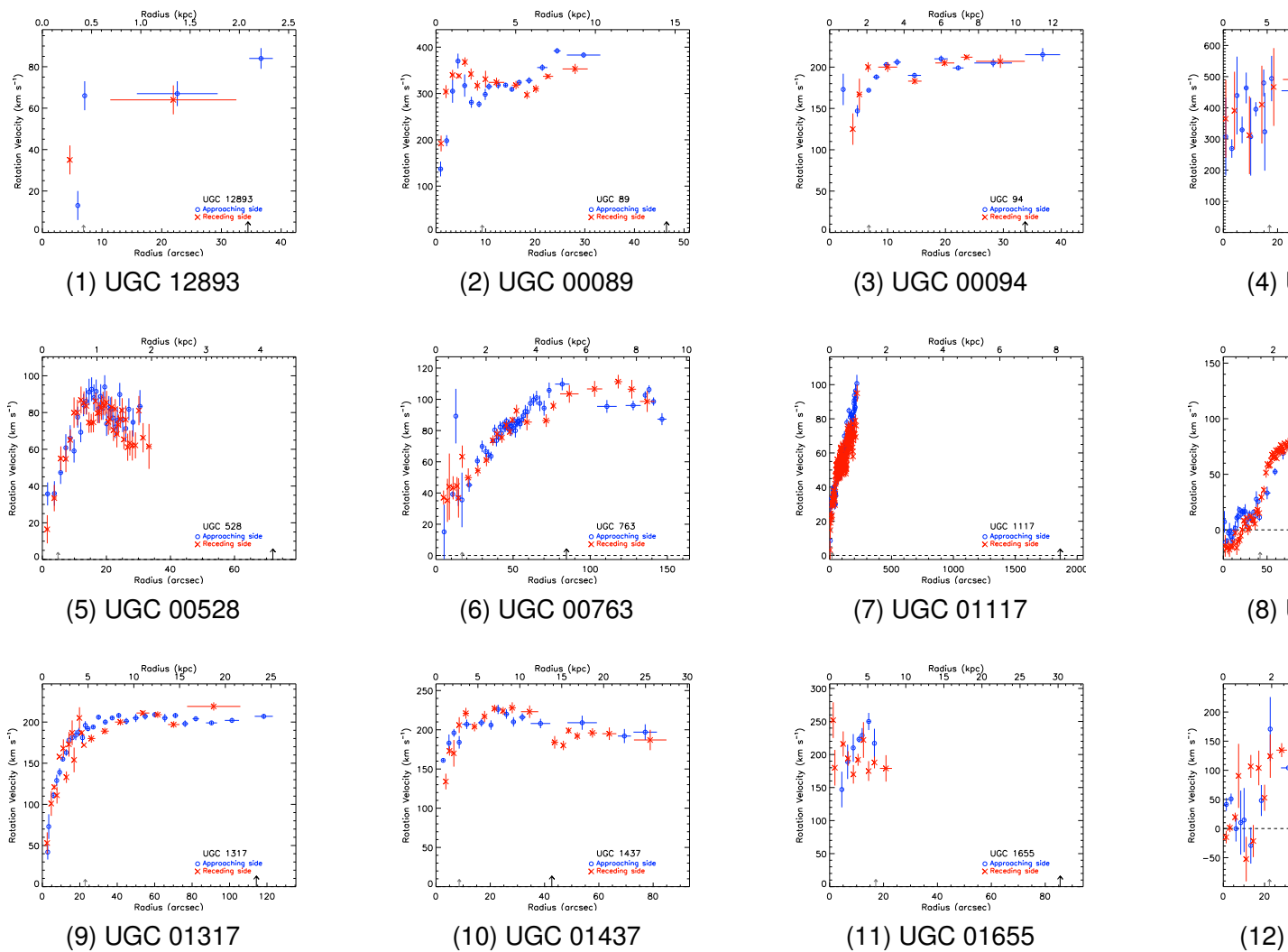

\caption{Courbes de rotation des galaxies GHASP. Les ronds bleus et les croix rouges identifient respectivement le c\^ot\'e qui s'approche et celui qui s'\'eloigne.  La fl\`eche noire indique le rayon optique ($D_{25}/2$, \citealp{de-Vaucouleurs:1995}), la fl\`eche grise indique le rayon de transition.}
\label{annexe_rcs}
\end{figure}
\clearpage
\noindent
\begin{center}
\textsc{Fig.} D.2: Disponible dans la version complète du manuscript \\ \url{http://tel.archives-ouvertes.fr/tel-00413769/fr/}.
\end{center}
\clearpage
\noindent
\begin{center}
\textsc{Fig.} D.2: Disponible dans la version complète du manuscript \\ \url{http://tel.archives-ouvertes.fr/tel-00413769/fr/}.
\end{center}
\clearpage
\noindent
\begin{center}
\textsc{Fig.} D.2: Disponible dans la version complète du manuscript \\ \url{http://tel.archives-ouvertes.fr/tel-00413769/fr/}.
\end{center}
\clearpage
\noindent
\begin{center}
\textsc{Fig.} D.2: Disponible dans la version complète du manuscript \\ \url{http://tel.archives-ouvertes.fr/tel-00413769/fr/}.
\end{center}
\clearpage
\noindent
\begin{center}
\textsc{Fig.} D.2: Disponible dans la version complète du manuscript \\ \url{http://tel.archives-ouvertes.fr/tel-00413769/fr/}.
\end{center}
\clearpage
\noindent
\begin{center}
\textsc{Fig.} D.2: Disponible dans la version complète du manuscript \\ \url{http://tel.archives-ouvertes.fr/tel-00413769/fr/}.
\end{center}
\clearpage
\noindent
\begin{center}
\textsc{Fig.} D.2: Disponible dans la version complète du manuscript \\ \url{http://tel.archives-ouvertes.fr/tel-00413769/fr/}.
\end{center}
\clearpage
\noindent
\begin{center}
\textsc{Fig.} D.2: Disponible dans la version complète du manuscript \\ \url{http://tel.archives-ouvertes.fr/tel-00413769/fr/}.
\end{center}
\clearpage
\noindent
\begin{center}
\textsc{Fig.} D.2: Disponible dans la version complète du manuscript \\ \url{http://tel.archives-ouvertes.fr/tel-00413769/fr/}.
\end{center}
\clearpage
\noindent
\begin{center}
\textsc{Fig.} D.2: Disponible dans la version complète du manuscript \\ \url{http://tel.archives-ouvertes.fr/tel-00413769/fr/}.
\end{center}
\clearpage
\noindent
\begin{center}
\textsc{Fig.} D.2: Disponible dans la version complète du manuscript \\ \url{http://tel.archives-ouvertes.fr/tel-00413769/fr/}.
\end{center}
\clearpage

\section{Cartes et profils de dispersion de vitesses des galaxies GHASP}
\noindent
\begin{tabular}{p{4.2cm}p{3.8cm}}
\centering
\includegraphics[width=3.7cm,trim=0.5cm 2.6cm 0.5cm 0.5cm, clip=true]{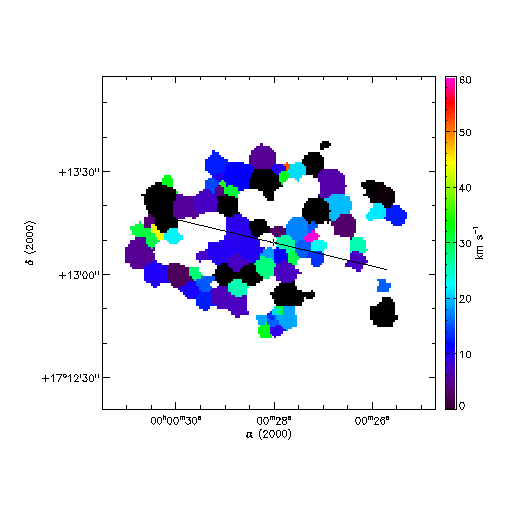} & \includegraphics[width=3.2cm,trim=0.9cm 0.cm 0.5cm 0.4cm, clip=true]{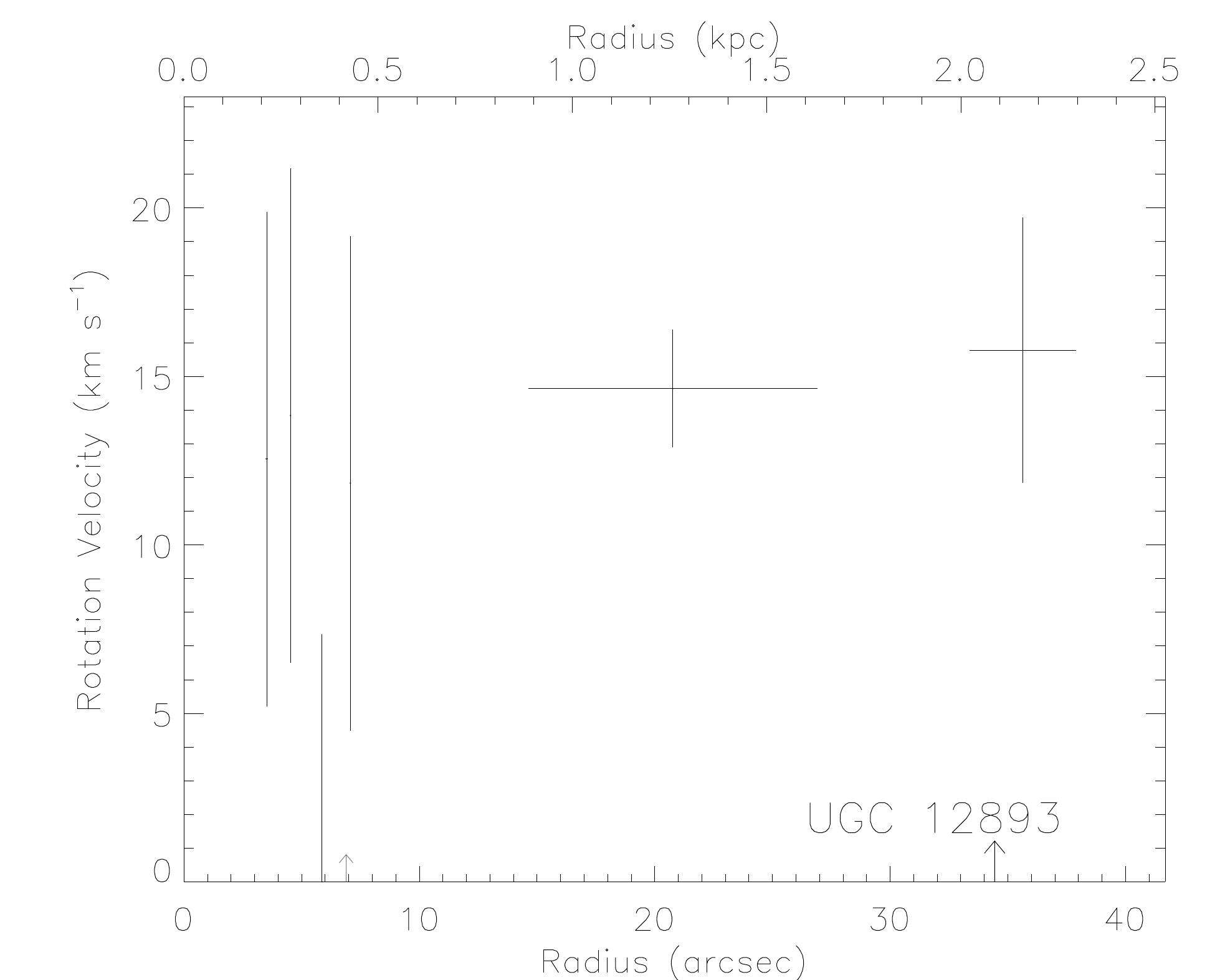}\\
\multicolumn{2}{c}{(1) UGC 12893}\\
\end{tabular}
\begin{tabular}{p{4.2cm}p{3.8cm}}
\centering
\includegraphics[width=3.7cm,trim=0.5cm 2.6cm 0.5cm 0.5cm, clip=true]{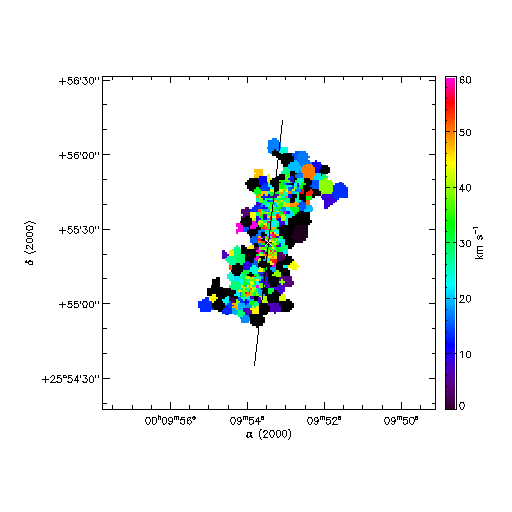} & \includegraphics[width=3.2cm,trim=0.9cm 0.cm 0.5cm 0.4cm, clip=true]{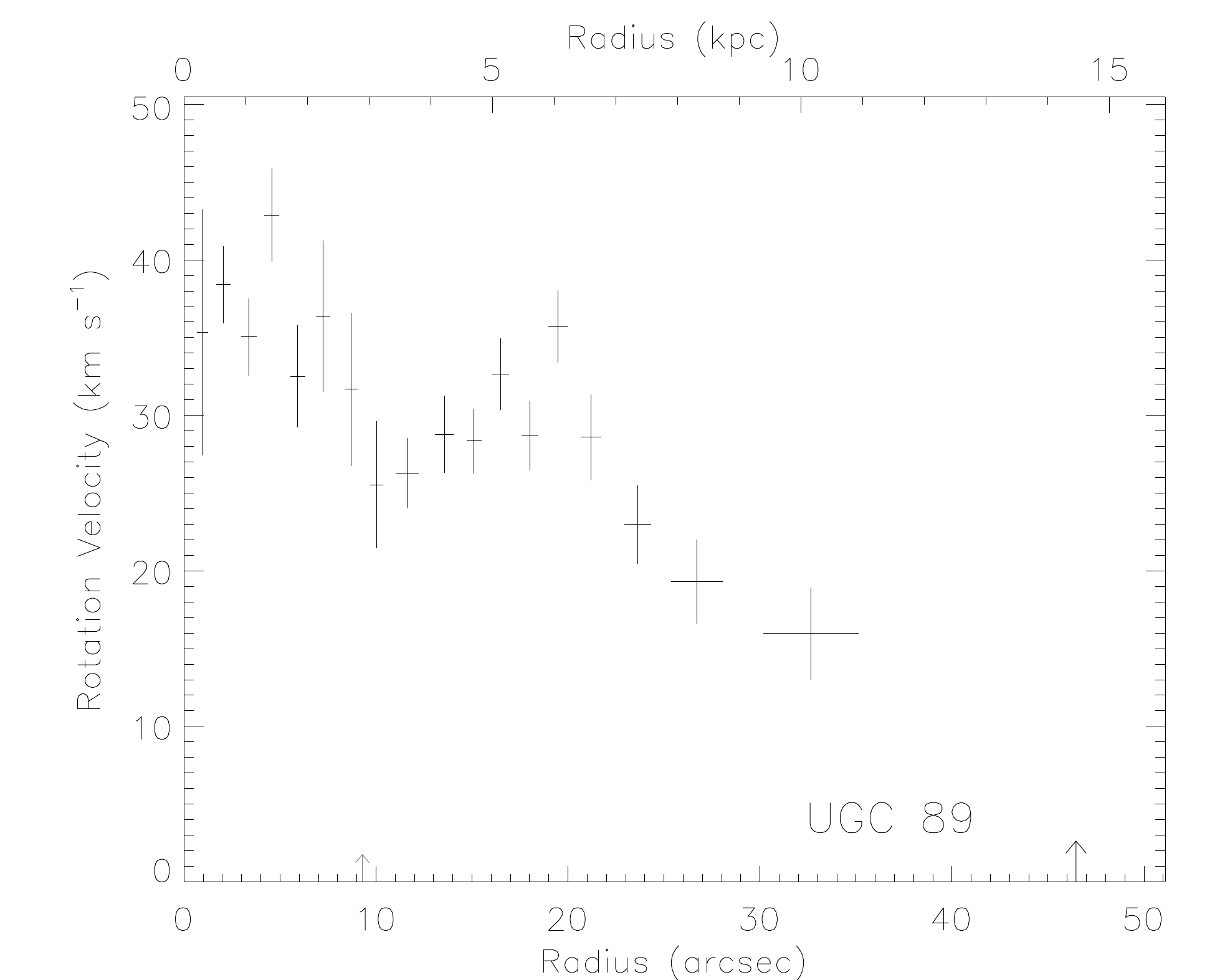}\\
\multicolumn{2}{c}{(2) UGC 00089}\\
\end{tabular}
\\
\begin{tabular}{p{4.2cm}p{3.8cm}}
\centering
\includegraphics[width=3.7cm,trim=0.5cm 2.6cm 0.5cm 0.5cm, clip=true]{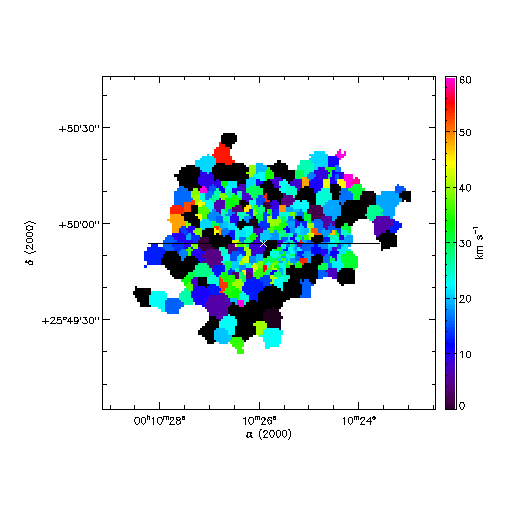} & \includegraphics[width=3.2cm,trim=0.9cm 0.cm 0.5cm 0.4cm, clip=true]{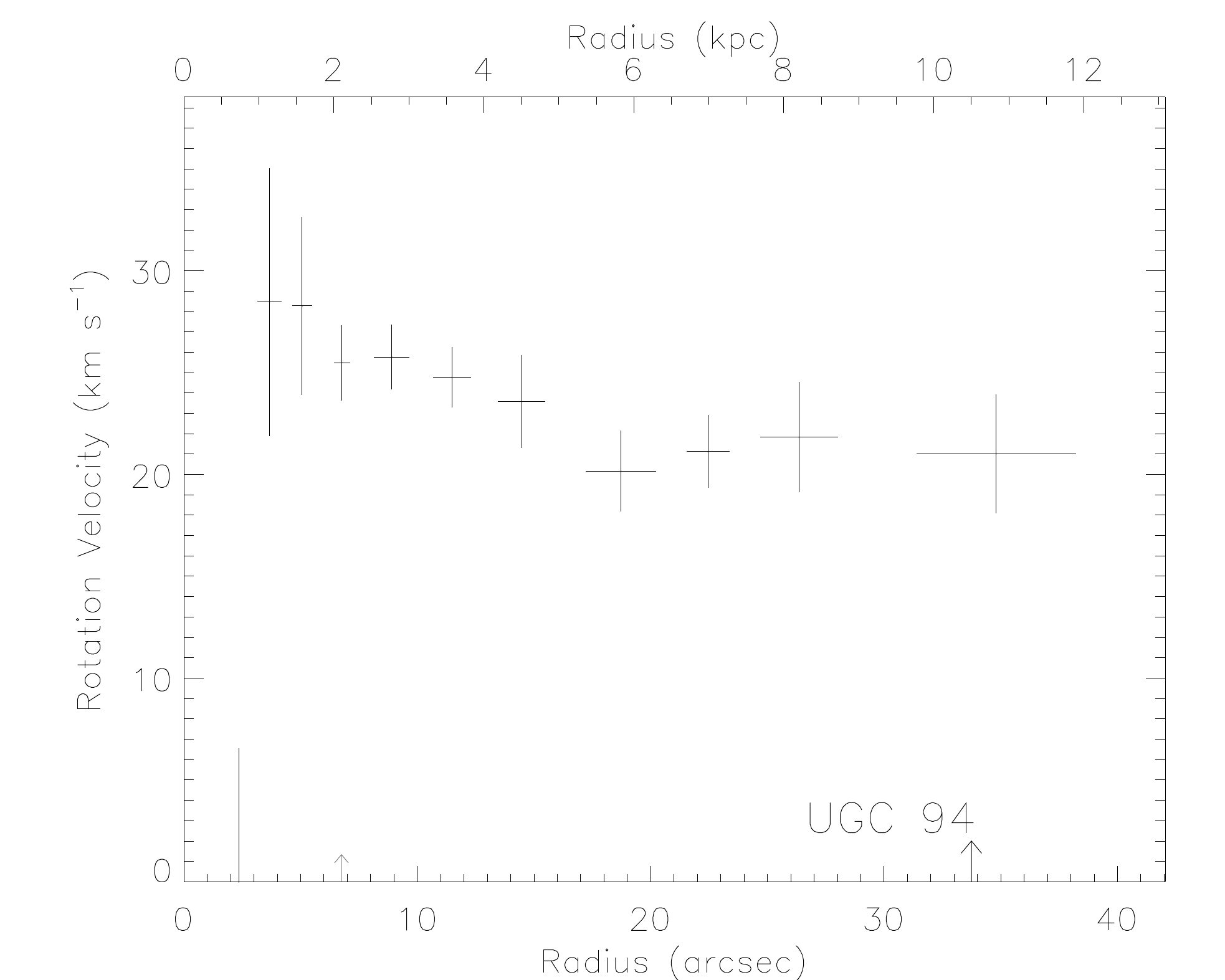}\\
\multicolumn{2}{c}{(3) UGC 00094}\\
\end{tabular}
\begin{tabular}{p{4.2cm}p{3.8cm}}
\centering
\includegraphics[width=3.7cm,trim=0.5cm 2.6cm 0.5cm 0.5cm, clip=true]{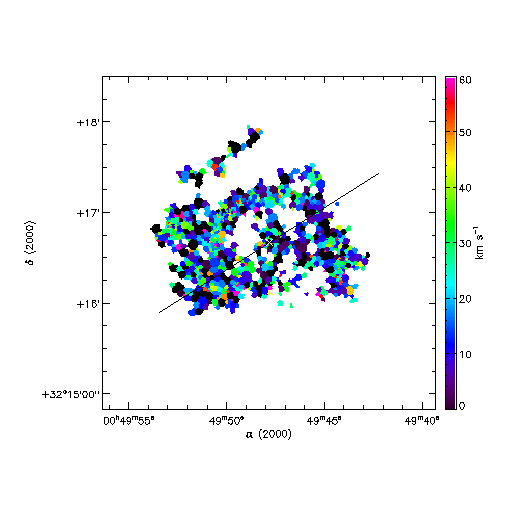} & \includegraphics[width=3.2cm,trim=0.9cm 0.cm 0.5cm 0.4cm, clip=true]{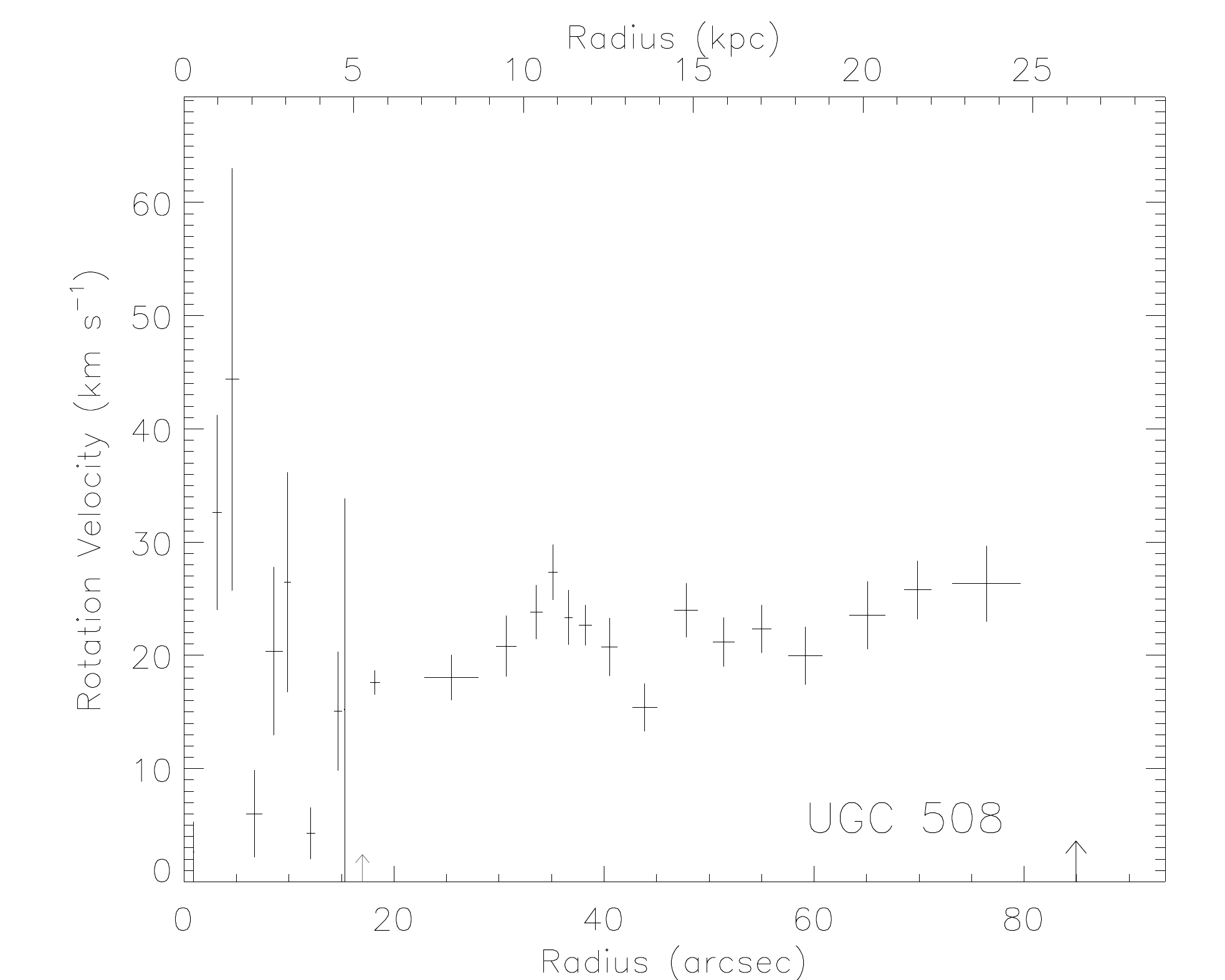}\\
\multicolumn{2}{c}{(4) UGC 00508}\\
\end{tabular}
\\
\begin{tabular}{p{4.2cm}p{3.8cm}}
\centering
\includegraphics[width=3.7cm,trim=0.5cm 2.6cm 0.5cm 0.5cm, clip=true]{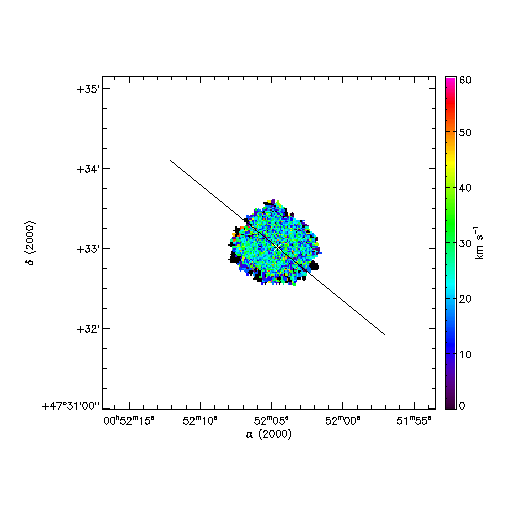} & \includegraphics[width=3.2cm,trim=0.9cm 0.cm 0.5cm 0.4cm, clip=true]{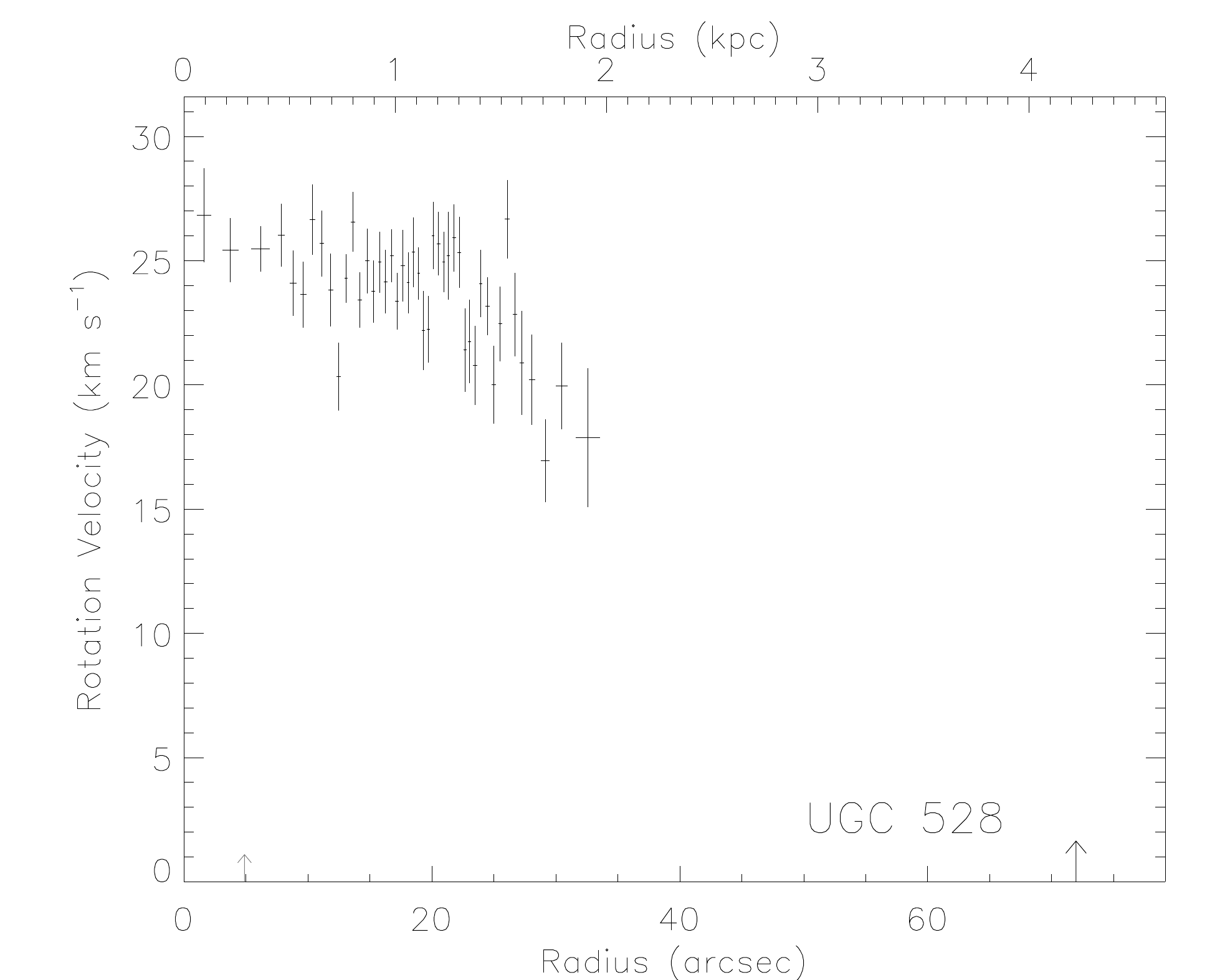}\\
\multicolumn{2}{c}{(5) UGC 00528}\\
\end{tabular}
\begin{tabular}{p{4.2cm}p{3.8cm}}
\centering
\includegraphics[width=3.7cm,trim=0.5cm 2.6cm 0.5cm 0.5cm, clip=true]{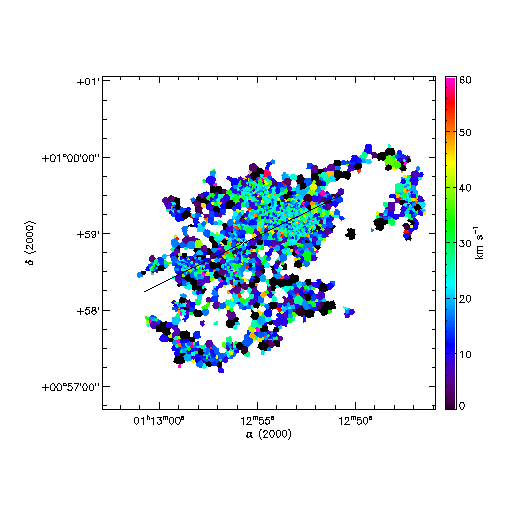} & \includegraphics[width=3.2cm,trim=0.9cm 0.cm 0.5cm 0.4cm, clip=true]{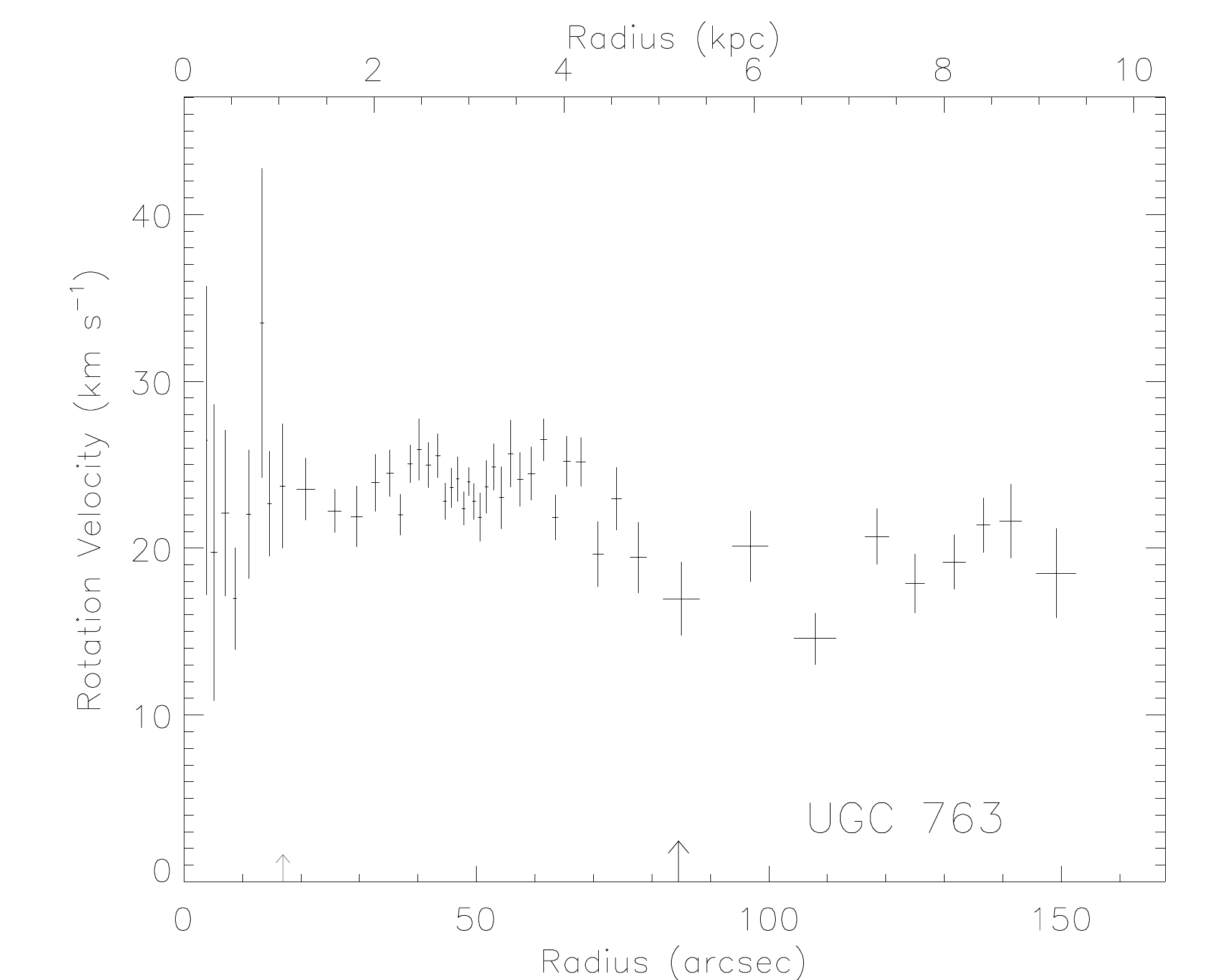}\\
\multicolumn{2}{c}{(6) UGC 00763}\\
\end{tabular}
\\
\begin{tabular}{p{4.2cm}p{3.8cm}}
\centering
\includegraphics[width=3.7cm,trim=0.5cm 2.cm 0.5cm 2.cm, clip=true]{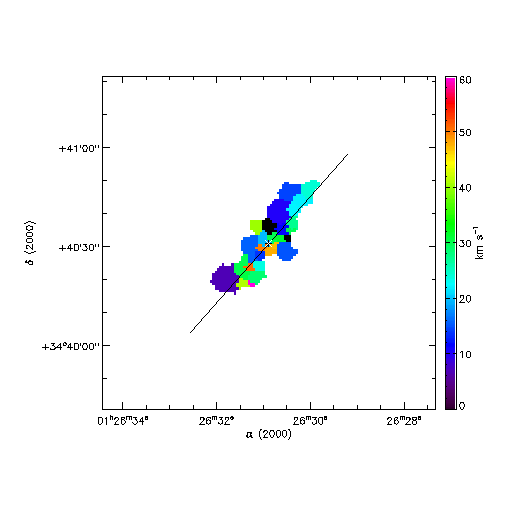} & \\
\multicolumn{2}{c}{(7) NGC 0542}\\
\end{tabular}
\begin{tabular}{p{4.2cm}p{3.8cm}}
\centering
\includegraphics[width=3.7cm,trim=0.5cm 2.6cm 0.5cm 0.5cm, clip=true]{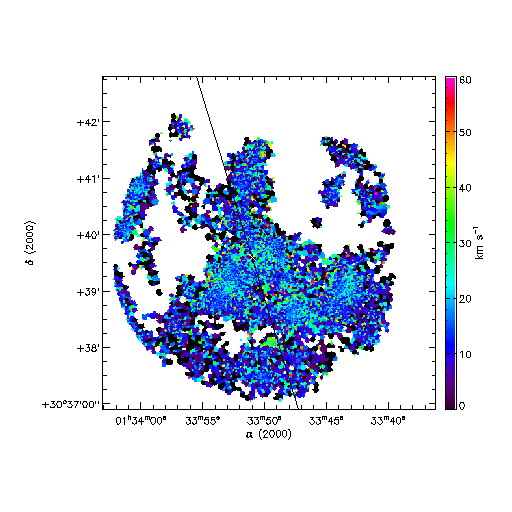} & \includegraphics[width=3.2cm,trim=0.9cm 0.cm 0.5cm 0.4cm, clip=true]{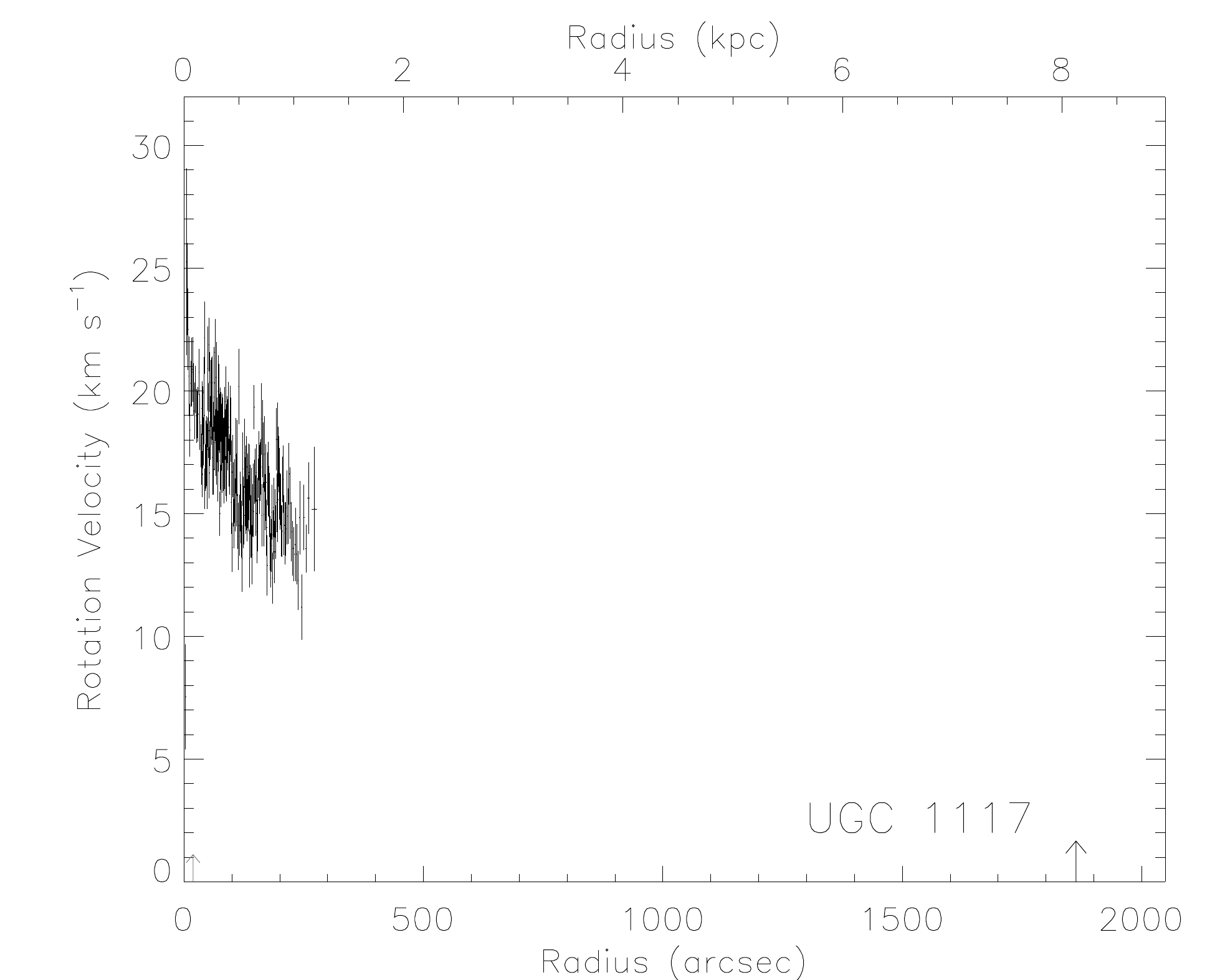}\\
\multicolumn{2}{c}{(8) UGC 01117}\\
\end{tabular}
\\
\begin{tabular}{p{4.2cm}p{3.8cm}}
\centering
\includegraphics[width=3.7cm,trim=0.5cm 2.cm 0.5cm 2.cm, clip=true]{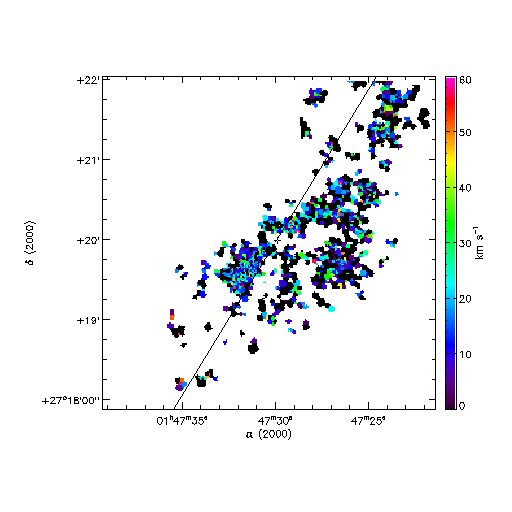} & \\
\multicolumn{2}{c}{(9) UGC 01249}\\
\end{tabular}
\begin{tabular}{p{4.2cm}p{3.8cm}}
\centering
\includegraphics[width=3.7cm,trim=0.5cm 2.6cm 0.5cm 0.5cm, clip=true]{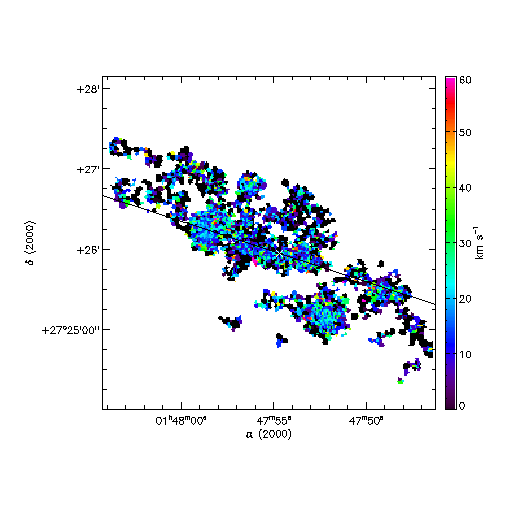} & \includegraphics[width=3.2cm,trim=0.9cm 0.cm 0.5cm 0.4cm, clip=true]{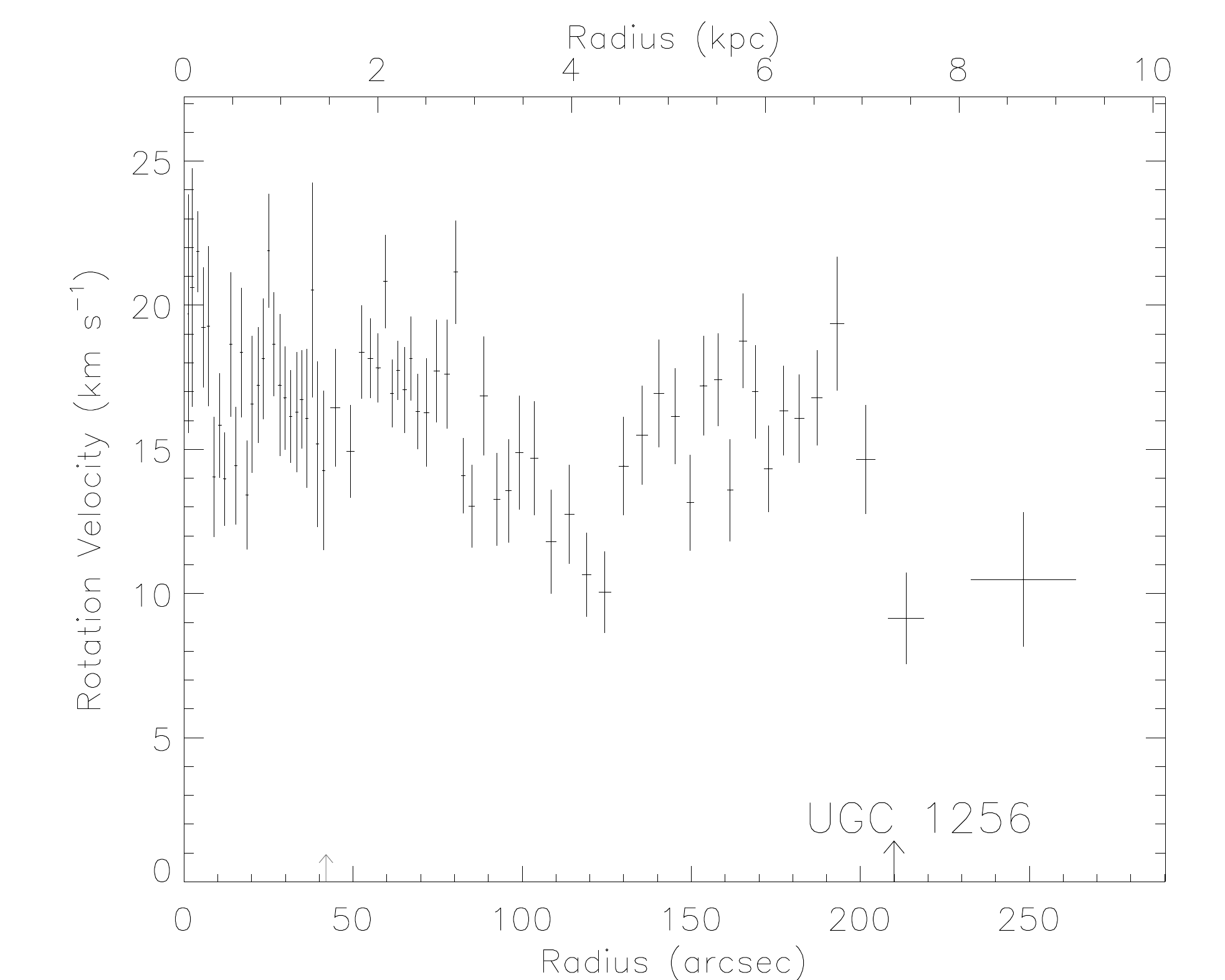}\\
\multicolumn{2}{c}{(10) UGC 01256}\\
\end{tabular}
\\
\begin{figure}[h]
\caption{Dispersion de vitesses des galaxies GHASP. \textbf{\`A gauche}: Carte de dispersion de vitesses. La double croix blanche et noire repr\'esente le centre cin\'ematique. La ligne noire repr\'esente le grand axe et s'arr\^ete au rayon optique ($D_{25}/2$, \citealp{de-Vaucouleurs:1995}). \textbf{\`A droite}: Profil radial de dispersion de vitesses. La fl\`eche noire indique le rayon optique, la fl\`eche grise indique le rayon de transition.}
\label{annexe_disp}
\end{figure}
\clearpage
\noindent
\begin{center}
\textsc{Fig.} D.3: Disponible dans la version complète du manuscript \\ \url{http://tel.archives-ouvertes.fr/tel-00413769/fr/}.
\end{center}
\clearpage
\noindent
\begin{center}
\textsc{Fig.} D.3: Disponible dans la version complète du manuscript \\ \url{http://tel.archives-ouvertes.fr/tel-00413769/fr/}.
\end{center}
\clearpage
\noindent
\begin{center}
\textsc{Fig.} D.3: Disponible dans la version complète du manuscript \\ \url{http://tel.archives-ouvertes.fr/tel-00413769/fr/}.
\end{center}
\clearpage
\noindent
\begin{center}
\textsc{Fig.} D.3: Disponible dans la version complète du manuscript \\ \url{http://tel.archives-ouvertes.fr/tel-00413769/fr/}.
\end{center}
\clearpage
\noindent
\begin{center}
\textsc{Fig.} D.3: Disponible dans la version complète du manuscript \\ \url{http://tel.archives-ouvertes.fr/tel-00413769/fr/}.
\end{center}
\clearpage
\noindent
\begin{center}
\textsc{Fig.} D.3: Disponible dans la version complète du manuscript \\ \url{http://tel.archives-ouvertes.fr/tel-00413769/fr/}.
\end{center}
\clearpage
\noindent
\begin{center}
\textsc{Fig.} D.3: Disponible dans la version complète du manuscript \\ \url{http://tel.archives-ouvertes.fr/tel-00413769/fr/}.
\end{center}
\clearpage
\noindent
\begin{center}
\textsc{Fig.} D.3: Disponible dans la version complète du manuscript \\ \url{http://tel.archives-ouvertes.fr/tel-00413769/fr/}.
\end{center}
\clearpage
\noindent
\begin{center}
\textsc{Fig.} D.3: Disponible dans la version complète du manuscript \\ \url{http://tel.archives-ouvertes.fr/tel-00413769/fr/}.
\end{center}
\clearpage
\noindent
\begin{center}
\textsc{Fig.} D.3: Disponible dans la version complète du manuscript \\ \url{http://tel.archives-ouvertes.fr/tel-00413769/fr/}.
\end{center}
\clearpage
\noindent
\begin{center}
\textsc{Fig.} D.3: Disponible dans la version complète du manuscript \\ \url{http://tel.archives-ouvertes.fr/tel-00413769/fr/}.
\end{center}
\clearpage
\noindent
\begin{center}
\textsc{Fig.} D.3: Disponible dans la version complète du manuscript \\ \url{http://tel.archives-ouvertes.fr/tel-00413769/fr/}.
\end{center}
\clearpage
\noindent
\begin{center}
\textsc{Fig.} D.3: Disponible dans la version complète du manuscript \\ \url{http://tel.archives-ouvertes.fr/tel-00413769/fr/}.
\end{center}
\clearpage
\noindent
\begin{center}
\textsc{Fig.} D.3: Disponible dans la version complète du manuscript \\ \url{http://tel.archives-ouvertes.fr/tel-00413769/fr/}.
\end{center}
\clearpage
\noindent
\begin{center}
\textsc{Fig.} D.3: Disponible dans la version complète du manuscript \\ \url{http://tel.archives-ouvertes.fr/tel-00413769/fr/}.
\end{center}
\clearpage
\noindent
\begin{center}
\textsc{Fig.} D.3: Disponible dans la version complète du manuscript \\ \url{http://tel.archives-ouvertes.fr/tel-00413769/fr/}.
\end{center}
\clearpage

\section{Base de données \FP}
\label{annexe_bddfp}

J'ai participé à l'élaboration de la base de données \FP, une collaboration entre le LAM, le G\'EPI et le LAE de Montréal.
Cette base de données est disponible à l'adresse \url{http://FabryPerot.oamp.fr/}. Elle est en cours de construction et sera compatible avec le projet VO.
A terme, elle regroupera toutes les données \FP~obtenues par les instruments présentés dans le chapitre \ref{instrumentation} de cette thèse. Elle mettra à disposition de la communauté scientifique les données réduites ainsi que les produits dérivés tels que les courbes de rotation, les diagrammes position-vitesse, les paramètres cinématiques déduits des analyses. Il sera également possible de récupérer les données brutes.
Les différents logiciels de réduction des données seront également disponibles.
Un outil de sélection de galaxies sur des critères de type morphologique, de magnitude, d'inclinaison, etc., est d'ores et déjà partiellement en place.
\par
Dans un deuxième temps, des outils élaborés seront proposés aux utilisateurs de la base de données. En particulier, un outil de projection des données à un décalage spectral voulu, avec résolutions et échantillonnages spatiaux et spectraux définis par l'utilisateur sera mis en place afin de généraliser les études du type de celle exposée dans la partie \ref{ghasp_highz}.  
\chapter{\'Etude des modèles de masse (chapitre \ref{ghasp_donnees})}
\chaptermarkannexe{\'Etude des modèles de masse (chapitre \ref{ghasp_donnees})}
\label{analyse_ghasp}
\minitoc
\textit{Cette annexe présente un article contenant les premiers résultats concernant l'étude de la distribution de masse des halos de matière sombre à partir des courbes de rotation d'une partie de l'échantillon GHASP. Cette étude s'insère dans les objectifs scientifiques du projet GHASP et est introduite dans la partie \ref{etude_dynamique}. \`A terme, l'ensemble de l'échantillon GHASP sera utilisé.}
\hl

\section{\underline{Article VIII:}~GHASP: an \Ha~kinematic survey of spiral and irregular galaxies - V. Dark matter distribution in 36 nearby spiral galaxies}
L'étude de la distribution de masse à partir de galaxies de l'échantillon GHASP a donné lieu à une publication dans un journal à comité de lecture\footnote{Monthly Notices of the Royal Astronomical Society} \citep{Spano:2008}. Cet article est présenté ici.
\par
Les résultats d'une étude de la distribution de masse sur $36$ galaxies spirales sont présentés. Ces galaxies ont été observées avec un interféromètre de Fabry-Perot dans le cadre du programme GHASP. L'obtention de champs de vitesses 2D de haute résolution en utilisant la raie \Ha~a pour objectif majeur de définir de manière détaillée la partie interne croissante des courbes de rotation, ce qui devrait permettre de mieux contraindre les paramètres de la distribution de masse. Lorsque des données HI (de plus faible résolution spatiale) étaient disponibles dans la littérature, elles ont été combinées aux vitesses \Ha. En combinant données cinématiques et photométriques, des modèles de masse ont été déterminés à partir de ces courbes de rotation en utilisant deux modèles de halo de matière sombre: une sphère isotherme (ISO) et un profil de Navarro-Frenk-White (NFW). Les résultats obtenus pour les galaxies ayant déjà fait l'objet d'études similaires par d'autres auteurs tendent à être concordants. Nos résultats mettent en évidence l'existence d'une densité de c\oe ur au centre des halos de matière sombre constante plutôt qu'un c\oe ur piqué et ce, quel que soit le type de galaxie de Sab à Im. Cela généralise à tous les types morphologiques le résultat déjà obtenu par d'autres auteurs à partir d'études de galaxies naines et de galaxies à faible brillance de surface, même s'il est encore nécessaire d'utiliser un échantillon plus important afin de conclure plus fermement. Quel que soit le modèle utilisé (ISO ou NFW), les halos de faible rayon de c\oe ur possèdent de plus grandes densités centrales, une fois encore indépendamment du type morphologique. Nous confirmons différentes lois d'échelle des halos, comme les corrélations entre le rayon de c\oe ur, la densité centrale des halos et la magnitude absolue des galaxies: les galaxies de faible luminosité ont un petit rayon de c\oe ur et une grande densité centrale. Nous trouvons que le produit de la densité centrale et du rayon de c\oe ur du halo est presque constant quel que soit le modèle et quelle que soit la magnitude absolue de la galaxie. Cela suggère que la densité de surface du halo est indépendante du type morphologique de la galaxie.

\includepdf[pagecommand={\pagestyle{headings}},scale=1.,offset=0 0,pages=-,
addtotoc={
1,subsubsection,2, Introduction,introghasp5,
2,subsubsection,2, L'échantillon, donneesghasp5,
4,subsubsection,2, Les données, dataghasp5,
6,subsubsection,2, Profils de densité et procédure d'ajustement des modèles, fitghasp5,
6,subsubsection,2, Résultats des modèles, resultsghasp5,
7,subsubsection,2, Discussion, discussionghasp5,
10,subsubsection,2, Résumé et conclusion, conclusionghasp5,
11,subsubsection,2, Annexe A: Meilleurs ajustements aux courbes de rotation, annexeghasp5
}
,addtolist={
3,table,{\underline{Article VIII}, Table 1: Data used for the analysis of our sample.},tab1_ghasp5,
5,table,{\underline{Article VIII}, Table 2: Results of the best fits of rotation curves with ISO and NFW models of halo profiles.},tab2_ghasp5,
7,figure,{\underline{Article VIII}, Figure 1: Minimum $\chi^{2}$ obtained with an ISO profile versus an NFW profile. For Figs 1-5 we used three different symbols for the main classes of morphological types of spiral galaxies (early, late and Magellanic).},fig1_ghasp5,
8,figure,{\underline{Article VIII}, Figure 2: Top panel: Central density (log) of DM versus the core radius (log) of DM for the ISO profile (ISO). Bottom panel: Same for the NFW profile. The solid lines are least-squares fits to our data (excluding the three open squares discussed in Section 6). The dashed line (top panel) is the least-squares fit extracted from fig. 3 of Kormendy \& Freeman (2004). The circles surround the points for which ISO is clearly the best model.},fig2_ghasp5,
9,figure,{\underline{Article VIII}, Figure 3: Top panel: central density (log) of dark halo versus absolute magnitude. Bottom panel: core radius (log) of dark halo versus absolute magnitude. Both are for the ISO models. The solid lines are least-squares fits for our data (excluding the three open squares discussed in Section 6). The dashed lines are the least-squares fits extracted from fig. 3 of Kormendy \&
Freeman (2004). The circles surround the points for which ISO is clearly the best model. },fig3_ghasp5,
9,figure,{\underline{Article VIII}, Figure 4: Central surface density (log) of DM (ISO model) as a function of absolute magnitude (given in Table 1, determined from $B_T$(0) found in the RC3 and our adopted distance). The solid line is a least-squares fit for our data (excluding the three open squares discussed in Section 6). The dashed line is the least-squares fit extracted from fig. 5 of Kormendy \& Freeman (2004). The circles surround the points for which ISO is clearly the best model.},fig4_ghasp5,
9,figure,{\underline{Article VIII}, Figure 5: Core radius (log) of the DM halo (ISO models) as a function of optical disc scalelength (top panel) and ($B-V$) colour (bottom panel). This last parameter, extracted from the RC3 catalog, could be found for 20 of our galaxies only. The solid lines are least-squares fits obtained with our data (excluding the three open squares discussed in Section 6). The dashed line on the top diagram is the least-squares fit from fig. 1 of Donato et al. (2004). On that diagram we have surrounded HSB, dS and LSB galaxies by circles of growing sizes.},fig5_ghasp5,
12,figure,{\underline{Article VIII}, Figures A: Best-fitting model for the rotation curve with ISO profile (left) and NFW profile (right-hand panel). The dots are for optical velocities, the open circles for HI velocities and the crosses for HI in the central part, not taken into account. The arrow on the X-axis indicates the disc scalelength.},figA_ghasp5
}]{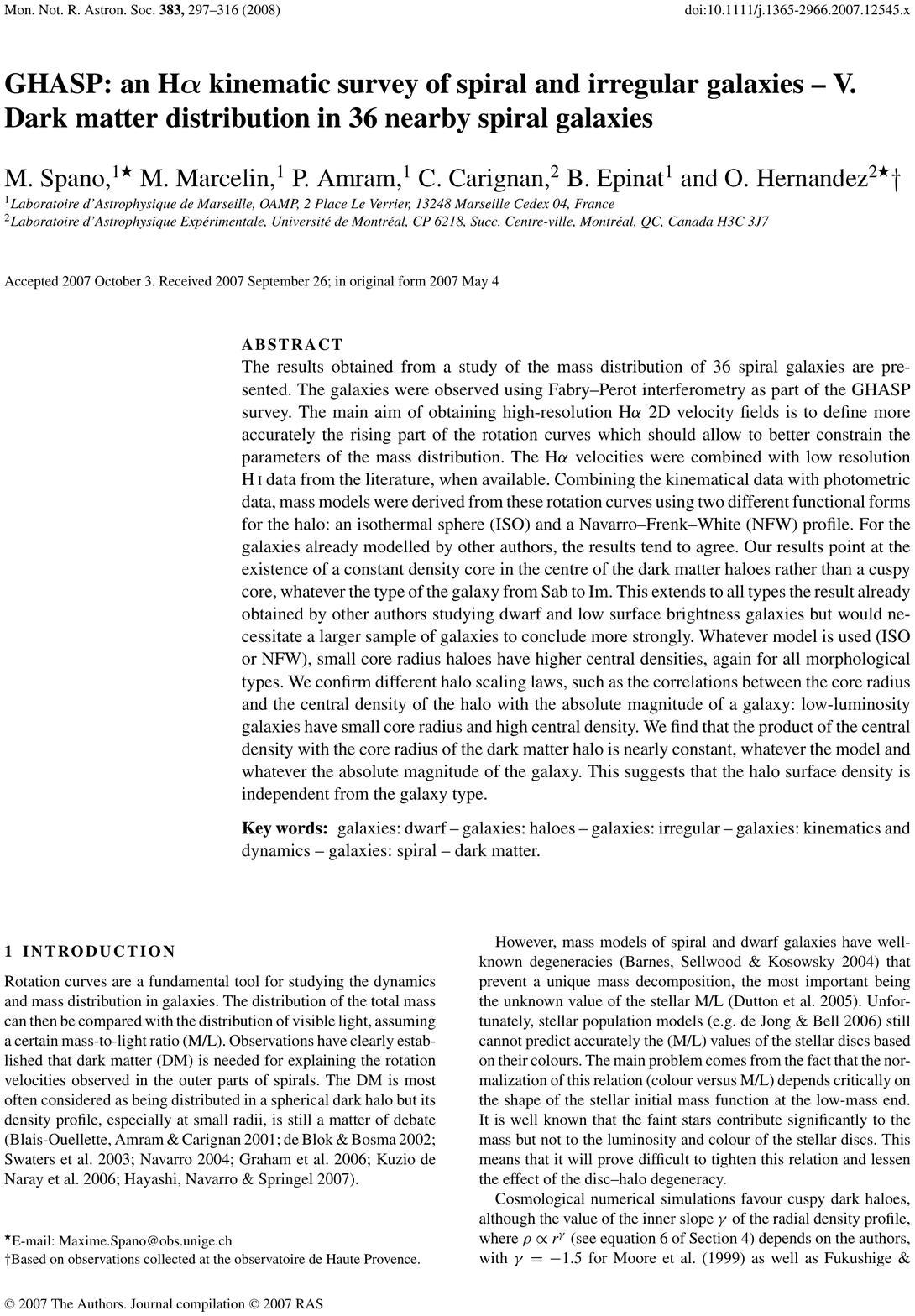}
\chapter{\'Etude de données à grand décalage spectral (chapitre \ref{etudes_highz})}
\chaptermarkannexe{\'Etude de données à grand décalage spectral (chapitre \ref{etudes_highz})}
\label{annexe_highz}
\minitoc
\textit{Cette annexe contient un article sur des observations à grand décalage spectral auquel j'ai contribué. Cet article présente des résultats concernant la relation de Tully-Fisher obtenus à partir d'observations de galaxies dont le décalage spectral est voisin de $0.6$ avec GIRAFFE. Quelques galaxies GHASP projetées de la manière décrite dans la partie \ref{ghasp_highz} ont été utilisées afin de vérifier les biais de la méthode de mesure des vitesses de rotation.}
\hl

\section{\underline{Article IX:}~IMAGES III. The evolution of the near-infrared Tully-Fisher relation over the last 6~Gyr}
\sectionmark{\underline{Article IX:}~IMAGES III. The evolution of the near-infrared Tully-Fisher relation over the last\\6~Gyr}

Le programme IMAGES (PI : F. Hammer), est un échantillon possédant actuellement $68$ galaxies observées à un décalage spectral $z\sim0.6$ par FLAMES/GIRAFFE dans le visible dont l'objectif est d’étudier l’assemblage de masses des galaxies depuis $z = 1$.
Une dizaine de galaxies de l'échantillon GHASP que j'ai projetées à un décalage spectral de $0.6$ dans les conditions d'observations de l'échantillon IMAGES (un pixel de $0.52''$ pour un seeing de $0.8''$) ont été utilisées lors d'une étude sur l'évolution de la relation de Tully-Fisher à ce décalage spectral. Cette étude a été publiée dans un journal à comité de lecture\footnote{Astronomy \& Astrophysics} \citep{Puech:2008}.
\par
\`A partir d'un échantillon représentatif de $65$ galaxies à raies d'émission ($W_0(OII)\ge15$ \AA) observées avec le spectrographe à multiples champs intégraux (multi-IFU) GIRAFFE sur le VLT, nous avons déterminé la relation de Tully-Fisher en bande $K$ (TFR) à $z\sim0.6$. Nous confirmons que la dispersion observée dans la TFR à $z\sim0.6$ est due à des galaxies possédant une cinématique anormale, et nous trouvons une forte corrélation entre la complexité de la cinématique des galaxies et la dispersion dans la TFR à laquelle elles contribuent. En ne considérant que les disques en rotation relaxés, la dispersion ainsi que la pente de la TFR ne semblent pas évoluer avec le décalage spectral. Nous détectons cependant une évolution du point zéro de la TFR en bande $K$ entre $z\sim0.6$ et $z=0$ qui impliquerait une augmentation de la luminosité des disques en rotation de $0.66\pm0.14$ mag entre $z\sim0.6$ et $z=0$. Les désaccords avec les résultats de Flores et al. (2006, A\&A, 455, 107) sont attribués à la fois à l'amélioration de la TFR locale et aux mesures plus précises des vitesses de rotation de l'échantillon distant. Les incertitudes s'expliquent globalement par la relativement basse résolution spatiale des données cinématiques.
Puisque la plupart des disques en rotation à $z\sim0.6$ semblent ne pas être sujets à de futurs épisodes de fusions, leur vitesse de rotation, qui est utilisée comme un traceur de la masse totale des galaxies, ne devrait pas varier de manière significative.
Si cette hypothèse est vérifiée, notre résultat implique que les disques en rotation observés à $z\sim0.6$ convertissent rapidement leur gaz en étoiles, de manière à doubler leur masse stellaire et être ainsi observés sur la TFR à $z=0$. Les disques en rotation observés sont en effet des galaxies à raies d'émission qui sont soit des pépinières d'étoiles (starbursts) soit des LIRGs,
ce qui implique qu'elles ont un fort taux de formation d'étoiles. Pour une fraction significative des disques en rotation, la majeure partie des étoiles se forment en $6$ à $8$ milliards d'années, ce qui est en bon accord avec les précédentes études concernant l'évolution de la relation masse-métallicité.

\includepdf[pagecommand={\pagestyle{headings}},scale=1.,offset=10 -15,pages=-,
addtotoc={
1,subsubsection,2, Introduction,introimages3,
2,subsubsection,2, Les données, donneesimages3,
3,subsubsection,2, Méthodologie, methodeimages3,
8,subsubsection,2, Relation de Tully-Fisher en bande K à $z\sim0.6$, tfimages3,
10,subsubsection,2, Discussion, discussionimages3,
12,subsubsection,2, Conclusion, conclusionimages3,
13,subsubsection,2, Annexe A: \'Evolution de la relation de Tully-Fisher en masse stellaire, annexeimages3
}
,addtolist={
4,table,{\underline{Article IX}, Table 1: Principle properties of the sample of $68$ galaxies used in this study, ordered by increasing RA (see text).},tab1_image3,
5,figure,{\underline{Article IX}, Figure 1:Illustration of the $\Delta V_{obs}$ correction method, for an early-type RC (upper panel, note that $V_{max} > V_{flat}$) and a late-type RC (bottom panel). The blue thick curves represent the input RC of the test, the black curves the best models (see text), and the dash-lines the two alternative RCs with non-optimal $r_t$ . We note that typical $r_t$ values lead to velocity gradients on spatial scales that are much smaller than the typical seeing of $\sim0.8$ arcsec.},fig1_images3,
6,figure,{\underline{Article IX}, Figure 2: Comparison between the $\Delta V_{model}$ values obtained in Monte-Carlo simulations of $100$ GIRAFFE data-cubes using a $0.8$ and $1$ arcsec seeing (see text). The black line is a linear fit fixing the intercept to zero. The residual dispersion is $\sim4.3$ km s$^{-1}$.},fig2_images3,
7,figure,{\underline{Article IX}, Figure 3: Examples of kinematical fitting of three $z \sim 0.6$ RD galaxies. From left to right: HST/ACS F775W image with the GIRAFFE IFU superimposed (from Paper I), observed VF (shown with a $5\times 5$ interpolation see Paper I), best modeled VF (shown with a $5\times 5$ interpolation), residual map between the observed and modeled VFs. Relatively large differences are found only close to the minor axis, where departure from pure circular motion is artificially exaggerated by projection effects (see a discussion of this effect in, e.g., Chemin et al. 2006).},fig3_images3,
7,figure,{\underline{Article IX}, Figure 4: Comparison between the $V_{flat}$ used as inputs to the Monte-Carlo simulations of $100$ GIRAFFE data-cubes, with the $V_0$ values obtained after using the method of correction described in Sect. 3.2. Black dots represent RCs generated using an arctan model (i.e., late-type like RCs), while open circles represent RCs typical of early-type galaxies. Black squares represent simulations where the rotation velocity is not sampled by the IFU (see text). The black line is a linear fit, and the dash lines represent the $1-\sigma$ residual dispersion $\sim17$ km s$^{-1}$. Blue stars represent real observations of local galaxies artificially redshifted to $z\sim0.6$ (see text).},fig4_images3,
7,figure,{\underline{Article IX}, Figure 5: Correction factors $\alpha$ used to correct $\Delta V_{obs}$ for RDs (blue dots; open blue dots represent the two RD+ galaxies). Small black dots correspond to Monte-Carlo simulations of $100$ GIRAFFE data-cubes in the same range of half-light radius, PA, inclination, rotation velocity, and RC gradient (see text). The horizontal dash line represents to the mean correcting factor of $1.25$ derived from Monte-Carlo simulations, while the $1-\sigma$ dispersion around the mean is shown in dash lines.},fig5_images3,
8,figure,{\underline{Article IX}, Figure 6: Monte-Carlo simulations of the velocity measurement accuracy (see Paper I, for details). The red line shows the corresponding bias (almost zero), while the black dashed lines shows the $1-\sigma$ error: $\sim12$ km s$^{-1}$ between $SNR = 3-5$, $\sim5$ km s$^{-1}$ between $SNR = 5-10$, and $\le 2$ for $SNR \ge 10$.},fig6_images3,
8,figure,{\underline{Article IX}, Figure 7: Evolution of the $K$-band TFR ($AB$ magnitudes). The completude limit $M_K \sim -20.14$ (corresponding to $M_J = -20.3$, see Sect. 2) is indicated by an horizontal dash line. Blue dots represent RDs (the two RD+ galaxies are represented with open blue dots), green squares PRs, and red triangles CK galaxies. The black line is the local TFR, while the blue dash-line represent a linear fit to the $z \sim 0.6$ TFR (see text).},fig7_images3,
8,table,{\underline{Article IX}, Table 2: Fits to the local and distant $K$-band TFRs, using $M_K (AB) = a + b \times log(V_{flat})$. },tab2_images3,
10,figure,{\underline{Article IX}, Figure 8: $K$-band TFR derived following the Flores et al. (2006) methodology, for the RD subsample. The black line is the local relation of Verheijen (2001), i.e., the one used as a reference by Flores et al. (2006). Note that $K$-band magnitudes in the distant sample have been converted into the Vega system using $M_K (Vega) = M_K (AB)-1.85$. For simplicity, we assume that the $K'$ magnitudes of Verheijen (2001) are roughly equivalent to those derived in the distant sample using the ISAAC $K$s filter. The blue line is the SDSS local TFR used as a reference in this study (converted into the Vega system). The blue dash line is a linear fit to the distant RDs, which has a zero point $0.4$ mag lower than the local one, fixing the slope to the local value. Open symbols represent galaxies from the new CDFS sample, while full symbols represent galaxies used in Flores et al. (2006).},fig8_images3,
11,figure,{\underline{Article IX}, Figure 9: Comparison between the rotation velocities $V_{F06}$ obtained using the method of Flores et al. (2006) (i.e., a constant correction factor of $1.2$) vs. rotation velocities $V_{flat}$ derived using the new method used in this paper, for the subsample of RDs. $V_{F06}$ underestimates $V_{flat}$ by $\sim11$\% on average. Open symbols represent galaxies from the new CDFS sample, while full symbols represent galaxies used in Flores et al. (2006).},fig9_images3,
13,figure,{\underline{Article IX}, Figure A.1: Evolution of the stellar-mass TFR in the RD subsample (the two RD+ galaxies are represented with open blue dots). The black line is the local smTFR, while the blue dash-line represents a linear fit to the $z\sim 0.6$ smTFR.},figa1_images3,
13,table,{\underline{Article IX}, Table A.1: Fits to the local and distant smTFRs, using $log(M_{stellar}/M_{\odot})=a+b\times log(V_{flat})$.},taba1_images3,
14,table,{\underline{Article IX}, Table A.2: Identified systematic uncertainties that could impact the shift of zero point between the local and the distant smTFRs. Systematic uncertainties on $V_{flat}$ have been converted into $M_{stellar}$ using Eq. (A.1). Negative values tend to reduce the shift of zero point, while positive values have the opposite trend.},taba2_images3,
14,figure,{\underline{Article IX}, Figure A.2: Histograms of log($M_{stellar}/L_K$) found in the local and distant samples of galaxies using the method of Bell et al. (2003). Both histograms have been re-centered using the median found in the local samples, which allows us to directly infer the evolution of log($M_{stellar}/L_K$) between $z \sim 0.6$ and $z=0$, i.e., $\sim0.625$ dex. Also shown, the evolution of log($M_{stellar}/L_K$) found by Drory et al. (2004) in a sample of intermediate-mass galaxies (black long-dashed line), and Arnouts et al. (2007) in blue star-forming galaxies (blue mixed-line) or independently of the color (black mixed-line).},figa2_images3
}]{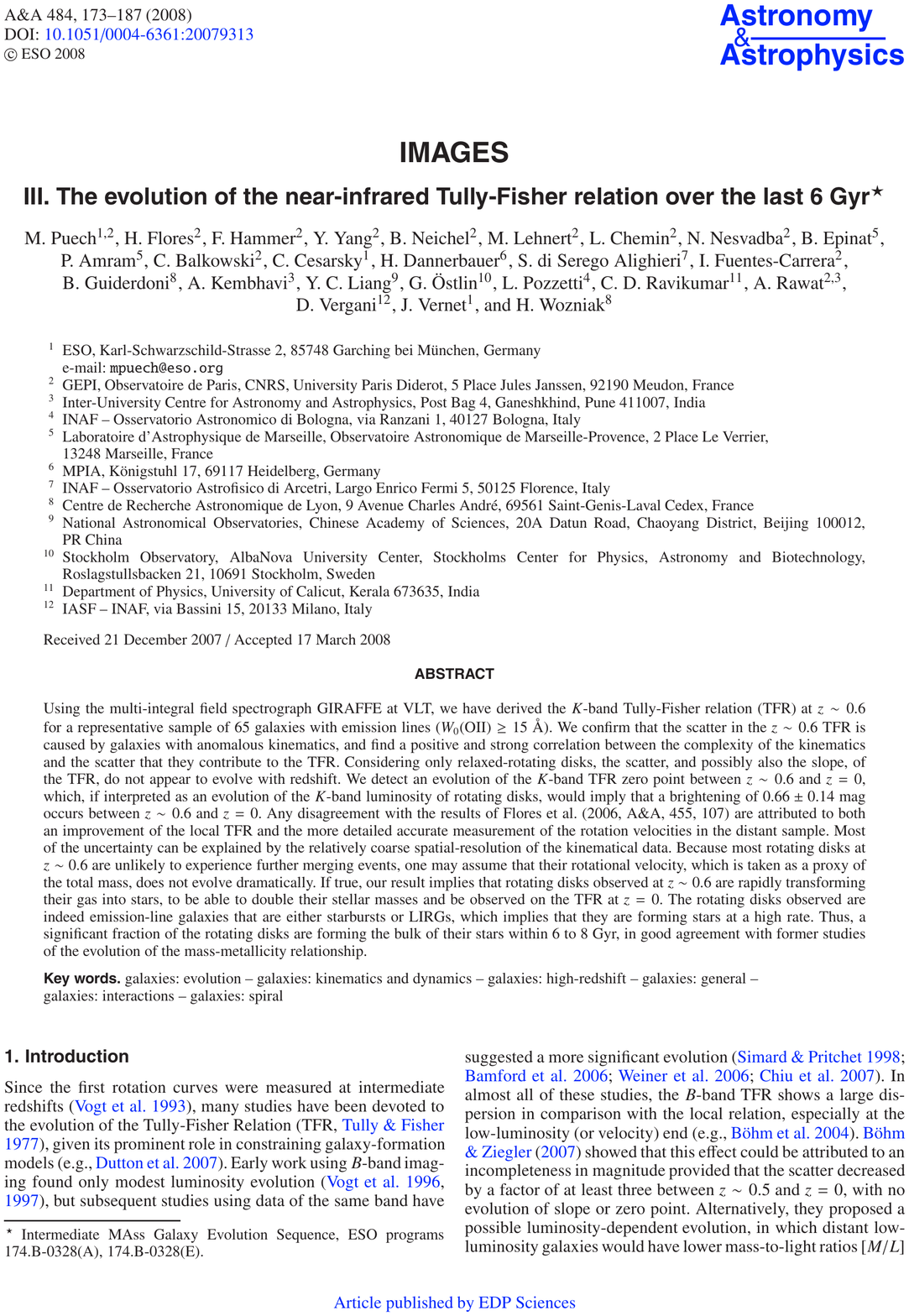}

\backmatter




\cleardoublepage
\newpage
\thispagestyle{empty}
\null
\newpage
\thispagestyle{empty}
\includepdf[fitpaper=true]{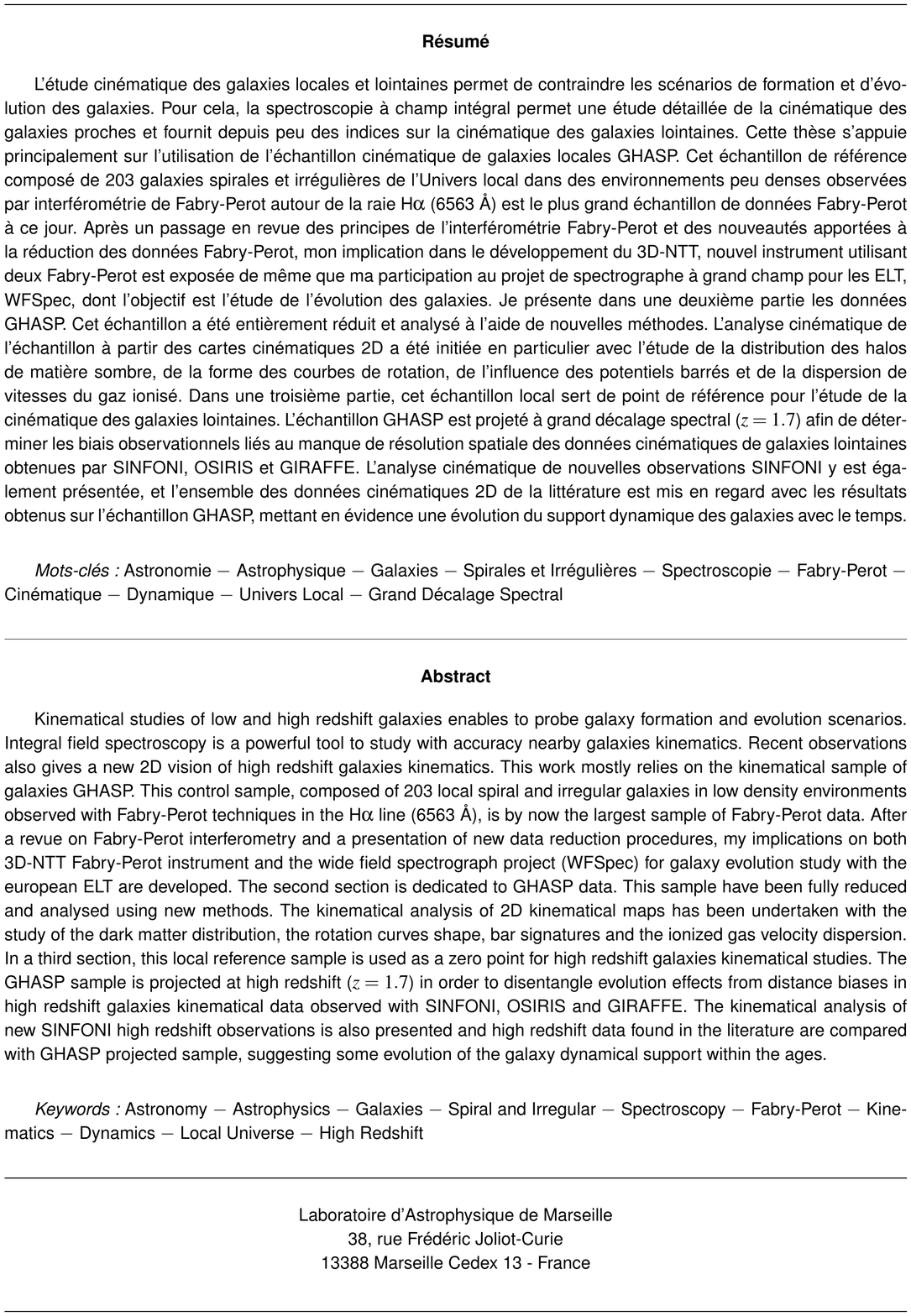}
\thispagestyle{empty}

\end{document}